\input harvmac
\input amssym.def
\input amssym
\baselineskip 14pt
\parskip 6pt

\def \si{\sigma}

\def \pr{\partial}
\def \d{{\rm d}}
\def \tr{{\rm tr }}

\def \by{{\bar y}}

\def\bm{{\bar m}}
\def \bY{{\bar Y}}
\def \bW{{\bar W}}
\def \bQ{{\bar Q}}

\def \bpsi{{\bar \psi}}
\def \bsi{{\bar \sigma}}
\def \bchi{{\bar \chi}}
\def \bphi{{\bar \phi}}

\def \l{\big \langle}
\def \r{\big \rangle}

\def \vep{\varepsilon}
\def \half{{\textstyle {1 \over 2}}}

\def \quar{{\textstyle {1 \over 4}}}
\def \ts{\textstyle}
\def\del{{\rm d}}
\def \A{{\cal A}}
\def \B{{\cal B}}
\def \C{{\cal C}}
\def \D{{\cal D}}

\def \G{{\cal G}}

\def \J{{\cal J}}
\def \K{{\cal K}}
\def \L{{\cal L}}
\def \M{{\cal M}}
\def \N{{\cal N}}
\def \O{{\cal O}}
\def \P{{\cal P}}

\def \R{{\cal R}}

\def \X{{\cal X}}

\def \Z{{\cal Z}}
\def\fd{{\frak d}}
\def\fe{{\frak e}}
\def\fN{{\frak r}}

\def \d{{\rm d}}

\def\uF{\bar{F}}
\def\uG{\bar{G}}
\def\uA{\bar{A}}
\def\uB{\bar{B}}
\def\uC{\bar{C}}
\def\uD{\bar{D}}
\def\uE{\bar{E}}
\def\uH{\bar{H}}
\def\ual{\bar{\alpha}}
\def\ube{\bar{\beta}}
\def\uga{\bar{\gamma}}
\def\ude{\bar{\delta}}
\def\uet{\bar{\eta}}
\def\uep{\bar{\epsilon}}
\def\hDel{\hat \Delta}
\def\wL{{\L}}
\def\hrho{{\tilde \rho}}

\def\hbet{\beta}
\def\hhbet{{\hat \beta}}
\def\hB{{\hat B}}
\def\cirk{\,{\raise1pt \hbox{${\scriptscriptstyle \circ}$}}\,}

\def \olr{{\raise6.5pt\hbox{$\leftrightarrow  \! \! \! \! \!$}}}

\font\ninerm=cmr9 \font\ninesy=cmsy9
\font\eightrm=cmr8 \font\sixrm=cmr6
\font\eighti=cmmi8 \font\sixi=cmmi6
\font\eightsy=cmsy8 \font\sixsy=cmsy6
\font\eightbf=cmbx8 \font\sixbf=cmbx6
\font\eightit=cmti8
\def\eightpoint{\def\rm{\fam0\eightrm}
  \textfont0=\eightrm \scriptfont0=\sixrm \scriptscriptfont0=\fiverm
  \textfont1=\eighti  \scriptfont1=\sixi  \scriptscriptfont1=\fivei
  \textfont2=\eightsy \scriptfont2=\sixsy \scriptscriptfont2=\fivesy
  \textfont3=\tenex   \scriptfont3=\tenex \scriptscriptfont3=\tenex
  \textfont\itfam=\eightit  \def\it{\fam\itfam\eightit}%
  \textfont\bffam=\eightbf  \scriptfont\bffam=\sixbf
  \scriptscriptfont\bffam=\fivebf  \def\bf{\fam\bffam\eightbf}%
  \normalbaselineskip=9pt
  \setbox\strutbox=\hbox{\vrule height7pt depth2pt width0pt}%
  \let\big=\eightbig  \normalbaselines\rm}
\catcode`@=11 %
\def\eightbig#1{{\hbox{$\textfont0=\ninerm\textfont2=\ninesy
  \left#1\vbox to6.5pt{}\right.\n@@space$}}}
\def\vfootnote#1{\insert\footins\bgroup\eightpoint
  \interlinepenalty=\interfootnotelinepenalty
  \splittopskip=\ht\strutbox %
  \splitmaxdepth=\dp\strutbox %
  \leftskip=0pt \rightskip=0pt \spaceskip=0pt \xspaceskip=0pt
  \textindent{#1}\footstrut\futurelet\next\fo@t}
\catcode`@=12 %
\def\today{\number\day\ \ifcase\month\or January\or February\or March\or April\or May\or June\or July\or
August\or September\or October\or November\or December\fi, \number\year}

\font \bigbf=cmbx10 scaled \magstep1

\lref\KS{
 Z.~Komargodski and A.~Schwimmer,
 On Renormalization Group Flows in Four Dimensions,
JHEP {1112}  (2011) 099, arXiv:1107.3987 [hep-th]\semi
Z.~Komargodski, The Constraints of Conformal Symmetry on RG Flows,
JHEP {1207} (2012) 069, arXiv:1112.4538 [hep-th].}

\lref\Luty{M.A. Luty, J. Polchinski and R. Rattazzi, 
The a-theorem and the Asymptotics of 4D Quantum Field Theory,
JHEP 1301 (2013) 152, arXiv:1204.5221 [hep-th].}
 
 \lref\ElvangST{
 H.~Elvang, D.Z.~Freedman, L.Y.~Hung, M.~Kiermaier, R.C.~Myers and S.~Theisen,
 On renormalization group flows and the a-theorem in 6d,
JHEP {1210} (2012) 011, arXiv:1205.3994 [hep-th]\semi
 H.~Elvang and T.M.~Olson,
 RG flows in d dimensions, the dilaton effective action, and the a-theorem,
JHEP {1303} (2013) 034, arXiv:1209.3424 [hep-th].}
 
\lref\Zam{
A.B. Zamolodchikov, Irreversibility of the Flux of the Renormalization 
Group in a 2D Field Theory, JETP Lett. 43 (1986) 730.}

\lref\CFL{
 A.~Cappelli, D.~Friedan and J.I.~Latorre,  C theorem and spectral 
 representation, Nucl.\ Phys.\ B{352} (1991) 616.}

\lref\TTa{H. Osborn and G.M. Shore, Correlation Functions of the Energy 
Momentum Tensor on Spaces of Constant Curvature, Nucl. Phys. B571 (2000) 287,  
hep-th/9909043.}

\lref\JackOne{
I.~Jack and H.~Osborn, 	
Two Loop Background Field Calculations For Arbitrary Background Fields, 
Nucl.\ Phys.\ B{207} (1982) 474.}
\lref\JackO{
 I.~Jack and H.~Osborn, General Background Field Calculations With Fermion 
Fields, Nucl.\ Phys.\ B{249} (1985) 472.}
\lref\Chet{
K.G.~Chetyrkin and M.F.~Zoller,
Three-loop $\beta$-functions for top-Yukawa and the Higgs 
self-interaction in the Standard Model, JHEP {1206} (2012) 033, 
arXiv:1205.2892 [hep-ph].}
\lref\Jack{
I.~Jack, D.R.T.~Jones and C.G.~North,
N=1 supersymmetry and the three loop anomalous dimension for the 
chiral superfield, Nucl.\ Phys.\ B{473} (1996) 308, hep-ph/9603386.}
\lref\FerreiraRC{P.M.~Ferreira, I.~Jack and D.R.T.~Jones,
The quasi-infra-red fixed point at higher loops,
Phys.\ Lett.\ B{392} (1997) 376, hep-ph/9610296.}

\lref\Fortin{
J.~-F.~Fortin, B.~Grinstein and A.~Stergiou, Scale without Conformal 
Invariance at Three Loops, JHEP 1208 (2012) 085, 
arXiv:1202.4757 [hep-th].}
\lref\FortinC{
J.~-F.~Fortin, B.~Grinstein and A.~Stergiou, 
Limit Cycles and Conformal Invariance, JHEP 1301 (2013) 184, 
arXiv:1208.3674 [hep-th].}
\lref\FortinS{
J.~-F.~Fortin, B.~Grinstein, C.W. Murphy and A.~Stergiou,
On Limit Cycles in Supersymmetric Theories, 
Phys. Lett. B719 (2013) 170, arXiv:1210.2718 [hep-th].}

\lref\Nak{Yu Nakayama, Supercurrent, Supervirial and Superimprovement,
Phys.\ Rev.\ D{87} (2013) 085005, arXiv:1208.4726 [hep-th].}

\lref\Nakthree{
Y.~Nakayama, Consistency of local renormalization group in d=3,
 Nucl. Phys. B879 (2014) 37,
arXiv:1307.8048 [hep-th].}

\lref\NakRev{Yu Nakayama, 	
A lecture note on scale invariance vs conformal invariance, $~~~~~$
arXiv:1302.0884 [hep-th].}

\lref\Wallace{
D.J.~Wallace and R.K.P.~Zia,
Gradient Properties of the Renormalization Group Equations in 
Multicomponent Systems,
Annals Phys.\  {92} (1975) 142.}
 
\lref\Cardy{
J.L. Cardy,  Is There a c Theorem in Four-Dimensions?,  {Phys. Lett.} 
B{215} (1988) 749.}
 
\lref\Analog{
H.~Osborn, Derivation of a Four-dimensional C theorem, 
Phys.\ Lett.\ B{222} (1998) 97\semi
I.~Jack and H.~Osborn,
Analogs For The C Theorem For Four-dimensional Renormalizable Field 
Theories, Nucl.\ Phys.\ B{343} (1990) 647.}
\lref\Weyl{H.~Osborn,
Weyl consistency conditions and a local renormalization group equation for 
general renormalizable field theories,
Nucl.\ Phys.\ B{363} (1991) 486.}

\lref\SusyA{
D.Z.~Freedman and H.~Osborn, Constructing a c function for SUSY gauge 
theories, Phys.\ Lett.\ B{432} (1998) 353, hep-th/980410.}

\lref\Papadodimas{K.~Papadodimas,
Topological Anti-Topological Fusion in Four-Dimensional Superconformal 
Field Theories, JHEP {1008} (2010) 118,
arXiv:0910.4963 [hep-th].}
\lref\Asnin{V.~Asnin,
On metric geometry of conformal moduli spaces of four-dimensional 
superconformal theories, JHEP {1009} (2010) 012,
arXiv:0912.2529 [hep-th].}

\lref\conform{
G.~Parisi, Conformal invariance in perturbation theory,
Phys.\ Lett.\ B{39} (1972) 643\semi
S.~Sarkar, Broken Conformal Ward Identities in Nonabelian Gauge 
Theories,
Phys.\ Lett.\ B{50} (1974) 499; Dimensional Regularization and Broken 
Conformal Ward Identities, Nucl.\ Phys.\ B{83} (1974) 108\semi
N.K. Nielsen, Conformal Ward Identities in Massive
Quantum Electrodynamics, Nucl. Phys. B65 (1973) 413; 
Gauge Invariance and Broken Conformal Symmetry,  Nucl. Phys. 
B97 (1975) 527.
}

\lref\Cgross{C.G. Callan and D.J. Gross, Fixed angle scattering in 
quantum field theory, Phys. Rev. D11 (1975) 2905.}

\lref\JackAJ{I.~Jack and D.R.T.~Jones,
RG invariant solutions for the soft supersymmetry breaking parameters,
Phys.\ Lett.\ B{465} (1999) 148, hep-ph/9907255.}

\lref\Sann{
 O.~Antipin, M.~Gillioz, E.~M{\o}lgaard and F.~Sannino,
The a theorem for Gauge-Yukawa theories beyond Banks-Zaks, 
Phys. Rev. D87 (2013) 125017, arXiv:1303.1525 [hep-th].}

\lref\Other{F. Baume, B. Keren-Zur, R. Rattazzi and L. Vitale, The
local Callan-Symanzik equation: structure and applications,
arXiv:1401.5983 [hep-th].}

\lref\BraunRP{
V.~M.~Braun, G.~P.~Korchemsky and D.~Mueller,
The uses of conformal symmetry in QCD,
Prog.\ Part.\ Nucl.\ Phys.\  {51} (2003) 311, hep-ph/0306057.
}

\lref\AnselmiAM{
D.~Anselmi, D.Z.~Freedman, M.T.~Grisaru and A.A.~Johansen,
Nonperturbative formulas for central functions of supersymmetric gauge theories,
Nucl.\ Phys.\ B{526} (1998) 543, hep-th/9708042\semi
D.~Anselmi, J.~Erlich, D.Z.~Freedman and A.A.~Johansen,
Positivity constraints on anomalies in supersymmetric gauge theories,
Phys.\ Rev.\ D{57} (1998) 7570, hep-th/9711035.
}
\lref\BarnesJJ{
E.~Barnes, K.A.~Intriligator, B.~Wecht and J.~Wright,
Evidence for the strongest version of the 4d a-theorem, via a-maximization 
along RG flows,
Nucl.\ Phys.\ B{702} (2004) 131, hep-th/0408156.
}
\lref\KutasovXU{
D.~Kutasov,
New results on the `a theorem' in four-dimensional supersymmetric field theory,
hep-th/0312098\semi
D.~Kutasov and A.~Schwimmer,
 Lagrange multipliers and couplings in supersymmetric field theory,
Nucl.\ Phys.\ B{702} (2004) 369, hep-th/0409029.
}

\lref\GrinsteinCKA{B.~Grinstein, A.~Stergiou and D.~Stone,
Consequences of Weyl Consistency Conditions,  JHEP 1311 (2013) 195,
arXiv:1308.1096 [hep-th].}

\lref\JackSK{
 I.~Jack and H.~Osborn,
Background Field Calculations in Curved Space-time. 1. 
General Formalism and Application to Scalar Fields,
Nucl.\ Phys.\ B{234} (1984) 331.
}

\lref\Mach{
M.E.~Machacek and M.T.~Vaughn,
Two Loop Renormalization Group Equations in a General Quantum Field Theory. 1. Wave Function Renormalization,
Nucl.\ Phys.\ B{222} (1983) 83.
}

\lref\Friedan{
  D.~Friedan and A.~Konechny,
Gradient formula for the beta-function of 2d quantum field theory,
J.\ Phys.\ A {43} (2010) 215401, arXiv:0910.3109 [hep-th]\semi
Curvature formula for the space of 2-d conformal field theories,
JHEP {1209} (2012) 113, arXiv:1206.1749 [hep-th]\semi
N.~Behr and A.~Konechny,
 Renormalization and redundancy in 2d quantum field theories,
arXiv:1310.4185 [hep-th].
}

\lref\Mincer{G. Gorishny, S.A. Larin, L.R. Surguladze \& F.K. Tkachov,
Mincer: Program for Multiloop Calculations in Quantum Field Theory for the Schoonschip System,
Comput. Phys. Commun. {55} (1989) 381\semi
S.A. Larin, F.V. Tkachov \& J.A.M. Vermaseren, The Form version
of Mincer, NIKHEF-H-91-18.
}

\lref\Form{J.A.M. Vermaseren, New Features of FORM, math-ph/0010025.
}

{\nopagenumbers
\rightline{DAMTP/13-53}
\rightline{arXiv:1312.0428[hep-th]}
\rightline{\today}
\vskip 1.5truecm
\centerline {\bigbf Constraints on RG Flow for Four Dimensional
Quantum Field Theories}
\vskip  6pt
\vskip 2.0 true cm
\centerline {I. Jack${}^*$ and H. Osborn${}^\dagger$}

\vskip 12pt
\centerline {${}^*$Department of Mathematical Sciences, University of 
Liverpool,
}
\centerline {Liverpool L69 3BX, UK}
\vskip 6pt

\centerline {${}^\dagger$Department of Applied Mathematics and Theoretical 
Physics,}
\centerline {Wilberforce Road, Cambridge, CB3 0WA, England}
\vskip 1.5 true cm

{\eightpoint
\parindent 1.5cm{

{\narrower\smallskip\parindent 0pt

The response of four dimensional quantum field theories to a Weyl 
rescaling of the metric in the presence of local couplings and which
involve $a$, the coefficient of the Euler density in the energy momentum
tensor trace on curved space, is reconsidered. Previous consistency 
conditions for the anomalous terms, which implicitly define a metric $G$ 
on the space of couplings and give rise to gradient flow like equations 
for $a$, are derived taking into account the role of lower dimension 
operators. The results for infinitesimal  Weyl rescaling are integrated 
to finite rescalings $e^{2\sigma}$ to a form which involves running 
couplings $g_\sigma$ and which interpolates between IR and UV fixed 
points. The results are also restricted to flat space where they give 
rise to broken conformal Ward identities. 

Expressions for the three loop Yukawa $\beta$-functions for a 
general scalar/fermion theory are obtained and the three loop 
contribution to the metric $G$ for this theory are also calculated. 
These results are used to check the gradient flow  equations to higher 
order than previously. It is shown that these are only valid when 
$\beta \to B$, a modified $\beta$-function, and that the equations provide  
strong constraints on the detailed form of the three loop Yukawa 
$\beta$-function. $\N=1$ supersymmetric Wess-Zumino theories are also 
considered as a special case. It is shown that the metric for the complex 
couplings in such theories may be restricted to a hermitian form.


\narrower}}
\medskip

\vfill
\line{\hskip0.2cm emails:${}^*${\tt dij@sune.amtp.liv.ac.uk} and
${}^\dagger${\tt ho@damtp.cam.ac.uk}
\hfill}
}


\eject
}
\pageno=1

\newsec{Introduction}

The paradigm shift in our understanding of quantum field theories due to
Wilson in the 1970's led to the view that quantum field theories
are not isolated objects but may be regarded as points on a manifold, with 
coordinates given by the couplings $\{g^I\}$, where
there is a natural flow under changes of the cut-off scale realising the
renormalisation  group.  The perturbative RG flow equations are just first
order equations determined by the $\beta$-functions $\beta^I(g)$, which are 
vector fields on the space of couplings. Even in this context
the global topology of such flows has been less certain, the simplest 
scenario arises when the flows link fixed points in the UV short distance 
limit to other fixed points in the large distance IR limit. At the fixed 
points the quantum field theory is scale invariant and moreover is
naturally expected to become a conformal field theory. However more 
complicated behaviours under RG flow, such as limit cycles or the flow 
becoming  chaotic, are also feasible. As was first suggested by 
Cardy \Cardy\ there may be additional constraints for unitary quantum
field theories in four dimensions due to the existence of a function $a(g)$
which has  monotonic behaviour under RG flow, or more minimally $a$ may
be defined at fixed points so that $a_{\rm UV} - a_{\rm IR}>0$. These
two scenarios are here described as the strong and weak $a$-theorem,
such a distinction was made in \NakRev.
If valid a strong $a$-theorem constrains the RG flow without assuming
any $UV$ completion although it requires the RG flow to be described by
linear equations involving $\beta$-functions.

The proposal of Cardy was for a four dimensional generalisation of the 
Zamolodchikov $c$-theorem, \Zam. This constrains the structure of two 
dimensional quantum field  theories and  has a simple elegant proof 
depending just  on the properties of the two point correlation 
function of the energy  tensor. The crucial positivity 
constraint arises from unitarity conditions applied to the two 
point function. No such approach works in four dimensions \CFL, \TTa\ 
but it was soon clear that only $a$, which is determined  by 
the topological term in the trace of the energy momentum tensor on 
curved space, is a viable candidate for a monotonic flow 
between fixed points. The energy momentum tensor two point function in 
conformal theories is determined by $c$, the coefficient of the square of the
Weyl tensor in the energy momentum tensor trace on curved space.

Much more recently Komargodski and Schwimmer \KS\ have described a 
proof of the four dimensional weak $a$-theorem which has been further 
analysed in \Luty\ with possible extensions to higher dimensions 
discussed in \ElvangST. This rests on coupling the theory to a 
dilaton and constructing an effective low energy field theory for 
the dilaton. The essential positivity requirement 
depends on positivity conditions arising from unitarity for  the four 
dilaton scattering amplitude. The starting point of the discussion in
\KS\ is the response of a conformal theory to a Weyl rescaling of 
the flat metric. The resulting expression determines the couplings 
of the dilaton introduced as a compensator for the local anomalous 
terms which arise under a Weyl rescaling and which have a coefficient 
proportional to $a$. The basic argument of Komargodski and Schwimmer 
is that coupling to a dilaton ensures a matching of these anomalies 
between the UV and IR fixed points.

However the results of \KS\ and also \Luty\ do not immediately extend
away from conformal fixed points. There is also no obvious connection
with a perturbative version of the strong $a$-theorem 
for four dimensional renormalisable quantum field theories. 
This was based on an analysis in terms of dimensional regularisation 
\Analog\ and also from Wess-Zumino consistency conditions for
the response of the theory on curved background to a Weyl rescaling 
of the metric \Weyl. Instead of a dilaton as in \KS\ the usual linear 
RG equations describing the response to a variation in the RG scale 
$\mu$
were extended to a local infinitesimal Weyl rescaling $\sigma(x)$ 
by allowing the couplings also to be local $g^I(x)$, with an arbitrary 
dependence on $x$. Local RG equations for variations of $\sigma(x)$ reduce 
to the conventional linear differential constraints for $\sigma$ and $g^I$ 
constant but contain additional local contributions depending on the 
derivatives of $g^I$, as well as the curvature. The consistency 
conditions arise from the abelian nature of the group 
of Weyl scale transformations. Such an approach has also
been extended to six dimensions in \GrinsteinCKA\ and three in \Nakthree.

In this paper we revisit some of the results in \Weyl, with an 
hopefully improved notation (although we apologise for alphabetical 
profligacy) and extensions.  The essential result is that there is a 
scalar function of the couplings ${\tilde A}(g)$  such that
\eqn\CT{
\del_g \, {\tilde A}(g) = \del g^I T_{IJ}(g) \beta^J(g) \, ,
}
where at a fixed point $\beta^I(g_*)=0$, $\quar {\tilde A}(g_*) =  a$. 
The symmetric part of $T_{IJ}$ defies a natural metric $G_{IJ}$ 
so that under RG flow
\eqn\strong{
\beta^I \pr_I\, {\tilde A} = G_{IJ} \beta^I \beta^J \, ,
}
Away from fixed points ${\tilde A}(g)$ is arbitrary up to
\eqn\Aarb{
{\tilde A}(g) \to {\tilde A}(g) + g_{IJ}(g)  \beta^I (g)\beta^J(g)\, ,
}
while correspondingly 
\eqn\Garb{
G_{IJ} \to G_{IJ} + \L_\beta g_{IJ} \, , \quad 
\L_\beta g_{IJ} = \beta^K\pr_K g_{IJ} + \pr_I \beta^K g_{KJ}
+ \pr_J\beta^K g_{IK} \, .
}
It is  then sufficient in order  to demonstrate the strong version of 
the $a$-theorem that $G_{IJ} + \L_\beta g_{IJ}$ is positive definite 
just for some particular  $g_{IJ}$.

In two dimensions positivity of the metric, up to the freedom in 
\Garb, flows from showing \Weyl\ that  $G_{IJ}+ \L_\beta g_{IJ} $, 
for suitable $g_{IJ}$, becomes the Zamolodchikov metric determined by 
the two point function $G_{IJ}(\mu^2 x^2)_{\rm Zam}= 
(x^2)^2\langle \O_I(x)\, \O_J(0) \rangle$, for 
$\{\O_I\}$ scalar operators dual to $\{g^I\}$, \Zam.  Variation
of $x^2$ in $G_{IJ\,\rm Zam}$ is equivalent to \Garb. However
the original analysis demonstrates \strong\ and does not directly 
imply \CT, see also \Friedan.

In four dimensions $G_{IJ}$ is related to  
$\langle \O_I\, T_{\mu\nu}\, \O_J \rangle$, although the precise 
connection is not fully clear and positivity, except at 
weak coupling when $G_{IJ}$ can be calculated or at a 
conformal fixed point, is however not apparent from a perturbative series
expansion. 

The consistency conditions such as \CT, obtained previously in \Weyl\
and discussed further in this work, are derived by considering the 
response to infinitesimal Weyl rescalings of the metric. We also
consider the response of the theory to finite Weyl rescalings of
the metric $\gamma_{\mu\nu} \to e^{2\sigma} \gamma_{\mu\nu}$.
The result is also expressed in terms of running couplings
$g_\sigma{\!\!}^I$ together with additional contributions also
depending explicitly on $\sigma$, involving derivatives up to
${\rm O}(\sigma^4)$, and containing $G_{IJ}$ and related functions
as well as derivatives of the couplings. For four dimensional theories
the final expression is quite involved but it extends the result
at a fixed point used as a starting point for the introduction of
a dilaton field in \KS\ and \Luty.

For four dimensional theories the local RG equations, from which \CT\
is derived, are essentially equivalent
to expressing the energy momentum tensor trace in terms of a basis
of scalar operators as well as contributions involving the curvature, 
defining $c$ and $a$, but also  scalars formed from derivatives of 
$g^I$. However even on flat space with constant $g^I$ there may be 
derivative terms so that
\eqn\trace{
\eta^{\mu\nu} T_{\mu\nu} = \beta^I(g) \O_I + 
\pr_\mu J_\upsilon{\!}^\mu \, .
}
Here $J_\upsilon{\!}^\mu$ is a current associated with an element 
$\upsilon$ of the Lie algebra of the symmetry group ${\rm G}_K$ of the 
kinetic terms of the theory. Such terms may arise at three loops in 
perturbative calculations for  scalar fermion theories \Fortin, 
\FortinC. A fixed point $\beta^I(g_*)=0$  would apparently give rise 
to scale but not conformally invariant theories if there is no 
redefinition of $T_{\mu\nu}$ which removes 
$\pr_\mu J_\upsilon{\!}^\mu$.
However the $\beta$-functions have an arbitrariness related
to the freedom to make transformations under ${\rm G}_K$ at the expense of
a redefinition of the couplings. This freedom cancels in \trace\ so 
that it can be rewritten as
 \eqn\traceB{
\eta^{\mu\nu} T_{\mu\nu} = B^I(g) \O_I  \, , 
}
where
\eqn\modB{
B^I(g) = \beta^I(g) - (\upsilon g)^I \, ,
}
so that if the couplings are not all invariant under ${\rm G}_K$ there 
may be a difference between $\beta^I $ and $B^I$.  If this 
possibility arises
\CT\ holds for $\beta^I \to B^I$ and hence the potential
strong $a$-theorem discussed here applies to the RG flow generated by 
$B^I$, and its vanishing, $B^I(g_*)=0$, at a fixed point defines a CFT.
The transformation from \trace\ to \traceB, in terms of the modified
$\beta$-functions as in \modB, assumes there are no anomalies in
$\pr_\mu J_\upsilon{\!}^\mu$. This should be the case in parity conserving
theories when $J_\upsilon{\!}^\mu$ is a vector current. 

The existence of  ${\tilde A}(g)$ satisfying \CT\ also requires 
integrability conditions which constrain the form of $\beta$-functions.
This was explored in \Analog\ and is investigated further in this 
paper, see also \Sann. The conditions require relations
between the coefficients appearing in $\beta$-functions at
different loop orders and which correspond to graphs of very
different topologies.

As an application of the results obtained and for the analysis
of the integrability constraints on $\beta$-functions we consider
here a model renormalisable scalar fermion theory with Yukawa and
quartic scalar couplings. Previously \Analog\ the various quantities
appearing in the consistency conditions were calculated to lowest
perturbative order for general theories including gauge fields. To
go beyond this requires three loop calculations. For complex 
scalars coupled to Weyl fermions imposing a $U(1)$ symmetry ensures
that the number of graphs necessary is ${\rm O}(10)$ rather than
${\rm O}(100)$, or more,  for a completely general scalar/fermion theory. 
We obtain results
for three loop anomalous dimensions and Yukawa $\beta$-functions without
calculating more than a couple of graphs by reducing this theory
to one describing the standard model top/Higgs coupling, recently obtained
by Chetyrkin and Zoller \Chet, and also a general $\N=1$ supersymmetric
scalar fermion theory when the relevant results have been known for some
time \Jack. The consistency conditions obtained here allow calculations
for $T_{IJ}$, initially defined in terms of a curved background, to be 
reduced to flat space calculations and we determine the three loop
contributions depending on the Yukawa couplings in the specific model
theory for which the three loop $\beta$-functions were obtained. The
result requires extracting the local divergences for two three-loop
vacuum diagrams. The results can be checked by reducing to supersymmetry
as a special case when much simpler superspace methods are possible.
As usual we use dimensional regularisation which may be problematic
at higher loop orders. These issues  are discussed in \Chet, but in
the absence of gauge fields here such problems appear to be 
irrelevant to the order considered here.

We consider in detail the application of these results to
$\N=1$ Wess-Zumino supersymmetric theories, extending the
discussion in \SusyA. For such theories the space of couplings is
naturally a complex manifold since they may be extended to chiral
or anti-chiral superfields. We show that three loop calculations 
demonstrate that the metric is hermitian to this order. 
Furthermore, when redefinitions as in \Garb\ are extended to the
supersymmetric case the assumption of a hermitian metric is preserved.
There is no all orders proof of hermiticity in the context of this paper, 
although for superconformal theories
related results have been obtained by Papadodimas \Papadodimas\
and Asnin \Asnin. The results for the metric can also
be expressed in K\"ahler form if allowance is made for potential
redefinitions of the couplings. 

Although this paper is quite lengthy each section is substantially
independent. In section 2 we rederive the local RG equations and
associated integrability conditions which follow by considering
the response to infinitesimal Weyl rescalings of the metric in 
theories in which the couplings are allowed to be local.  In section
3 the infinitesimal transformations  are integrated to obtain the change 
in the vacuum energy functional $W$ under finite rescalings. The 
results depend on running couplings $g_\sigma{\!}^I$ and 
provide an interpolation between UV and IR fixed points. In
section 4 we restrict the equations to flat space and broken conformal
symmetry. This context is sufficient to allow the metric $G_{IJ}$, 
which is initially defined for curved space backgrounds, to be
recovered just from flat space calculations.

The scalar fermion theory used as an illustration is introduced
in section 5 and the various $\beta$-functions and anomalous
dimensions listed. In particular three loop results for the Yukawa
$\beta$-functions and also the anomalous dimensions for this
theory are obtained, primarily using previous calculations and
also the restriction to the supersymmetric case. In section 6 we
analyse the RG equations for this theory. It is shown how they
impose non-trivial consistency conditions on the coefficients 
which are present in the general expansions for the $\beta$-functions
and associated anomalous dimensions. In particular it is shown that
at three loop order it is necessary to take account of \modB\ for
\CT\ to be valid. The result for $\upsilon$ at this order is
in agreement with the detailed three loop calculations of  
Fortin {\it et al} \FortinC\ for scalar fermion theories. In section 7
we restrict to supersymmetric theories and demonstrate the
consistency of a hermitian metric. The results are compared
with expressions when $a$-maximisation is extended away
from superconformal fixed points by introducing Lagrange 
multipliers and also the possibility of
a K\"ahler form for the metric is discussed. Sections 8 and 9
describe how the metric  and related quantities can be determined by
flat space calculations using dimensional regularisation. Section 8
discusses the general formalism for renormalisable theories with
local couplings and sets up the required RG equations. Section 9
applies these methods to the scalar/fermion theory and determines the
additional necessary field independent counterterms to three loops. 
These determine the metric and, specialised to the supersymmetric case,
show that it is hermitian to this order.

There are four appendices containing further calculational details.
Appendix A analyses how particular contributions to the anomalous 
dimensions in supersymmetric theories which are proportional to
transcendental numbers can be extended to determine the related contributions
to the metric and also $a$. Appendix B contains further details on 
the derivation of local RG equations in the context of dimensional
regularisation. The RG equations are extended to allow for special conformal 
transformations as well as the usual variations of scale.
The methods used here to obtain the three loop
counterterms for Yukawa theories with dimensional regularisation are
described in Appendix C and are also extended to four loops for
scalar theories in Appendix D.

\newsec{Local RG Equations and Integrability Conditions}

As was demonstrated in \Weyl, and more recently in \KS, non-trivial 
constraints on the RG flow in quantum field theories can be obtained 
by considering the response to infinitesimal local  Weyl rescalings of 
the metric of the form
\eqn\rmet{
\delta_\sigma \gamma_{\mu \nu} = 2 \sigma \, \gamma_{\mu \nu} \, ,
}
when the theory is extended to an arbitrary curved space background. 
Conformally invariant theories are invariant under such rescalings up
to local conformal anomalies induced by the non vanishing of the energy
momentum tensor on curved space.
Equations for the response to such Weyl rescalings for quantum 
field theories not at conformal fixed points may be obtained if 
the couplings are extended to arbitrary local functions and at the same time
as \rmet\ there is a flow in the space of local couplings.
The resulting equations are then an extension of the standard linear 
equations  which determine the RG flow in terms of the usual
$\beta$-functions and are realised by restricting to constant $\sigma$
as well as  constant couplings. Choosing  couplings $\{g^I\}$, which
are coordinates for a manifold $\M_g$,
the local RG equations obtained in \Weyl\ by assuming the  quantum
field theories are extended to arbitrary $g^I(x)$ as well as 
$\gamma_{\mu\nu}(x)$ are then generated in four
dimensions by the functional differential operator
\eqn\defD{
\Delta_\sigma = \int \d^4 x \, \sigma \, \bigg ( 2\gamma_{\mu \nu}
{\delta \over \delta \gamma_{\mu \nu}} + {\hbet}^I
{\delta \over \delta g^I} \bigg ) \, ,
}
where the $\beta$-functions, which are contravariant vectors on $\M_g$,  
have in general a linear contribution
\eqn\Bhat{
{\hbet}^I(g) = - (d- \Delta_I) g^I +  {\rm O}(g^2) \, .
}
In \Bhat, in the present context,  the spatial dimension $d=4$ and
$\Delta_I$  is the scale dimension of the operator $\O_I$, which is dual to
$g^I$, at the critical point when all $g^J\to 0$. Initially we restrict for 
simplicity to just marginal operators with  $\Delta_I = 4$, as for 
renormalisable theories when $g^J=0$ is the free theory.

Acting on the vacuum energy functional $W[\gamma_{\mu\nu},g^I]$, 
$\Delta_\sigma$ gives zero up to a residual local contribution, depending just 
on $\gamma_{\mu\nu},g^I$ and their derivatives at $x$,  so that
\eqn\DelW{\eqalign{
\Delta_\sigma \, 16\pi^2 W = {}& - \int d^4x \sqrt{-\gamma} \; \sigma \,\Big (
- C\, F + \quar A \, G + {\ts{1\over 72}} B \, R^2 + E^{\mu\nu} \, G_{IJ}
\pr_\mu g^I \pr_\nu g^J  \cr
\noalign{\vskip -6pt}
&\hskip 2.8cm {}+ {\ts {1\over 6}} R \, 
\big ( E_I \nabla^2 g^I + F_{IJ} \pr^\mu g^I \pr_\mu g^J
\big ) - X \Big ) \cr
&{}- 2  \int d^4x \sqrt{-\gamma} \, \pr_\mu \sigma \; \Big (
 E^{\mu\nu} \, W_I \pr_\nu g^I + {\ts {1\over 6}} R \, H_I \pr^\mu g^I
+ Y^\mu \Big ) \cr
&{}-  \int d^4x \sqrt{-\gamma} \, \nabla^2 \sigma \; \Big ( 
{\ts {1\over 6}} R \, D + Z \Big ) \, ,
}
}
where the curvature terms, apart from the Ricci scalar $R$, are
\eqn\defFGE{
F = C^{\mu\nu\sigma\rho} C_{\mu\nu\sigma\rho} \, , \quad
G= \quar \epsilon_{\mu\nu\sigma\rho}\epsilon^{\alpha\beta\gamma\delta}
R^{\mu\nu}{\!}_{\alpha\beta} R^{\sigma\rho}{\!}_{\gamma\delta}  \, , \quad
E^{\mu\nu} = R^{\mu\nu} - \half \gamma^{\mu\nu} \, R \,,
}
so that $G$ is the Euler density and $E^{\mu\nu}$ is the Einstein tensor. 
With the normalisations in \DelW
\eqn\CAfree{\eqalign{
C_{\rm free} = {}& {\ts {1\over 40}} \big ( {\ts{1\over 3}} n_S + n_W + 4n_V \big ) \, , \cr
A_{\rm free} = {}& {\ts {1\over 90}} \big (  n_S +{\ts {11\over 2}} n_W + 62n_V \big ) \, , 
}
}
for $n_S$ real scalars, $n_W$ Weyl fermions and $n_V$ vectors.
The remaining terms in \DelW, $X,Y^\mu, Z$, are independent of the curvature
and involve just the local couplings $g^I$ and their derivatives. $X,Y^\mu,Z$ 
therefore remain on restriction to flat space and can be decomposed in the form
\eqnn\XYZ$$\eqalignno{
X(g) = {}& \half \, A_{IJ}\, \nabla^2 g^I \nabla^2 g^J +
B_{IJK}\, \nabla^2 g^I \pr^\mu g^J \pr_\mu g^K +
 \half \, C_{IJKL} \,
\pr^\mu g^I \pr_\mu g^J \, \pr^\nu g^K \pr_\nu g^L \, ,\cr
Y^\mu( g)  = {}&  S_{IJ}\, \pr^\mu g^I \nabla^2 g^J 
+ T_{IJK}\,  \pr^\mu g^I \, \pr^\nu g^J \pr_\nu g^K \, , \cr
Z( g) = {}&  U_I\,  \nabla^2 g^I + V_{IJ}\, 
\pr^\mu g^I \pr_\mu g^J \, . & \XYZ
}
$$
Clearly $G_{IJ},F_{IJ},F_{IJ},V_{IJ}$ are symmetric while
$B_{IJK}=B_{I(JK)}, \, T_{IJK}=T_{I(JK)}$ and $C_{IJKL}=C_{(IJ)(KL)}
=C_{(KL)(IJ)}$.
The notation in \DelW\ and \XYZ\ is an adaptation of that
in \Weyl, with suitable modifications to ensure later simplifications. 
$G_{IJ},A_{IJ},S_{IJ}$ are covariant tensors under a redefinition of the 
couplings $g^I \to h^I(g)$  while $E_I,W_I,H_I,U_I$ are vectors. 
Since  $\nabla^2 g^I \to \pr_J h^I \nabla^2 g^J + 
\pr_J \pr_K h^I \pr^\mu g^J \pr_\mu g^K$, the transformation of
$B_{IJK}, C_{IJKL}$ under such
a change in the couplings contains additional inhomogeneous terms.
If $A_{IJ}$ is invertible $X$ may be written as
\eqn\ConD{\eqalign{
X =  \half \, A_{IJ}\, \D^2 g^I \D^2 g^J + \half \, {\hat  C}_{IJKL} \,
 \pr^\mu g^I \pr_\mu g^J \, \pr^\nu g^K\! \pr_\nu g^L \, ,
}
} 
where $\D^2 g^I$ is defined by
\eqn\defDg{
\D^2 g^I = \nabla^2 g^I + B^I{\!}_{JK}\, \pr^\mu  g^J \pr_\mu g^K \, , \quad
B^I{\!}_{JK}=(A^{-1})^{IL}B_{LJK} \, ,
}
with $B^I{\!}_{JK}$ acting as a connection on $\M_g$.
In \ConD \ ${\hat C}_{IJKL}= C_{IJKL} -  B^M{\!}_{IJ} B_{MKL}$  which then
also transforms as a tensor under redefinitions of the couplings.

Defining the energy momentum tensor and local operators $\O_I$ by
\eqn\defTO{
2 {\delta \over \delta \gamma^{\mu\nu}(x)} W = - \sqrt{-\gamma(x)} \, \langle
T_{\mu\nu}(x) \rangle \, , \qquad
{\delta \over \delta g^I(x) } W = - \sqrt{-\gamma(x)} \, \langle
\O_I (x) \rangle \, ,
}
the result \DelW\ then encodes the standard form for the trace anomaly
\eqn\tranom{
16 \pi^2 \big ( \gamma^{\mu\nu} \langle T_{\mu\nu} \rangle
- {\hbet}^I  \langle \O_I \rangle \big ) \big |_{\pr g = 0} =
C \, F - \quar A  \, G  - {\ts{1\over 72}} B \, R^2 
- {\ts {1\over 6}} D \, \nabla^2 R\, .
}

The crucial consistency conditions arise from the fact that the group of
local Weyl transformations is abelian so that
\eqn\comD{
\big [ \Delta_\sigma, \Delta_{\sigma'} \big ] = 0 \, .
}
Using, under Weyl rescalings  of the metric as in \rmet, 
\eqn\varFG{\eqalign{
\delta_\sigma F = {}& - 4 \sigma\, F \, , \quad 
\delta_\sigma G = - 4 \sigma\, G + 8\,  E^{\mu\nu}\nabla_\mu \nabla_\nu 
\sigma \, , \quad \delta_\sigma R = - 2 \sigma\, R - 6 \nabla^2 \sigma \, ,\cr
 \delta_\sigma  E^{\mu\nu} = {}& - 4 \sigma\, E^{\mu\nu} 
-  2 (\nabla^\mu\nabla^\nu - \gamma^{\mu\nu} \nabla^2) \sigma\, ,  \ \qquad
 \delta_\sigma \nabla^2 = - 2 \sigma\, \nabla^2 
+ 2 \pr_\mu \sigma \, \nabla^\mu  \, , 
}
}
then the curvature dependent terms arising from imposing \comD\ give
the integrability condition
\eqn\raG{
\pr_I A = G_{IJ} {\hbet}^J - \L_\hbet W_I \, ,
}
and relations which determine the $R$ dependent terms
\eqn\brel{
B =  E_I \hbet^I - \L_\hbet D \, , 
}
and
\eqn\EFHdet{\eqalign{
E_I = {}& - A_{IJ} \beta^J - \L_\hbet U_I \, ,\cr
F_{IJ} = {}& G_{IJ} - B_{KIJ} \hbet^K
-  U_K\, \pr_I\pr_J \hbet^K -  \L_\hbet V_{IJ} \, , \cr
H_I = {}&  S_{IJ}\hbet^J - {\tilde U}_I \, , \qquad
{\tilde U}_I \equiv U_I + \pr_I\hbet^J U_J + V_{IJ} \hbet^J \, ,
}
}
together with the condition
\eqn\EFHrel{
{\tilde E}_I \equiv E_I + \pr_I \hbet^J E_J + F_{IJ} \hbet^J =  \L_\hbet H_I \, .
}
Further relations which constrain $W_I,G_{IJ}$ are
\eqn\WS{
\pr_{[I}W_{J]} = - {\tilde S}_{[IJ]} \, ,
}
defining
\eqn\Stil{
{\tilde S}_{IJ} \equiv S_{IJ} + \pr_J \beta^K S_{IK} + T_{IJK} \beta^K \, ,
}
and
\eqn\GArel{
G_{IJ} - \L_\hbet S_{IJ}=  {\tilde A}_{IJ} \equiv
A_{IJ} + \pr_I \hbet^K A_{KJ}  + B_{JIK} \hbet^K  \, , 
}
and also a consistency  relation involving the derivative of
$G_{IJ}$ which can be simplified to
\eqn\relGBC{\eqalign{
\Gamma^{(G)}{\!}_{IJK} -  \L_\beta T_{IJK} -  S_{IL} \pr_J\pr_K \hbet^L 
 =   B_{IJK} + \pr_I \hbet^L  B_{LJK} + C_{ILJK} \hbet^L \, , 
}
}
for
\eqn\Chr{
\Gamma^{(G)}{\!}_{IJK}  = \half \big ( \pr_J G_{IK} + \pr_K G_{IJ} -
\pr_I G_{JK} \big ) \, ,
}
the Christoffel connection formed from $G_{IJ}$. The constraint 
\EFHrel\ follows by combining \EFHdet\ with \GArel. 

In  the above relations $\L_\hbet$ is the Lie derivative determined
by $\hbet^I$ so that
\eqn\Lie{
\L_\hbet W_I = \hbet^J \pr_J \, W_I + \pr_I \hbet^J \, W_J \, , \qquad
\L_\hbet D = \hbet^J\pr_J D \, ,
} 
with obvious extensions for $ \L_\hbet S_{IJ}, \L_\hbet V_{IJ}$,
analogous to $\L_\beta g_{IJ}$ in \Garb.  $\L_\hbet$ preserves 
tensorial properties under redefinitions of the couplings $g^I \to h^I(g)$. 

In \EFHrel, \Stil\ and \GArel\ ${\tilde E}_I, {\tilde S}_{IJ}, {\tilde A}_{IJ}$ are  tensors.
The relations \EFHdet\ and also \relGBC\ are not manifestly
invariant under redefinitions  but covariance can be verified  by 
combining different identities. The result  for $F_{IJ}$
is thus equivalent to
\eqn\GFrela{\eqalign{
G_{IJ} = {}& \big ( F_{IJ} - \pr_{(I} E_{J)} \big ) +
\big  ( B_{KIJ}-   \Gamma^{(A)}{\!}_{KIJ} \big )\hbet^K  
+  \L_\hbet \big ( V_{IJ} - \pr_{(I} U_{J)} -  A_{IJ} \big ) \, , 
}
}
where the three terms each  transform as a tensor.
\GArel\ determines $G_{IJ}$, which is later used as a metric on $\M_g$,
in terms of flat space results. It may be recast as
\eqn\GAA{
G_{IJ} = A_{IJ} - \half \, \beta^K \D_K A_{IJ} +
 \L_\beta \big ( S_{(IJ)} + \half A_{IJ} \big ) \, ,
}
where
\eqn\DLAV{\eqalign{
\D_K A_{IJ} =  \pr_K A_{IJ} - B_{JKI} - B_{IKJ} \, .
}
}

From \Stil\ and \GArel\ we may obtain
\eqnn\ABtil$$\eqalignno{
& G_{IJ} +  \pr_J\beta^K G_{IK} + \Gamma^{(G)}{\!}_{IJK} \beta^K =
G_{IJ} + \half \L_\hbet G_{IJ} - \half \big ( \pr_I (G_{JK} \beta^K) 
- \pr_J ( G_{JK} )\beta^K ) \big ) \cr 
& \qquad {}= \L_\hbet {\tilde S}_{IJ} + {\tilde G}_{IJ}  \, ,  & \ABtil 
}
$$
for 
\eqn\Gtil{
{\tilde G}_{IJ} = ( \delta_I {\!}^K + \pr_I \beta^K) (\pr_J{\!}^L + \pr_J \beta^L)
A_{KL} + 2 \,  ( \delta_{(I} {\!}^L + \pr_{(I} \beta^L) B_{LJ)K} \beta^K 
+ C_{IKJL} \beta^K \beta^L \, .
}

The essential variation and RG equations \CT\ and \strong\ follow 
directly from \raG\ for
\eqn\ATT{
{\tilde A} = A + W_I \beta^I \, , \qquad
T_{IJ} = G_{IJ} + \pr_I W_J - \pr_J W_I \, .
}

The coefficients in \DelW\  have an intrinsic arbitrariness induced
by adding to $W$ local terms of the same form as in \DelW\  for $\sigma$
a constant. This freedom gives an equivalence
\eqn\equivW{\eqalign{
W_I \sim {}& W_I -  \pr_I a + g_{IJ} \hbet^J \, , \cr
H_I \sim {}& H_I + e_I +  \pr_I \beta^J e_J +  f_{IJ} \hbet^J \, , \cr
S_{IJ} \sim {}& S_{IJ} +  g_{IJ} - a_{IJ} -  \pr_I \hbet^K a_{JK} - 
b_{JIK} \hbet^K \, , \cr
T_{IJK} \sim {}& T_{IJK} +  \Gamma^{(g)}{\!}_{IJK} - b_{IJK}  
- \pr_I\hbet^L b_{LJK} - c_{ILJK}\hbet^L  \, ,  \cr
D \sim {}& D -  b + e_I \hbet^I \, , \cr
U_I \sim {}&  U_I - e_I - a_{IJ} \hbet^J \, , \cr
V_{IJ} \sim {}& V_{IJ} +  g_{IJ} - f_{IJ} - b_{KIJ} \hbet^K\, , \cr
F_{IJ} \sim {}& F_{IJ} + \L_\hbet f_{IJ} + \pr_I\pr_J \hbet^K e_K \, , \cr
B_{IJK} \sim {}& B_{IJK} + 
\L_\hbet b_{IJK} + \pr_J\pr_K \hbet^L a_{IL} \, , \cr
C_{IJKL}\sim {}& C_{IJKL} +  \L_\hbet c_{IJKL} + \pr_I\pr_J \hbet^M \, b_{MKL}
+ \pr_K \pr_L \hbet^M b_{MIJ} \, ,
}
}
as well as
\eqn\equiva{
(A,B,C,E_I,G_{IJ},A_{IJ}) \sim (A,B,C,E_I,G_{IJ},A_{IJ}) 
+ \L_\hbet( a,b,c,e_I,g_{IJ},a_{IJ}) \, .
} 
With the definition \Stil\ then from \equivW
\eqn\Stran{\eqalign{
{\tilde S}_{IJ} \sim {}& {\tilde S}_{IJ} - 
\pr_{[I} \big ( g_{J]K} \beta^K \big ) + g_{IJ} + \half \, \L_\beta g_{IJ} \cr
&{}- ( \delta_I{\!}^K + \pr_I \beta^K)( \delta_J{\!}^L + \pr_J \beta^L) a_{KL}
- 2 \, ( \delta_{(I}{\!}^L + \pr_{(I}\beta^L) b_{LJ)K} \beta^K \cr
&{}- c_{ILJK} \, \beta^L \beta^K \, .
}
}
As a consequence of \equivW\ we may set, if $ \delta_I{\!}^J + \pr_I \beta^J$ is
invertible,
\eqn\Szero{
S_{(IJ)} = T_{IJK} = D = U_I = V_{IJ} = 0 \, .
}

To describe the RG flow of  four dimensional  quantum field theories it is 
necessary to take into  account contributions to the basic equations
corresponding to relevant operators, in addition to just the marginal 
operators with couplings $\{g^I\}$. 
These may induce modifications of the consistency conditions
obtained above for the RG flow. We first consider vector operators. A general
analysis may be obtained by extending the global symmetry
group of the kinetic terms ${\rm G}_K$ to a local symmetry by 
introducing background gauge fields $a_\mu(x) \in {\frak g}_K$,
the Lie algebra corresponding to ${\rm G}_K$, and extending all
derivatives to covariant derivatives $D_\mu = \pr_\mu + a_\mu$.
The symmetry extends to the full quantum field theory if,
for any $\omega \in {\frak g}_K$, the couplings $g^I $ and $a_\mu$ transform
as
\eqn\transga{\eqalign{
\delta_\omega g^I(x) = {}&- (\omega g)^I(x) = - \omega^I{\!}_J(x) g^J(x) \, , \cr
\delta_\omega a_\mu(x) = {}&  D_\mu \omega(x) =  \pr_\mu \omega(x) + 
[a_\mu(x) ,\omega(x) ] \, .
}
}
where $\omega^I{\!}_J$ belongs to the appropriate representation of 
${\frak g}_K$ acting on the couplings $\{g^I\}$. Under such variations
 $\delta_\omega \hbet^I(g) = \omega^I{\!}_J \hbet^J(g)$.
The corresponding covariant derivative acting on the couplings is then
\eqn\deg{
 D_\mu g^I = \pr_\mu g^I +(a_\mu g)^I \, , \qquad
(a_\mu g)^I =   a_\mu{\!}^I{\!}_J g^J \, ,
}
with the curvature as usual 
\eqn\cura{
f_{\mu \nu} = \pr_\mu a_\nu - \pr_\nu a_\mu + [ a_\mu , a_\nu ] \, .
}
The generator of local ${\rm G}_K$ transformations as in \transga\ is then
\eqn\Delom{
\Delta_\omega = \int \d^4 x \; \bigg (  D_\mu \omega \cdot 
{\delta \over \delta a_\mu} - (\omega g)^I {\delta \over \delta g^I}
\bigg ) \, , \qquad \big [ \Delta_\omega , \Delta_{\omega'} \big ]
= \Delta_{[\omega,\omega']} \, ,
}
with $\cdot $ denoting an  invariant scalar product on ${\frak g}_K$.

The introduction of background gauge fields $a_\mu$, so that
now we take $W[\gamma_{\mu\nu},g^I,a_\mu]$, allows
\defTO\ to be extended to define local vector   currents by
\eqn\defJ{
{\delta \over \delta a_{\mu}(x)} W 
= - \sqrt{-\gamma(x)} \, \langle J^\mu(x)  \rangle \, , 
\qquad  J^\mu \in {\frak g}_K \, .
}
For this paper we assume manifest background gauge invariance so that 
\eqn\back{
\Delta_\omega W = 0 \, ,
}
although in general there can be anomalies which involve $\epsilon$-tensor
contributions. If present there would be additional consistency 
conditions.  If \back\ holds then from the definition \defJ\ the current $J^\mu$
satisfies the conservation equation
\eqn\conserv{
\omega \cdot D_\mu  \langle J^\mu  \rangle = - (\omega g)^I 
\langle \O_I \rangle \, , \qquad \omega \in {\frak g}_K \, .
}

Under Weyl rescalings of the metric there are  additional
contributions to the functional differential operator in \defD\
involving $a_\mu$ given by
\eqn\defDa{
\Delta_{\sigma,a} = \int \d^4 x  \; \big ( \sigma \, 
\rho_I    D_\mu g^I
- \pr_\mu \sigma  \, \upsilon   \big ) \cdot 
{\delta \over \delta a_\mu}  \, , \quad \rho_I(g), \upsilon(g) \in
{\frak g}_K \, .
}
Assuming \defJ\ then \defDa\ implies \trace.
We assume that manifest covariance under ${\rm G}_K$ is maintained so
that, for all $\omega\in {\frak g}_K$,
\eqn\covD{
\big [ \Delta_\omega, \Delta_\sigma\big ]  =
\big [ \Delta_\omega, \Delta_{\sigma,a} \big ]  =  0 \, ,
}
which implies
\eqn\covarG{
(\omega g)^J\pr_J \hbet^I = (\omega \hbet)^I \, , \quad
(\omega g)^J\pr_J \rho_I + \rho_J \, \omega^J{\!}_I = [\omega, \rho_I] \, ,
\quad   (\omega g)^J\pr_J \upsilon = [\omega, \upsilon]  \, ,
}
In this case \defDa\ can be equivalently  expressed as 
\eqn\defDar{
\Delta_{\sigma,a} = \int \d^4 x  \; \big ( \sigma \, 
{\hrho}_I  D_\mu g^I 
- D_\mu ( \sigma  \, \upsilon )  \big ) \cdot 
{\delta \over \delta a_\mu} \, , \qquad {\hrho}_I
= \rho_I + \pr_I \upsilon  \, ,
}
since, using \covarG,
\eqn\diffv{
D_\mu \upsilon = \pr_\mu \upsilon + [ a_\mu , \upsilon ] = \pr_I \upsilon\, 
D_\mu g^I \, .
}

For general quantum theories it is also necessary to 
consider the extra contributions arising from operators $\{\O_M\}$
with canonical dimension two. The associated couplings  $\{M\}$
 are mass terms belonging to the dual space $V_M$.
 The vacuum self energy now extends to a functional
 $W[\gamma_{\mu\nu},g^I,a_\mu, M]$. The action of gauge
 transformations in \transga\ now extends also to
 $\delta_\omega M(x) = M(x) \omega_M(x) - {\bar \omega}_M(x) M(x)$
 for $\omega_M,{\bar \omega}_M$ belonging to appropriate 
 representations of ${\frak g}_K$. There is also a corresponding 
 additional term in $\Delta_\omega$ in \Delom\ which requires that
  \conserv\ is extended to
 \eqn\conservM{
\omega \cdot D_\mu  \langle J^\mu  \rangle = - (\omega g)^I 
\langle \O_I \rangle - (  M\omega_M - {\bar \omega}_M M ) \cdot
\langle \O_M \rangle \, .
}
for ${\delta \over \delta M} W 
= - \sqrt{-\gamma} \, \langle \O_M \rangle$
and  $\cdot$ also denoting
the natural scalar product on $V_M \times V_M{\!}^*$

As for $g^I$  local RG equations require extension to arbitrary
$M(x) \in V_M$.
In \DelW\ describing the response to Weyl rescalings of the metric, 
besides $\Delta_{\sigma,a}$, it is 
necessary also to include the additional term
\eqn\defDM{\eqalign{
\Delta_{\sigma,M} = - \int \d^4 x  \; \Big ( & \sigma  \, 
\big ( (2 -  {\gamma}_M )  M +
{\ts{1\over 6}} R \, \eta + \delta_I  \, D^2 g^I 
+ \epsilon_{IJ}\, D^\mu g^I D_\mu g^J  \big ) \cr 
\noalign{\vskip - 4pt}
&{} + 2\, \pr_\mu \sigma  \, \theta_I  D^\mu g^I
+ \nabla^2 \sigma \, \tau   \Big ) \cdot 
{\delta \over \delta M}   \, ,
}
}
where $\eta, \delta_I,\epsilon_{IJ}= \epsilon_{JI}, \theta_I,
 \tau \in V_M$ 
and  ${ \gamma}_M :V_M\to V_M$ . 
\covD\ is extended to $ [ \Delta_\omega, \Delta_{\sigma,M} ]  =  0$.

The requirement that
 \eqn\comDa{
\big [ \Delta_\sigma + \Delta_{\sigma,a} + \Delta_{\sigma,M}, 
\Delta_{\sigma'} + \Delta_{\sigma',a} + \Delta_{\sigma',M}\big ] = 0 \, ,
}
imposes further consistency conditions which follow by using
\eqn\varDg{\eqalign{
(  \Delta_\sigma + \Delta_{\sigma,a}  ) D_\mu g^I = {}& \pr_\mu \sigma\,  B^I
+ \sigma \, D_\mu g^J \big ( \pr_J B^I + ({\hrho}_J g)^I \big )
+ \sigma\,  (\upsilon D_\mu g)^I \, , \cr
\noalign{\vskip 2pt}
(  \Delta_\sigma + \Delta_{\sigma,a}  ) D^2 g^I = {}& \nabla^2 \sigma \, 
 B^I + 2 \, \pr_\mu \sigma \, D_\mu g^J \Psi_J{}^I 
+ \sigma \, D^\mu g^J D_\mu g^K  \, \Omega_{JK}{}^I \cr
&{} + \sigma \big ( - 2 \, D^2 g^I + D^2 g^J \big ( \pr_J B^I 
+ ({\hrho}_J g)^I \big ) +  (\upsilon D^2 g)^I \big ) \, , 
}
}
with $B^I$ the modified $\beta$-function defined in \modB, $\hrho_I$ as
in  \defDar,   and 
\eqn\deflam{\eqalign{
\Psi_J{}^I = {}& \delta_J{}^I + \pr_J B^I + \half ( {\hrho}_J  g)^I \, ,\cr
\Omega_{JK}{}^I = {}& \pr_J\pr_K B^I + (\pr_{(J} {\hrho}_{K)} g)^I 
+ 2 \,  ({\hrho}_{(J})^I{}_{K)} \, .
}
}
The Lie derivative defined by  \Lie\  is also extended to
ensure that it transforms covariantly under ${\rm G}_K$ rotations so that
$\L_\beta \, W_I \to \wL_{B,\hrho} \, W_I$ where
\eqn\exLie{
\wL_{B,\hrho} \, W_I = \L_B W_I + ( {\hrho}_I  g)^J W_J \, .
}
Hence $\wL_{B-(\omega g),\hrho+\partial \omega} \, W_I 
= \wL_{B,\hrho} \, W_I - (\omega g)^J \pr_J W_I - W_J \, \omega^J{}_I
= \wL_{B,\hrho} \, W_I$ for $\omega\in {\frak g}_K$.

With these results, and  $\L_{\upsilon g} \upsilon = 0$, the condition
\comDa\ requires
\eqn\Brho{
{\hrho}_I B^I = 0 \, , 
}
and
\eqn\comDM{
\eta =  \delta_I B^I  - (\L_B - \gamma_M) \tau \, .
}
which determines $\eta$, and
\eqn\derel{
\Psi_I{\!}^J \delta_J + \epsilon_{IJ} B^J 
= \big  (\wL_{B,\hrho} - \gamma_M \big ) \theta_I  \, .
}
The property \Brho\ ensures that the extended Lie derivative commutes
with contraction with $B^I$ so that in \exLie\
$B^I \, \wL_{B,\hrho} \, W_I = \L_B ( B^I W_I)$. Furthermore,  we then,
with the definitions in \deflam, obtain
\eqn\comLPsi{
\big [ \wL_{B,\hrho} \, , \, \Psi_I{}^J\big ] = \Omega_{I K}{}^J  \, B^K\, .
}

The functional differential operators in \defDa\ and \defDM\
are essentially arbitrary up to variations arising from purely
local contributions which automatically maintain the
consistency conditions \Brho\ and \comDM. Such variations can be
generated by
\eqn\genD{\eqalign{
\delta \,\big ( \Delta_\sigma+ \Delta_{\sigma,a} + \Delta_{\sigma,M} \big )
= {}& \big [ \D ,  \Delta_\sigma+ \Delta_{\sigma,a} + \Delta_{\sigma,M} \big ] \, ,
\cr
\delta \, \big ( \Delta_\sigma+ \Delta_{\sigma,a} + \Delta_{\sigma,M} \big ) W
= {}& \D \, \big ( \Delta_\sigma+ \Delta_{\sigma,a} + \Delta_{\sigma,M} \big ) W \, ,
}
}
for any local functional differential operator $\D$. Choosing
\eqn\varrho{
\D = \int \d^4x \; r_I D_\mu g^I \cdot {\delta \over \delta a_\mu} \, ,
\qquad r_I(g) \in {\frak g}_K \, ,
}
gives
\eqn\vrhores{\eqalign{
& \delta {\hrho}_I = (r_I g)^J {\hrho}_J 
- ({\hrho} _I g)^J r_J + (\pr_I r_J -\pr_J r_I)B^J \,, \quad
\delta \upsilon = r_I  B^I \, , \cr
& \delta B^I = - B^J (r_J g)^I \, , \qquad
\delta \delta_I = (r_I g)^J \delta_J \, , \qquad
\delta \theta_I = (r_I g)^J \theta_J \, , \cr
& \delta \epsilon_{IJ} =   (r_I g)^K \epsilon_{KJ} + (r_J g)^K \epsilon_{IK}
+ (\pr_{(I} r_{J)} g )^K \delta_K + 2\, \delta_K (r_{(I})^K{}_{J)}  \, .
}
}
From this it follows that $\delta \big ( \wL_{B,\hrho} \, \theta_I \big ) =  ( r_I g)^J 
\wL_{B,\hrho} \,  \theta_J$.
In a similar fashion we may obtain
\eqn\vMres{\eqalign{
\delta \eta = {}& (\L_B - \gamma_M ) h \, ,  \qquad \delta \tau
= - h + d_I B^I  \, , \qquad
\delta \theta_I =  \Psi_I{\!}^J d_J + e_{IJ} B^J \, , \cr
\delta \delta_I = {}& \big (\wL_{B,\hrho} - \gamma_M\big ) d_I \, , \qquad
\delta \epsilon_{IJ} = \big  (\wL_{B,\hrho} \, - \gamma_M \big ) e_{IJ}
+ \Omega_{IJ}{}^K \, d_K \, ,
}
}
for $h, d_I,e_{IJ} \in V_M$. In consequence we may set $\tau=0$.

The essential equation \DelW\ is modified so that 
\eqnn\modW$$\eqalignno{
( \Delta_\sigma +{}&  \Delta_{\sigma,a} + \Delta_{\sigma,M})  \, 16\pi^2 W \cr
= {}& - \int d^4x \sqrt{-\gamma} \; \sigma \,\Big (
- C\, F + \quar A  \, G + {\ts{1\over 72}} B \, R^2 + E^{\mu\nu} \, G_{IJ}
D_\mu g^I D_\nu g^J  \cr
\noalign{\vskip -6pt}
&\hskip 2.8cm {}+ {\ts {1\over 6}} R \, 
\big ( E_I  D^2 g^I + F_{IJ} D^\mu g^I D_\mu g^J + I \cdot M
\big ) - X \Big ) \cr
&{}- 2  \int d^4x \sqrt{-\gamma} \, \pr_\mu \sigma \; \Big (
E^{\mu\nu} \, W_I D_\nu g^I + {\ts {1\over 6}} R \, H_I D^\mu g^I
+ Y^\mu \Big )
\, , & \modW
}
$$
where for simplicity the part involving $\nabla^2 \sigma$ is dropped
since the relevant terms can be set to zero by adding local 
contributions to $W$. In \modW\ $I\in V_M{\!}^*$ and $X,Y$ now have
additional terms involving $f$ and $M$,
\eqn\rgLa{\eqalign{
X( g ,a,M)  = {}& {\ts{1\over 2}} A_{IJ}\, D^2 g^I D^2 g^J
+  B_{IJK}\, D^2 g^I \, D^\mu g^J \! D_\mu g^K \cr
&{} + \half 
C_{IJKL}\, D^\mu g^I D_\mu g^J \, D^\nu g^K\! D_\nu g^L \cr
& {}+  \quar \,  f^{\mu \nu} \cdot \beta_f \cdot  f_{\mu \nu} 
+ \half \, M \cdot \beta_M \cdot  M 
+ f^{\mu \nu} \cdot P_{IJ} \, D_\mu g^I D_\nu g^J \cr
&{} + J_{ I} \cdot M\, D^2 g^I + K_{IJ} \cdot M \, D^\mu g^I D_\mu g^J  \, , \cr
\noalign{\vskip 2pt}
Y^\mu( g,a,M)  = {}&  S_{IJ}\, D^\mu g^I D^2 g^J 
+ T_{IJK}\,  D^\mu g^I \, D^\nu g^J D_\nu g^K \cr
&{} +  f^{\mu\nu}\cdot Q_I D_\nu g^I 
+ L_I \cdot M D^\mu g^I\, , \cr
}
}
for $P_{IJ}=-P_{JI}, Q_I \in {\frak g}_K$, $J_I,K_{IJ}=K_{JI},L_I \in V_M{\!}^*$.

The presence of the additional terms in \rgLa, together with the extension
$\Delta_\sigma \to \Delta_\sigma +  \Delta_{\sigma,a} + \Delta_{\sigma,M}$
leads to modifications of the previous consistency conditions 
together with some further necessary relations. In general 
$\beta^I \to B^I$, assuming G${}_K$-covariance as in \covarG\ with
additionally
\eqn\covarGr{
(\omega g)^K\pr_K  G_{IJ} + G_{KJ}\omega^K{}_I + 
G_{IK} \omega^K{}_J =0 \, ,
}
etc, but there are further required changes. To avoid too much 
complication we focus on the results related to the variation of $A$.
The basic equation \raG\  becomes
\eqn\raGa{
\pr_I A = G_{IJ} B^J - \wL_{B,\hrho} \, W_I  \, .
}
Taking into account
\eqn\varf{\eqalign{
\Delta_{\sigma,a} f_{\mu\nu} = {}& \sigma \big ( [\upsilon, f_{\mu\nu}]
+ (f_{\mu\nu}g)^I {\hrho}_I + ( \pr_I {\hrho}_J -
 \pr_J {\hrho}_I ) D_\mu g^I D_\nu g^J \big ) \cr
 &{}+ \pr_\mu\sigma \, {\hrho}_I \, D_\nu g^I - 
  \pr_\nu\sigma \, {\hrho}_I \, D_\mu g^I   \, ,
} 
}
and defining
\eqn\Shat{ 
{\hat S}_{IJ} = \Psi_{J}{\!}^K  S_{I K} + T_{IJK} B^K + L_J \cdot \theta_I
+ \half \, \hrho_J \cdot Q_I \, ,
} 
then instead of \WS, 
\eqn\WSa{
\pr_{[I}W_{J]}  = - {\hat S}_{[IJ]} \, . 
}

There are also extra relations from terms involving $f_{\mu\nu}$ which give,
for any  $\omega \in {\frak g}_K $,
\eqna\WQ$$\eqalignno{
(\omega g)^I W_I = {}&  - \omega \cdot Q_I B^I \, , &\WQ{a}  \cr
(\omega g)^J G_{IJ} ={}& -  \omega \cdot \wL_{B,\hrho} \, Q_I  
+ \omega \cdot P_{IJ} B^J 
- \half \, \omega \cdot \beta_f \cdot  {\hrho}_I \, ,  & \WQ{b}
}
$$
where \exLie\ is extended, as in \Other,
\eqn\LieQ{
\omega \cdot \wL_{B,\hrho} \, Q_I = \omega \cdot ( \L_B Q_I 
( {\hrho}_I  g)^J Q_J \big ) (\omega g)^J {\hrho}_J \cdot Q_I \, ,
}
so that $\wL_{B-(\omega g),\hrho+\partial \omega} \, Q_I 
= \wL_{B,\hrho} \, Q_I - (\omega g)^J \pr_J Q_I - Q_J \, \omega^J{}_I
+ [ \omega , Q_I ] = \wL_{B,\hrho} \, Q_I$.  
From \WQ{a}\ 
\eqn\rWQ{
({\hrho}_I g)^J W_J  = - {\hrho}_I \cdot Q_J B^J \, ,
}
so that the essential result \raGa\ can still be rewritten in the succinct 
form \CT
\eqn\ATB{
\pr_I  {\tilde A}  = T_{IJ} B^J \, ,
}
where ${\tilde A},T_{IJ}$ are now defined, using \Brho, by an extension
of \ATT\ to
\eqn\tAT{
{\tilde A} = A + W_I B^I \, , \qquad
T_{IJ} = G_{IJ} + 2\, \pr_{[I} W_{J]}  +  2\, {\hrho}_{[I} \cdot Q_{J]} \, .
}
Furthermore from \WQ{b}, in conjunction with \WQ{a},
\eqn\rgG{\eqalign{
(\omega g)^I G_{IJ} B^J = {}& - \omega \cdot B^I \pr_I (Q_J B^J) 
- (\omega g)^I {\hrho}_I \cdot Q_J B^J \cr
= {}& (\omega g)^I \big (  B^J \pr_J W_I +  ( {\hrho}_I g)^J W_J  
\big ) + (\omega B^I) W_I 
= (\omega g)^I \wL_{B,\hrho} \, W_I  \, ,
}
}
which ensures that \raGa\ implies $(\omega g)^I \pr_I A=0$.  

The consistency conditions also generate additional relations for the
terms in \modW, \rgLa\ containing $M$ which take the form. 
\eqn\relI{
I + J_I B^I = 0 \, ,
}
and
\eqn\relJKL{\eqalign{
{\tilde J}_I  + \wL_{B,\hrho} \, L_I + L_I\cdot  \gamma_M  = 
\theta_I \cdot \beta_M \, , \qquad
{\tilde J}_I \equiv \Psi_I{\!}^J J_J + K_{IJ} B^J  \, .
}
}
The relation \GArel,  determining $G_{IJ}$, now becomes
\eqn\GArela{\eqalign{
G_{IJ} =  {\tilde A}_{IJ}  + \wL_{B,\hrho} \, S_{IJ} 
- J_ J \cdot \theta_I - L_I \cdot \delta_J  \, , \quad
{\tilde A}_{IJ} = \Psi_I{\!}^K A_{KJ} + B_{JIK} B^K \, , 
}
}
which gives rise to a modification of  \GAA,
\eqn\GArelb{\eqalign{
G_{IJ} = {}&  
A_{IJ} - \half \big ( (\hrho_{(I}g)^K A_{J)K} +  B ^K \D_K A_{IJ} \big ) \cr
& {} + \wL_{B,\hrho} \big ( S_{(IJ)} + \half A_{IJ} \big )
- J_ {(I} \cdot \theta_{J)} - L_{(I} \cdot \delta_{J)} \, ,
}
}
with the definition of $ \D_K A_{IJ} $ unchanged from \DLAV.
Also \relGBC\ becomes
\eqn\relGBCa{\eqalign{
\Gamma^{(G)}{\!}_{IJK} = {}&  \Psi_I{\!}^L B_{LJK}  + C_{ILJK} B^L  + 
\Omega_{JK}{}^L\, S_{IL} + \wL_{B,\hrho} \, T_{IJK} \cr
&{} + \big ( \pr_I {\hrho}_{(J} -  \pr_{(J} {\hrho}_{I} \big ) \cdot 
Q_{K)} - {\hrho}_{(J} \cdot P_{K)I}  
- K_{JK} \cdot \theta_I - L_I \cdot \epsilon_{JK}  \, .
}
}

By considering $\Psi_J{}^K  G_{IK} + \Gamma^{(G)}{\!}_{IJK} B^K$ we may obtain
\eqn\RelGS{
G_{IJ} + \half \, \L_B \, G_{IJ} - \half \, Z_{IJ} = {\hat G}_{IJ} + 
\wL_{B,\hrho} \, {\hat S}_{IJ} - \theta_J \cdot \beta_M \cdot \theta_I 
+ \quar \, \hrho_J \cdot \beta_f \cdot \hrho_I \, , 
}
for
\eqn\GZhat{\eqalign{
{\hat G}_{IJ} = {}& \Psi_I{}^K \Psi_J{}^L A_{KL} 
+ 2\, \Psi_{(I}{\!}^K B_{KJ)L} B^L + C_{IKJL} B^K B^L\cr
& \hskip 1.1cm {} -2\, L_{(I} \cdot (\wL_{B,\hrho} - \gamma_M) \theta_{J)} 
- \L_B \hrho_{(I} \cdot Q_{J)} \, , \cr
\noalign{\vskip 2pt}
Z_{IJ} = {}& \pr_I \big (G_{JK} B^K \big ) - \pr_J \big ( G_{IK} B^K \big ) 
- \big ( (\pr_I \hrho_J - \pr_J \hrho_I) g \big ){}^K W_K \, .
}
}
Instead of \ATB\ we may then write
\eqn\ATBhat{
\pr_I  {\hat A}  = {\hat T}_{IJ} B^J \, , \qquad {\hat A} = A + W_I B^I
+ \half \, G_{IJ} B^I B^J \, ,
}
with
\eqn\That{\eqalign{
{\hat T}_{IJ} = {}& {\hat G}_{IJ} +
\wL_{B,\hrho} \, {\hat S}_{IJ} - \theta_J \cdot \beta_M \cdot \theta_I
+ \quar \, \hrho_J \cdot \beta_f \cdot \hrho_I  \cr
&{} + 2 \, \pr_{[I} \big ( W_{J]} + G_{J]K} B^K \big ) 
+  2\, {\hrho}_{[I} \cdot Q_{J]} 
-\half \, \big ( (\pr_I \hrho_J - \pr_J \hrho_I) g)^K W_K \, .
}
}

The equivalence relations \equivW, \equiva\
also extend to the more general case with additional terms stemming
from the presence of $a_\mu,M$. In particular
the essential equation \ATB\ is arbitrary up to the equivalence relations
given by
\eqn\equivA{\eqalign{
{\tilde A}\sim {} & {\tilde A} + g_{IJ} B^I B^J \, , \quad 
G_{IJ} \sim G_{IJ} + \wL_{\beta,\rho} \,  g_{IJ} 
=  G_{IJ} + \wL_{B,\hrho}  \, g_{IJ} \, , \cr 
W_I \sim {}& W_I + g_{IJ}B ^J \, , \qquad
\omega \cdot Q_I \sim \omega \cdot Q_I -  g_{IJ} (\omega g)^J \, , \ \
\omega \in {\frak g}_K \, .
}
}
Using ${\hat S}_{IJ} \sim {\hat S}_{IJ} + g_{IJ} + \half \, \L_B \, g_{IJ}
- \pr_{[I} \big (g_{J]K} B^K \big ) $ 
we may verify that \WSa\ and \RelGS\ are invariant. For simplicity we may
choose  ${\hat S}_{(IJ)} = 0$.
There is also an additional arbitrariness 
arising from local contributions to $W$ involving $f_{\mu\nu}$ such as
\eqn\equivp{\eqalign{
Q_I \sim{}&  Q_I +  p_{IJ} B^J \, , \quad
\omega\cdot  P_{IJ} \sim \omega \cdot P_{IJ} 
+ \omega \cdot \wL_{B,\hrho}  \, p_{IJ} + (\omega g)^K 
\hrho_K \cdot p_{IJ} \, , \cr
C_{ILJK} \sim {}& C_{ILJK} +
 ( \pr_L \hrho_{(J} - \pr_{(J} \hrho_L ) \cdot p_{IK)} +
 ( \pr_I \hrho_{(J} - \pr_{(J} \hrho_I ) \cdot p_{LK)} \, , \cr
 T_{IJK} \sim {}& T_{IJK} - \hrho_{(J} \cdot p_{IK)} \, , \qquad\qquad
 p_{IJ} = - p_{JI}  \in {\frak g}_K \, .
}
}
This gives in \tAT\ $T_{IJ} \sim T_{IJ} + 2 \, \hrho_{[I} \cdot p_{J]K} B^K$
so that $T_{IJ} B^J$ is invariant. From local terms containing $M$ we have
further
\eqn\equivj{\eqalign{
J_I \sim {}& J_I + \wL_{B,\hrho}  \, j_I + j_I \cdot \gamma_M \, , \quad
K_{IJ} \sim K_{IJ} + \Omega_{IJ}{}^K j_K \, , \quad
L_I \sim L_I - \Psi_I{\!}^J j_J \, , \cr
E_I \sim {} & E_I + j_I \cdot \eta \, , \quad A_{IJ} \sim  A_{IJ}
- 2\, j_{\smash{(I}} \cdot \delta_{J)} \, , \quad
B_{IJK} \sim B_{IJK} - j_I \cdot \epsilon_{JK} \, , \cr
S_{IJ} \sim {}& S_{IJ} + j_J \cdot \theta_I \, , \qquad \qquad
j_I \in V_M {\!}^* \, .
}
}
For consistency with omitting $\nabla^2 \sigma$ terms in \modW\ it
is necessary to impose $j_IB^I = 0$. ${\hat S}_{IJ}$ in \Shat\ 
and ${\hat G}_{IJ},Z_{IJ}$ in \GZhat\ are invariant under \equivp\ 
and \equivj.

\newsec{Integration of Weyl Scaling}

The consistency conditions obtained in the previous section
are obtained as integrability conditions for the response to
local Weyl rescalings  of the metric. Here we describe how
results for the vacuum energy functional $W[\gamma_{\mu\nu},g^I]$ 
for finite  rescalings of the metric can be obtained. 

For simplicity we focus initially on two dimensional quantum field 
theories.
With the functional differential operator $\Delta_\sigma$ given
by the corresponding form to \defD\  in two dimensions the basic
equation   \DelW\ becomes
\eqn\DelWt{
\Delta_\sigma \, 2\pi W = \int \d^2x \sqrt{-\gamma} \; \Big (\sigma \,\big (
 C\, R  -   G_{IJ}\, \pr^\mu g^I \pr_\mu g^J  \big )
 - 2 \pr_\mu \sigma \, W_I \, \pr^\mu g^I \Big ) \, ,
 }
for $C(g),G_{IJ}(g),W_I(g)$ depending on  the couplings $g^I$.
The consistency conditions flowing from \comD\ are just \Weyl
\eqn\ctwo{
\pr_I C = G_{IJ} \beta^J - \L_\beta W_I \, ,
}
which is essentially identical to the four dimensional result
given in \raG.

To integrate \DelWt\ we define $g_\sigma{\!\!}^I$ by
\eqn\rung{
{\d \over \d \sigma} g_\sigma{\!\!}^I = \beta^I(g_\sigma) \, , \qquad
g_0{\!}^I = g^I \, ,
}
where such running couplings depending on $\sigma(x)$ were 
discussed in \Fortin.
With this definition \DelWt\ directly implies, for arbitrary $\delta \sigma(x)$,
\eqn\DelWs{\eqalign{
\delta_\sigma\, & 2\pi 
W [ e^{2\sigma}\gamma_{\mu\nu} , g_\sigma{\!\!}^I ]\cr
 = {}& \!\!
\int \! \d^2x \sqrt{-\gamma} \; \Big (\delta \sigma \big (
 C(g_\sigma) \, (R -2 \nabla^2 \sigma)  -   G_{IJ}(g_\sigma) \, \pr^\mu 
 g_\sigma{\!\!}^I \pr_\mu g_\sigma{\!\!}^J  \big )
 -2 \pr_\mu \sigma \, W_I(g_\sigma)\, \pr^\mu g_\sigma{\!\!}^I \Big ) \, ,
} 
}
where on the right hand side the dependence on $\sigma$ is explicit. To
integrate this we first define ${\breve C}(\sigma)$ by
\eqn\defCc{
{\d \over \d \sigma}  {\breve C}(\sigma) = C(g_\sigma) \, , \qquad 
{\breve C}(0) = 0 \, ,
}
and then \DelWs, using \ctwo\ with the condition $G_{IJ} = G_{JI}$, gives
\eqnn\trialW$$\eqalignno{
\delta_\sigma \, \bigg ( 2\pi 
W [ e^{2\sigma} &\gamma_{\mu\nu} , g_\sigma{\!\!}^I ]
- \int \! \d^2x \sqrt{-\gamma} \; \Big ({\breve C}(\sigma) \, R 
+ \big ( C(g_\sigma) - W_I (g_\sigma) \beta^I(g_\sigma) \big ) 
\, \pr^\mu \sigma \pr_\mu \sigma \Big ) \bigg ) \cr
= &{}- \int \! \d^2x \sqrt{-\gamma} \;  \delta \sigma \,
G_{IJ}(g_\sigma) \,{\bar  \pr}^\mu g_\sigma {\!\!}^I 
{\bar  \pr}_\mu g_\sigma {\!\!}^J\cr
\noalign{\vskip -2pt}
&{}-  2 \int \! \d^2x \sqrt{-\gamma} \; \big ( \pr_\mu \delta \sigma \, 
W_I (g_\sigma) + \delta\sigma \,  \pr_\mu \sigma \, \L_\beta 
W_I(g_\sigma) \big ) \, {\bar \pr}^\mu g_\sigma{\!\!}^I \, , & \trialW
}
$$
where we define
\eqn\prg{
{\bar  \pr}_\mu g_\sigma {\!\!}^I = {\pr}_\mu g_\sigma {\!\!}^I -
\beta^I(g_\sigma) \, \pr_\mu \sigma \, .
}
Noting that
\eqn\sigg{
\delta_\sigma  {\bar  \pr}_\mu g_\sigma {\!\!}^I =
\delta \sigma \, \pr_J \beta^I(g_\sigma) \,  
{\bar  \pr}_\mu g_\sigma {\!\!}^J \, ,
}
does not involve $\pr_\mu \delta \sigma$ and defining 
${\breve G}_{IJ}(\sigma)$ by the solution to the differential
equation
\eqn\defGc{
{\d \over \d \sigma}  {\breve G}_{IJ}(\sigma) 
+ \pr_I \beta^K(g_\sigma)  {\breve G}_{KJ}(\sigma) 
+ \pr_J \beta^K(g_\sigma)  {\breve G}_{IK}(\sigma) 
= G_{IJ}(g_\sigma) \, , \qquad {\breve G}_{IJ}(0) = 0 \, ,
}
then we may finally obtain
\eqn\solW{\eqalign{
 2\pi \big ( W [ e^{2\sigma}\gamma_{\mu\nu} , g_\sigma{\!\!}^I ] -
W [ \gamma_{\mu\nu} , g^I ] \big ) =  \int \! \d^2x \sqrt{-\gamma} \; 
{\cal W} \, ,
}
}
where
\eqn\Wsol{\eqalign{
{\cal W} = {}& {\breve C}(\sigma) \, R 
+ {\tilde  C}(g_\sigma) \, \pr^\mu \sigma \pr_\mu \sigma 
-  {\breve G}_{IJ}(\sigma)
\; {\bar  \pr}^\mu g_\sigma {\!\!}^I {\bar  \pr}_\mu g_\sigma {\!\!}^J
- 2  \, W_I (g_\sigma) \,  {\pr}^\mu g_\sigma{\!\!}^I \pr_\mu \sigma \, ,
}
}
for ${\tilde C} = C + W_I \beta^I$. 

The differential equations \defCc\ and \defGc\ may be formally solved
as an expansion  in $\sigma$, noting that $f(g_\sigma) =
\exp( \sigma \L_\beta ) f(g)$,  in the form
\eqn\power{
{\breve C}(\sigma) = \big ( \exp( \sigma \L_\beta ) - 1 \big ) \L_\beta {\!}^{-1}
 C(g) \, , \quad
 {\breve G}_{IJ} (\sigma) =
 \big ( \exp( \sigma \L_\beta ) - 1 \big ) \L_\beta {\!}^{-1}  G_{IJ}(g) \, ,
}
which  gives rise to results  corresponding to those in \Other. The behaviour
for large $\sigma$ is less apparent in this expression.

The result \solW\ with \Wsol\ provides an interpolation of the anomalous 
contributions to the self energy functional $W$ between UV fixed points as 
$\sigma\to \infty$ and IR fixed points as $\sigma\to - \infty$
assuming  $g_\sigma{\!\!}^I$ is on a RG trajectory linking
to fixed points satisfying $\beta^I (g_*) = 0$. If this holds
then asymptotically ${\breve C}(\sigma) \sim C(g_*) \sigma$ and if the
fixed point is a surface ${\cal M}_{g_*}$ in the space of couplings, 
corresponding to exactly marginal operators, then on ${\cal M}_{g_*}$
$\pr_I C(g_*) =0 $ since then $\pr_I \beta^J(g_*)=0$.

It is also of interest to rewrite \solW\ to determine the response
to just a Weyl rescaling of the metric which can be achieved
by letting $\gamma_{\mu\nu} \to e^{-2\sigma} \gamma_{\mu\nu}$.
Apart from anomalous terms arising from ${\cal W}$ the Weyl
rescaling is realised by introducing the running couplings $g_\sigma{\!\!}^I$
since \solW\ and \Wsol\ give
\eqn\solWQ{
 2\pi \big (
W [e^{-2\sigma} \gamma_{\mu\nu} , g^I ] -  
W [\gamma_{\mu\nu} , g_\sigma{\!\!}^I ] \big ) 
=  \int \! \d^2x \sqrt{-\gamma} \;  {\cal W}' \, , \quad
 {\cal W}' = 2 \, \pr^\mu \sigma \pr_\mu {\breve C}
- {\cal W} \, .
}
To complete this result  it is necessary to determine 
$ \pr_\mu {\breve C}$. In general ${\breve C}(\sigma)$, determined by \defCc,
depends also the initial $g^I$. It is convenient to let $g^I \to g_\sigma{\!}^I$, 
${\breve C}= {\breve C} (\sigma,g_\sigma)$ and then 
${\del \over \del \sigma} = {\pr \over \pr \sigma} + \beta^I(g_\sigma)\pr_I$.
Hence $\pr _\mu {\breve C} =  {\pr \over \pr \sigma } {\breve C} \, \pr_\mu \sigma
+ \pr_I {\breve C} \, \pr_\mu  g_\sigma{\!}^I =
C(g_\sigma) \, \pr_\mu \sigma 
+ \pr_I {\breve C} \, {\bar \pr_\mu}  g_\sigma{\!}^I $. 
From  \defCc, \defGc\ with \ctwo\ we may  obtain
\eqn\recCG{\eqalign{
{\d \over \d \sigma} \big ( & \pr_I {\breve C}(\sigma)-  
{\breve G}_{IJ}(\sigma) \beta^J(g_\sigma) + W_I (g_\sigma) \big ) \cr
\noalign{\vskip -3pt}
& {} + \pr_I \beta^K(g_\sigma) \big ( \pr_K {\breve C}(\sigma)-
{\breve G}_{KJ}(\sigma) \beta^J(g_\sigma) + W_K (g_\sigma) \big ) 
= 0 \, .
}
}
This has the solution, with the necessary boundary conditions at $\sigma=0$,
\eqn\solCG{
\pr_I {\breve C}(\sigma)-
{\breve G}_{IJ}(\sigma) \beta^J(g_\sigma) + W_I (g_\sigma) 
= {\breve W}_I(\sigma )
}
so long as
\eqn\solM{
{\d \over \d \sigma} {\breve W}_I (\sigma) + \pr_I \beta^J(g_\sigma) 
{\breve W}_J (\sigma) = 0 \, ,
\qquad {\breve W}_I (0) = W_I(g) \, .
}
It is easy to check that ${\breve W}_I (\sigma) {\bar \pr}_\mu g_\sigma{\!}^I 
= W_I(g) \pr_\mu g^I$, ${\breve W}_I (\sigma)\beta^I(g_\sigma) = W_I(g)\beta^I(g)$.
With these results \solCG\ gives, since
$\pr_\mu = \pr_\mu\sigma {\del \over \del \sigma}  +  {\bar \pr}_\mu g_\sigma{\!}^I  \pr_I$,
\eqn\CWG{
\pr_\mu {\breve C} (\sigma)  = C(g_\sigma)  \, \pr_\mu \sigma
- W_I(g_\sigma)  \, {\bar \pr}_\mu g_\sigma{\!}^I  + 
{\breve G}_{IJ}(\sigma)  \, {\bar \pr}_\mu g_\sigma{\!}^I \beta^J 
+  W_I(g)\, \pr_\mu g^I \, .
}
Subject to \CWG, \solW\ and \Wsol\ then entails in \solWQ
\eqn\solWP{\eqalign{
 {\cal W}' = {}& -{ \breve C}(\sigma) \, R 
+  {\tilde C}(g)  \, \pr^\mu \sigma \pr_\mu \sigma 
+  {\breve G}_{IJ}(\sigma) \,
 { \pr}^\mu g_\sigma  {\!\!}^I { \pr}_\mu g_\sigma {\!\!}^J 
+ 2 \, W_I(g) \, \pr^\mu g^I \pr_\mu \sigma\, ,
}
}
where the result has been simplified  by using
\eqn\Cint{
{\tilde C}(g_\sigma) - {\tilde C}(g) = {\breve G}_{IJ}(\sigma) \beta^I (g_\sigma)
\beta^J(g_\sigma) = \int_0^\sigma \!\!\! \d t \; 
G_{IJ}(g_t) \beta^I (g_t) \beta^J(g_t) \, .
}
This follows from $\beta^I \pr_I {\tilde C} =
G_{IJ} \beta^I \beta^J$  which may be integrated, with 
the definition \defGc, to give \Cint.
Assuming $G_{IJ}(g')\beta^I(g')\beta^J(g')>0$
for all $g'{}^I\in (g^I, g_\sigma{\!\!}^I)$ then from \Cint\
${\tilde C}(g_\sigma) < {\tilde C}(g) $ for $\sigma<0$. 

A similar analysis may be extended to four dimensions starting from 
\DelW. For simplicity we impose $D=U_I=V_{IJ}=0$, as in \Szero, 
although $S_{IJ},T_{IJK}$ are not restricted initially.
The integrability conditions \brel\ and \EFHdet\ then become
\eqn\BEFH{
B =  E_I \hbet^I \, , \quad
E_I = - A_{IJ} \beta^J \, ,\quad
F_{IJ} = G_{IJ} - B_{KIJ} \hbet^K \, , 
\quad H_I =   S_{IJ}\hbet^J  \, .
}
\varFG\ extends to finite Weyl rescalings of the metric to give in four
dimensions
\eqn\varFGER{\eqalign{
F_\sigma  = {}& e^{- 4 \sigma}\,  F \, , \cr
G_\sigma = {}& e^{- 4 \sigma}  \big ( G + 8\,  E^{\mu\nu}\nabla_\mu \nabla_\nu
\sigma  \cr
\noalign{\vskip -4pt}
&\hskip 0.9cm {} - 4 \, \nabla^2( \pr^\mu \sigma \, \pr_\mu \sigma ) 
+ 8 \, \nabla^\mu ( \pr_\mu \sigma \, \nabla^2  \sigma) + 
8 \, \nabla^\mu ( \pr_\mu \sigma \, \pr^\nu  \sigma \, \pr_\nu \sigma  )  
\big ) \, ,\cr
E_\sigma{\!} ^{\mu\nu} = {}& e^{- 4 \sigma}\big (  E^{\mu\nu}
-  2 (\nabla^\mu\nabla^\nu - \gamma^{\mu\nu} \nabla^2) \, \sigma 
+2 \, \pr^\mu \sigma \, \pr^\nu \sigma + \gamma^{\mu \nu}\,  
\pr^\lambda \sigma \, \pr_\lambda \sigma \big ) \, , \cr 
R_\sigma = {}& e^{- 2 \sigma} \big (  R - 6 \, \nabla^2 \sigma - 
6 \, \pr^\mu \sigma \, \pr_\mu \sigma \big ) \, , \qquad 
\nabla_\sigma{\!}^2 = e^{- 2 \sigma} \big (  \nabla^2
+ 2 \, \pr^\mu \sigma \, \pr_ \mu \big )   \, .
}
}
It is also important in this case to extend \prg\ defining
\eqn\defDelg{
\Delta g_\sigma {\!\!}^I =  \nabla^2 g_\sigma {\!\!}^I -
\beta^I(g_\sigma) \, \nabla^2\sigma - 2 \,
\pr_J \beta^I(g_\sigma) \, \pr^\mu g_\sigma {\!\!}^J \pr_\mu \sigma 
+ \beta^J(g_\sigma)\, \pr_J \beta^I(g_\sigma)\, 
\pr^\mu \sigma \pr_\mu  \sigma  \, ,
}
such that, analogous to \sigg,
\eqn\sigD{
\delta_\sigma  \Delta g_\sigma {\!\!}^I =
\delta \sigma \, \pr_J \beta^I(g_\sigma) \,
\Delta g_\sigma {\!\!}^J + \delta \sigma \, \pr_J \pr_K 
\beta^I(g_\sigma)\, {\bar \pr}^\mu g_\sigma {\!\!}^J
{\bar \pr}_\mu g_\sigma {\!\!}^K \, ,
}
and for $g^I \to h^I$,
$ \Delta g_\sigma {\!\!}^I \to \pr_J h^I \Delta g_\sigma {\!\!}^J +             
\pr_J \pr_K h^I {\bar \pr}^\mu g_\sigma {\!\!}^J {\bar \pr}_\mu 
g_\sigma {\!\!}^K$. 

Using \DelW\ it follows that the local anomalous response to
Weyl rescaling can be written as  
\eqn\DelWff{
\delta_\sigma\, 16\pi^2 
W [ e^{2\sigma}\gamma_{\mu\nu} , g_\sigma{\!\!}^I ]
 =  \int \! \d^4x \sqrt{-\gamma} \; \A \, ,
} 
where $\A$ is determined by \DelW\ in conjunction with \varFGER.
Even with \Szero\ the general form is lengthy. Only the final expression
is of possible interest but we include below some intermediate steps in case
of any desire to verify the calculational details. For the curvature 
dependent terms, using \raG, \EFHrel\ as well as $B=E_I\beta^I$,
\eqnn\Acurve$$\eqalignno{
\A_{\rm curvature} ={}& \delta \sigma \big ( 
C(g_\sigma) \, F - {\ts{1\over 4}} A(g_\sigma) \, G - {\ts{1\over 72}}
E_I(g_\sigma) \beta^I(g_\sigma)  R^2   - E^{\mu\nu} \,
G_{IJ}(g_\sigma)\,  {\bar \pr}_\mu g_\sigma {\!\!}^I
{\bar \pr}_\nu g_\sigma {\!\!}^J \big ) \cr
&{}+ E^{\mu\nu} \, \delta_\sigma \Big ( \big (
 A(g_\sigma)  - W_I(g_\sigma) \beta^I(g_\sigma) \big )
\pr_\mu \sigma  \pr_\nu \sigma - 2 \, W_I(g_\sigma) \,
{\bar \pr}_\mu g_\sigma {\!\!}^I \pr_\nu \sigma \Big ) \cr
\noalign{\vskip 0pt}
&{}- {\ts{1\over 6}} R \; \delta_\sigma \Big ( 
 H_I(g_\sigma) \big ( 2  \,
{\bar \pr}^\mu g_\sigma {\!\!}^I \pr_\mu \sigma 
+ \beta^I(g_\sigma) \, \pr^\mu \sigma \pr_\mu \sigma \big ) \Big )  \cr
&{}- {\ts{1\over 6}} R \; \delta \sigma
\big ( E_I(g_\sigma) \Delta g_\sigma {\!\!}^I  + F_{IJ}(g_\sigma)\,
{\bar \pr}^\mu g_\sigma {\!\!}^I {\bar \pr}_\mu g_\sigma {\!\!}^J 
\big ) 
\,  .
& \Acurve
}
$$
There  are also contributions which remain  on flat space
and are independent of ${\bar \pr} g_\sigma$
\eqn\AA{\eqalign{
\A_A ={}&  \delta_\sigma \Big ( A(g_\sigma) \, \big ( \nabla^2\sigma\,  
 + \half \,  \pr^\mu \sigma \pr_\mu \sigma\big  ) \pr^\nu \sigma
\pr_\nu \sigma  \Big ) \cr
&{}+ \delta\sigma \,  \pr_I A(g_\sigma) \big ( \nabla^2  g_\sigma {\!\!}^I 
+ 2 \, \pr^\mu  g_\sigma {\!\!}^I \pr_\mu \sigma (\nabla^2 \sigma 
+ \pr^\nu \sigma \pr_\mu \sigma ) \big ) \cr
&{}+ \delta\sigma \,  \pr_I \pr_J A(g_\sigma)\, \pr^\mu  g_\sigma {\!\!}^I 
 \pr_\mu g_\sigma {\!\!}^I 
-  \delta\sigma \,   \L_\beta  A(g_\sigma)(\nabla^2 \sigma +
 \half \,  \pr^\nu \sigma \pr_\mu \sigma )\pr^\nu \sigma  \pr_\nu \sigma \cr
 &{}+2 \, \pr_\mu \delta\sigma \,  \pr_I A(g_\sigma)\, \pr^\mu  g_\sigma {\!\!}^I \, 
 \pr^\nu \sigma \pr_\nu \sigma \, .
}
}
The remaining contributions in \DelWff\ are also curvature 
independent but involve ${\bar \pr} g_\sigma$ in a non-trivial 
fashion. From the $R$ dependent terms
\eqn\EFHR{\eqalign{
\A_{EFH} ={}& 
\delta \sigma \Big ( E_I(g_\sigma) \big ( \Delta  g_\sigma {\!\!}^I 
+ \half \, \beta^I(g_\sigma) \big ( \nabla^2 \sigma + 
\pr^\mu \sigma \pr_\mu \sigma ) \big ) + F_{IJ}(g_\sigma)\,
 {\bar \pr}^\mu g_\sigma {\!\!}^I {\bar \pr}_\mu g_\sigma {\!\!}^J \cr
\noalign{\vskip -3pt}
& \hskip 0.7cm {} + \L_\beta H_I(g_\sigma) \, \big (2\,
{\bar \pr}^\mu g_\sigma {\!\!}^I \pr_\mu \sigma 
+ \beta^I(g_\sigma) \, \pr^\mu\sigma \pr_\mu \sigma \big )
\Big ) \, (\nabla^2 \sigma + \pr^\nu \sigma \pr_\nu \sigma ) \cr
&{} + 2\, \pr_\mu \delta \sigma \, H_I(g_\sigma) \big ( 
{\bar \pr}^\mu g_\sigma {\!\!}^I + \beta^I(g_\sigma)  \pr^\mu \sigma \big )\, 
(\nabla^2 \sigma + \pr^\nu \sigma \pr_\nu \sigma ) \, .
}
}
For the terms involving $W_I$, including those arising from $A$
using \raG\ in \AA,
\eqnn\Weq$$\eqalignno{
\A_W = {}& - \delta_\sigma \Big ( W_I(g_\sigma) \big ( 
\Delta  g_\sigma {\!\!}^I \, \pr^\nu \sigma \pr_\nu \sigma 
+ 2\, {\bar \pr}^\mu g_\sigma {\!\!}^I \pr_\mu \sigma \, 
( \nabla^2 \sigma + \pr^\nu \sigma \pr_\nu \sigma ) \big )  \cr
\noalign{\vskip -3pt}
& \hskip 0.8cm {} + \pr_J W_I(g_\sigma) \, 
{\bar \pr}^\mu g_\sigma {\!\!}^I {\bar \pr}_\mu g_\sigma {\!\!}^J \,
\pr^\nu \sigma \pr_\nu \sigma 
+ W_I(g_\sigma) \beta^I(g_\sigma)  \big ( 2 \, \nabla^2 \sigma  + 
{\ts {3\over 2}} \, \pr^\mu \sigma \pr_\mu \sigma \big )
\pr^\nu \sigma \pr_\nu \sigma \cr
&  \hskip 0.8cm  {}+ 2\, \pr_{[I} W_{J]}(g_\sigma)
\, {\bar \pr}^\mu g_\sigma {\!\!}^I \beta^J(g_\sigma) \,  \pr_\mu \sigma \,
\pr^\nu \sigma \pr_\nu \sigma 
 \cr \noalign{\vskip -3pt} & \hskip 0.8cm {} 
+ \L_\beta W_I(g_\sigma) \big ( 2 \, 
{\bar \pr}^\mu g_\sigma {\!\!}^I  +
\beta^I(g_\sigma) \pr^\mu \sigma \big ) \pr_\mu \sigma 
\, \pr^\nu \sigma \pr_\nu \sigma \Big ) \cr
& \!\! {}- 4 \, \pr_\mu  \delta\sigma \, \pr_{[I} W_{J]}(g_\sigma) 
\big ( {\bar \pr}^\mu g_\sigma {\!\!}^I {\bar \pr}^\nu g_\sigma {\!\!}^J
+ \beta^I(g_\sigma) \pr^\mu  \sigma {\bar \pr}^\nu g_\sigma {\!\!}^J
+ \half {\bar \pr}^\mu g_\sigma {\!\!}^I \beta^J(g_\sigma) \pr^\nu \sigma
\big ) \pr_\nu \sigma . & \Weq
}
$$
In a similar fashion the corresponding contributions containing 
$G_{IJ}$, including contributions from $F_{IJ}$ in \EFHR\ and from \AA\
with \raG\
may be written, noting that $\pr_{(I}\beta^K G_{J)K} + \Gamma_{(IJ)K} \beta^K
 = \half \L_\beta G_{IJ}$, as
\eqnn\Geq$$\eqalignno{
\A_G = {}& - \delta_\sigma \Big (
G_{IJ}(g_\sigma)\, \big ( 
{\bar \pr}^\mu g_\sigma {\!\!}^I {\bar \pr}^\nu g_\sigma {\!\!}^J \,
\pr_\mu \sigma \pr_\nu \sigma 
- \half \, {\bar \pr}^\mu g_\sigma {\!\!}^I {\bar \pr}_\mu g_\sigma {\!\!}^J
\pr^\nu \sigma \pr_\nu \sigma  \big ) \, \cr
\noalign{\vskip -6pt}
& \hskip 2.6  cm {} -
\quar \, G_{IJ}(g_\sigma)\, \beta^I(g_\sigma) \beta^J(g_\sigma)
\pr^\mu \sigma \pr_\mu \sigma \, \pr^\nu \sigma \pr_\nu \sigma
\Big ) \cr
&{}\! \! - \delta \sigma \, \Big ( G_{IJ}(g_\sigma)\big  ( 
\Delta g_\sigma {\!\!}^J + \beta^J (g_\sigma ) (\nabla^2 \sigma 
+ \pr^\mu \sigma \pr_\mu \sigma )\big )  + \Gamma^{(G)}{\!}_{IJK}(g_\sigma) 
{\bar \pr}^\mu g_\sigma {\!\!}^J {\bar \pr}_\mu g_\sigma {\!\!}^K \Big ) 
\cr 
\noalign{\vskip -2pt} 
&\hskip 5cm {} \times \big ( 2\, {\bar \pr}^\nu g_\sigma {\!\!}^I \pr_\nu \sigma 
+ \beta^I(g_\sigma) \, \pr^\nu \sigma \pr_\nu \sigma \big ) \cr
&{}-   \delta\sigma \, \half \big ( G_{IJ}(g_\sigma)  + 
 \pr_J \beta^K(g_\sigma) G _{IK}(g_\sigma) 
 + \Gamma^{(G)}{\!}_{IJK}(g_\sigma)\beta^J (g_\sigma ) \big ) \cr
\noalign{\vskip - 1pt}
&\hskip 1 cm {}   
\times \big (   2\, {\bar \pr}^\mu g_\sigma {\!\!}^I \pr_\mu \sigma  
+ \beta^I(g_\sigma)\,  \pr^\mu \sigma \pr_\mu \sigma \big ) 
\big (   2\, {\bar \pr}^\nu g_\sigma {\!\!}^J \pr_\nu \sigma  
+ \beta^J(g_\sigma)\,  \pr^\nu \sigma \pr_\nu \sigma \big ) 
 \, .&  \Geq
}
$$
For the corresponding result containing $S_{IJ},T_{IJK}$ we include also  the 
terms arising from from $H_I$ in \EFHR\ and from $\pr_{[I} W_{J]}$ in \Weq.
Using \WS\ we obtain, with ${\tilde S}_{IJ}$ given by \Stil,
\eqnn\Seq$$\eqalignno{
\A_S = {}& - \delta_\sigma \Big ( \big (
S_{IJ}(g_\sigma)\, \Delta  g_\sigma {\!\!}^J + T_{IJK}(g_\sigma)\,
{\bar \pr}^\mu g_\sigma {\!\!}^J {\bar \pr}_\mu  g_\sigma {\!\!}^K \big )  
\big ( 2 {\bar \pr}^\nu g_\sigma {\!\!}^I    
\pr_\nu \sigma  +  \beta^I(g_\sigma) \, \pr^\nu \sigma \pr_\nu \sigma \big )\cr
\noalign{\vskip -2pt}
&\hskip 0.7cm {}+ \half \, {\tilde S}_{IJ} (g_\sigma) \,
 \big (   2\, {\bar \pr}^\mu g_\sigma {\!\!}^I \pr_\mu \sigma
+ \beta^I(g_\sigma)\,  \pr^\mu \sigma \pr_\mu \sigma \big )
\big (   2\, {\bar \pr}^\nu g_\sigma {\!\!}^J \pr_\nu \sigma
+ \beta^J(g_\sigma)\,  \pr^\nu \sigma \pr_\nu \sigma \big )\Big ) \cr
&{}+ \delta \sigma \, \Big(  \L_\beta S_{IJ}(g_\sigma) 
 \big ( \Delta g_\sigma {\!\!}^J + \beta^J (g_\sigma ) (\nabla^2 \sigma 
+ \pr^\mu \sigma \pr_\mu \sigma )\big  ) +\L'{\!}_\beta  T_{IJK}(g_\sigma) 
{\bar \pr}^\mu g_\sigma {\!\!}^J {\bar \pr}_\mu g_\sigma {\!\!}^K \Big )
\cr
\noalign{\vskip -2pt} 
&\hskip 5cm {} \times 
\big (   2\, {\bar \pr}^\mu g_\sigma {\!\!}^I \pr_\mu \sigma  
+ \beta^I(g_\sigma)\,  \pr^\mu \sigma \pr_\mu \sigma \big ) 
\cr
&{}+  \delta\sigma \, \half  \L_\beta {\tilde S}_{IJ}(g_\sigma) \, 
\big (   2\, {\bar \pr}^\mu g_\sigma {\!\!}^I \pr_\mu \sigma  
+ \beta^I(g_\sigma)\,  \pr^\mu \sigma \pr_\mu \sigma \big ) 
\big (   2\, {\bar \pr}^\nu g_\sigma {\!\!}^J \pr_\nu \sigma  
+ \beta^J(g_\sigma)\,  \pr^\nu \sigma \pr_\nu \sigma \big ) 
\, . \cr
& &\Seq
}
$$
The expressions \Geq\ and \Seq\ combine so that we may use
\GArel\ and  \relGBC\ and also \ABtil\ so that
the remaining terms, with the results in \EFHR\ applying \BEFH, become
\eqn\ABCeq{\eqalign{
\A_{ABC} = \delta \sigma \Big ( & 
\half \, A_{IJ} (g_\sigma )\, \Delta  g_\sigma {\!\!}^I \Delta   
g_\sigma {\!\!}^J +
B_{IJK}(g_\sigma)\, \Delta  g_\sigma {\!\!}^I \, {\bar \pr}^\mu 
g_\sigma {\!\!}^J{\bar  \pr} _\mu g_\sigma {\!\!}^K \cr
\noalign{\vskip -4pt}
&{} +  \half \, C_{IJKL}(g_\sigma )\,  {\bar \pr}^\mu 
g_\sigma {\!\!}^I {\bar \pr}_\mu 
g_\sigma {\!\!}^J\,  {\bar \pr}^\nu   g_\sigma {\!\!}^K
{\bar \pr}_\nu  g_\sigma {\!\!}^L \ \Big ) \, .
}
}

With these results it is then possible to extend \solW\ to 
four dimensions in the form
\eqn\solWf{
 16\pi^2 \big ( 
W [ e^{2\sigma}\gamma_{\mu\nu} , g_\sigma{\!\!}^I ] -
W [ \gamma_{\mu\nu} , g^I ] \big ) =  \int \! \d^4x \sqrt{-\gamma} \; 
{\cal W} \, , 
}
with ${\cal W}$ a local function expressible as sum of contributions $
{\cal W}_1, {\cal W}_2,{\cal W}_3$. The curvature dependent terms are
contained in 
\eqn\Wres{\eqalign{
{\cal W}_{1} = {}& {\breve C}(\sigma) \, F
-  {\ts{1\over 4}} {\breve A}(\sigma) \, G \cr
\noalign{\vskip 2pt}
&{}+  {\tilde  A}(g_\sigma)  
\big ( E^{\mu\nu} \pr_\mu \sigma  \pr_\nu \sigma  
+ \nabla^2\sigma\,  \pr^\mu \sigma \pr_\mu \sigma + 
\half \,  \pr^\mu \sigma \pr_\mu \sigma\,  \pr^\nu \sigma
\pr_\nu \sigma \big ) \cr
\noalign{\vskip 2pt}
&{} 
- {\breve G}_{IJ}(\sigma)\,  ( E^{\mu\nu} 
+ \gamma^{\mu\nu}  {\ts{1\over 6}} R ) \, 
{\bar \pr}_\mu g_\sigma {\!\!}^I {\bar \pr}_\nu
g_\sigma {\!\!}^J   - 2 \, W_I(g_\sigma) \,
E^{\mu\nu} {\pr}_\mu g_\sigma {\!\!}^I \pr_\nu \sigma  \, , 
}
}
where $ {\breve C}(\sigma) , {\breve A}(\sigma) $ are defined analogously
to \defCc\ and $ {\breve G}_{IJ}(\sigma)$ is again given by \defGc. \Wres\
is an evident extension of the two dimensional result in \Wsol\ with 
$\gamma^{\mu\nu} \to E^{\mu\nu}$. The additional terms involving $G,W$,
after some simplification, are given by
\eqn\WresG{\eqalign{
{\cal W}_{2} = {}&  
-\quar \,  G_{IJ}(g_\sigma) \big (  2 {\pr}^\mu
g_\sigma {\!\!}^I  \pr_\mu \sigma 
- \,\beta^I(g_\sigma) \pr^\mu \sigma \pr_\mu  \sigma \big )
\big ( 2 \pr^\nu g_\sigma {\!\!}^J \pr_\nu \sigma 
- \beta^J(g_\sigma)  \pr^\nu \sigma \pr_\nu \sigma \big )\cr
&{} +  \half \, G_{IJ}(g_\sigma) 
{\pr}^\mu g_\sigma {\!\!}^I {\pr}_\mu g_\sigma {\!\!}^J 
\, \pr^\nu \sigma \pr_\nu  \sigma  \cr 
&{} - \big ( W_I (g_\sigma) \, \nabla^2 g_\sigma {\!\!}^I 
+  \pr_I W_J (g_\sigma)  \, {\pr}^\mu g_\sigma {\!\!}^I
{\pr}_\mu g_\sigma {\!\!}^J \big ) \, 
\pr^\nu \sigma \pr_\nu \sigma \cr
\noalign{\vskip 2pt}
&{} - 2\,  W_I (g_\sigma)\, {\pr}^\mu g_\sigma {\!\!}^I 
\pr_\mu \sigma \, 
( \nabla^2 \sigma + \pr^\nu \sigma \pr_\nu \sigma ) \, .
}
}
The remaining contributions to ${\cal W}$ 
imposing,  by a choice of $a_{IJ}$ in \Stran,
\eqn\Srel{
{\tilde S}_{(IJ)}( g) = 0 \, ,
}
then reduce to
\eqn\WresABC{\eqalign{
{\cal W}_{3} = &{}  -  \big ( S_{IJ}(g_\sigma) \big ( \nabla^2 g_\sigma {\!\!}^J 
+ 2 \, \pr^\mu g_\sigma {\!\!}^J \pr_\mu  \sigma
+ ( {\ts {1\over 6}}R - \nabla^2 \sigma -\pr^\mu \sigma \pr_\mu \sigma ) \beta^I(g_\sigma)\big )\cr
\noalign{\vskip -2pt}
&\hskip 2.8cm {} + T_{IJK}(g_\sigma)\, {\pr}^\mu g_\sigma {\!\!}^J {\pr}_\mu  g_\sigma {\!\!}^K \big )
\big (2\, { \pr}^\nu g_\sigma {\!\!}^I \pr_\nu \sigma -  \beta^I(g_\sigma)
 \, \pr^\nu \sigma \pr_\nu \sigma \big ) \cr
&{}+ \half \,{\breve A}_{IJ} (\sigma )\, 
\hDel  g_\sigma {\!\!}^I  \hDel  g_\sigma {\!\!}^J +
{\breve B}_{IJK}(\sigma)\, \hDel  g_\sigma {\!\!}^I \, {\bar \pr}^\mu 
g_\sigma {\!\!}^J{\bar  \pr} _\mu g_\sigma {\!\!}^K 
\cr \noalign{\vskip 2pt} &{} 
+  \half \, {\breve C}_{IJKL}(\sigma )\,  {\bar \pr}^\mu 
g_\sigma {\!\!}^I {\bar \pr}_\mu 
g_\sigma {\!\!}^J\,  {\bar \pr}^\nu   g_\sigma {\!\!}^K
{\bar \pr}_\nu  g_\sigma {\!\!}^L \, .  
}
}
In \WresABC\ $R$ dependent terms have been absorbed in a
redefinition of $\Delta  g_\sigma$,
\eqn\hD{
\hDel g_\sigma {\!\!}^I = \Delta  g_\sigma {\!\!}^I 
+  \beta^I(g_\sigma)  \,  {\ts {1\over 6}} R \, ,
}
which satisfies the corresponding equation to \sigD.

In  \WresABC\ ${\breve A}_{IJ}(\sigma)$ is defined similarly to
${\breve G}_{IJ}(\sigma)$ in  \defGc\ while
${\breve B}_{IJK}(\sigma)$ is determined by
\eqn\defBc{\eqalign{
{\d \over \d \sigma}&   {\breve B}_{IJK}(\sigma) 
+ \pr_I \beta^L(g_\sigma)  {\breve B}_{LJK}(\sigma) 
+ \pr_J \beta^L(g_\sigma)  {\breve B}_{ILK}(\sigma) \cr
&{} + \pr_K \beta^L(g_\sigma)  {\breve B}_{IJL}(\sigma) + \pr_J \pr_K
\beta^L(g_\sigma)  {\breve A}_{IL}(\sigma) 
= B_{IJK}(g_\sigma) \, ,  \quad {\breve B}_{IJK}(0) = 0 \, ,
}
}
with a corresponding equation for ${\breve C}_{IJKL}(\sigma )$. Just as in 
\power\ there is a formal solution
\eqn\powerB{\eqalign{
{\breve B}_{IJK}(\sigma) = {}& 
 \big ( \exp( \sigma \L_\beta ) - 1 \big ) \L_\beta {\!}^{-1}
\big ( B_{IJK}(g) - \pr_J \pr_K
\beta^L(g)  \, \L_\beta{\!}^{-1} A_{IL}(g) \big ) \cr
\noalign{\vskip  1 pt}
&{} +  \big ( \exp( \sigma \L_\beta ) - 1 \big ) \L_\beta {\!}^{-1}
\big (\pr_J \pr_K \beta^L(g)\big ) \,  \L_\beta{\!}^{-1} A_{IL}(g)  \, .
}
}
By obtaining analogous equations to \recCG\
the relations \GArel\ and \relGBC\ imply
\eqn\GSrel{\eqalign{
{\breve G}_{IJ}(\sigma) -  S_{IJ}(g_\sigma) +{\breve S}_{IJ}(\sigma) =  {}& 
\Psi_I{}^K (g_\sigma) {\breve A}_{KJ}(\sigma)  + 
{\breve B}_{JIK}(\sigma) \hbet^K(g_\sigma)  \, ,  \cr
{\breve \Gamma}_{IJK}(\sigma) -  T_{IJK} (g_\sigma) +{\breve T}_{IJK} (\sigma) = {}&
\Psi_I{}^L (g_\sigma) 
{\breve B}_{LJK}(\sigma)+ {\breve C}_{ILJK}(\sigma)\beta^L(g_\sigma) 
\, ,
}
}
with ${\breve S}_{IJ}, {\breve T}_{IJK}$ defined similarly to ${\breve W}_I$ in \solM\
and for  $\Psi_I{\!}^J(g) = \delta_I{\!}^J + \pr_I \beta^J(g)$.
${\breve \Gamma}_{IJK} $ satisfies \defBc\ with ${\breve B}_{IJK} \to
{\breve \Gamma}_{IJK} $, $B_{IJK} \to \Gamma^{(G)}{\!}_{IJK} $ and 
${\breve A}_{IL} \to {\breve G}_{IL}$ and as a consequence
\eqn\GGam{
{\breve \Gamma}_{IJK}  = \Gamma^{({\breve G})}{\!}_{IJK} \, ,
}
with $\Gamma^{({\breve G})}$ defined in terms of ${\breve G}_{IJ}$ as in \Chr.
As a consequence of \Srel\ we have from \GSrel
\eqn\GGABC{\eqalign{
{\breve G}_{IJ}(\sigma) + \half \L_\beta {\breve G}_{IJ}(\sigma) = {}& 
\Psi_I{\!}^K(g_\sigma) \Psi_J{\!}^L(g_\sigma) {\breve A}_{KL}(\sigma) \cr
&{} + \Psi_J{\!}^L(g_\sigma){\breve B}_{LIK}(\sigma) \beta^K(g_\sigma)
+  \Psi_I{\!}^L(g_\sigma){\breve B}_{LJK}(\sigma) \beta^K(g_\sigma)\cr
&{} +{\breve C}_{ILJK}(\sigma )  \beta^K(g_\sigma) \beta^L(g_\sigma) \, .
}
}

Applying  \GSrel\ in \WresG\ we may use
\eqnn\STin$$\eqalignno{
 \big ( {\breve S}_{IJ}(\sigma) \, \Delta g_\sigma {\!\!}^J 
+ {\breve T}_{IJK}(\sigma)\,
{\bar \pr}^\mu g_\sigma {\!\!}^J {\bar \pr}_\mu  g_\sigma {\!\!}^K \big )
{\bar \pr}_\nu  g_\sigma {\!\!}^I
= {}& \big ( {S}_{IJ}(g) \, \nabla^2 g^J + {T}_{IJK}(g)\,
{\pr}^\mu g^J {\pr}_\mu  g^K \big ) {\pr}_\nu  g^I \, , \cr
{\breve S}_{IJ}(\sigma) \, {\bar \pr}_\nu  g_\sigma {\!\!}^I \beta^J(g_\sigma) = {}& 
S_{IJ}(g) \, \pr_\nu g^I \beta^J(g) \, , & \STin
}
$$
and similarly for ${\bar \pr}_\nu  g_\sigma {\!\!}^I \to \beta^I(g_\sigma)$.
By applying ${\pr \over \pr \sigma}$ to \defGc\ so that
it becomes a homogeneous equation,  we may obtain
\eqn\LieG{
\L_\beta {\breve G}_{IJ}(\sigma)\, {\bar \pr}^\mu g_\sigma{\!\!}^I {\bar \pr}_\mu g_\sigma{\!\!}^J
 = G_{IJ} (g_\sigma) \,
{\bar \pr}^\mu g_\sigma{\!\!}^I {\bar \pr}_\mu g_\sigma{\!\!}^J
- G_{IJ}(g) \, \pr^\mu g^I \pr_\mu g^J \, ,
}
and also, as in \Cint,
\eqn\AArel{
 {\tilde A}(g_\sigma) -   {\tilde A}(g)= {\breve G}_{IJ}(\sigma) \,
 \beta^I(g_\sigma)  \beta^J(g_\sigma) \, .
 }

Starting from \solWf, with \Wres, \WresG, \WresABC, and letting 
$\gamma_{\mu\nu} \to e^{-2\sigma} \gamma_{\mu\nu}$
then, similarly to \solWQ, 
\eqn\solWQf{
16 \pi^2  \big ( W [e^{-2\sigma} \gamma_{\mu\nu} , g^I ] -  
W [\gamma_{\mu\nu} , g_\sigma{\!\!}^I ] \big ) 
=  \int \! \d^4x \sqrt{-\gamma} \;  {\cal W}' \, , 
}
For an IR fixed point so that  $g_\sigma{\!\!}^I \to g_*{\!\!}^I$ as 
$\sigma \to - \infty$ then, assuming $W [\gamma_{\mu\nu} , g_\sigma{\!\!}^I ] 
\to W [\gamma_{\mu\nu} , g_*{\!\!}^I ]$ smoothly, ${\cal W}'$ determines the dependence
on $\sigma$ in the neighbourhood of the fixed point.

To determine ${\cal W}'$ we use \varFGER\ for $\sigma \to -\sigma$
and the corresponding equation to \CWG\ and discard total derivatives
as appropriate. Writing ${\cal W}' = {\cal W}'{\!}_1 + {\cal W}'{\!}_2 + {\cal W}'{\!}_3$
the result, using \GSrel, \GGABC,  \STin, \LieG, \AArel, is
\eqn\Wresf{\eqalign{
{\cal W}'{\!}_1 ={}&   -{ \breve C}(\sigma) \, F +
\quar {\breve A}(\sigma) \, G \cr
&{}+ {\tilde A}(g) \, \big ( E^{\mu\nu} \pr_\mu \sigma \pr_\nu \sigma
- \nabla^2\sigma\,  \pr^\mu \sigma \pr_\mu \sigma +
\half \,  \pr^\mu \sigma \pr_\mu \sigma\,  \pr^\nu \sigma
\pr_\nu \sigma \big )  \cr
&{} +  {\breve G}_{IJ}(\sigma) \,
 (E^{\mu\nu}+ \gamma^{\mu\nu} {\ts {1\over 6}}R) \,  
{ \pr}_\mu  g_\sigma  {\!\!}^I { \pr}_\nu g_\sigma {\!\!}^J 
+ 2\, W_I(g) \, E^{\mu\nu} \pr_\mu g^I \pr_\nu \sigma \, ,
}
}
and
\eqn\WresGf{\eqalign{
{\cal W}'{\!}_2 =
&{}  - \quar \, G_{IJ}(g) \big ( 2  {\pr}^\mu g^I \pr_\mu \sigma 
+  \beta^I(g) \pr^\mu \sigma \pr_\mu  \sigma\big )
\big ( 2{ \pr}^\nu g^J  \pr_\nu \sigma 
+ \beta^J(g) \pr^\nu \sigma \pr_\nu  \sigma \big ) \cr
&{} +  \half \, G_{IJ}(g) {\pr}^\mu g^I {\pr}_\mu g^J \, \pr^\nu \sigma \pr_\nu  \sigma  \cr
&{} - \big ( W_I (g) \, \nabla^2 g^I 
+  \pr_I W_J (g)  \, {\pr}^\mu g^I
{\pr}_\mu g^J \big ) \, 
\pr^\nu \sigma \pr_\nu \sigma \cr
&{}- 2 \, W_I(g) \, \pr^\mu g^I \pr_\mu \sigma \, 
(\nabla^2 \sigma - \pr^\nu \sigma  \pr_\nu \sigma) \, , 
 } 
 }
and
\eqn\WresABCf{\eqalign{
{\cal W}'{\!} _{3} = {}&  \big ( S_{IJ}(g)  \big ( \nabla^2 g^J
- 2\, \pr^\mu g^J \pr_\mu \sigma +
({\ts {1\over 6}}  R + \nabla^2 \sigma - \pr^\mu \sigma \pr_\mu \sigma) \beta^J(g)\big ) \cr
\noalign{\vskip -2pt}
& \hskip 3.5cm {} + T_{IJK}(g) \pr^\mu g^J \pr_\mu g^K \big ) 
 \big ( 2 \pr^\nu g^I \pr_\nu \sigma + \beta^I(g) \pr^\nu \sigma
\pr_\nu \sigma \big ) \cr
& {} -\half \,  {\breve A}_{IJ}(\sigma )\, 
\big (\nabla^2  g_\sigma {\!\!}^I + {\ts {1\over 6}}R \,\beta^I(g_\sigma) \big )
\big (\nabla^2 g_\sigma {\!\!}^J + {\ts {1\over 6}}R \,\beta^J(g_\sigma) \big )
\cr \noalign{\vskip 2pt}  &{} 
- {\breve B}_{IJK}(\sigma)\; \big (\nabla^2 g_\sigma {\!\!}^I 
+ {\ts {1\over 6}}R \, \beta^I(g_\sigma) \big ) \, 
{\pr}^\mu g_\sigma {\!\!}^J{  \pr} _\mu g_\sigma {\!\!}^K 
\cr \noalign{\vskip 2pt} &{} 
- \half \, {\breve C}_{IJKL}(\sigma )\,  { \pr}^\mu 
g_\sigma {\!\!}^I { \pr}_\mu 
g_\sigma {\!\!}^J\,  { \pr}^\nu   g_\sigma {\!\!}^K
{\pr}_\nu  g_\sigma {\!\!}^L \, .
}
}
$ {\cal W}'{\!} _{1}, {\cal W}'{\!} _{2} ,{\cal W}'{\!} _{3}$ may also be obtained
from ${\cal W}_1, {\cal W}_2,{\cal W}_3$ by letting $g_\sigma \to g$ and then
$\sigma\to-\sigma$. The contributions involving ${\breve G}_{IJ}$, as well as
${\breve A}_{IJ}, {\breve B}_{IJK},  {\breve C}_{IJKL}$ depend on the RG
trajectory linking $g$ and $g_\sigma$, for variations arising from \equivW, \equiva\
the associated freedom becomes a difference of contributions from the end
points of the RG flow.

These expressions simplify if we assume that the $x$-dependence in 
$g_\sigma$ arises only from $\sigma$, so that in solving \rung\
$g^I$ is a constant. In this case we may take $\pr_\mu  g_\sigma {\!\!}^I =
 \beta^I(g_\sigma)\pr_\mu \sigma$, $\nabla^2  g_\sigma {\!\!}^I 
 + {\ts {1\over 6}}R \,\beta^I(g_\sigma) 
=  \beta^I(g_\sigma) {\ts {1\over 6}}{\breve R}  + \Psi_J{}^I (g_\sigma)\beta^J(g_\sigma)
\pr^\mu \sigma \pr_\mu \sigma$ for ${\ts {1\over 6}} {\breve R} = {\ts {1\over 6}} R + \nabla^2 \sigma
- \pr^\mu \sigma \pr_\mu \sigma$ and then
\eqn\Wfull{\eqalign{
{\cal W}' ={}&   -{ \breve C}(\sigma) \, F +
\quar {\breve A}(\sigma) \, G \cr
&{} +  {\tilde A}(g_\sigma)  \, ( E^{\mu\nu} \pr_\mu \sigma \pr_\nu \sigma 
-\nabla^2 \sigma \, \pr^\mu \sigma \pr_\mu \sigma 
+  \half  \,  \pr^\mu \sigma \pr_\mu \sigma\,  \pr^\nu \sigma
\pr_\nu \sigma )  \cr
&{}  -  \quar \, G_{IJ} (g_\sigma)\beta^I(g_\sigma)  \beta^J(g_\sigma) \, 
  \pr^\mu \sigma \pr_\mu \sigma\,  \pr^\nu \sigma\pr_\nu \sigma \cr
&{} + {\ts {1\over 6}} {\breve R} \, S_{IJ}(g_\sigma) 
\beta^I(g_\sigma)  \beta^J(g_\sigma) \, \pr^\mu \sigma \pr_\mu \sigma
- \half ( {\ts {1\over 6}} {\breve R} )^2 \, {\breve A}_{IJ}(\sigma) 
\beta^I(g_\sigma)  \beta^J(g_\sigma) \, .
}
}
Of course at a fixed point with a vanishing beta function this coincides with 
the result used in \Luty\ for $\sigma \to \tau$ and a similar expression was
obtained in \Other.

Although lengthy, and tedious to obtain, the extended result \solWQf,
with \Wresf, \WresGf, \WresABCf, is still relatively simple and
potentially allows for the analysis of dilaton couplings away
from conformal fixed  points.\foot{If all $\beta$ terms are set to zero
in \DelW\ and the various conditions for integrability are implemented
along with \Szero\ then \DelW\ becomes
$$\eqalign{
\noalign{\vskip -4pt}
\Delta_\sigma \, 16\pi^2 W = {}& \int d^4x \sqrt{-\gamma} \; \sigma \,\Big (
C\, F - \quar \, A \, G + \half \, G_{IJ} ( \D^2 g^I \D^2 g^J - 2\,
(E^{\mu\nu} + {\ts {1\over 6}}R \, \gamma^{\mu\nu}) \, \pr_\mu g^I \pr_\nu g^J 
) \cr
\noalign{\vskip -6pt}
&\hskip 2.2cm {}+  \half \, {\hat  C}_{IJKL} \,
 \pr^\mu g^I \pr_\mu g^J \, \pr^\nu g^K\! \pr_\nu g^L \Big ) \cr
\noalign{\vskip -4pt}
& {}- 2 \int d^4x \sqrt{-\gamma} \; \pr_\mu \sigma \, \Big (
E^{\mu\nu} W_I \pr_\nu g^I - \pr_{[I} W_{J]} \,
\pr^\mu g^I \nabla^2 g^J \Big ) \,,
}
$$
where $ \D^2 g^I$ is defined as in \defDg\ with $A_{IJ} \to G_{IJ}, \,
B_{IJK} \to \Gamma^{(G)}{\!}_{IJK}$ and we must also impose $\pr_I A = 0$.
This may be integrated straightforwardly to give
$ 16 \pi^2 ( W [e^{2\sigma} \gamma_{\mu\nu} ] - W [\gamma_{\mu\nu} ] )
=  \int \! \d^4x \sqrt{-\gamma} \;  {\cal W}_{\rm FP}$ where
$$\eqalign{
\noalign{\vskip -6pt}
{\cal W}_{\rm FP} = {}& \sigma \Big ( C \, F - \quar \, A \, G 
+ \half \, G_{IJ} ( \D^2 g^I \D^2 g^J - 2\,
(E^{\mu\nu} + {\ts {1\over 6}}R \, \gamma^{\mu\nu}) \, 
\pr_\mu g^I \pr_\nu g^J ) 
\cr \noalign{\vskip -6pt} &\hskip 2.2cm {}
+  \half \, {\hat  C}_{IJKL} \,
 \pr^\mu g^I \pr_\mu g^J \, \pr^\nu g^K\! \pr_\nu g^L \Big ) \cr
\noalign{\vskip - 2 pt} &
{}+ A \, ( E^{\mu\nu} \pr_\mu \sigma \pr_\nu \sigma + \nabla^2 \sigma\, 
\pr^\mu \sigma \pr_\mu \sigma + \half \, \pr^\mu \sigma \pr_\mu \sigma \,
\pr^\nu \sigma \pr_\nu \sigma ) \cr
\noalign{\vskip - 2 pt} &
{}- G_{IJ} ( \pr^\mu g^I \pr^\nu g^J \, \pr_\mu \sigma \pr_\nu \sigma 
- \half \, \pr^\mu g^I \pr_\mu g^J \, \pr^\nu \sigma \pr_\nu \sigma )  \cr
\noalign{\vskip - 2 pt} & - 2 \, W_I \, E^{\mu\nu} \pr_\mu g^I \pr_\nu \sigma
+ 2 \,  \pr_{[I} W_{J]} \, \pr^\mu g^I \nabla^2 g^J \, \pr_\mu \sigma  \cr
\noalign{\vskip - 2 pt} & 
{}- 2\, W_I \,\pr^\mu g^I \pr_\mu \sigma \, (\nabla^2 \sigma + 
\pr^\nu \sigma \pr_\nu \sigma ) - ( W_I \nabla^2 g^I +
\pr_I W_J \pr^\mu g^I \pr_\mu g^J ) \, \pr^\nu \sigma \pr_\nu \sigma 
\, .
}
$$
This result is relevant at a fixed point when $\{g^I\}$ are the couplings
for exactly marginal operators and so parameterise the moduli space. 
The terms proportional to $G_{IJ}$ can be expressed in terms of the 
Riegert operator, a conformally covariant 4th order differential operator 
acting on dimensionless scalars.
On the moduli space $A$ is constant, whereas $C$ may vary, and we 
expect, since $(\omega g)^I W_I=0$,  $W_I=\pr_I f$ for some scalar $f$, 
and so by virtue of the freedom in \equivW\ we may then set $W_I=0$.} 
Setting the curvature terms to zero
and $\sigma \to \tau$ \Wresf\ becomes part of the Lagrangian
determining couplings of scalar fields $\O_I$ to the dilaton $\tau$
in the dilaton effective action. The results used in \KS\ and \Luty\
depend also on imposing additional boundary conditions whose generalisation
is less apparent.

\newsec{Broken Conformal Symmetry}

The results obtained in section 2 depend on extending the quantum
field theory to a curved space background. In this section we show
how a subset of the consistency relation equations can be defined
by restricting to flat space and considering broken conformal symmetry.
These are derived by considering diffeomorphisms, as well as
Weyl rescalings. Their intersection defines the conformal group

In general quantum field theories on curved space,
within appropriate regularisation schemes, are invariant under 
diffeomorphisms. This may be expressed, for arbitrary smooth 
$v^\mu(x)$, as
\eqn\diffW{
\int \d^4 x \, \bigg ( \L_v \gamma_{\mu \nu}
{\delta \over \delta \gamma_{\mu \nu}} + v^\mu \pr_\mu g^I
{\delta \over \delta g^I} \bigg ) W = 0 \, ,
}
where
\eqn\Lv{
\L_v \gamma_{\mu\nu} = \nabla_\mu v_\nu + \nabla_\nu v_\mu\, .
}
Conformal Killing vectors  satisfy
\eqn\Killv{
\nabla_\mu v_\nu + \nabla_\nu v_\mu = 2 \, \sigma_v \, \gamma_{\mu\nu} \, ,
}
and for any such conformal Killing vector   acting on $W$ we may take from 
\diffW\ and \defD
\eqn\DDV{
\Delta_{\sigma_v} \to \Delta_v
= \int \d^4 x \, \big ( -  v^\mu \pr_\mu g^I +\sigma_v \,
\hbet^I \big ) {\delta \over \delta g^I} \, .
}
Defining the commutator of two diffeomorphisms by
\eqn\comv{
 [ v , v' ] ^\mu = v^\nu \pr_\nu \, v'^\mu - v'^\nu \pr_\nu \, v^\mu \, ,
 }
\Killv\ implies
\eqn\dsig{
v^\mu \pr_\mu \, \sigma_{v'} - v'^\mu \pr_\mu \, \sigma_{v} 
= \sigma_{[v,v']} \, .
}
It is then easy to verify that, from the definition \DDV,
\eqn\comvD{
\big [ \Delta_v , \Delta_{v'} \big ] = \Delta_{[v,v']} \, .
}

On flat space the solutions of \Killv,  for $\nabla_\mu\to \pr_\mu,
\, \gamma_{\mu\nu} \to \eta_{\mu\nu}$, are of course the usual
conformal Killing vectors
\eqn\solKv{
v^\mu(x) = a^\mu + \omega^\mu{}_\nu x^\nu +  \lambda x^\mu 
+ b^\mu x^2 - 2 \, x^\nu b_\nu \, x^\mu \, , \qquad
\sigma_v(x) = \lambda - 2 \, x^\mu b_\mu \, ,
}
for $\omega_{\mu\nu}=- \omega_{\nu\mu}$. Combining \diffW\ with 
\DelW\ gives a condition on the flat space vacuum energy
functional $W[g^I]$ which reduces, since $\pr^2 \sigma_v=0$, to
\eqn\ConW{
\Delta_v \, 16\pi^2 W =
\int \d^4 x \; \big ( \sigma_v \, X - 2 \, \pr _\mu \sigma_v \,
Y^\mu \big ) \, ,
}
for $X(g),Y^\mu(g)$ given by \XYZ, albeit $\nabla^2 \to \pr^2$.
In $Y^\mu$, since $S_{(IJ)}\, \pr^\mu g^I \pr^2 g^J = 
S_{(IJ)} \big ( \pr^\nu ( \pr^\mu g^I \pr_\nu g^J) - \half \, \pr^\mu
( \pr^\nu g^I \pr_\nu g^J) \big )$ and $\pr_\nu \pr_\mu \sigma_v = 0$,
the symmetric part of $S_{IJ}$ may be dropped.        
\ConW\ expresses broken conformal symmetry\foot{Broken conformal
Ward identities were first discussed at the same time as the usual
RG equations \conform\ but in \Cgross\ `appear to be useless'.
For other approaches see \BraunRP.}, valid so long
as the couplings are local functions of $x$. 

Linear conditions
on correlation functions for the operators $\O_I$, which reduce
to standard RG equations for $v^\mu(x) = \lambda x^\mu$ and
$g^I$ constant,  can be obtained
from
\eqn\commDg{
\bigg [ \Delta_v , {\delta \over \delta g^I(x)}  \bigg ] =
- v^\mu(x) \, \pr_\mu {\delta \over \delta g^I(x)}  - \sigma_v(x) 
\Big ( d \, \delta_I{\!}^J + \pr_I \hbet^J\big (g(x)\big ) \Big )
{\delta \over \delta g^J(x)}   \, ,
}
with $d=4$ here again. With the definition \defTO\ then
\eqn\WardO{
16\pi^2 \big ( \Delta_v  \langle \O_I \rangle + 4 \, \sigma_v 
\langle \O_I \rangle + \sigma_v \, \pr_I \beta^J \langle \O_J \rangle
+ v^\mu \pr_\mu \langle \O_I \rangle \big ) = \A_I \, ,
}
for 
\eqn\defAO{
\A_I = - {\delta \over \delta g^I} \int \del^4 x \; \big ( \sigma_v \, X
- 2 \, \pr_\mu \sigma_v \, Y^\mu \big ) \, .
} 

To impose \comvD\ making use of \dsig\ we note that
\eqnn\comZY$$\eqalignno{
\Delta_v & \int \d^4 x \; 
\big ( \sigma_{v'} \, X - 2 \, \pr _\mu \sigma_{v'} \,
Y^\mu  \big ) - \Delta_{v'} \int \d^4 x \; 
\big (\sigma_v \, X  - 2 \, \pr _\mu \sigma_v \, Y^\mu \big )\cr 
= {}& \int \d^4 x \; \big ( \sigma_{[v,v']} \, X - 2 \, 
\pr _\mu \sigma_{[v,v']}  \, Y^\mu  \big )  
+  \int \d^4 x \; \big ( 2\,  k_\mu \,  K^\mu
 + 4  \;  l_{\mu\nu}\, L^{\mu\nu} \big ) \, , \cr
& k_\mu = \sigma_{v'}\, \pr_\mu \sigma_v 
- \sigma_{v}\, \pr_\mu \sigma_{v'} \,   , \qquad 
l_{\mu\nu} =  \pr_\mu \sigma_{v'}\, \pr_\nu \sigma_v
- \pr_\mu \sigma_{v}\, \pr_\nu \sigma_{v'} = 8\, b'{\!}_{[\mu}\, b_{\nu]}
\, , & \comZY
}
$$
for
\eqnn\resKL$$\eqalignno{
K^\mu = {} & \big ( 
A_{IJ} + \pr_I \hbet^K A_{JK} + B_{JIK} \hbet^K + \L_\hbet S_{IJ} \big )
\pr^\mu g^I \pr^2 g^J \cr
&{}+ \big ( B_{IJK} + \pr_I \hbet^L  B_{LJK}
+ C_{ILJK} \hbet^L  + S_{IL} \pr_J\pr_K \hbet^L + \L_\hbet T_{IJK} \big )
\pr^\mu g^I \pr^\nu g^J \pr_\nu g^K \, , \cr
L^{\mu\nu} = {}& - \big ( S_{[IJ]} - \pr_{[I} \hbet^K S_{J]K} + 
T_{[IJ]K} \hbet^K \big ) \pr^\mu g^I \pr^\nu g^J \, . & \resKL
}
$$
Hence \comvD\ is satisfied, assuming \ConW, if the terms involving $K^\mu$ and
$L^{\mu\nu}$ in \comZY\ vanish. As $\sigma_v$ is just linear in $x$ the 
conditions in this case do not require either $K^\mu$ or $L^{\mu\nu}$ to be 
zero. For the term involving $l_{\mu\nu}$, since this is a constant, it  is 
necessary 
and sufficient only that $L^{\mu\nu}$ is a total derivative so that we require
\eqn\resL{
L^{\mu\nu} = \pr^{[\mu} \big ( W_I \pr^{\nu]} g^I \big ) \, ,
}
for some $W_I$, which is then equivalent to the result \WS\ for 
$\pr_{[I}W_{J]}$ and hence $W_I$ is determined in terms of $S_{IJ},T_{IJK}$ 
up to the freedom $W_I \sim W_I - \pr_I a$.
For the term containing $k_\mu$  in \comZY\ then since
\eqn\derl{
\pr_{(\nu} k_{\mu)} = 0 \, ,
}
it is sufficient to require
\eqn\resK{
K^\mu = \pr_\nu \big ( G_{IJ} \pr^\mu g^I  \pr^\nu g^J \big ) 
- \half \, \pr^\mu \big ( G_{IJ} \pr^\nu g^I  \pr_\nu g^J \big ) \, , \quad 
G_{IJ}=G_{JI} \, ,
}
choosing the relative coefficients to match the form of $K^\mu$ in \resKL. 
Combining \resKL\ and \resK\ is equivalent to \GArel\ and  \relGBC\ 
with the definition \Chr.

Although restricting to broken conformal symmetry on flat space does
not directly determine $A$, which plays the role of a $c$-function, the
relations defining $W_I$ and $G_{IJ}$ are sufficient to reconstruct
the critical result \raG. Using \GArel\ and \relGBC
\eqnn\intGW$$\eqalignno{
\pr_{[I} \big ( G_{J]K} \hbet^K\big )  ={}& \pr_{[I} \hbet^K G_{J]K} -
\Gamma^{(G)}{\!}_{[IJ]K} \hbet^K \cr
={}&  \pr_{[I} \hbet^K A_{J]K} - B_{[IJ]K}\hbet^K \cr
\noalign{\vskip -2pt}
&{} + \pr_{[I} \hbet^K 
\L_\hbet S_{J]K} - S_{[I\,L} \pr_{J]}\pr_K \hbet^L \hbet^K
- \L_\hbet T_{[IJ]K} \hbet^K \cr
= {}& G_{[IJ]}-A_{[IJ]} - 
\L_\hbet \big ( S_{[IJ]} + T_{[IJ]K} \hbet^K \big ) + \pr_{[I} \hbet^K 
\L_\hbet  S_{J]K}  - S_{[I\,L} \pr_{J]}\pr_K \hbet^L \hbet^K \cr
= {}& \L_\hbet \big ( \pr_{[I} W_{J]} \big ) = \pr_{[I} \L_\hbet W_{J]} \, ,
&\intGW
}
$$
using \WS\ and $G_{[IJ]}=A_{[IJ]}=0$. 
\intGW\ is the necessary condition for the integrability of \raG\
so that $A$ may be calculated in terms of the flat space quantities 
$A_{IJ},B_{IJK}, S_{IJ},T_{IJK}$ up
to a $g$-independent constant.

If in \resKL
\eqn\varST{
\Delta S_{IJ} = g_{IJ} \, , \qquad \Delta T_{IJK} = \Gamma^{(g)}{\!}_{IJK} \, ,
}
then
\eqn\varKL{\eqalign{
\Delta K^\mu = {}& \L_\beta g_{IJ} \, \pr^\mu g^I \pr^2 g^J 
+ \big ( \pr_K \L_\beta g_{IJ} - \half \,  \pr_I \L_\beta g_{JK} \big )
\, \pr^\mu g^I \pr^\nu g^J \pr_\nu g^K \, , \cr
\Delta L^{\mu\nu} = {}& \pr_{[I} \big ( g_{J]K} \beta^K \big ) \,
\pr^\mu g^I \pr^\nu g^J\, ,
}
}
and it is easy to see that this implies
\eqn\varGW{
\Delta G_{IJ} =  \L_\beta g_{IJ} \, , \qquad \Delta W_I = g_{IJ} \beta^J \, ,
}
in accord with \equivW\ and \equiva.

At a fixed point, assuming 
\eqn\fixD{
\pr_I \beta^J \big |_{g=g_*} = - (4 -\Delta_I ) \delta_I{\!}^J \, ,
}
then with \defTO\ the identity \ConW\ requires, by considering
$ {\delta \over \delta  g^I(x)} {\delta \over \delta g^J(y)}$ and then 
restricting to constant couplings,
\eqn\twop{\eqalign{
\big ( v^\mu (x) \pr_{\mu x} + \Delta_I \, \sigma_v(x)&   +
v^\mu (y) \pr_{\mu y} + \Delta_J \, \sigma_v(y) \big )
 \big \langle \O_I(x) \, \O_J(y) \big \rangle \cr
= {}& {1\over 16\pi^2} \, 
A_{IJ} \, \pr_x{\!}^2 \pr_y{\!}^2 \big ( \sigma_v(x) \delta^4 (x-y) \big ) \, .
}
}
There is a potential term involving $S_{IJ}$ but this cancels for $S_{IJ}
= - S_{JI}$. The conformal identity \twop\ has a solution only for
$\Delta_I = \Delta_J = \Delta $ when
\eqn\twopsol{\eqalign{
& \big \langle \O_I(x) \, \O_J(y) \big \rangle =  {C_{IJ} \over
( (x-y)^2 )^{\Delta} } - {1\over 16\pi^2} \, 
{A_{IJ} \over 2(4-\Delta )} \, \pr_x{\!}^2 \pr_y{\!}^2  \delta^4 (x-y) \cr
&{} = {1\over \Delta -4} \, (\pr^2)^3 \bigg ( {C_{IJ} \over 64 (\Delta -3)^2
(\Delta - 2)^2 (\Delta - 1)} \, {1  \over ( (x-y)^2 )^{\Delta-3} } 
- {A_{IJ} \over 2 (8 \pi^2)^2}  \, {1\over (x-y)^2} \bigg )\, .
}
}
For this to be well defined for $x\approx y$ we must have
$(2\pi^2)^2 C_{IJ} = 24\, A_{IJ} + {\rm O}(\Delta-4)$.

With the definition \tranom\ and restricting to flat space then
$\langle T^{\mu\nu} \rangle$ satisfies
\eqna\contr $$\eqalignno{
\pr_\mu \langle T^{\mu\nu} \rangle + \pr^\nu g^I \, \langle \O_I \rangle =
{}& 0 \, ,    & \contr{a} \cr
16\pi^2 \big ( \eta_{\mu\nu}  \langle T^{\mu\nu} \rangle - \beta^I
\langle \O_I \rangle \big ) = {}& X + 2 \, \pr_\mu Y^\mu \, , & \contr{b}
}
$$
and also, with $\Delta_v$ as in \DDV, a corresponding broken conformal identity
\eqn\WardT{\eqalign{
& 16\pi^2 \big ( \Delta_v  \langle T^{\mu\nu} \rangle  + 6 \, \sigma_v  \, 
\langle T^{\mu\nu} \rangle + \L_v  \langle T^{\mu\nu} \rangle  \big ) =
\A^{\mu\nu} \, , \cr
& \L_v  \langle T^{\mu\nu} \rangle  = v^\rho \pr_\rho 
\langle T^{\mu\nu} \rangle - \pr_\rho v^\mu \, \langle T^{\rho\nu} \rangle
- \pr_\rho v^\nu \, \langle T^{\mu\rho} \rangle \, ,
}
}
where
\eqn\defAT{\eqalign{
\A^{\mu\nu} = {}& 2\, {\delta \over \delta \gamma_{\mu\nu}} \int \del^4 x \; 
\sqrt{-\gamma}\, 
\big ( \sigma \, X - 2 \, \pr_\alpha \sigma \, Y^\alpha \big )  
\Big |_{\gamma_{\mu\nu} \to \eta_{\mu\nu}, \sigma\to \sigma_v}  \cr
&{}+ {\ts {1\over 3}} (\eta^{\mu\nu} \pr^2 - \pr^\mu \pr^\nu  ) 
\big ( \sigma_v ( E_I \pr^2 g^I + F_{IJ} \pr^\alpha g^I \pr_\alpha g^J ) + 
2\, \pr_\alpha \sigma_v \,H_I \pr^\alpha g^I \big ) \cr
&{}+ \D^{\mu\nu\sigma\rho } \big ( \sigma_v \,  
G_{IJ} \pr_\sigma g^I \pr_\rho  g^J 
+ 2\, \pr_\sigma \sigma_v \, W_I \pr_\rho  g^I \big ) \, ,
}
}
with $ \D^{\mu\nu\sigma\rho } $ defined so that
\eqn\defDabc{
\D^{\mu\nu\sigma\rho } f_{\sigma\rho } = \pr^2 f^{\mu\nu} + \eta^{\mu\nu} \, 
\pr^\sigma\pr^\rho    f_{\sigma\rho } -
2 \, \pr^{(\mu} \pr_\sigma f^{\nu)\sigma} + 
(\pr^\mu \pr^\nu - \eta^{\mu\nu} \pr^2 ) 
\eta^{\sigma\rho }  f_{\sigma\rho } \, ,
}
for any $f_{\sigma\rho }=f_{\rho \sigma}$.
 
The form for $\A^{\mu\nu}$ in \WardT\ is constrained by \contr{a,b} in 
conjunction with \WardO. Using $\pr_\mu (\L_v + 6 \sigma_v ) T^{\mu\nu} = 
(\L_v + 6 \sigma_v ) \pr_\mu T^{\mu\nu} 
+ \pr^\nu \sigma_v \, \eta_{\sigma\rho}T^{\sigma\rho}$ we may obtain
from \contr{a}
\eqn\AAid{
\pr_\mu \A^{\mu\nu} + \pr^\nu g^I \, \A_I = \pr^\nu \sigma_v \, 
( X + 2\, \pr_\mu Y^\mu) \, ,
}
and from $\eta_{\mu\nu}  (\L_v + 6 \sigma_v ) T^{\mu\nu} =
(v^\rho \pr_\rho + 4 \sigma_v )  \eta_{\mu\nu} T^{\mu\nu}$ 
from \contr{b}
\eqn\AAAid{
\eta_{\mu\nu} \A^{\mu\nu} - \beta^I \A_I = \big (
\Delta_v + 4 \, \sigma_v + v^\mu \pr_\mu \big ) ( X + 2\, \pr_\nu Y^\nu) \, .
}
\AAAid\ constrains the additional derivative terms in \defAT\ as it reduces to
\eqn\EFHGid{\eqalign{
& \pr^2 \big ( \sigma_v ( E_I \pr^2 g^I + F_{IJ} \pr^\mu g^I \pr_\mu g^J ) 
+ 2\, \pr_\mu \sigma_v \, H_I \pr^\mu g^I \big ) \cr
&{} -2  (\eta^{\mu\nu} \pr^2 - \pr^\mu \pr^\nu  ) 
 \big ( \sigma_v \,  G_{IJ} \pr_\mu g^I \pr_\nu g^J 
+ 2\, \pr_\mu \sigma_v \, W_I \pr_\nu g^I \big ) \cr
&{}= - \pr^2  \big ( \sigma_v ( A_{IJ} \pr^2 g^I \beta^J + B_{KIJ} 
\pr^\mu g^I \pr_\mu g^J  \beta^K ) - 2\, \pr_\mu \sigma_v \,  
S_{IJ} \pr^\mu g^I \beta^J \big )\cr
&\hskip 0.6 cm {} +8\,  \pr_\mu \big ( \pr_\nu \sigma_v \, {\tilde S}_{[IJ]} \,
\pr^\mu g^I \pr^\nu g^J \big ) \cr
&\hskip 0.6 cm {} + 2\, \big ( \pr_\mu \, \sigma_v + (\pr_\mu \sigma_v ) \big ) 
\big ( ({\tilde A}_{IJ} + \L_\beta S_{IJ} ) \pr^\mu g^I \pr^2 g^J \cr
& \hskip 4.1 cm +   ({\tilde B}_{IJK} + \L_\beta T_{IJK}  + \pr_J\pr_K \beta^L
S_{IL} ) \pr^\mu g^I \pr^\nu g^J \pr_\nu g^K \big ) \, ,
}
}
for ${\tilde S}_{IJ}, {\tilde A}_{IJ},{\tilde B}_{IJK}$ as in \Stil, 
\GArel, \relGBC. Since
\eqn\DeGW{\eqalign{
(\eta^{\mu\nu}& \pr^2 - \pr^\mu \pr^\nu  )
\big ( \sigma_v \,  G_{IJ} \pr_\mu g^I \pr_\nu g^J
 + 2\, \pr_\mu \sigma_v \, W_I \pr_\nu g^I \big ) \cr
 ={}& \half \, \pr^2 \big ( \sigma_v \, G_{IJ} \pr^\mu g^I \pr_\mu g^J \big )
 + 4\, \pr_\mu \big ( \pr_\nu \sigma_v \, \pr_{[I} W_{J]} \pr^\mu g^I 
\pr^\nu g^J \big ) \cr
&{}- \big ( \sigma_v \pr_\mu + 2(\pr_\mu \sigma_v ) \big )
\big ( G_{IJ}  \pr^\mu g^I \pr^2 g^J  + \Gamma^{(G)}{\!}_{IJK} 
\pr^\mu g^I \pr^\nu g^J \pr_\nu g^K \big ) \, ,
}
}
\EFHGid\ reduces to the consistency relations \EFHdet, \WS, \GArel\ and  
\relGBC. Hence the broken conformal identity \WardT, with \defAT\ may be used
to define $G_{IJ},W_I$ and also $E_I,F_{IJ},H_I$ just in terms of 
correlation functions involving the energy momentum tensor on flat space. 

The relations \contr{a,b} and \WardT\ which are expressed in terms of local
couplings can be translated into equivalent constraints on various 
correlation functions involving the energy momentum tensor and with $g^I$
constant.  We describe here the simplest results for the three point function 
$\langle T^{\mu\nu}(x) \, \O_J(y) \, \O_K(z) \rangle $ in 
the conformal limit assuming \fixD\ with $\Delta_J =\Delta_K= \Delta $. 
In this case we can drop contributions arising from $H_I, S_{IJ},W_I$. 
Suppressing the argument $x$ the conformal
Ward identity becomes
\eqn\threep{\eqalign{
16\pi^2 \big ( \L_v + 6 \, \sigma_v + 
v^\mu (y) \pr_{\mu y} + \Delta \, \sigma_v(y)&   +
v^\mu (z) \pr_{\mu z} + \Delta \, \sigma_v(z) \big )
 \big \langle T^{\mu\nu} \, \O_J(y) \, \O_K(z) \big \rangle \cr
= {}& \, \A^{\mu\nu}_{JK} ( y,z) \, ,
}
}
with
\eqnn\AJK$$\eqalignno{
 \A^{\mu\nu}_{JK} ( y,z) = {}& A_{JK} \big ( 2 \, 
\pr^{(\mu} \delta_y \; \pr^{\nu)} ( \sigma_v \, \pr^2 \delta_z) + 
2\, \pr^{(\mu} ( \sigma_v \, \pr^2 \delta_y) \; \pr^{\nu)} \delta_z \cr
& \qquad {}- \eta^{\mu\nu} \, \big (  
\pr^{\rho} \delta_y \; \pr_{\rho} ( \sigma_v \, \pr^2 \delta_z) +
\pr^{\rho} ( \sigma_v \, \pr^2 \delta_y) \;\pr_{\rho} \delta_z
+ \sigma_v \, \pr^2 \delta_y \; \pr^2 \delta_z \big ) \big ) \cr
\noalign{\vskip 2pt}
{}& + {\ts{1\over 3}} (\eta^{\mu\nu} \pr^2 - \pr^\mu \pr^\nu) 
\big ( \sigma_v \big ( (4-\Delta ) \, A_{JK} ( \pr^2 \delta_y \, \delta_z
+ \delta_y \, \pr^2 \delta_z) 
+ 2 \, G_{JK} \, \pr^\rho \delta_y \, \pr_\rho \delta_z \big ) \big ) \cr
\noalign{\vskip 2pt}
&{} + 2\, G_{JK} \, \D^{\mu \nu \sigma  \rho } \big ( 
\sigma_v \, \pr_{(\sigma} \delta_y \, \pr_{\rho)} \delta_z  \big ) \, ,
& \AJK
}
$$
for $\delta_y \equiv \delta^4(x-y), \, \delta_z \equiv \delta^4(x-z)$
and where we have let $E_I \to - A_{IJ} \beta^J, \, F_{IJ} \to G_{IJ}$.
Corresponding to \contr{a,b} we have
\eqn\threepW{\eqalign{
& \pr_\mu \big \langle T^{\mu\nu} \, \O_J(y) \, \O_K(z) \big \rangle 
- \pr^\nu \delta_y \, \big \langle \O_J \, \O_K(z) \big \rangle
- \pr^\nu \delta_z \, \big \langle \O_J (y) \, \O_K \big \rangle 
=  0 \, , \cr
& 16\pi^2 \big ( \eta_{\mu \nu} 
\big \langle T^{\mu\nu} \, \O_J(y) \, \O_K(z) \big \rangle \cr
& \quad {}+
(\Delta -4) \, \delta_y \,  \big \langle \O_J(y) \, \O_K(z) \big \rangle  +
(\Delta -4) \, \delta_z \,  \big \langle \O_J(y) \, \O_K(z) \big \rangle
\big ) = A_{JK} \, \pr^2 \delta_y \,  \pr^2 \delta_z \, .
}
}
It is again somewhat non-trivial to check consistency of \threep\
and \threepW, the necessary condition reduces to
\eqn\GAJK{
G_{JK} = (\Delta - 3) \, A_{JK} \, ,
}
which is equivalent to \GArel\ in the conformal limit.

\newsec{Beta functions for Scalar Fermion Theory}

We consider as an example for the application of the general consistency
relations a general scalar fermion field theory
involving $n_\psi,n_\chi$  two component chiral spinor fermion fields 
$\psi,\chi$, of opposite chirality, and $n_\phi$ complex scalars $\phi_i$,
$i=1,\dots n_\phi$, with a Lagrangian of the form
\eqn\Lag{
\L = - \pr \bphi^i \cdot \pr \phi_i - \bpsi \, i \sigma \cdot \pr \, \psi
- \bchi \, i \bsi \cdot \pr \, \chi - {\bchi} \, m(\phi)  \, \psi  - 
{\bpsi} \,  \bm( \bphi) \, \chi - V(\bphi, \phi )  \, ,
}
where $ \sigma{\cdot a}\, \bsi{\cdot a} = - a^2 \, 1$, 
$\tr_\sigma(\sigma{\cdot a}\, \bsi{\cdot b}) = - 2 \, a{\cdot b}$ with
$\cdot$ in this context denoting contraction of Lorentz indices. In \Lag\ we 
assume
\eqn\Vphi{
m(\phi) = y^i \phi_i + \mu \, , \ \  \bm(\bphi) = \bphi^i \,\by_i  
+ {\bar \mu} \, , \quad
V(\bphi,\phi) = \quar \,\lambda_{ij}{\!}^{kl}\, \bphi^i\bphi^j\phi_k\phi_l 
+ {\rm O}(\phi^2\bphi, \phi\bphi^2) \, .
}
The Yukawa coupling $y^i$ is  a $n_\chi \times n_\psi$ matrix and 
$\by_i = (y^i)^\dagger$.  Also $(\lambda_{ij}{\!}^{kl})^* = \lambda_{kl}{\!}^{ij}$.
For $n_\chi = n_\psi$ \Lag\ can be re-expressed in 
terms of four component Dirac fermions. The Lagrangian \Lag\ has a 
$U(1)\times U(1)$ symmetry for the dimension four interactions under
\eqn\Sym{
\psi \to e^{i\theta} \psi \, , \qquad \chi \to e^{i\tau} \chi \, ,
\qquad \phi_i \to e^{i(\tau-\theta)} \phi_i \, .
}
This is sufficient to significantly
reduce the number of Feynman diagrams at each loop order.

The $\beta$-functions associated with the couplings $y,\lambda$ in $\L$ 
can be expressed as
\eqn\bbPV{\eqalign{
\beta_y{\!}^i = {}& {\tilde \beta}_y{\!}^i + \gamma_\chi \, y^i + y^i \, 
\gamma_\psi + y^j \gamma_{\phi\,j}{}^i \, , \cr
\beta_V = {}& {\tilde \beta}_V 
+ V^j \gamma_{\phi\,j}{}^i \phi_i + \bphi^i \gamma_{\phi\,i}{}^j V_j \, ,
}
}
for 
\eqn\Vder{
V^j = {\pr \over \pr \phi_j}V \, , \qquad V_j = {\pr \over \pr \bphi^j}V \, .
}
In addition
\eqn\conjy{
\beta_{\by i} = (\beta_y{\!}^i)^\dagger \, .
}

In giving results for $\beta$ and related functions  it is convenient to 
rescale
\eqn\rescale{
\lambda_{ij}{\!}^{kl} \to 16\pi^2 \lambda_{ij}{\!}^{kl}\, , \quad 
y^i \to 4\pi\, y^i \, , \quad \by_i \to 4\pi\, \by _i \, ,
}
thereby removing factors of $1/16\pi^2$ which arise at each loop order.
The anomalous dimension matrices at one and two loops are given by
\eqn\onel{
\gamma_\chi{\!}^{\,(1)} = \half \, y^j \by_j \, , \qquad
\gamma_\psi{\!}^{\,(1)} = \half \, \by_j y^j \, , \qquad
\gamma_\phi^{\ (1)}{\!}_{j}{}^i = \tr(\by_j y^i ) \, ,
}
and
\eqn\twol{\eqalign{
\gamma_\chi{\!}^{\,(2)} ={}& - {\ts{1\over 8}} \, y^i \by_j \, y^j \by_i 
-{\ts{3\over 4}} \, \tr (y^j \by_i) \, y^i \by_j \, , \cr
\gamma_\psi{\!}^{\,(2)} = {}& - {\ts{1\over 8}} \, \by_i\, y^j \by_j\, y^i 
- {\ts{3\over 4}} \, \tr (y^j \by_i) \, \by_j y^i  \, , \cr
\gamma_\phi^{\ (2)}{\!}_{j}{}^i = {}& \quar \, \lambda_{jk}{}^{mn} 
\lambda_{mn}{}^{ki} - {\ts {3\over 4}} \, \big (\tr(\by_j\, y^k \by_k\, y^i ) 
+ \tr(\by_k\, y^k\,  \by_j \, y^i ) \big ) \, .
}
}
The $\beta$-functions are then given by \bbPV\ with \JackO
\eqn\bpert{\eqalign{
{\tilde \beta}_y^{\, (1)i} = {}& 0 \, , \qquad
{\tilde \beta}_y^{\, (2)i} =  2 \, y^j \by_k\, y^i \by_j\, y^k - 2 \,
\lambda_{jk}{}^{li}\,  y^j \by_l\, y^k \, , \cr
{\tilde \beta}_V^{\, (1)} = {}& \half \, V_{rs}V^{rs} - 2 \,
\tr( m \,  \bm  \, m \,  \bm )  \, , \cr
{\tilde \beta}_V^{\, (2)} = {}& - \half \, V_{rst}V^{rs}{}_u V^{tu}
- 2\,\tr(\by_i\, y^j ) \,\big (  V^{ik}V_{kj} + V_j{}^k  V_k{}^{i} \big ) \cr
&{}+ 2 \, \tr(y^k\, \bm\, y^l \bm ) \, V_{kl} 
+ 2\,  \tr(\by_k\, m\, \by_l\, m ) \, V^{kl} \cr  
&{} + 2 \big( \tr ( y^k \by_k \, m\, \bm \, m\, \bm )
+ \tr ( \by_k \, y^k \bm \, m\, \bm \, m ) 
+ 2 \, \tr (y^k \bm \, m\, \by_k \, m \, \bm ) \big )  \, , 
}
}
where $a_r b^r = a^i b_i + a_i b^i$ and $V_{rs},V_{rst}$  are defined by
obvious extensions of \Vder. In consequence $\half \, V_{rs}V^{rs} 
= V_{ij}V^{ij} + V_i{}^j V_j{}^i$.

Two special cases are of particular interest. Assuming 
$n_\chi = r , \, n_\psi = r n , \, n_\phi = n $  we require
\eqn\Vsp{
m(\phi) \psi = y\, \phi_i\psi^i \, , \quad \bpsi\, m(\bphi) = 
\by \, \bpsi_i \bphi^i \, , \quad
V(\bphi,\phi) = \half \lambda \, (\bphi^i \phi_i )^2 \, ,
}
and there is then a manifest $U(n)$ symmetry (for the scalar couplings
the symmetry extends to $O(2n)$), with 
$\chi,\bchi$  singlets, and the  couplings reduce to 
just $\lambda,y,\by$. In the above formulae
\eqn\lyy{
\lambda_{ij}{\!}^{kl} \to \lambda ( \delta_i{\!}^k \delta_j{\!}^l +
\delta_i{\!}^l \delta_j{\!}^k ) \, , \qquad
y^i \, \by_j\to \by y \, \delta_j{\!}^i \, , \qquad
\by_i y^i \to \by y \, 1_n \, .
}
The anomalous dimensions are no longer matrices and from the above we get
\eqn\anom
{\eqalign{
\gamma_\psi{\!}^{\,(1)} = {}& \half \, \by y \, , \qquad
\gamma_\chi{\!}^{\,(1)} = \half n  \, \by y \, , \qquad
\gamma_\phi^{\ (1)} =  r \, \by y \, ,\cr
\gamma_\psi{\!}^{\,(2)} = {}& -{\ts{1\over 8}}(6r+n)\,(\by y)^2 \, , \qquad
\gamma_\chi{\!}^{\,(2)} = -{\ts{1\over 8}}(6r+1)n\, (\by y)^2  \, , \cr
\gamma_\phi^{\ (2)} = {}& (n+1) 
\big ( \half \lambda^2 - {\ts{3\over 4}}r \, (\by y)^2 \big ) \, , 
}
}
with $r$ arising from the trace due to additional fermion degrees of 
freedom. Furthermore, from \bpert
\eqnn\betyl
$$\eqalignno{
{\tilde \beta}_y{\!}^{\,(1)} = {}& 0 \, , \qquad
{\tilde \beta}_y{\!}^{\,(2)} = 2 \big ( (\by y)^2 - (n+1) \, \by y \, 
\lambda \big ) y\, , \qquad {\tilde \beta}_\lambda{\!}^{\,(1)} =
2(n+4) \, \lambda^2 - 4r \, (\by y)^2 \, , \cr
\noalign{\vskip 2pt}
{\tilde \beta}_\lambda{\!}^{\,(2)}  = {}& - 4 (5n+11) \lambda^3 
-4(n+4) \, \lambda^2 \by y + 8 r \, \lambda (\by y)^2 + 4(n+3)r \, 
(\by y)^3 \, ,  &  \betyl
}
$$
where now
\eqn\bbbt{
\beta_y = {\tilde \beta}_y + 
( \gamma_\chi + \gamma_\psi + \gamma_\phi)\,y \, ,
\qquad \beta_\lambda = {\tilde \beta}_\lambda + 4 \gamma_\phi \, 
\lambda \, .
}
Combining \betyl\ and \anom\ for $n=2$ reproduces standard model results
in \Chet.\foot{Assuming \lyy\ the detailed relation with the results of
\Chet\  at each loop order $\ell$ is given by 
$\beta_\lambda {\!}^{(\ell)}  |_{n=2} = 4\, \beta_\lambda{\!}^{(\ell)}
 |_{\lambda\to{1\over 2} \lambda,\, g_s =0}$,
 $\beta_y {\!}^{(\ell)}  |_{n=2} = 2\, \beta_{y_t}{\!}^{(\ell)}
 |_{\lambda\to{1\over 2} \lambda,\,  g_s =0}$,
 $\gamma_\psi {\!}^{(\ell)}  |_{n=2} =  \gamma^t_{2,L}{\!}^{(\ell)}
 |_{\lambda\to{1\over 2} \lambda,\,  g_s =0}$,
$\gamma_\chi {\!}^{(\ell)}  |_{n=2} =  \gamma^t_{2,R}{\!}^{(\ell)}
 |_{\lambda\to{1\over 2} \lambda,\, g_s =0}$ and
 $\gamma_\phi {\!}^{(\ell)}  |_{n=2} =  \gamma^\Phi_{2,L}{\!}^{(\ell)}
 |_{\lambda\to{1\over 2} \lambda,\  g_s =0}$ where $\by=y=y_t$ and $r=d_R$.
}

The other special case corresponds to $\N=1$ supersymmetry. This is 
achieved by letting $n_\psi = n_\phi = n_C $ and imposing
\eqn\susy{
\bchi \to {\tilde \psi} = \psi^T C \, , \qquad \chi \to {\tilde \bpsi} 
= - C^{-1}\bpsi^T \, ,
}
with $C^T=-C$ $C\bsi C^{-1} = - \si^T$, and then rescaling 
$\psi,\bpsi$ to achieve a canonical kinetic term. $\phi_i,\psi_i$
and $\bphi^i, \bpsi^i $ form $n_C$ chiral supermultiplets and a general 
renormalisable $\N=1$ supersymmetric Lagrangian is achieved by letting
\eqnn\red$$\eqalignno{
V(\bphi,\phi) = {}& u^i(\phi)\,{\bar u}_i(\bphi)  \, ,  \quad 
m^{ij}(\phi) = u^{i,j}(\phi) = m^{j i}(\phi) \, , \quad 
\bm_{i j}(\bphi) = {\bar u}_{i,j}(\bphi ) = \bm_{ji}(\bphi) \, , \cr
Y^{ijk} = {}& u^{i,jk} = Y^{(ijk)}\, , \qquad \bY_{ijk} = {\bar u}_{i,jk} 
= \bY_{(ijk)}\, , \qquad \lambda_{ij}{}^{kl} = \bY_{ijm} Y^{mkl} \, . &\red
}
$$
\susy\ is compatible with \Sym\  if $\tau = -\theta$ so that
$U(1)\times U(1)\to U(1)_R$ corresponding to the usual $R$-symmetry.
Standard supersymmetry results based on superspace ensure that the 
$\beta$-functions are determined in terms of the anomalous dimension
\eqn\bsusy{\eqalign{
\beta_Y^{\, ijk} = {}& Y^{ljk}\gamma_{\,l}{}^i + Y^{ilk}\gamma_{\,l}{}^j + 
Y^{ijl}\gamma_{\,l}{}^k \, , \qquad
\beta_{\bY \, ijk} = \gamma_i{}^l\, \bY_{ljk} +  \gamma_j{}^l\, \bY_{ilk} 
+  \gamma_k{}^l \, \bY_{ijl} \, . \cr}
}
Hence with the definitions \bbPV
\eqn\bYV{
{\tilde \beta}_Y = 0 \, , \qquad {\tilde \beta}_V(\phi,\bphi) =
2\, u^i(\phi)\, \gamma_i{}^j \, {\bar u}_j (\bphi)  \, .
}
The results for anomalous dimensions and beta functions for \Lag\ with 
\red\ 
reduce to the supersymmetric form so long as the coefficient of all traces,
which each correspond to a fermion loop, have an additional
coefficient $\half$. This reflects the restriction \susy. Then we have
\eqn\restrict{
\gamma_{\psi\,i}{}^j = \gamma_{\phi\,i}{}^j = \gamma_i{}^j \, , \qquad
\gamma_\chi{\!}^j{}_i = {\bar \gamma}^{\,j}{}_i \, ,  }
With the modification of the trace coefficients the results \onel\ 
and \twol\ are compatible with \restrict\ for
\eqn\otl{
\gamma^{(1)}{\!}_i{}^j = \half \, (\bY Y)_i{}^j \, , \qquad
\gamma^{(2)}{\!}_i{}^j = - 
\half \, \bY_{ikl} \, (Y \bY)^l{}_m \, Y^{mkj} \, .
}
The results in \anom\ and \betyl\ also correspond to a single field
supersymmetric theory for $n=1$, $r=\half$ if $\lambda = \half\, \by y$.

At three-loop order the general expressions for the anomalous dimensions 
are restricted to correspond to one particle irreducible graphs and have 
the form for the fermions
\eqn\threel{\eqalign{
\gamma_\chi{\!}^{\,(3)} ={}& a \, y^i \by_j \, y^j \by_k\, y^k \by_i 
+ b \, y^i \by_j \, y^k \by_k\, y^j \by_i + 
c \, y^i \by_j \, y^k \by_i \, y^j \by_k \cr
&{} + d\, y^i \by_j \, \lambda_{im}{}^{kl}
\lambda_{kl}{}^{mj}  + e\, y^i \by_k\, y^j \by_l \, \lambda_{ij}{}^{kl} \cr
&{} + f  \, \big ( \tr (y^j \by_k \, y^k  \by_i) + 
\tr (y^k \by_k \, y^j  \by_i) \big ) \, y^i \by_j  \cr
&{} + g \, \tr (y^j \by_i) \, y^i \by_k\, y^k \by_j  +
h \, \tr (y^j \by_i) \, y^k \by_j\, y^i \by_k \cr 
&{} + i \,  \tr (y^j \by_k) \,  \tr (y^k \by_i) \, \, y^i \by_j \, , \cr
\gamma_\psi{\!}^{\,(3)} = {}& a \, \by_i \, y^j \by_j \, y^k\by_k \, y^i 
+ b\, \by_i \, y^j \by_k \, y^k\by_ j \, y^i 
+ c \, \by_i\, y^j \by_k \, y^i \by_j \, y^k \cr
&{} + d\, \by_j\, y^j \, \lambda_{im}{}^{kl}
\lambda_{kl}{}^{mj}  + e\, \by_k \, y^i \by_l \, y^j \, \lambda_{ij}{}^{kl} \cr
&{} + f \, \big ( \tr (y^j \by_k \, y^k  \by_i) +  
\tr (y^k \by_k \, y^j  \by_i) \big ) \, \by_j\,  y^i \cr
&{} + g \, \tr (y^j \by_i) \, \by_j\, y^k \by_k\, y^i  +
h \, \tr (y^j \by_i) \, \by_k\, y^i \by_j\, y^k \cr
&{} + i \,  \tr (y^j \by_k) \,  \tr (y^k \by_i) \, \, \by_j\,  y^i \, ,
}
}
and for the scalar field
\eqn\threels{\eqalign{
\gamma_\phi^{\ (3)}{\!}_{j}{}^i = {}& 
a'\,\big (  \lambda_{jk}{}^{mn} \lambda_{mn}{}^{pq} \lambda_{pq}{}^{ki} 
+ 4 \, \lambda_{jk}{}^{mn} \lambda_{ml}{}^{kp} \lambda_{np}{}^{li} \big ) \cr
&{} + b'\, \big (  \lambda_{jk}{}^{mn} \lambda_{mn}{}^{li} + 
2 \, \lambda_{jm}{}^{ln} \lambda_{k n }{}^{m i} \big ) \, \tr(y^k \by_l ) \cr
&{}  + c'\,  \big ( \tr(\by_j\,y^k \by_l\, y^m ) \, \lambda_{km}{}^{li} +
\lambda_{jl}{}^{km} \, \tr (\by_k \, y^l \,\by_m \, y^i ) \big ) \cr
&{} + d' \, \big ( \tr(\by_j \, y^k \by_k \, y^l \, \by_l \, y^i ) +
\tr(\by_k \, y^k \by_l \, y^l \, \by_j \, y^i )  \big ) \cr
&{} + e' \, \big ( \tr(\by_j \, y^k \by_l \, y^l \, \by_k \, y^i )  +
\tr(\by_k \, y^l \by_l \, y^k  \, \by_j \, y^i ) \big ) \cr
&{} + f' \, \tr(\by_k \, y^k \by_j \, y^l\, \by_l \, y^i ) + g' \,
\tr(\by_k \, y^l \by_j \, y^k \by_l \, y^i ) \cr
&{}+ h' \, \big ( \tr(\by_j\, y^k \by_l \, y^i )+  \tr(\by_l\, y^k\, \by_j \, y^i ) 
\big )  \, \tr(y^l\, \by_k) \, .
}
}
The individual contributions in \threel\ and \threels\ are all hermitian
except for those involving the coefficient $c'$ where the two
terms are hermitian conjugates. Furthermore, the expressions are
constrained by  $\gamma_\chi \leftrightarrow \gamma_\psi$
and $\gamma_\phi^{\ (3)}{\!}_{j}{}^i \to \gamma_\phi^{\ (3)}{\!}_{i}{}^j$ 
for $y^i \leftrightarrow \by_i, \, \lambda_{ij}{}^{kl} \to \lambda_{kl}{}^{ij}$
everywhere.

Restricting to the $U(n)$ case given by \lyy
\eqnn\threeln$$\eqalignno{
\gamma_\chi{\!}^{\,(3)} = {}& n \big ( a + n\, b + c+ r (n+1)  \, f + 
r \, (g +h ) + r^2 \, i \big ) (\by y)^3 + 
n(n+1)\big ( 2d  \, \lambda^2 \by y + e \, \lambda (\by y)^2 \big ) \, ,\cr
\noalign{\vskip 2pt}
\gamma_\psi{\!}^{\,(3)} = {}& \big ( n^2 \, a + n\, b + c+ r(n+1) \, f 
+ r n \, (g + h)  + r^2 \, i \big ) (\by y)^3 
+ (n+1)\big ( 2d \, \lambda^2 \by y + e  \, \lambda (\by y)^2 \big ) \, , \cr
\gamma_\phi^{\ (3)} = {}& 2(n+1) \big ( 2(n+4)\, a' \, \lambda^3 + 3r \, b' \,
\lambda^2 \by y + r \, c'\, \lambda (\by y)^2 \big ) \cr
&{}+ r \big ( (n^2+1) \,d' + 2n \, e' + n\, f' + g' + r(n+1)\, h' \big ) 
(\by y)^3 \, .  &  \threeln
}
$$
Comparing with \Chet\ for $n=2$ we may obtain
\eqn\abc{
a = - {\ts{5\over 32}} \, , \quad 2b+c = - {\ts{7\over 8}} +
{\ts{3\over 2}} \zeta(3) \, ,\quad
d = - {\ts{11\over 32}} \, , \quad e=f=1 \, , \quad 
g + h = {\ts{5\over 16}} \, , \quad i = - {\ts{3\over 8}} \, ,
}
and
\eqn\abco{
a' = - {\ts{1\over 16}} \, , \quad b ' = - {\ts{5\over 16}} \, , \quad
c' = {\ts{5\over 4}} \, , \quad h'=2 \, , \quad 5d'+4e'+2f'+g' = 
- {\ts{25\over 16}} + 3 \zeta(3) \, .
}
The graphs associated with $a',b',c'$ were calculated in \Fortin, 
the numerical values given are consistent with \abco\ if an additional
factor of 2 for fermion loops is supplied due to the absence of a 
symmetry factor here.

In the supersymmetric case given by \red\ 
there are four independent terms \Jack\ so that
\eqn\sthree{\eqalign{
\gamma^{(3)}{\!}_{i}{}^j = {}& 
\bY_{ikl} \big ( A \, (Y \bY)^l{}_m (Y\bY)^m{}_n  + C \, 
Y^{lmp} \, (\bY Y)_p{}^q \,\bY_{qmn}  \, \big ) Y^{nkj} \cr
&{}+  \bY_{ikl} \big ( B\,  (Y \bY)^k{}_m (Y\bY)^l{}_n 
+ D\, Y^{kps} \, Y^{lqr} \, \bY_{prm} \, \bY_{qsn} \,\big )  Y^{mnj}  \, .
}
}
From \threel\ and \threels
\eqn\susyth{\eqalign{
A = {}& a+ \quar i  = a'+d' \, ,
\hskip 2.55cm  B = \half g = \half(b'+f') \, , \cr
C = {}& b+d+f + \half h = b'+e'+\half h' \, , \quad 
D = c+e = 2a'+c'+ \half g' \, .
}
} 
According to \Jack
\eqn\ABCD{
A = - \quar \, , \qquad B= -{\ts{1\over 8}} \, , \qquad C= 1  \, , \qquad 
D = {\ts{3\over 2}} \zeta(3) \, .
}
This resolves the freedom  present in \abc\ by requiring in addition
\eqn\abcn{
b = {\ts {1\over 16}} \, , \qquad c = -1 + {\ts{3\over 2}} \zeta(3) \, , \qquad
g = - \quar \, , \qquad  h = {\ts {9\over 16}} \, ,
}
with two additional linear constraints on the coefficients also satisfied.
If the results for $a',b',c',h'$  in \abco\ are used in \susyth\ with \ABCD\
then
\eqn\prim{
d' = -{\ts{3\over 16}} \, , \qquad e' = {\ts{5\over 16}} \, , \qquad f'=
{\ts{1\over 16}} \, , \qquad g' = -2 + 3 \zeta(3) \, .
}
With these values $5d'+4e'+2f'+g'$ is compatible with \abco\  providing
a further check.

In a similar fashion we may write
\eqnn\ythree$$\eqalignno{
{\tilde \beta}_y{\!\!}^{\,(3)i} = {}& \alpha \, y^j\by_k \, y^l \, 
\lambda_{jl}{}^{mn}\lambda_{mn}{}^{ki} + \beta \,  y^j\by_k\, y^l 
\big ( \lambda_{jm}{}^{ni}\lambda_{nl}{}^{km}  +  
\lambda_{jm}{}^{kn}\lambda_{nl}{}^{mi} \big ) \cr
&{}+ \gamma\,  \big ( \tr(y^j\by_m) \, y^m\by_l\, y^k 
+  \tr(y^k\by_m) \, y^j \by_l y^m \big ) \lambda_{jk}{}^{li} 
+ \delta \, \tr(\by_l y^m) \, y^j \by_m\, y^k \, \lambda_{jk}{}^{li} \cr
&{}+ \epsilon \, \big ( y^k \by_m\, y^m \by_j\, y^l + 
y^k \by_j \, y^m\by_m y^l \big )\,  \lambda_{kl}{}^{ji} 
+\eta \, \big ( y^m \by_j \, y^k \by_m \, y^l + y^k \by_m \, y^l\by_j \, y^m \big )
\, \lambda_{kl}{}^{ji} \cr
&{}+ \zeta \, y^k \by_m \, y^i \by_n \, y^l \, \lambda_{kl}{}^{mn} \cr
&{}+ \iota \, \big ( y^j \by_l \, y^i \by_k \, y^l 
+ y^l \by_k \, y^i \by_l \, y^j \big )  \tr(y^k \by_j ) 
+ \kappa \, \big ( y^j \by_k \, y^l + y^l \by_k \, y^j \big ) \, 
\tr ( \by_j \, y^k \by_l \, y^i ) \cr
&{} + \mu \, \big ( y^k \by_j \, y^j \by_l \, y^i \by_k \, y^l +
y^k \by_l \, y^i \by_k \, y^j \by_j \, y^l \big ) \cr
&{} + \nu \, \big ( y^k \by_l \, y^j \by_j \, y^i \by_k \, y^l +
y^k \by_l \, y^i \by_j \, y^j \by_k \, y^l \big ) \cr
&{}+ \theta \, \big ( y^j \by_k \, y^l \by_j \, y^i \by_l \, y^k +
y^k \by_l \, y^i \by_j \, y^l \by_k \, y^j \big ) \, . & \ythree
}
$$
This reduces to
\eqn\ythreered{\eqalign{
{\tilde \beta}_y{\! \!}^{\,(3)} =  {}& (n+1) \big ( 2 \alpha + (n+3) \beta \big )
\, \lambda^2 \by y \, y + (n+1) (2\gamma+\delta) \, r \lambda(\by y)^2 \, y \cr
&{}+ (n+1) \big ( (n+1) \epsilon 
+ 2 \eta + \zeta\big ) \, \lambda(\by y)^2 \, y \cr
&{}+ \big ( 2 \iota + (n+1) \kappa \big ) r \, (\by y)^3 \, y 
+ (n+1) ( \mu+\nu+\theta) \, (\by y)^3 \, y \, .
}
}
Comparing with \Chet
\eqn\sola{
2 \alpha + 5 \beta = 8 \, , \ \ 2 \gamma+ \delta = 5 \, , \ \
3 \epsilon + 2 \eta + \zeta = {\ts {15\over 2}} \, , \ \
2 \iota + 3\kappa = -2 \, , \ \ \mu+\nu+\theta = - 6 \, .
}
In the supersymmetric case then ${\tilde \beta}_y{\!\! }^{\,(3)i} = 0$ 
requires
\eqn\susyB{
\alpha+\half \delta + 2 \nu = 0 \, , \quad
\half \gamma + \epsilon + \half \iota + \mu =0 \, , \quad
\beta + \eta + \half \zeta + \half \kappa + \theta = 0 \, .
}
Each term in \ythree\ corresponds to a particular Feynman graph. 
By calculating the relevant integrals corresponding to individual 
graphs we found
\eqn\resthree{
\alpha = {\ts{3\over 2}} \, , \quad \beta = \gamma= 1 \, , \quad 
\delta = 3 \, , \quad \epsilon = \half \, , \
\quad \eta = \zeta = 2 \, ,
}
which are consistent with the first three relations in \sola.
In \Fortin\ those graphs corresponding to $\alpha,\beta, \gamma,
\delta,\epsilon, \eta$ were also calculated, the numbers quoted for each
graph  appear to be in accord with the coefficients in \resthree\ up to 
factors of 2 which are a consequence of the different symmetry factors for the
theory considered here. By using \sola\
and also \susyB\ with \resthree\ it is easy to obtain
\eqn\resthreeb{
\iota = - 1 \, , \quad \kappa = 0 \, , \quad \mu = -\half \, , \quad
\nu = - {\ts {3\over 2}} \, , \quad \theta = - 4 \, ,
}
so that the three-loop Yukawa  beta function for the theory described
the Lagrangian \Lag\ is fully determined.

\newsec{Gradient Flow Properties}

Based on the results for the scalar fermion $\beta$-functions we explore
at low loop order the constraints arising from the flow equation \CT.
Here we initially neglect the distinction between the standard perturbative
$\beta$-function and the modified $B$-function given by \modB. 
If $T_{IJ} = G_{IJ}$ is symmetric and $G_{IJ}$ is positive definite
then \CT\ defines a gradient flow. 
For purely scalar theories a gradient flow was
postulated and investigated by Wallace and Zia \Wallace, who
showed how $G_{IJ}$ may be found by diagrammatic arguments to
quite high loop order. In general an antisymmetric part in $T_{IJ}$ is
necessary to ensure \CT\ remains valid under  the equivalence relations
\equivA\ which correspond to the freedom in \Aarb\ and \Garb.

We assume here the lowest order results found in \Analog\ determining
$G_{IJ}$. Applied to the theory defined by \Lag,
so that $g^I= \{ y^i , \by_i , \lambda_{ij}{}^{kl}\}$, then 
at two-loop order 
\eqn\Tab{
T_{IJ}{\!}^{(2)} \, \del g^I \del ' g^J = 
G_{IJ}{\!}^{(2)} \, \del  g^I \del ' g^J = 
{\ts{1\over 3}} \,\big ( \tr( \del y^i \,\del' \by_i) + 
\tr( \del \by_i \, \del' y^i )\big )  \, ,
}
for $\del g^I =  \{\del  y^i , \del \by_i , \del \lambda_{ij}{}^{kl}\} , \
\del' g^I =  \{ \del' y^i , \del' \by_i , \del' \lambda_{ij}{}^{kl}\} $. 
With the one-loop result
for $\beta_y{\!}^i$ given by \bbPV\ and \onel
\eqn\Cthree{
{\tilde  A}^{(3)} = {\ts {1\over 12}} \, \big ( \tr( \by_i \, y^i \by_j \, y^j ) +
\tr ( y^i \by_i \, y^j \by_j ) \big ) + {\ts {1\over 6}} \, \tr ( \by_i \, y^j ) \, 
\tr ( \by_j \, y^i ) \, .
}

At the next order the three-loop contribution to $T_{IJ}$ must be
of the general  form
\eqn\Tabn{\eqalign{
T_{IJ}{\!}^{(3)} \del g^I \del' g^J = {}& 
{\ts {1\over 24}} \, \del \lambda_{ij}{}^{kl} \, \del' \lambda_{kl}{}^{ij}\cr
&{}+ \Big ( \ual  \, \big (  \tr ( \del \by_i\,  \del' y^i \, \by_j \, y^j ) +
 \tr (\del \by_i \, y^j  \by_j \, \del'  y^i ) \big )   \cr 
 \noalign{\vskip -3pt}
&\quad \  {} + \ube \, \big (  \tr ( \del \by_i\, \del' y^j \, \by_j \, y^i ) + 
 \tr (\del \by_i \, y^i \by_j \, \del' y^j ) \big )   \cr
& \quad \ {} + \uga \, \big (   \tr ( \del \by_i\, y^i\, \del' \by_j \, y^j ) 
+  \tr (\del \by_i \, y^j \, \del' \by_j \, y^i ) \big )  \cr
& \quad \ {} + \ude \, \tr( \del \by_i \, \del' y^j ) \, \tr (\by_j \, y^i )
+ \uet \, \tr( \del \by_i \, y^j ) \, \tr (\by_j \, \del' y^i )\cr
 \noalign{\vskip -3pt}
&\quad \ {} 
+ \uep \; \tr(\del \by_i \, y^j) \, \tr (\del' \by_j \, y^i )
\hskip 2cm {} + \hbox{conjugate} \Big ) \, ,
}
}
where the first term was calculated in \Analog. The remaining terms
correspond to three-loop vacuum diagrams, with one and two fermion
loops, with two vertices selected. The result is also required
to be invariant under conjugation when $y\leftrightarrow \by$.
Although this is not imposed the expression \Tabn\ 
is symmetric under $\del g^I\leftrightarrow \del' g^I$ so that 
at this order $T_{IJ}{\!}^{(3)} = G_{IJ}{\!}^{(3)}$.

The real coefficients $\ual, \ube, \uga, \ude, \uet, 
\uep $ in \Tabn\ have not been determined hitherto. Without explicit 
determination the  integrability conditions necessary
for \CT\ provide constraints on these coefficients and also on the 
$\beta$-functions themselves, as was also demonstrated to 
two-loop order in \Analog. 
The dependence of ${\tilde A}^{(4)}$ on $\lambda$ is determined in 
terms of $\beta_\lambda{\!}^{(1)}$ and then this fixes the
$\lambda$-dependent terms  in $\beta_y{\!}^{(2)}$. Using the results 
for $\beta_\lambda{\!}^{(1)}$ 
\eqnn\betl$$\eqalignno{
\beta_\lambda{\!}^{(1)}{\!}_{ij}{}^{kl} = {}& \lambda_{ij}{}^{mn}
\lambda_{mn}{}^{kl} + 4 \, \lambda_{m(i}{}^{n(k} \lambda_{j)n}{}^{l)m}
+ 2 \, \tr\big (\by_{(i}\, y^m\big ) \, \lambda_{j)m}{}^{kl} + 2 \,
\lambda_{ij}{}^{m(k} \, \tr \big (\by_m \, y^{l)} \big ) \cr
&{}- 8 \, \tr\big (\by_{(i} \, y^{(k} \, \by_{j)} \, y^{l)} \big ) \, ,
& \betl}
$$
and $\beta_y{\!}^{(2)}$ from  \twol\ and \bpert\ in \CT, with \Tab\ and
\Tabn, requires the three integrability conditions on $\ual,\ube,
\uga,\ude,\uet ,\uep$
\eqn\integ{
2(\ube + \uga) = 4\ual +{\ts{1\over 6}} 
= 2\ual + \ude + \half  =   \uet + \uep  \, .
}
Subject to these conditions
\eqnn\Cfour$$\eqalignno{
{\tilde A}^{(4)} = {}& {\ts{1\over 72}} \, \big (
\lambda_{ij}{}^{kl} \lambda_{kl}{}^{mn}\lambda_{mn}{}^{ij} 
+ 4 \,
\lambda_{ij}{}^{kl} \lambda_{km}{}^{in}\lambda_{ln}{}^{jm}\big ) \cr
&{} + {\ts {1\over 12}}  \, \lambda_{ij}{}^{kl}\, \tr( \by_l \, y^m) \, 
\lambda_{km}{}^{ij}  -{\ts {1\over 3}} \, \lambda_{ij}{}^{kl} 
\tr ( \by_k \, y^i \by_l \, y^j )\cr
&{}+ {\ts{ 2\over 9}} \, \tr( \by_i \, y^j \by_k \, y^i \by_j \, y^k )
+ {\ts{1\over 72}} \, \big ( 
\tr ( \by_i \, y^i \, \by_j \, y^j\, \by_k \, y^k)
+ \tr ( y^i \by_i \, y^j \by_j \, y^k \by_k ) \big ) \cr
&{} -{\ts {1\over 6}}  \big ( \tr ( y^i \by_i \, y^k \by_j ) 
+ \tr ( \by_i \, y^i \, \by_j\,
y^k ) \big ) \, \tr(\by_k \, y^j )
- {\ts {1\over 18}} \, \tr ( \by_i \, y^j ) \, \tr(\by_j \, y^k ) \, 
\tr(\by_k \, y^i ) \cr
&{}+ 2\ual \, \tr \big ( \beta_y{\!}^{(1)i} 
\beta_\by{\!}^{(1)}{\!}_i \big ) \, .   & \Cfour
}
$$

Precise results for $G^{(3)}{\!}_{IJ}$ can be obtained in terms of 
flat space calculations by applying \GArelb, noting that $\D_K A_{IJ}$ 
is zero at three loops. This gives, with the aid of  results from 
section 9,
\eqn\Resabc{
\ual = - {\ts{13\over 72}} \, , \quad 
\ube =  -{\ts{5\over 18}} \, , \quad
\uga = 0  \, , \quad \ude = - {\ts{25\over 36}} \, , \quad
\uet = -{\ts{7\over 18}} \, , \quad \uep = - {\ts {1\over 6}} \, .
} 
These of course satisfy \integ. The freedom associated with \equivA\ 
corresponding to letting
${\tilde A} \to {\tilde A} + z \;  \tr  ( \beta_y{\!}^i \beta_{\by\, i } )$
is realised at this order by
\eqn\equivabc{
\ual \sim \ual + \half z \, , \quad \ube \sim \ube + z \, , \quad
\ude \sim \ude + z \, , \quad \uet \sim \uet + 2 z \, ,
}
under which \integ\ is invariant.
In this case we have correspondingly
\eqn\varW{
W_I{\!}^{(3)}\d g^I  \sim W_I {\!}^{(3)}\d g^I  + 
\d \, \quar z  \big ( \tr(\by_i \, y^j\by_j \, y^i) + \tr ( y^i \by_j \, y^j \by_i )
+2\, \tr(\by_i\, y^j) \, \tr(\by_j \, y^i ) \big ) \, .
}

Higher order results become
more involved. At the next order the metric
for the purely scalar couplings has the general form
\eqn\Gfoura{\eqalign{
G_{IJ}{\!}^{(4)} \, \del  g^I \del g^J \big |_{\lambda\lambda}  = {} & 
\uG \,  \big ( \lambda_{ij}{}^{mn}\,\del  \lambda_{mn}{}^{kl} 
\, \del \lambda_{kl}{}^{ij} + 4 \,
  \lambda_{im}{}^{kn}\,\del  \lambda_{jn}{}^{l m}  \,
 \del \lambda_{kl}{}^{ij}  \big ) \, ,
}
}
where $\uG$ is essentially arbitrary due to the freedom in \Garb\
but has been calculated in a minimal subtraction scheme below.
The $\lambda$-terms do not generate any consistency conditions, in
accord with \Wallace, giving
\eqnn\Afive$$\eqalignno{
{\tilde A}^{(5)}\big |_{\lambda}  = {}& {\ts{1\over 96}} \, 
\lambda_{ij}{}^{kl} \lambda_{kl}{}^{mn}\lambda_{mn}{}^{pq} \lambda_{pq}{}^{ij} 
- {\ts{1\over 12}}  \, \lambda_{ij}{}^{kl}
\big (  \lambda_{kl}{}^{mn}  \lambda_{mp}{}^{iq} \lambda_{nq}{}^{jp} + 
\lambda_{km}{}^{jn} \lambda_{np}{}^{iq}\lambda_{lq}{}^{mp}\big ) \cr&{}
+ {\ts {1\over 4}}\uG  \; \beta_\lambda{\!}^{(1)}{\!}_{ij}{}^{kl}
\beta_\lambda{\!}^{(1)}{\!}_{kl}{}^{ij} \, . & \Afive
}
$$

With the results for $\beta$-functions in the previous section we
may extend these results to include mixed scalar Yukawa contributions
for the theory defined by \Lag. There is then an additional four loop
contribution so that instead of \Gfoura
\eqn\Gfourb{\eqalign{
G_{IJ}{\!}^{(4)} \, \del  g^I \del g^J \big |_{\lambda\lambda}  = {} & 
\uG \,  \big ( \lambda_{ij}{}^{mn}\,\del  \lambda_{mn}{}^{kl} 
\, \del \lambda_{kl}{}^{ij} + 4 \,
\lambda_{im}{}^{kn}\,\del  \lambda_{jn}{}^{l m}  \,
\del \lambda_{kl}{}^{ij}  \big )\cr
&{}+ \uH \,  \del  \lambda_{ij}{}^{kl} \, \tr(\by_l \, y^m) \, 
\, \del \lambda_{km}{}^{ij} \, .
}
}
In addition we assume
\eqn\Tfoura{\eqalign{
T_{IJ}{\!}^{(4)} \, \del g^I \del'  g^J \big |_{\lambda y}  = {} & 
\uA \,  \del  \lambda_{ij}{}^{kl }\, \lambda_{kl}{}^{im} \, 
\tr(\by_m \, \del' y^j ) + \uB \,  \del  \lambda_{ij}{}^{kl }\, 
\tr(\by_l\, \del' y^m ) \,  \lambda_{km}{}^{ij} \cr
&{}+ \uC \,   \del  \lambda_{ij}{}^{kl }\, \tr(\by_k \, y^i \by_l \, \del' y^j ) \, ,
}
}
with a corresponding result for 
$T_{IJ}{\!}^{(4)} \, \del  g^I \del'  g^J \big |_{\lambda \by}$. 
In terms of \Gfourb\ and \Tfoura,  using the one and two  loop 
$\beta$-functions from the previous section,
\eqn\Afivea{\eqalign{
{\tilde A}^{(5)}\big |_{\lambda y \by}  = {}& 
{\ts {2\over 3}} \,\lambda_{ij}{}^{kl }\, 
\tr ( \by_k \, y^m \by_l\, y^i\,  \by_m \, y^j ) \cr 
&{}+ ( \uC + {\ts {1\over 3}}) \, \lambda_{ij}{}^{kl }\, \big ( 
\tr ( \by_m \, y^m \by_k\, y^i\, \by_l \, y^j )  
+ \tr ( y^m \by_m \,y^i\,  \by_k\, y^j\, \by_l  )
\big ) \cr 
&{}+ ( \uC + {\ts{2\over 3}} ) \, \lambda_{ij}{}^{kl }\, \big (  \tr( \by_k \, y^m) \, 
\tr ( \by_m \, y^i \, \by_l\, y^j )  + 
\tr (\by_k \,y^i \, \by_l\, y^m )\, \tr( \by_m \, y^j ) 
\big ) \cr 
&{}+ {\ts{1\over 6}} \, 
\lambda_{ij}{}^{mn} \lambda_{mn}{}^{kl} \,
\tr (\by_k\, y^i \, \by_l \, y^j ) \cr
&{}+ \half (\uA +\uB - {\ts{1\over 8}} ) \, 
\lambda_{ik}{}^{lm} \lambda_{lm}{}^{kj} \, \big (
\tr (\by_j \, y^i \, \by_n\, y^n) + \tr (\by_j \, y^n\by_n \, y^i ) \big ) \cr 
&{}+ (\uA +\uB - {\ts{1\over 12}}  ) \, \lambda_{ik}{}^{lm} \lambda_{lm}{}^{kj} \, 
\tr (\by_j \, y^n ) \,  \tr (\by_n \, y^i ) \big ) \cr 
&{} -{\ts{1\over 12}} \, \big ( \lambda_{ij}{}^{mn} \lambda_{mn}{}^{kl} 
+ 2\, \lambda_{im}{}^{nl} \lambda_{jn}{}^{mk} \big ) \,
\tr (\by_k\, y^i ) \,  \tr (\by_l \, y^j ) \cr
&{} - {\ts{1\over 12}} \, 
\big ( \lambda_{ij}{}^{mn} \lambda_{mn}{}^{pq}\lambda_{pq}{}^{jk} 
+ 4 \, \lambda_{ij}{}^{mn} \lambda_{mp}{}^{jq}\lambda_{nq}{}^{pk} \big ) 
\, \tr(\by_k\, y^i ) \cr
&{} + {\ts {1\over 4}}\uG  \; \beta_\lambda{\!}^{(1)}{\!}_{ij}{}^{kl}
\beta_\lambda{\!}^{(1)}{\!}_{kl}{}^{ij} \big |_{\lambda y \by}  \, . 
}
}
There is one integrability constraint which is used to eliminate $\uH$,
\eqn\intGF{
\uH = 2 \, \uG - {\ts {1\over 6}} \, .
}

The result \Afivea\ may be used to constrain  $\lambda$ contributions
to $ \beta_y{\!}^{(3)}$ by considering $\del_\by {\tilde A}^{(5)}$.
For generality we must include further possible $\lambda$-dependent 
terms in $T_{IJ}{\!}^{(4)}$ for which the relevant contributions are
\eqn\Tfourb{\eqalign{
T_{IJ}{\!}^{(4)} \, \del g^I \del'  g^J \big |_{\by \lambda}  = {} & 
\uA' \,  \tr( \del\by_i\,  y^m)\,  \lambda_{mj}{}^{kl }\, \del' \lambda_{kl}{}^{ji} \, 
+ \uB'  \, \tr( \del \by_k  \,  y^m ) \,  \lambda_{ij}{}^{kl }\, 
\del'  \lambda_{l m }{}^{ij} \cr
&{}+ \uC'  \, \tr( \del \by_i \, y^k \by_j \, y^l )  \,   \del'  \lambda_{kl}{}^{ij }\, ,
}  
}
and
\eqn\Tfourc{\eqalign{
T_{IJ}{\!}^{(4)} \, \del g^I \del'  g^J \big |_{\by y}  = {} & 
\uD \,  \tr( \del\by_i\,  \del' y^j )\,  \lambda_{jm}{}^{kl }\,  \lambda_{kl}{}^{mi} \cr
&{} + \uE \, \big ( \tr( \del \by_i \, \del'  y^k \, \by_j \, y^l )  
+ \tr( \del \by_i \, y^k \by_j \, \del'  y^l ) \big ) \,   \lambda_{kl}{}^{ij }\, ,\cr
T_{IJ}{\!}^{(4)} \, \del g^I \del'  g^J \big |_{\by \by}  = {} & \uF\, 
\tr( \del \by_i \, y^k \, \del' \by_j \,  y^l )  \,   \lambda_{kl}{}^{ij }\,  .
}
}
If $T_{IJ}{\!}^{(4)} $ is symmetric then $\uA'=\uA, \, \uB' = \uB, \, \uC' = \uC$.

At this order it is necessary to take into account the potential necessity
of modifying the perturbative $\beta$-function as in \modB. For the
theory defined by \Lag
\eqn\upL{
\upsilon = - \upsilon^\dagger = \big  \{ \upsilon_{\phi \,i}{}^j , 
\upsilon_\psi, \upsilon_\chi \big  \} \, ,
}
and $(\upsilon g)^I$ is obtained by using, for any $\upsilon \in {\frak g}_K$,
\eqn\upg{\eqalign{
& (\upsilon \, y)^i =  \upsilon_{\chi} y^i - y^i \upsilon_{\psi}
- y^j  \upsilon_{\phi\,j}{}^i \, , \quad
(\upsilon \, \by)_i = \upsilon_{\psi} \by_i - \by_i \upsilon_{\chi}
+  \upsilon_{\phi\, i}{}^j \by_j  \, , \cr
& (\upsilon \, \lambda)_{ij}{}^{kl} = 
\upsilon_{\phi\,i}{\!}^m\lambda_{mj}{}^{kl}
+  \upsilon_{\phi\,j}{\!}^m \lambda_{im}{}^{kl} -
\lambda_{ij}{}^{ml}  \upsilon_{\phi\,m}{\!}^k
- \lambda_{ij}{}^{km}  \upsilon_{\phi\,m}{\!}^l \, .
}
}
At three loops all contributions to $\gamma_\phi^{\ (3)}{\!}_{j}{}^i,
\gamma_\chi{\!}^{\,(3)} , \gamma_\psi{\!}^{\,(3)} $ in \threel, \threels\
are separately hermitian except the terms involving $c'$ in \threels.
Hence there is a  unique three loop possibility
\eqn\threeup{
\upsilon_\phi^{\ (3)}{\!}_{j}{}^i =
u \big ( \tr(\by_j\,y^k \by_l\, y^m ) \, \lambda_{km}{}^{li} -
\lambda_{jl}{}^{km} \, \tr (\by_k \, y^l \,\by_m \, y^i ) \big ) \, .
}

Applying  \CT\ for $\del_\by {\tilde A}^{(5)}$ given by  \Afivea\
 requires combining \Tfourb\ with $\beta_\lambda{\!}^{(1)}$ and \Tfourc\ with 
 $\beta_y{\!}^{(1)}$, $\beta_\by {\!}^{(1)}$.
Using also \Tab\ in conjunction with the $\lambda$ dependent contributions 
to the three-loop Yukawa beta functions given by \ythree,   \threel, \threels\
and \Tabn\ for $\uga=0$, combined with the corresponding
two loop results determined by \bpert\ and \twol, then to ${\rm O}(\lambda)$
\eqn\conla{
{\ts {1\over 3}} \, \eta = {\ts {1\over 3}} \, \zeta = {\ts {2\over 3}}\, ,
}
and
\eqn\conla{\eqalign{
{\ts {1\over 3}} \, e - 2 \, \ube = {}& - 2 \, \ual + \half \, \uE = {\ts {1\over 3}} \, 
\epsilon + \half (\uE+\uF)  = \uC + {\ts {1\over 3}} \, , \cr
\noalign{\vskip 3pt}
{\ts {1\over 3}} \, (c'+u) - 2 \, \uet  - 8 \, \uB' = {}&
{\ts {1\over 3}} \, \delta + \uC' + \uF =  - 2 \, \ude + \uC'  \cr
=  {\ts {1\over 3}} \, (c'-u)  - 2 \, \uep - 8 \, \uA' = {}& {\ts {1\over 3}} \, \gamma 
+ \uC'  +\uE  =  \uC + {\ts {2\over 3}}  - 8 \, \uG \, .
}
}
To ${\rm O}(\lambda^2)$
\eqn\conlal{\eqalign{
{\ts {1\over 3}} \, \beta + 2\, \uC' = {}& - 16 \, \uG \, , \cr
{\ts {1\over 6}} \, \alpha + \half \, \uC'  = {} & {\ts {1\over 6}} - 4 \, \uG \, ,\cr
{\ts {1\over 6}} \, b' + \half ( \uA' + \uB') = {}& - {\ts {1\over 12}} + \uG  \, , \cr
{\ts {2 \over 3}} \, d + \half \, \ube = {}& 
\half \, \ual + \uD = \uA + \uB - {\ts {1\over 8}} \, ,\cr
\quar \, \uet +\uB' + \uD = {}& \quar ( \ude + \uep)  +\uA'  = 
\uA + \uB - {\ts {1\over 12}} + \uG \, ,
}
}
and  to ${\rm O}(\lambda^3)$
\eqn\conlall{
{\ts {1\over 3}} \, a' +\uA' +\uB' =  - {\ts {1\over 12}} +  2 \, \uG  \, .
}

The coefficient  of $\uG$ is arbitrary as expected since \conla, \conlal,
\conlall\ are invariant under
\eqn\invG{
\uG \to \uG + \xi \, , \quad \uA' \to \uA' + \xi \, , \quad \uB' \to \uB'  + \xi \, ,
\quad \uC' \to \uC' - 8 \xi \, .
}
as this corresponds to the freedom
${\tilde A} \to {\tilde A} +   {\ts {1\over 4}}\xi   \; \beta_{\lambda \, ij}{}^{kl}
\beta_{\lambda\,kl}{}^{ij}$. Furthermore,
\eqn\BBfivea{\eqalign{
\big ( \tr  ( \beta_y{\!}^{(1)i} &
\beta_\by{\!}^{(2)}{\!}_i  ) +  \tr ( \beta_y{\!}^{(2)i} 
\beta_\by{\!}^{(1)}{\!}_i )  \big ) \big |_{\lambda y \by}  \cr
= {}& 
- 2 \, \lambda_{ij}{}^{kl }\, \big ( 
\tr ( \by_m \, y^m \by_k\, y^i\, \by_l \, y^j )  
+ \tr ( y^m \by_m \,y^i\,  \by_k\, y^j\, \by_l  ) \cr 
\noalign{\vskip -2pt}
&\hskip1.52cm {} +   \tr( \by_k \, y^m) \,  \tr ( \by_m \, y^i \, \by_l\, y^j )  + 
\tr (\by_k \,y^i \, \by_l\, y^m )\, \tr( \by_m \, y^j ) 
\big ) \cr 
&{}+ \quar \, 
\lambda_{ik}{}^{lm} \lambda_{lm}{}^{kj} \, \big (
\tr (\by_j \, y^i \, \by_n\, y^n) + \tr (\by_j \, y^n\by_n \, y^i ) +
2\, \tr (\by_j \, y^n ) \,  \tr (\by_n \, y^i ) \big ) \, ,
}
}
so that letting 
${\tilde A} \to {\tilde A} + z \; \tr  ( \beta_y{\!}^i \beta_{\by\, i } )$
corresponds in \Afivea\ to
\eqn\varABC{
\uA + \uB \to \uA + \uB + \half z \, , \qquad \uC \to \uC - 2 z \, .
}
The consistency constraints \conla, \conlal, \conlall\ are then invariant if, 
along with \equivabc, at the same time
\eqn\varABDEF{
\uA' \to \uA' + \quar z \, , \quad \uB' \to \uB'-\quar z \, , \quad \uD 
\to \uD + \quar z \, , \quad \uE \to \uE -2z \, , \quad \uF \to \uF - 2 z \, .
}

The conditions \conlal, \conlall\ entail various constraint equations for the
coefficients appearing in the general expressions for the three-loop Yukawa 
$\beta$-function and associated anomalous dimensions. Together with
\conla\ the full list is
\eqn\constraint{\eqalign{
\eta = \zeta =2 \, , \quad & 2\alpha - \beta = 2 \, , \quad
\delta+\gamma - 2\epsilon - \beta = 2 \, ,  \quad a'-b' = \quar \, , \cr
& 2 c' - \beta + \gamma - 2 e - 16d = 6 \, .
}
}
Reassuringly these relations are in accord with the results \abc, \abco\  and \resthree.
In addition
\eqn\resu{
u = - \half \gamma - e - 8d = {\ts {5 \over 4}} \, .
}
This demonstrates that the RG equations such as \CT\ hold only for the modified 
$\beta$-function determined by a non zero $\upsilon$ as in \threeup. 
The coefficient appears to  be exactly  in accord with that determined
by Fortin {\it et al} \FortinC\ by explicit three loop calculation for a general 
scalar fermion theory.\foot{They considered couplings to real scalars 
and there was also a purely Yukawa contribution to $\upsilon_\phi{\!}^{(3)}$
which is absent in the model discussed here.} It is interesting to note that
$u=c'$. There are also constraints on the three-loop metric given by \Tabn\ with
$\uga=0$
\eqn\conTab{\eqalign{
2 \ual -  \ude = {}& {\ts{1\over 6}}(- 2 \epsilon + \delta ) = {\ts {1\over 3}} \, , \cr
\ube -  \ude =  {}& {\ts{1\over 12}}( \beta + 2e + 2) =  {\ts {5\over 12}} \, , \cr
2 \ube -  \uep -  \uet ={}&   {\ts {1\over 3}}( e - c ' - 4 a') = 0  \, ,
}
}
which are equivalent to \integ, and so \conTab\ provides an additional 
confirmatory check on the three loop results obtained in section 5.

From \conlal, \conlall\
\eqn\resABC{
\uA + \uB = \ual - {\ts{1\over 16}} \, , \quad \uC = - 4\ual -  {\ts{1\over 6}} \, 
}
so that ${\tilde A}^{(5)} |_{\lambda y \by}$ is determined in \Afivea\ up
to the freedom of choice for $\uG$ and that corresponding to \varABC.
We also have $\uA'+\uB' = 2 \uG - {1\over 16}, \, \uC' = - 8 \uG - {1\over 6}$
so there is the potentiality of a symmetric $T_{IJ}{\!}^{(4)}$
if we take $\ual = 2 \uG$ but this need not be true in general
renormalisation schemes (with dimensional regularisation $\ual = - {7\over 72},
\, \uG = - {7\over 216}$).

\newsec{Supersymmetry}

For supersymmetric theories with just $\N=1$ 
supersymmetry there are further constraints 
which simplify many details  significantly. The
results obtained in \Analog\ were restricted to
supersymmetric field theories previously in \SusyA.
Here the analysis is extended to a general  $\N=1$ Wess-Zumino 
supersymmetric scalar fermion theory, which may be obtained from 
\Lag\ by imposing \susy, \red, to a higher order. 
Such a  theory can of course be rewritten in terms of $n_C$ chiral and
corresponding conjugate anti-chiral  superfields. 
The local couplings may also be extended so that $Y^{ijk},\bY_{ijk}$ for 
this theory are also chiral, anti-chiral superfields.
Divergences which arise in a perturbative expansion are cancelled
by counterterms which are integrals of local polynomials 
in the fields and couplings of dimension two over full $\N=1$ superspace. This 
restriction crucially ensures  that $\beta$-functions for  $Y^{ijk},\bY_{ijk}$
are determined in terms of just the anomalous dimension matrix $\gamma$
as in \bsusy\ but further conditions on the functions which are present in
local RG equations also arise. The various RG functions are 
further constrained by assuming manifest $U(n_C)$ symmetry.

The formalism of section 2 can be adapted to this case by taking
\eqn\gYYb{
g^I = ( Y^{ijk},\bY_{ijk}) \, , \qquad
(\omega g)^I =\big  ( -(Y*\omega)^{ijk}, (\omega * \bY) _{ijk} \big ) \, , 
\quad \omega_i{}^j \in {\frak g}{ \frak l}(n_C, {\Bbb C}) \, ,
}
where
\eqn\omegast{\eqalign{
(Y * \omega)^{ijk} \equiv  {}&
Y^{ljk}\omega_{\,l}{}^i + Y^{ilk}\omega_{\,l}{}^j + 
Y^{ijl}\omega_{\,l}{}^k \, , \cr
(\omega * \bY)_{ijk}  \equiv {}&
\omega_i{}^l\, \bY_{ljk} +  \omega_j{}^l\, \bY_{ilk} 
+  \omega_k{}^l \, \bY_{ijl} \, . }
}
With this notation  the result for the Yukawa supersymmetric $\beta$-functions
\bsusy\  becomes\foot{More generally we may have $\beta_Y   =  
Y * \gamma , \, \beta_{\bY} = {\bar \gamma} *  \bY$. This form is 
preserved under  transformations 
$Y^{ijk} \to Y^{lmn}G_l{}^i G_m{}^j G_n{\!}^k = Y'{}^{ijk}$,
$\bY_{ijk} \to {\bar  G}_i{}^l {\bar  G}_j{}^m
{\bar G}_k{}^n  \bY_{lmn}  = \bY'{\!}_{ijk}$ for $G\in Gl(n_C, {\Bbb C})$.
In this case
 $\beta'{\!}_Y   =  Y' * \gamma' , \,
\beta'{\!}_{\bY} = {\bar \gamma}'*  \bY' $ with
$\gamma' = G^{-1} \gamma\,  G + G^{-1} \dot G, \,
{\bar \gamma}{}'=  {\bar G}\, {\bar \gamma}\,  {\bar G}{}^{-1} 
+{\dot {\bar G}} \, {\bar G}{}^{-1}$  for
$\dot G = ( \beta_Y\cirk \pr_Y + \beta_\bY \cirk \pr_\bY ) G$
and similarly for ${\dot {\bar G}} $.
For $U(n_C)$ transformations ${\bar G} = G^{-1}$. 
Requiring then
$\dot G + \half (\gamma - {\bar \gamma}) G = 0$ ensures $\gamma' = 
{\bar \gamma}'$ so the general case can be reduced to $\gamma={\bar \gamma}$
by virtue of $U(n_C)$ symmetry.
}
\eqn\bsusyst{
\beta_Y   =   Y * \gamma \, , \qquad   
\beta_{\bY} = \gamma *  \bY \, .
}
To avoid explicit indices where possible we also define, in this section
and Appendix A,  a scalar 
product ${}\cirk{}$ on Yukawa couplings so that  for instance
$Y\cirk \bY = Y^{ijk}\bY_{ijk}$. 

Besides the $\beta$-functions other expressions appearing
in the equations of section 2 are determined
in terms of the anomalous dimension matrix $\gamma$.
Based on a superspace framework  Fortin {\it et al}  \FortinS\ 
showed that $\rho_I$ to all orders  is given by
(a related result is given in Appendix C of \Nak)
\eqn\Susyall{
\big ( \rho_I(g)\,  \del g^I \big ){}{}_i{}^j  
= - \del_Y  \gamma_i{}^j + \del_\bY \gamma_i{}^j \, ,
}
for $\del_Y = \del Y \cirk \pr_Y , \, \del_\bY = \del \bY \cirk \pr_\bY $.
In a similar fashion to the derivation of \Susyall\ we may also obtain in 
\defDM\ results which are determined just in terms of $ \gamma_i{}^j$,
\eqn\SusyMall{
\big (\delta_I(g)\, \del g^I \big ){}_i{}^j  =0 \, , \qquad
\big ( \epsilon_{IJ} (g)\,  \del g^I  \del g^J \big ){}_i{}^j 
=  2 \,  \del_\bY \del_Y^{\vphantom g} \gamma_i{}^j \, ,
}
The result \Susyall\ implies $\rho_I(g)\, g^I = 0$
which in turn ensures that in the supersymmetric case
\eqn\susyup{
\upsilon = 0 \, .
}
Thus there is no modification of the $\beta$-function as in \modB.  
The necessary constraint \Brho\  on $\rho_I$ applied to  \Susyall\ 
requires
\eqn\betrho{
\beta_Y \cirk \pr_Y  \gamma_i{}^j  = \beta_\bY \cirk \pr_\bY  \gamma_i{}^j \, .
}
This is a special case of the identity, for any $\omega_i{}^j$, 
\eqn\conga{
\big ( (\omega * \bY ) \cirk \pr_\bY - (Y*\omega ) \cirk \pr_Y\big )
 \, \gamma_i{}^j   = \big [ \omega , \gamma] {}_i{}^j  \, ,
}
taking $\omega \to \gamma$.
The result \conga\ was obtained  in \JackAJ\foot{
See eq. (A.7).}  and is a consequence of 
$\gamma_i{}^j (Y, \bY) $ transforming as a $(1,1)$ tensor
under $U(n_C)$  with $\omega = - {\bar \omega} \in {\frak u}(n_C)$,
the associated Lie algebra.

In the supersymmetric theory the equation \CT\  is assumed to now
take the form
\eqn\ST{\eqalign{
\del_Y {\tilde A} =  {}& \half \big ( \del Y \cirk T \cirk \beta_\bY + 
\beta_Y \cirk K \cirk \del Y \big ) \, , \qquad K^T = -K \, , \cr
\del_\bY {\tilde A} =  {}& \half \big ( \beta_Y \cirk {\bar T} \cirk \del \bY 
+  \del \bY \cirk {\bar K}
\cirk \beta_\bY \big ) \, , \qquad  {\bar K}^T = - {\bar K}  \, ,
}
}
so that 
\eqn\SAG{
( {\beta_Y}\cirk \pr_Y + \beta_{\smash \bY}\cirk  \pr_{\smash{\bY}}  ) 
{\tilde A} = \beta_Y \cirk G \cirk \beta_{\smash \bY} \, ,  \qquad
G = \half ( T +{\bar T} ) \, .
}
By $U(n_C)$ invariance 
$\big ( {\beta_Y}\cirk \pr_Y - \beta_{\smash \bY}\cirk
\pr_{\smash{\bY}} \big ) {\tilde A} =  0$ so for consistency we should
require $\beta_Y \cirk T \cirk \beta_{\smash \bY} =
\beta_Y \cirk {\bar T} \cirk \beta_{\smash \bY}$. $T,K$  may 
be determined by perturbative calculations but from the perspective
of just analysing the integrability conditions flowing from \ST\ and 
using known results for $\beta$-functions, as is
considered mainly in this section, there  is an ambiguity such that 
$T,K$ satisfy the equivalence relations
 \eqn\varTK{
T \sim T + T' \, , \ K \sim K+K' \quad \hbox{if} \quad
 \del Y \cirk  T' \cirk \beta_\bY + 
\beta_Y \cirk  K' \cirk \del Y = 0 \, .
}

The result \tAT\ constrains the form of $K$ and $T-{\bar T}$.
Writing
\eqn\WQeq{
W_I \d g^I =\half (  \d Y \! \cirk \bW + W \cirk \d \bY) \, , \qquad 
Q_I \d g^I = \half ( \d Y \! \cirk \bQ - Q \cirk \d \bY) 
\in {\frak{u}}(n_C)\, ,
}
then
\eqn\relTK{\eqalign{
\del' Y \! \cirk  K \cirk \del Y = {}& \d'  Y\!\cirk  \d_Y \bW
- \tr \big (\d' Y \! \cirk \bQ \; \d_Y \gamma\big  ) 
- \del' Y \leftrightarrow \del Y \, , \cr
\del Y \cirk  \half ( T - {\bar T})   \cirk \del \bY = {}& 
\d_Y W \cirk \d \bY + \tr \big ( \d_Y \gamma \; Q \cirk \d \bY \big )
- \hbox{conjugate} \, .
}
} 
The relation \WQ{a} requires
\eqn\relWQ{
3\, (\bY W) - 3\, (\bW Y) =   Q \cirk \beta_\bY - \beta_Y \cirk \bQ \, ,
}
 defining $(\bY W), (\bW Y) \in {\frak{ g l}}(n_C, {\Bbb C}) $ by
\eqn\WYr{
( Y * \omega ) \cirk \bW  = 3\, \tr \big ( (\bW Y) \, \omega \big )  \, 
\, , \quad W \cirk (\omega * \bY ) = 3\, \tr\big ( ( \bY W) \, \omega \big )
 \, , \quad \omega  \in {\frak{ g l}}(n_C, {\Bbb C}) \, .
}
If $W \cirk \d \bY$ corresponds to a $\ell$-loop vacuum graph then 
$(\bY W)$ may be represented by an associated $(\ell-1)$-loop graph with two 
external lines. For any $Q', \bQ'$ such that
\eqn\varQ{
 Q' \cirk \beta_\bY = \beta_Y \cirk  \bQ' \, ,
 }
 then $Q\sim Q + Q', \, \bQ \sim \bQ + \bQ'$ since  \relTK\  ensures
 that the  corresponding  $T',K'$ satisfy \varTK. Up to this equivalence
 \relWQ\ determines $Q,\bQ$ in terms of $W,\bW$.
 
The RG flow equations \ST\ are invariant under 
\eqn\varA{
\Delta {\tilde A} = \beta_Y \cirk g \cirk \beta_\bY 
+ ( {\beta_Y}\cirk \pr_Y + \beta_{\smash \bY}\cirk  \pr_{\smash{\bY}} ) a\, , 
\qquad g = {\bar g} \, ,
}
when
\eqn\varKT{\eqalign{
\del' Y \cirk \Delta K \cirk \del Y = {}& 2 \, \del' Y \cirk g \cirk
 (\del_Y \gamma * \bY) + \del' Y \cirk \del_Y g \cirk \beta_\bY
 - \del' Y \leftrightarrow \del Y \, , \cr
 \del Y \cirk \Delta T \cirk \del \bY = {}& 
2 \, \d_Y \big  ( (Y * \gamma) \cirk g \big )  \cirk \del \bY +
2 \, \del Y \cirk g \cirk ( \del_\bY \gamma * \bY )  \cr
&{} + \del Y \cirk
\big ( {\beta_Y} \! \cirk \pr_Y + \beta_{\smash \bY}\!\cirk  \pr_{\smash{\bY}} \big ) 
g \cirk \del \bY 
- \del Y \cirk \del_\bY g \cirk \beta_\bY - \beta_Y \cirk 
\del_Y g \cirk \del \bY \, .
 }
 }
For $\Delta {\bar K}, \, \Delta {\bar T}$ given by the conjugate  
equations to \varKT\  then $\Delta G = \half (\Delta T + \Delta {\bar T})$
is therefore
\eqnn\varGS$$\eqalignno{
\del Y \cirk \Delta G \cirk \del \bY = {}& \del Y \cirk
 \big ( {\beta_Y}\cirk \pr_Y + \beta_{\smash \bY}\cirk  \pr_{\smash{\bY}} \big ) 
g \cirk \del \bY  + ( \del Y * \gamma ) \cirk g  \cirk \del \bY
 + \del Y \cirk g \cirk (\gamma * \del \bY )  \cr
&{} + 2 \,( Y * \del_Y \gamma ) \cirk g \cirk \del \bY + 
2 \, \del Y \cirk g \cirk ( \del_\bY \gamma * \bY ) \, . & \varGS
}
$$
\varKT\ and \varGS\ correspond exactly to the freedom in 
\equivA\ assuming \Susyall\ and demonstrate that it is consistent
to require that  $G$ defines  a hermitian metric for supersymmetric
theories. Corresponding to this freedom there are  associated variations
in $W_I, Q_I$ given by
\eqn\equivWQ{\eqalign{
\Delta W \! \cirk \d \bY ={}&  \beta_Y \cirk g \cirk \d \bY - 2\, \d_\bY a \, ,
\hskip 1.45cm 
\d Y \! \cirk \Delta \bW =  \d Y \cirk g \cirk \beta_\bY - 2\,  \d_Y a \, , \cr
\Delta Q \cirk \d \bY = {}& - 3 \, ( g \cirk \d \bY \, Y )
+ \beta_Y \cirk p \cirk \d \bY \, , \ \
\d Y\! \cirk \Delta \bQ = -  3 \, (\bY \, \d Y \! \cirk g) +
\d Y \! \cirk p \cirk \beta_\bY  \, ,
} 
}
with $ ( g \cirk \d \bY \, Y ), (\bY \, \d Y \! \cirk g) $ defined similarly
to \WYr\ and $\d Y\! \cirk p \cirk \d \bY \in  {\frak{ g l}}(n_C, {\Bbb C})$. 
These results ensure that \relTK\  is compatible with \varKT, variations
in $Q,\bQ$ arising from $p$ satisfy \varQ.
We may also verify the invariance of \relWQ, so long as
$(Y* \omega ) \cirk \pr_Y a = (\omega *\bY) \cirk \pr_\bY a $.

There is also freedom corresponding essentially to a choice of scheme.
For this we consider variations
\eqn\varAt{
\delta {\tilde A} = -(Y*h) \cirk \pr_Y {\tilde A} = - (h * \bY)\cirk \pr_\bY
{\tilde A} \, ,
}
for arbitrary $h_i{}^j(Y,\, \bY)$. We assume that there is a corresponding
variation in $\gamma$ of the form 
\eqn\varbet{
\delta \beta_Y = Y * \delta \gamma \, ,\qquad \delta \beta_\bY =
\delta \gamma * \bY \, ,
}
for
\eqn\varga{
\delta \gamma = \beta_\bY \cirk \pr_\bY h - (Y*h) \cirk \pr_Y \gamma \, .
}
This expression for $\delta \gamma$ 
 may be rewritten in various equivalent forms using \conga\ for
$\omega \to h$ or for $\omega\to\gamma, \, \gamma\to h$.  In
consequence $\delta \gamma^\dagger = \delta \gamma$ if 
$h^\dagger =  h , \gamma^\dagger = \gamma$ and also if $h$ corresponds
to a 1PI graph then so does $\delta \gamma$ as well.
Assuming \varAt\ and \varbet, \varga\ the essential equations
\ST\ are invariant if
\eqn\varKTn{\eqalign{
\del' Y \! \cirk \delta K \cirk \del Y = {}& 
-  \del' Y \cirk\big  ((Y*h)\cirk \pr_Y K \big )  \cirk \d Y\cr
&{} -  \del'{\!}_Y (Y*h)  \cirk K \cirk \d Y 
 - \del' Y \cirk K \cirk d_Y (Y*h)  \, , \cr
 \del Y \cirk \delta T \cirk \del \bY = {}& 
-  \d Y  \cirk \big  ((Y*h)\cirk \pr_Y T \big )  \cirk \del \bY \cr
& {} -  \del_Y (Y*h)  \cirk T \cirk \d  \bY 
 - \del Y \cirk T  \cirk d_{\smash \bY} h * \bY   \, .
}
}
since then $2\, \d_Y \delta {\tilde A} = 
 \del Y \cirk T \cirk  \delta \beta_{\smash \bY} 
+  \delta \beta_Y \cirk K \cirk \del Y +
\del Y \cirk \delta T \cirk \beta_{\smash \bY} + 
\beta_Y \cirk \delta K \cirk \del Y$.

The basic equations \ST\  may be verified using perturbative results. 
For  convenience we adopt a notation where the one and two
loop contributions to the anomalous dimension $\gamma$ in \otl\
are given by $\gamma^{(1)} = \half \, (\bY Y), \ \gamma^{(2)} = 
- \half \, (\bY Y \bY Y)$.
Restricting the metric \Tab\ to the supersymmetric case gives
\eqn\susyG{
\del Y \cirk T^{(2)}\cirk  \del \bY  =
{\ts{1\over 3}} \, \d Y \cirk \d \bY = {\ts{1\over 3}} \, 
\tr \big ( (\d \bY \d Y ) \big )  \, ,
}
and in general
\eqn\susyGt{
\del Y \cirk T^{(3)} \cirk \del \bY   =
a \; \tr \big (  (\d \bY \d Y ) \, ( \bY Y ) \big ) +
b \; \tr \big (  (\d \bY  Y ) \, ( \bY \d Y ) \big )  \, ,
}
where we note that $\tr \big ( ( \bY_1 Y_2 \bY_3 Y_4  ) \big ) = 
\tr \big (  (\bY_1  Y_4 ) \, ( \bY_3 Y_2 ) \big )$.
To this order $K,{\bar K}=0$ and $T={\bar T}=G$. For integrability
we require
\eqn\intab{
2 a -b = - \half \, ,
}
which accords with the constraints for supersymmetric theories described in 
\SusyA.\foot{
In terms of the parameters in \SusyA\ $\alpha= 2a,
\beta=2b,\gamma=0$.} If we let
\eqn\gone{
\d Y \cirk g^{(2)} \cirk \d \bY =  z  \, \d Y \cirk \d \bY \, ,
}
then \varKT\ gives at this order $\Delta K=0$ and $\Delta T$ is determined
by
\eqn\varab{
\Delta a = 3z \, , \qquad \Delta b = 6 z \, ,
}
under which \intab\ is  invariant.
Integration of \CT\  subject to \intab\ then gives
\eqn\susyA{\eqalign{
{\tilde  A}^{(3)} = {}& {\ts{1\over 8}} \, \tr \big ( 
(\bY Y)^2 \big )  \, , \cr
{\tilde  A}^{(4)} = {} & {\ts{ 1\over 24}}  \, \tr \big ( (\bY Y)^3 \big ) 
+ {\ts{1\over 3}} a \, 
\beta_Y{\!}^{(1)} \cirk \beta_\bY{\!}^{(1)}  \, .
}
}
Reducing the results in \Tabn\ requires $a={1\over 12}+2\ual +{1\over 2}\ude, \,
b= 2 \ube + {1\over 2}\uet$ and hence from \Resabc
\eqn\Resab{
a = - {\ts {5\over 8}} \, , \qquad
b = -  {\ts {3\over 4}}  \, ,
}
which of course satisfy \intab.

This discussion can be extended to the next order using as input
the form of the three-loop $\gamma$ given by \sthree. It is
convenient to summarise this in the form
\eqn\sthreea{
\gamma^{(3)}
 = A \, \gamma_A + B \, \gamma_B + C\, \gamma_C + D\, \gamma_D \, ,
 }
 where the coefficients $A,B,C,D$ are given in \ABCD. However there
 is potential scheme-dependence since if in \varga\  we take 
 $h = v\, (\bY Y \bY Y)$ and $\beta_\bY\to \beta_\bY{\!}^{(1)},
 \gamma \to \gamma^{(1)}$ then
 \eqn\varAB{
 \delta A = v \, , \qquad \delta B = \half v \, .
}
From \ABCD\ it is evident that we may use this freedom to set
$A=B=0$ but $C$ is scheme-independent.
At this order there are three relevant connected 1PI vacuum graphs with
different topologies and to determine $ T^{(4)}$ it is necessary
to choose for each graph one $Y$ vertex and one $\bY$ vertex 
in inequivalent ways.
The number of possible terms multiply but this procedure gives the
general expression
\eqn\Tfour{\eqalign{
\del Y \cirk T^{(4)} \cirk \del \bY   = {}& a_1 \; 
\tr \big (  (\d \bY \d Y) \, (\bY Y \bY Y ) \big ) 
+  a_2 \; \tr \big (  (\d \bY Y) \, (\bY Y \bY \d Y) \big ) \cr
& {} +  a_3 \; \tr \big (  (\d \bY Y) \, (\bY \d Y \, \bY Y ) \big ) 
+  a_4 \; \tr \big (  (\bY Y) \, (\d \bY \, Y \bY \d Y ) \big ) \cr
& {} +  a_5 \; \tr \big (  (\bY \d Y) \, (\d \bY \, Y \bY Y ) \big ) \cr
&{}+ b_1 \; \tr \big (  (\d \bY \d Y ) \, ( \bY Y )^2 \big ) 
+ b_2 \; \tr \big (  (\d \bY Y ) \, ( \bY \d Y ) \, (\bY Y )  \big ) \cr
&{}+ b_3 \; \tr \big (  (\d \bY Y ) \, ( \bY Y ) \, (\bY \d Y )  \big ) \cr
&{}+ c \;
\d \bY_{ikl} \, \d Y^{kps} \, Y^{lqr
} \, \bY_{prm} \, \bY_{qsn}  Y^{mni} \, . 
}
}
In this case $\tr \big ( (\bY_1 Y_2) \, ( \bY_3 Y_4 \bY_5 Y_6  ) \big ) =
\tr \big ( (\bY_5 Y_4) \, ( \bY_3 Y_2 \bY_1 Y_6  ) \big )$. At four loops 
there may also be contributions to $K$ in \ST\ so that, following a similar
prescription as for $T^{(4)}$ but choosing two $Y$ vertices and 
antisymmetrising, there are two possible terms
\eqn\Kfour{\eqalign{
\del' Y \cirk K^{(4)} \cirk \del Y   = {}& e \;
\tr \big (  (\bY \d' Y) \, (\bY Y \bY \d Y ) \big ) + f \;
\tr \big (  ( \bY \d' Y ) \, ( \bY \d Y ) \, (\bY Y )  \big ) \cr
&{} - \del' Y \leftrightarrow \del Y \, .
}
}
At this order $T$ and ${\bar T}$ are also no longer necessarily equal
since
\eqn\aat{
\d Y \cirk  {\bar T}^{(4)} \cirk \d \bY = 
\del Y \cirk T^{(4)} \cirk \del \bY \big |_{a_2 \leftrightarrow a_5} \, .
}

It is easy to see that, by virtue of \equivWQ,  we may take
$Q^{(2)}, W^{(3)} \to 0$. At the next order there may be non-trivial
$Q,W$. If we allow only contributions corresponding to connected
diagrams then it is sufficient to assume
\eqn\WQfour{
\d Y \cirk \bW^{(4)} 
= - \quar e \, \tr \big ( (\bY Y) \, (\bY Y \bY \d Y ) \big ) \, , \qquad
\d Y \cirk \bQ^{(3)}  = e \,  (\bY Y \bY \d Y ) \, ,
}
where the coefficients are related by imposing  \relWQ\ and  using  \relTK\ 
which  relates \WQfour\ to  \Tfour\ and \Kfour\ if $a_2 = a_5,  \, f=0$.

At five loops ${\tilde A}^{(5)}$ is determined  in terms of 
the five connected vacuum diagrams for this theory at this order.
The relevant contributions can be written in the general form
\eqn\Afive{\eqalign{
2\, {\tilde A}^{(5)} = {}& X_1\; \tr \big (  (\bY Y)^2 \, (\bY Y \bY Y ) \big )
+ X_2 \; \tr \big ( (\bY Y \bY Y )^2 \big ) + 
X_3 \; \tr \big (  (\bY Y) \, \gamma_B  \big ) \cr
&{} + X_4\; \tr \big (  (\bY Y)^4 \big ) + 
X_5 \; \tr \big (  (\bY Y) \, \gamma_D  \big ) \, ,
}
}
where $\gamma_B, \gamma_D$ are explicitly defined by \sthree.
Using \ST\ for $\d_Y {\tilde A}^{(5)}$ we may then obtain,
for arbitrary values for $A,B,C,D$ in \sthreea,
\eqn\integfive{\eqalign{
X_1 ={} & \half (a_2+e) + a_4 + b_1 = \half \, a_5 + b_2 + f - \half a \cr
= {}& \half (a_1 -e)  + b_3 -  f - \half b 
= \half (a_3+a_5-e ) + A \, , \cr
X_2 ={} & \half ( a_1 + a_2 + e  - a) = \half ( a_3 -b + C )  \, , \cr
X_3 = {}& \half \, a_4 = {\ts {1\over 6}} ( a_5 -e + 2 B ) \, ,  \cr
X_4 = {}& {\ts {1\over 8}} (b_1+b_2+b_3) \, ,  \qquad\qquad 
X_5 = \half \, c = D \, .
}
}
For each term in \Afive\  integrability conditions arise whenever
the number of inequivalent $Y$ vertices in the associated graph
is greater than one. The equations \integfive\ are invariant under
\eqn\varabef{\eqalign{
& a_1\to a_1 + \mu \, , \ \ a_2 \to a_2 - \mu - \nu \, , \ \ a_5 \to
a_5 + \nu \, , \quad e \to e + \nu  \, , \ \ f\to f + \omega \, , \cr
& b_1 \to b_1 + \half \mu \, , \ \ b_2 \to b_2 - \half \nu
- \omega \, , \ \ b_3 \to b_3 - \half \mu + \half \nu + \omega \, ,
}
}
which correspond to variations satisfying \varTK\ for one loop $\beta_Y,\beta_\bY$.
The freedom in \varabef\ in part can be realised by changes in $Q,{\bar Q}$ 
satisfying \varQ. As a consequence,
even setting $K^{(4)}=0$, ${\tilde A}^{(5)}$ does not determine $T^{(4)}$.

If we take 
\eqn\susygt{
\del Y \cirk g^{(3)} \cirk \del \bY   =
x \; \tr \big (  (\d \bY \d Y ) \, ( \bY Y ) \big ) +
y \; \tr \big (  (\d \bY  Y ) \, ( \bY \d Y ) \big )  \, ,
}
then \varKT, with one loop results for $\gamma$, generates
\eqn\varabxy{\eqalign{
& \Delta a_1 = 2x \, , \ \ \Delta a_2 = x + 3y \, , \ \ 
\Delta a_3 = 4y \, , \ \ \Delta a_4 = 2x \, , \ \ \Delta a_5 = 3x+ y \, , \cr 
& \Delta b_1 = 2x \,  , \ \ \Delta b_2 = 2x+y \, , \ \  \ \Delta b_3 = 3y
\, , \ \ \Delta e = -3x + y \, , \ \ \Delta f = - \half(x-y) \, ,
}
}
so that $\Delta X_1 = 3x +2y, \ \Delta X_2 =2y, \ \Delta X_3 = x, \ \Delta X_4
= \half (x+y)$. Corresponding to \gone, along with \varab, we have in addition
\eqn\varz{
\Delta a_1 = \Delta a_2 = \Delta a_5 = \Delta e = - 3z \, , \qquad
\Delta a_3 = -6z \, ,
}
which entails $\Delta X_1 = -3z, \ \Delta X_2  = - 6z$. There is one
invariant under \varabxy\ and \varz
\eqn\XXXX{
2\, X_1 - X_2  - 4 \, X_3 - 4\, X_4 = \half \, A - B - \quar \, C = - \quar  \, ,
}
imposing the numerical results in \ABCD. The freedom in \varabxy\ may be used
to set $\del' Y \cirk K^{(4)} \cirk \del Y =0$.

The results for $T$ in \susyG, \susyGt\  and \Tfour\ determine the
metric $G$  at each order. It is of interest to consider whether this
is K\"ahler so that
\eqn\Kahler{
\d Y \! \cirk G \cirk \d \bY = \d_Y \d_{\smash \bY} F \, .
}
It is possible to construct $F$ so long as the freedom
due to variations as in \varKT\ and \varKTn, or equivalently \varga,
are allowed for. From \susyG, \susyGt
\eqn\Grest{
F^{(2)} = {\ts{1\over 3}} \, \tr \big ( (\bY Y) \big ) \, , \qquad
F^{(3)} =-  {\ts{1\over 4}} \, \tr \big ( (\bY Y) ^2\big ) \, ,
}
 if we use \varab\ to set 
\eqn\resab{
a=b = - \half \, .
}
At the next order the general expression has the form
\eqn\Grest{
F^{(4)} ={\hat a} \,  \tr \big ( (\bY Y)(\bY Y\bY Y)  \big ) +
 {\ts {1\over 3}}{\hat b}\,  \tr \big ( (\bY Y) ^3 \big )
 +{\ts {2\over 9}} D \, \tr \big (\gamma_D \big )  \, ,
} 
and then \Tfour\ and \Kahler\ require
\eqn\abK{
a_1=a_3 = 2{\hat a} \, , \ a_4 = {\hat a} \, , \
a_2 = 2{\hat a}  + \lambda \, , \ a_5 = 2{\hat a} -\lambda \, , \
b_1=b_2=b_3={\hat b} \, .
} 
for arbitrary $\lambda$ since $G^{(5)}$ depends only on $a_2+a_5$.
Imposing the conditions in \integfive\ 
is possible only by  choosing a scheme with  $A=B=0,C=1$ and then
${\hat a},{\hat b}$ as well as $e,f$ are determined so that 
\eqn\resabef{
 {\hat a} = {\hat b} =1\, , \qquad
e = -1 -\lambda \, , \quad  f =  \quar + \half \lambda \, ,
}
giving $X_1={5\over 2}, \, X_2 = {7\over 4}, \, X_3 = \half , \, 
X_4= {\ts {3\over 8}}$. 

For $\N=1$ supersymmetric theories there is, at critical points with
vanishing $\beta$-functions, an exact expression for $a$
\AnselmiAM\ in terms of the anomalous dimension matrix $\gamma$
or alternatively  the $R$-charge $R= {2\over 3}(1+\gamma)$.
Introducing terms linear in $\beta$-functions there is 
a corresponding expression which is valid  away from critical points
and this can then
be shown to  satisfy many of the properties associated with the
$a$-theorem \refs{\BarnesJJ, \KutasovXU}.
For the theory considered here, with $n_C$ chiral scalar multiplets,
these results take the form
\eqn\Aexact{
{\tilde A} = {\ts{1\over 12}}\, n_C - \half \, \tr(\gamma^2) + 
{\ts {1\over 3}} \, \tr(\gamma^3) + \Lambda \cirk \beta_\bY 
+  \beta_Y \cirk H \cirk \beta_\bY \, ,
}
where we require\foot{In \BarnesJJ\ and 
\KutasovXU\ $\Lambda$ plays the role of a Lagrange multiplier
enforcing constraints on the $R$-charges. At lowest order the
result for $\Lambda$ and also the metric $G$ 
obtained in \BarnesJJ\ are equivalent, up to matters
of definition and normalisation, with those obtained later here and
in \susyG.} 
\eqn\GaMB{
\Lambda \cirk \beta_{\smash\bY} = \beta_Y \cirk {\bar \Lambda} \, , 
\qquad H = {\bar H} \, .
}
$\Lambda$ is determined in \Aexact\ up to terms which may
be absorbed in $H$ so that 
$ \Lambda \cirk \d \bY \sim \Lambda \cirk \d \bY
+ \beta_Y \cirk g \cirk \d \bY $.
Assuming the result \Aexact\ for $\tilde A$ satisfies \ST\ then
$H$ is arbitrary as a consequence
of  \varA. 

However $\Lambda$ is constrained by imposing \ST.
Defining $(\bY\Lambda )_i{}^j $ in a similar fashion to \WYr, then
\eqn\Dgam{\eqalign{
\d_Y  \big ( - \half \, \tr(\gamma^2) + 
{\ts {1\over 3}} \, \tr(\gamma^3)
 +  \Lambda \cirk \beta_\bY \big )
 = {}& \tr \big (  \d_Y \gamma  \; \big ( 
3\, (\bY\Lambda )  - \gamma + \gamma^2 \big ) \big  ) \cr
&{} + \big ( \d_Y \Lambda \big )  \cirk \beta_\bY  \, .
}
}
Hence if $\Lambda$ is required to obey\foot{More generally if
$3\, (\bY \Lambda ) = \gamma - \gamma^2 + \Theta \cirk \beta_\bY 
+  [\Xi , \, \beta_Y \!  \cirk \pr_Y \gamma ] $, 
$\d' Y \! \cirk K \cirk \d Y  = 
 \tr  ( \Xi \;[ \d'{\!}_Y \gamma , \d_Y \gamma ]  ) $. Such a term
 can be removed by considering changes as in \varTK.}
\eqn\detL{
3\, (\bY \Lambda ) = \gamma - \gamma^2 + \Theta \cirk \beta_\bY 
\, , \qquad
\Theta\cirk \d \bY \in {\frak{ g l}}(n_C, {\Bbb C}) \, , 
}
then \Aexact, excluding the $H$ term, satisfies \ST\ if we take
\eqn\resGK{\eqalign{
\half \, \d Y \! \cirk T \cirk \d \bY = {}&
\tr \big ( \d_Y \gamma \; \Theta \cirk \d \bY \big ) + 
 \d_Y \Lambda \cirk  \d \bY + \half \, \d Y \! \cirk T' \cirk \d \bY \, ,\cr 
\d' Y \! \cirk K \cirk \d Y  = {}& 0 \, , \qquad
\d Y \! \cirk T' \cirk \beta_\bY  = 0 \, .
}
}
A related result, with effectively $\Theta=0$, is contained in
\BarnesJJ. For supersymmetric theories,
satisfying \detL\ is consequently essentially equivalent to 
requiring \ST, although terms involving  $\Theta$
are necessary at higher orders. The relations \detL\ and \resGK\
are not invariant under variations of $g$
as in  \GaMB\ and so this freedom is no longer present.

Since $\gamma$ is hermitian a corollary of \detL\ is that $\Lambda,\Theta$
must satisfy
\eqn\relLaTh{
3\, (\bY \Lambda ) - 3 \, ({\bar \Lambda} Y ) =  \Theta \cirk \beta_\bY
- \beta_Y \cirk {\bar \Theta} \, .
}
This is essentially identical to \relWQ\ and suggests a relation between
$\Lambda,\Theta$ and $W,Q$ but a precise connection is as yet unclear.
 
For variations as in \varAt\ and \varga\ then compatibility with \Aexact\
requires
\eqn\varLa{
\delta \Lambda \cirk \d \bY = - (Y*h) \cirk \pr_Y \Lambda \cirk \d \bY
 + \delta' \Lambda \cirk \d \bY  \, ,
}
where $ \delta' \Lambda$ satisfies, assuming \detL,
\eqn\varLap{
\delta' \Lambda \cirk \beta_{\smash \bY} = -  \beta_{\smash \bY} \cirk
S \cirk \beta_{\smash \bY} \, , \qquad
\d \bY \cirk S \cirk \d \bY = 
\tr \big ( \d_\bY h \; \Theta\cirk \d \bY \big ) \, .
}
Furthermore, \detL\ is also invariant if
\eqn\varTh{
\delta \Theta\cirk \d \bY  = - (Y*h) \cirk \pr_Y \Theta  \cirk \d \bY
- \d_\bY h + \d_\bY h \; \gamma + \gamma \; \d_\bY h 
- \Theta  \cirk ( \d_\bY h * \bY ) + \delta' \Theta \cirk \d \bY \, ,
}
so long as
\eqn\varThpt{
3\, (\bY \delta'\Lambda) =  \delta' \Theta \cirk \beta_{\smash \bY} \, .
}
This can be solved subject to \varLap\ by taking
\eqn\solLT{
\delta ' \Lambda \cirk \d \bY = - \beta_{\smash \bY} \cirk
S \cirk \d  \bY \, , \qquad 
\delta' \Theta \cirk \d \bY = - 3 \, (\bY \, \d \bY \cirk S ) \, .
}
Using \varLa,\varTh,\solLT\ in \resGK\ generates variations
in agreement with \varKTn\ up to contributions which may be absorbed
in $T'$. Such variations generate terms in $\Theta$ which are 1PR. 
Also we may show $\delta ( \Lambda \cirk \beta_{\smash \bY}
- \beta_Y \cirk {\bar \Lambda}) =0$ subject to
$( \beta_{\smash \bY} \cirk \pr_{\smash \bY} - \beta_Y \cirk \pr_Y) h
= [ \gamma , h ]$.

The perturbative results obtained here for $\tilde A$ may be expressed in 
the form \Aexact, although this can require additional constraints on $\gamma$
beyond those required for integrability of \ST. As was already shown in 
\SusyA\ the low order results in \susyA,
with the one and two loop expressions for $\gamma$ in \otl,
can be expressed in the form \Aexact. At lowest order it is necessary that
\eqn\GHone{
\Lambda^{(2)} \cirk \d \bY = {\ts {1\over 6}} \, Y \! \cirk \d \bY
\quad \Rightarrow \quad 3\, (\bY \Lambda^{(2)} ) = \gamma^{(1)}   \, .
}
In general at the next order we may take
\eqn\GHtwo{
\Lambda^{(3)} \cirk \d \bY = \lambda \,  
\tr \big ( ( \d \bY Y ) \, (\bY Y) \big ) \, , \qquad 
\Theta^{(2)} \cirk \d \bY = \theta \, (\d \bY Y) \, .
}
In this case 
\eqn\Latwo{
3\, (\bY \Lambda^{(3)})  - \Theta^{(2)} \! \cirk \beta_\bY{\!}^{(1)} 
= (\lambda - \half \theta) \,  \big ( 2 (\bY Y \bY Y ) + (\bY Y)^2 \big ) \, .
}
Equating this to $\gamma^{(2)} - \gamma^{(1)2}$, in accord with \detL,
requires
\eqn\resl{
\lambda - \half \theta = - {\ts{1\over 4}}  \, ,
}
and determines $\gamma^{(2)}= - \half (\bY Y \bY Y) $ just as in \otl.
Using \GHtwo\ in \resGK\ is compatible with \susyGt\ for
\eqn\ablt{
a = 2\lambda =-\half + \theta \, , \qquad 
b = 2 \lambda + \theta = - \half +  2 \theta  \, .
}
For ${\tilde A}^{(4)}$ given by  \susyA, \Aexact\ is then satisfied 
with $H^{(2)}=0$. 

At the next order there are several terms which may contribute to
 $\Lambda^{(4)}$ and $\Theta^{(3)}$ in \detL. The general form is 
\eqn\Lfour{\eqalign{
\Lambda^{(4)} \cirk \d \bY = {}& 
\alpha  \, \tr \big ( ( \bY Y ) \, (\d \bY Y \bY Y ) \big )
+  \beta \, \tr \big ( ( \d \bY Y ) \, (\bY Y \bY Y ) \big ) \cr
&{}+ \gamma \, \tr \big ( ( \d \bY Y ) \, (\bY Y)^2 \big ) 
+ D \, {\ts {1\over 9}} d_\bY \tr \big ( \gamma_D \big ) \, , \cr
\Theta^{(3)} \! \cirk \d \bY = {}& \sigma \, (\d\bY Y \bY Y ) + \tau \,
(\bY Y \d \bY Y ) + \mu \, (\d \bY Y) ( \bY Y ) + 
\nu \, (\bY Y) ( \d \bY Y ) \, .
}
}
Imposing now
\eqn\Lathree{\eqalign{
3\, (\bY \Lambda^{(4)}) - \Theta^{(3)}\! \cirk \beta_\bY{\!}^{(1)} 
 - \Theta^{(2)}\! \cirk \beta_\bY{\!}^{(2)} =
 \gamma^{(3)}-  \gamma^{(1)} \gamma^{(2)} - \gamma^{(2)} \gamma^{(1)}\, , 
}
}
determines $\gamma^{(3)}$ with 
\eqn\fourres{
A = 2 \gamma - \half (\sigma+\tau) \, , \qquad
B = \alpha - \half \sigma \, , \qquad C = 2 \beta - \tau + \theta \, ,
}
and from the 1PR contributions the additional relations
\eqn\fouresa{
\beta - \half \sigma - \nu = \quar \, , \qquad
2 \alpha - \mu + \half \theta = \quar \, , \qquad \gamma -
\half (\mu+\nu) = 0 \, .
}
These equations require the constraint on $\gamma^{(3)}$
\eqn\conABC{
A-2B-\half C  = - \half \,  ,
} 
which is satisfied by the calculated results \ABCD.

For simplicity we may assume $\Theta^{(2)}$ is restricted to just
1PI contributions so that $\mu= \nu = \gamma=0$. Then 
using \Lfour\  with \fourres, \fouresa\ in  \resGK\ gives contributions to 
$T^{(4)},K^{(4)}$ of the form \Tfour, \Kfour\ with
\eqn\resabc{\eqalign{
&{} a_1=a_2 = 2\beta = {\ts{3\over 4}} - 2B-\half \theta  \, , \quad a_3 = 
2\beta+\tau = \half - 2A \, , \quad 
a_4=2\alpha = \quar -\half \theta \, , \cr 
&{} a_5 = 2\beta -\theta = {\ts{3\over 4}} - 2B - {\ts {3\over 2}}  \theta \, , \qquad
 b_1 =  \ b_2 = \ b_3 = 0 \, , \quad e = f =0 \, ,
}
}
which is compatible with \integfive\ for 
$X_1 = {\ts{5\over 8}}-B- {3\over 4}\theta, \, 
X_2 = 1-2B-\half\theta,  \, X_3={\ts{1\over 8}}- \quar \theta$ and 
$ X_4 = 0$ so long as $a,b$ satisfy \ablt. With these results we may check
\eqn\Afivech{\eqalign{
{\tilde A}^{(5)} = {}& - \tr \big ( \gamma^{(1)} \gamma^{(3)} \big ) - \half \, 
\tr \big ( \gamma^{(2)2} \big ) + \tr \big ( \gamma^{(1)2} \gamma^{(2)} \big )  \cr
&{}+ \Lambda^{(2)} \! \cirk \beta_\bY{\!}^{(3)}  + \Lambda^{(3)} \! \cirk \beta_\bY{\!}^{(2)} 
+ \Lambda^{(4)} \! \cirk \beta_\bY{\!}^{(1)} \, ,
}
}
as required by \Aexact\ to this order with $H=0$. The results for
$\Lambda$ may be expressed 
also in the form
\eqn\Wlam{\eqalign{
 \Lambda^{(2)} \cirk \d \bY = {}& \d_\bY \, {\ts{1\over 6}} \tr \big ( (\bY Y)\big )\, , \qquad
\Lambda^{(3)} \cirk \d \bY = \d_\bY \, \half \lambda\,  \tr \big ( (\bY Y)^2\big ) \, , \cr
\Lambda^{(4)} \cirk \d \bY = {}& 
(\alpha - \half \beta)  \, \tr \big ( ( \bY Y ) \, (\d \bY Y \bY Y ) \big )\cr
&{} + \d_\bY \big ( \half  \beta \, \tr \big ( (  \bY Y ) \, (\bY Y \bY Y ) \big )
+ {\ts {1\over 9}}  D \,  \tr \big ( \gamma_D \big ) \big )  \, .
}
}

At higher orders the number of potential constraints increases when
the number of inequivalent lines of a $(\ell+1)$-loop vacuum graph, 
related to the number of terms  in $\gamma^{(\ell)}$, becomes larger 
than the number of inequivalent
vertices, which are related to possible contributions to $\Lambda^{(\ell+1)}$.
The calculations of \FerreiraRC\ for $\gamma^{(4)}$ in terms of $Y,\bY$ 
correspond to 11 distinct graphs which are related to 6 5-loop
vacuum graphs giving 13 possible $\Lambda^{(5)}$. However the number of
independent terms in  $\gamma^{(4)}$ may be reduced by considering
redefinitions as in \varga\ with $h\propto \gamma_A,\gamma_B,\gamma_C,\gamma_D$
and letting $\beta_\bY \to \beta_\bY{\!}^{(1)}, \, \gamma\to \gamma^{(1)}$. By
taking $h= {3\over 4} \zeta(4) \, \gamma_D$ all terms, corresponding
to non planar graphs which contain the $\gamma_D$ subgraph, involving $\zeta(4)$
in the expression given in \FerreiraRC\ are generated by \varga. There are 7 
planar graphs relevant for $\gamma^{(4)}$ and applying \detL\ in conjunction 
with lower order contributions gives one relation, 
which is invariant under changes of scheme and is analogous to \conABC,  
amongst the coefficients. This is satisfied by results of \FerreiraRC.

Some calculations checking the validity of the  essential equations \ST\ 
or \detL\ at each loop order when new transcendental numbers appear
 are also undertaken in Appendix A.

\newsec{Renormalisation with Local Couplings}

The results derived in section 2 can be specialised to
renormalisable quantum field theories when
the metric $G_{IJ}$ and other quantities may be
calculated in a perturbative loop expansion on a curved
space background. Within the framework of dimensional
regularisation with minimal subtraction on flat space
there is also a precise prescription for determining 
quantities, such as $S_{IJ}$ and $W_I$, which are initially
defined in terms of contributions involving $\pr_\mu \sigma$, 
in terms of the $\sigma$-independent counterterms,
necessary for a finite theory, which  are simple poles in
$\vep = 4-d$.

To demonstrate this we consider initially
a generic  renormalisable quantum field theory described by
a Lagrangian density $\L$  formed from  fields  $\Phi$  and their 
conjugates ${\bar \Phi}$ depending on local couplings $\{g^I(x)\}$  
 for a complete set of marginal operators $\{\O_I(x)\}$.
 For renormalisability $\L$ must contain background gauge fields
$\{a_\mu(x)\}$   and   local couplings $\{ M(x)\}$
for all relevant dimension two operators, corresponding to
contributions to $\L$ of the form $\L_{M} = - {\bar \Phi} \, M \, \Phi$.
In  $\L$ the kinetic terms, which are bilinear in the scalar/fermion
fields $\Phi$ and their conjugates ${\bar \Phi}$ and have the form
$\L_{K} = - {\bar \Phi} \, \K(\pr) \, \Phi$, are invariant under 
a maximal symmetry group ${\rm G}_K$  where, for any ${\rm g}\in {\rm G}_K$, 
$\Phi \to {\rm  g} \, \Phi$ and ${\bar \Phi} \to {\bar \Phi}\, {\bar {\rm g}}$ 
we require ${\bar {\rm g}}\, {\rm g} =1$,
$ {\bar {\rm g}}\,  \K(\pr) \,{\rm g} =  \K(\pr)$. 
For infinitesimal transformations corresponding to the associated Lie 
algebra ${\frak g}_K$ then for 
$\omega \in {\frak g}_K$, $\omega + {\bar \omega}=0$.
In general ${\rm G}_K$ is not simple but is a product of $U(n)$'s or $O(n)$'s. 
The symmetry ${\rm G}_K$ extends to the complete action
$\L$ if the couplings are also transformed appropriately, 
so that for any $\omega \in {\frak g}_K$ then $\delta g^I$ is given by \transga.
A local symmetry  ${\rm G}_K$  is obtained as usual by replacing
all derivatives   in $\K(\pr)$ by appropriate covariant derivatives
$ D_\mu = \pr_\mu + a_\mu$ for $a_\mu(x) \in {\frak g}_K$. In 
general then $ \L(\Phi,{\bar \Phi}, g , a , M) $.

As usual a finite quantum field theory in a perturbative expansion 
obtained from $\L$
is achieved at each order by adding appropriate local counterterms
 $\L_{\rm c.t.}$. As well as counterterms involving $\Phi,{\bar \Phi}$
 with $x$-dependent couplings, additional local
contributions independent of the fields involving  contributions
containing $\prod_i \pr^{m_i} g^{I_i}$ with  $\sum_i m_i \le 4$ and
also $f_{\mu\nu}$ as defined in \cura, are 
also necessary. Assuming an invariant regularisation then
all derivatives of the couplings are extended to covariant derivatives, 
$\pr_\mu g^I \to D_\mu g^I$, as in \deg.  RG equations are  obtained
by assuming that $\L$ is such that the bare Lagrangian generating
a finite perturbation expansion order by order is
\eqn\bareL{\eqalign{
\L_0 ={}&  \L(\Phi,{\bar \Phi}, g , a , M) + 
\L_{\rm{c.t.}}(\Phi,{\bar \Phi}, g, a , M) \cr
={}&   \L(\Phi_0,{\bar \Phi}_0, g_0 , a_{0} , M_0 )
  - {1\over 16\pi^2}\, \X(g , a, M) \, .
}
}
$\X$ includes all the extra field independent counterterms and is
arbitrary up to total derivatives.
Assuming dimensional regularisation with
minimal subtraction, then in a loop expansion
\eqn\bareloop{
\L_{\rm{c.t.}}(\Phi,{\bar \Phi}, g, a , M)^{(\ell)} =
\sum_{r=1}^\ell \L_{\rm{c.t.}}(\Phi,{\bar \Phi}, g, a , M)^{(\ell)}_r \,
{1\over \vep^r} \, ,
} 
so that $\X$ contains just poles in $\vep$.

The RG flow equations which are  considered here are obtained from
\eqn\rgflow{\eqalign{
\bigg ( \vep \, \sigma - \D_\sigma - \D_{\smash{\sigma,\Phi,{\bar \Phi}}}
- (2-\vep) \, \pr_\mu \sigma \, D^{\mu} & g^I {\pr \over \pr D^2 g^I } \bigg  ) 
\L(\Phi_0,{\bar \Phi}_0, g_0 , a_{0} , M_0 )\cr
\noalign{\vskip -2pt} 
{}= {}&   \pr^\mu \bigg ( \pr_\mu \sigma \,  T \cdot {\pr \over \pr M} \, 
\L(\Phi_0,{\bar \Phi}_0, g_0 , a_{0} , M_0 ) \bigg ) \, , }
}
where $\sigma$ is linear in $x$, of the same form as $\sigma_v$ in \solKv,
and the right hand side for $T\in V_M$ is a potential total derivative 
contribution when $\sigma$ is not constant which can be neglected in 
the subsequent discussion.
In \rgflow\ $\D_\sigma, \D_{\sigma,\Phi,{\bar \Phi}} $ are derivatives defined by
\eqn\defDsi{\eqalign{
\D_\sigma ={}&  \sigma {\hat \beta}^I  \cdot {\pr \over \pr g^I} 
+ \big ( \sigma  \rho_I D_\mu g^I - \pr_\mu \sigma\,  \upsilon \big ) \cdot 
{\pr \over \pr a_\mu } \cr
&{}+ \big ( \sigma  ( { \gamma}_M M -
 \delta_I  \, D^2 g^I - \epsilon_{IJ}\, D^\mu g^I D_\mu g^J)
-  2\, \pr_\mu \sigma\,  \theta_I D^\mu g^I   \big ) \cdot 
{\pr \over \pr M} \, , \cr
\D_{\sigma,\Phi,{\bar \Phi}} ={}& 
\big  ( \sigma \,  (\half \vep - \gamma) \, \Phi \big )\cdot  {\pr \over \pr \Phi}
+ \big  (\sigma\,  {\bar \Phi}\, (\half \vep - {{\bar \gamma}} ) \big )\cdot
{\pr \over \pr {\bar \Phi}}  \, .
}
}
Here $\D_\sigma, \D_{\sigma,\Phi,{\bar \Phi}}$
act on local functions of $g^I,a_\mu,M, \Phi,{\bar \Phi}$ and their derivatives so that
for instance acting  on $f(g(x),\pr_\mu g(x))$,
${\scriptstyle h}\cdot {\pr \over \pr g }  = {\scriptstyle h(x)} 
{\pr \over \pr g(x) } + {\scriptstyle \pr_\mu h(x)} {\pr \over \pr \pr_\mu g(x) } $. 
The action of $\D_\sigma$ is then equivalent to the corresponding contributions
to the functional derivative operator 
$\Delta_\sigma + \Delta_{\sigma,a} + \Delta_{\sigma,M}$ defined by \defD, 
\defDa\ and \defDM\ although $\beta^I \to {\hat \beta}^I$.
A derivation of \rgflow\ is sketched in Appendix B.

For the marginal couplings $g^I$ \Bhat\ becomes
\eqn\Bhatr{
\hhbet^I(g) = - \vep \, k_I  g^I + \beta^I(g) \, ,
}
and minimal subtraction ensures that $\beta^I(g) $ is independent of $\vep$.
In a loop expansion
\eqn\loops{
\big ( 1 + {\ts {\sum_I}}\,  k_I g^I \cdot \pr_I - \half  \Phi\cdot \pr_\Phi -
\half  {\bar\Phi}\cdot \pr_{\bar\Phi} \big ) \L_{\rm{c.t.}}{\!}^{(\ell)}
= \ell \,  \L_{\rm{c.t.}}{\!}^{(\ell)} \, .
}

Amongst the counterterms in $\L+ \L_{\rm c.t.}$  for constant $g^I$ the 
quadratic kinetic terms are in general  modified just  by the introduction of 
an appropriate matrix 
${Z}(g) = {\bar Z}(g) = 1 + {\rm O}(g)$, $\L_{K} \to  
- {\bar \Phi} \,  Z \, \K(\pr) \, \Phi$. This determines the anomalous 
dimension matrices $\gamma(g), {\bar \gamma}(g)$ for the fields 
$\Phi,{\bar \Phi}$ in \rgflow\ through
\eqn\Zrg{
{\hat \beta}^I(g){\pr \over \pr g^I}\, Z(g) =
{\bar\gamma}(g) \, Z(g) + Z(g) \, \gamma(g) \, .
}
At $\ell$ loops, with $Z^{(\ell)}$ expanded as in \bareloop, \Zrg\ 
requires $ \gamma^{(\ell)} +{\bar \gamma}^{(\ell)}
= - \ell \, Z^{(\ell)}_1$.
The standard prescription determines $ \gamma^{(\ell)}(g)$ 
by assuming
${\bar \gamma}(g)=\gamma(g)$ so that the eigenvalues are real.
In obtaining RG equations describing the RG flow it is necessary to 
factorise $Z$, 
\eqn\ZZZ{
Z = {\bar \Z} \, \Z \, ,
}
so that  in \bareL
\eqn\barep{
\Phi_0 = \Z \Phi \, , \qquad {\bar \Phi}_0 = {\bar \Phi} {\bar \Z} \, .
}
The factorisation in \ZZZ\ has an essential arbitrariness generated by 
infinitesimal variations $\delta \Z = \omega \, \Z,\ \delta {\bar \Z}=  
{\bar \Z}\,{\bar \omega}= - {\bar \Z} \, \omega$ for $\omega \in  {\frak g}_K$. 
The RG equations for $\Z$ then take the form, from \Zrg,
\eqn\rgg{
{\hat \beta}^I(g){\pr \over \pr g^I}\, \Z(g) = \omega(g) \, \Z(g)
+ \Z(g) \, \gamma(g) \, , \qquad \omega(g) \in {\frak g}_K  \, .
}
Assuming ${\bar \gamma}=\gamma$ and taking $\Z^{(1)}= \half Z^{(1)}$,
$\Z^{(2)}= \half Z^{(2)}- {1\over 8} Z^{(1)}{}^2$ then combining \Zrg\ and
\rgg\ gives $\omega^{(2)} = \quar [ \gamma^{(1)} , Z^{(1)} ] = 0 $ but
$\omega^{(3)} = \quar [ \gamma^{(2)} , Z^{(1)} ] +  \quar 
[ \gamma^{(1)} , Z^{(2)} ]$ may be non zero. It is possible to choose
$\Z$ so that in \rgg\ $\omega=0$ but then ${\bar \gamma}\ne \gamma$
in general.

In \bareL
\eqn\azero{
a_{0 \mu} = a_\mu + \fN_I D_\mu g^I \, , \qquad  
\fN_I \in {\frak g}_K \, , 
}
is determined so that all terms involving derivatives of 
$\Phi$ or ${\bar \Phi}$  in $\L_{\rm {c.t.}}$ are absorbed 
by letting $\Phi, {\bar \Phi} \to \Phi_0, {\bar \Phi}_0$
and $D_\mu \Phi, D_\mu {\bar \Phi} \to D_{0\mu}\Phi_0, 
D_{0\mu}{\bar \Phi}_0$ with  $D_{0 \mu} = \pr_\mu +  a_{0 \mu}$.
Hence $\L_{K\, 0} = - {\bar \Phi}_0 \, \K(D_0) \, \Phi_0$ up to
total derivatives. 
The RG equation from \rgflow\ then requires from \rgg
\eqn\rga{
\D_\sigma a_{0\mu}= - D_{0\mu} ( \sigma \, \omega ) 
= - \pr_\mu  ( \sigma \, \omega ) - \sigma\,  [ a_{0\mu} , \omega ]  \, .
}

The resulting equations from the terms in \rga\ proportional to
$\sigma$ and $\pr_\mu \sigma$ become
\eqn\rgN{
\wL_{\smash{\hB,\hrho}}\,  \fN_I  + \hrho_I = \pr_I (\upsilon - \omega)
+ [  \fN_I \, , \, \upsilon - \omega ] \, , 
}
for 
\eqn\Bhat{
\hB^I = {\hat \beta}^I - (\upsilon g)^I \, ,
}
and
\eqn\upeq{
 \fN_I  \hB^I = \upsilon - \omega \, .
}
Assuming $ \fN_I , \omega$ contain only poles in $\vep$,
so that $ \fN_I = \sum_{n\ge 1} \fN_{I,n}\, \vep^{-n}$,  the
${\rm O}(1)$ terms in \rgN\ and \upeq\ determine $\hrho_I,\upsilon $ 
\eqn\Nloop{
\hrho_I =  {\ts {\sum_J}}\,  k_J g^J 
(\pr_J \, \fN_{I,1} - \pr_I \,  \fN_{J,1} ) \, , \qquad
\upsilon =-  {\ts {\sum_I}}\,  \fN_{I ,1} k_I g^I\, .
}
Since $\sum_I \hrho_I  \, k_I g^I = 0$ then contracting \rgN\ with $\hB^I$ and
using \upeq\ shows that these equations require
\eqn\conBp{
\hrho_I  \hB^I = \hrho_I  B^I = 0 \, ,
}
in agreement with \Brho.

The counterterms contained in $M_0$, where 
$\L_{M\, 0} = - {\bar \Phi}_0\,  M_0\, \Phi_0$, have  the general form
\eqn\Mzero{
 M_0 = Z_M  \big ( M  - \fd_I D^2 g^I - \fe_{IJ}D^\mu g^I D_\mu g^J\big ) \, ,
}
with $\fd_I, \fe_{IJ} \in V_M , \, Z_M : V_M \to V_M$. 
\rgflow\ then implies
\eqn\rgM{
\bigg ( \D_\sigma 
+ (2-\vep) \, \pr_\mu \sigma \, D^{\mu} g^I {\pr \over \pr D^2 g^I } \bigg )M_0
=  \sigma \, \big [ \omega,\,  M_0 \big ] \, .
}
This decomposes into
\eqn\gamm{
\hhbet^I {\pr \over \pr g^I} Z_M - [ \omega , \, Z_M ] = - Z_M \, \gamma_M \, ,
}
which determines $\gamma_M{\!}^{(\ell)} = \ell \, Z_{M\,1}^{(\ell)}$, and
\eqn\gmm{\eqalign{
- \big ( \wL_{\smash { \hB, \hrho}} - \gamma_M \big )\,  \fd_I 
= {}& \delta_I \, ,\cr
- \big ( \wL_{\smash { \hB , \hrho} } - \gamma_M \big ) \, \fe_{IJ} -
{\hat \Omega}_{IJ}{}^K  \fd_K = {}& \epsilon_{IJ} \, , \cr
- {\hat \Psi}_I{}^J \fd_J -\fe_{IJ} \, \hB^J = {}& \theta_I \, ,
}
}
for ${\hat \Psi}_I{}^J = (1-\half \vep) \, \delta_I{}^J + \pr_I \hB^J + 
\half ( {\hrho}_I  g)^J$ and ${\hat \Omega}_{IJ}{}^K$ as in \deflam\
with $B\to \hB$. \gmm\ then determines the $\vep$ independent
$\delta_I, \epsilon_{IJ}$ and
\eqn\deftheta{
\theta_I = (k_I + \half ) \, \fd_{I,1} + {\ts {\sum_J}}\, \fe_{IJ,1} k_J g^J \, .
}
By virtue of \conBp, \comLPsi\ also 
extends to $\big [ \wL_{\smash{\hB,\hrho}} \, , \, {\hat \Psi}_I{}^J \big ] = 
{\hat \Omega}_{I K}{}^J  \hB^K$ so that we may obtain directly from
\gmm\  the finite relation
\eqn\deth{
\big  (\wL_{\smash{\hB,\hrho}} - \gamma_M \big ) \theta_I 
= {\hat \Psi}_I{\!}^J \delta_J + \epsilon_{IJ} \hB^J \, ,
}
for which the 
${\rm O}(\vep^0)$ contribution is identical to \derel\ while the
${\rm O}(\vep)$ terms equivalently determine $\theta_I$ in terms
of $\delta_I, \epsilon_{IJ}$.

The additional field independent local counterterms in \bareL\
may be reduced, by discarding total derivatives, to the form
\eqn\counter{\eqalign{
\X( g, a , M) 
= {}& {\ts{1\over 2}} \A_{IJ}\, D^2 g^I D^2 g^J 
+ \B_{IJK}\, D^2 g^I \,  D^\mu g^J D_\mu g^K \cr
&{} + \half \, 
\C_{IJKL}\, D^\mu g^I D_\mu g^J \, D^\nu g^K D_\nu g^L \cr
&{} + \quar \, f^{\mu \nu} \cdot \L_f \cdot f_{\mu \nu} 
+\half \,  M \cdot \L_M \cdot M  
+ f^{\mu \nu} \cdot \P_{IJ} \, D_\mu g^I D_\nu g^J \cr
&{} +\J_{I}\cdot  M\,  D^2 g^I + \K_{IJ} \cdot M\,  D^\mu g^I D_\mu g^J  \, .
}
}
Assuming this expression  the flat space 
contributions $X,Y$ in \modW\ are determined through the RG equation
\eqnn\rgeq$$\eqalignno{
\bigg (  \vep \, \sigma&{}  - \D_\sigma  - (2-\vep) \, 
\pr_\mu \sigma \, D^{\mu} g^I {\pr \over \pr D^2 g^I } \bigg ) \X(g, a , M)  
- \sigma \, X( g ,a,M) + 2\, \pr_\mu \sigma  \, Y^\mu( g,a,M) \cr
= {}& - 2\, \pr_\mu \sigma \, \big ( \pr_\nu ( \G_{IJ} \, D^\mu g^I D^\nu g^J )
- \half \, \pr^\mu (  \G_{IJ} \, D^\nu g^I D_\nu g^J ) \big ) \cr
= {}& - 2\, \pr_\mu \sigma \, \big ( \G_{IJ}  D^\mu  g^I D^2 g^J 
- \G_{IJ}  (f^{\mu \nu}g)^I D_\nu g^J + \Gamma^{(\G)}{\!}_{IJK}
D^\mu g^I  D^\nu g^J D_\nu g^K \big ) \, , & \rgeq
}
$$
allowing on the right hand side a total derivative which generates
terms of the same form as in $\X$ and $X,Y^\mu$ as given by \rgLa. 
To obtain \rgeq\ we assume that $\G_{IJ} = \G_{JI}$
satisfies $(\omega g)^K \pr_K \G_{IJ} + \G_{KJ}\omega^K{\!}_I +
\G_{IK} \omega^K{\!}_J =0$. The contributions in \rgeq\ arising from 
$\G_{IJ}$ are the same form as the terms in $Y^\mu$ which involve
$S_{(IJ)}, T_{IJK}, Q_I$ so $\vep$-independent contributions to $\G_{IJ}$
give rise to a corresponding ambiguity in $Y^\mu$. This freedom
is removed by requiring that $\G_{IJ}$ contains only poles in $\vep$.

Decomposing \rgeq\ we find for the $M$-dependent terms
\eqn\rgMM{\eqalign{
\big ( \vep - \wL_{\smash { \hB, \hrho}}\big )\, \J_I  - \J_I \cdot \gamma_M 
+ \delta_I \cdot \L_M = {}& J_I \, ,\cr
\big ( \vep - \wL_{\smash { \hB , \hrho} }\big ) \, \K_{IJ}  - 
\K_{IJ} \cdot \gamma_M - {\hat \Omega}_{IJ}{}^K  \J_K +
\epsilon_{IJ} \cdot \L_M = {}& K_{IJ} \, , \cr
{\hat \Psi}_I{}^J \J_J + \K_{IJ} \, \hB^J - \theta_I \cdot \L_M
= {}& L_I \, ,
}
}
which determine $J_I,K_{IJ},L_I$ so that
\eqn\resLJK{
L_I = - (k_I+\half ) \,  \J_{I,1}
- {\ts {\sum_J}} \, \K_{IJ ,1} k_J g^J \, .
}  
Using \deth, and in a similar fashion,
assuming
\eqn\betmm{
\big (\vep - \wL_{\smash{\hB,\hrho}}\big ) \L_M - \gamma_M \cdot \L_M
- \L_M \cdot \gamma_M = \beta_M \, ,
}
\rgMM\ requires for consistency 
$(\vep - \wL_{\smash{\hB,\hrho}}) L_I - L_I \cdot \gamma_M 
= {\hat \Psi}_I{\!}^J J_J + K_{IJ} \hB^J - \theta_I \cdot \beta_M$
which is equivalent to \relJKL. For the contributions involving $f^{\mu\nu}$
\rgeq\ reduces to
\eqn\rgf{\eqalign{
\omega\cdot \big ( \vep - \wL_{\smash { \hB, \hrho}}\big )\, \P_{IJ}
- (\omega g)^K \hrho_K \cdot \P_{IJ}
- \half \, \omega \cdot \L_f \cdot (\pr_I \hrho_J - \pr_J \hrho_I )
= {}& \omega \cdot P_{IJ} \, , \cr
\omega\cdot \big ( \vep - \wL_{\smash { \hB, \hrho}}\big )\, \L_f \cdot \omega'
- \omega \cdot \L_f \cdot (\omega' g)^K \hrho_K 
- (\omega g)^K \hrho_K \cdot \L_f \cdot \omega' 
= {}& \omega \cdot \beta_f \cdot \omega' \, , \cr
{} - \omega \cdot \P_{IJ} \hB^J + \half \, \omega \cdot \L_f \cdot \hrho_I
+ \G_{IJ} (\omega g)^J = {}& \omega \cdot Q_I \, .
}
}
To obtain \rgf\ we presume G${}_K$ covariance as in \covD\ to ensure
$\hhbet, \rho \to \hB , \hrho$ so that for instance
$ \omega\cdot \wL_{\smash { \hB, \hrho}} \, \P_{IJ} =
\omega\cdot \wL_{\smash { \hhbet, \rho}} \, \P_{IJ} - [ \omega , \upsilon ]
\cdot \P_{IJ}$. From \rgf\ 
\eqn\resQ{
Q_I = {\ts {\sum_J}}\,  \P_{IJ,1} k_J g^J \, ,
}
and also from \rgf\ we may obtain, 
using $-  (\pr_I \hrho_J - \pr_J \hrho_I ) \hB^J = \wL_{\smash { \hB, \hrho}}\,
\hrho_I - (\hrho_I g)^J \hrho_J$,
the finite relation
\eqn\conrgf{
\omega\cdot \big ( \vep - \wL_{\smash { \hB, \hrho}}\big )\, Q_I 
- (\omega g)^J \hrho_J \cdot Q_I = - \omega \cdot P_{IJ} \hB^J 
+ \half \, \omega \cdot \beta_f \cdot \hrho_I +  G_{IJ} (\omega g)^J \, ,
}
assuming
\eqn\GGrel{
\big ( \vep - \wL_{\smash { \hB, \hrho}}\big )\, \G_{IJ} = G_{IJ} \, ,
}
with $G_{IJ}$ $\vep$-independent. For $\vep\to 0$ \conrgf\ is just \WQ{b}.
Directly from \rgf
\eqn\relQG{
\omega \cdot Q_I \hB^I = (\omega g)^I \G_{IJ} \hB^J \, .
}
Since \resQ\ ensures that $\sum_I Q_I k_I g^I = 0 $ so that $ Q_I \hB^I 
= Q_I B^I$, \WQ{b} is satisfied if
\eqn\defW{
W_I = - \G_{IJ} \hB^J \quad \Rightarrow \quad W_I = {\ts \sum_J}\,
\G_{IJ,1} k_J g^J \, .
}

The remaining equations arising from the decomposition of \rgeq\ are then
\eqn\rgeqA{\eqalign{
& \big ( \vep - \wL_{\smash{{\hB},\hrho}} \big  ) \A_{IJ}
 + 2\, \J_{(I}\cdot \delta_{J)}= A_{IJ} \, , \cr
& \big ( \vep - \wL_{\smash{{\hB},\hrho}} \big  ) \B_{IJK}
- {\hat \Omega}_{JK}{}^L \A_{IL}
 + \J_{}\cdot \epsilon_{JK} + \K_{JK} \cdot \delta_I= B_{IJK}  \, , \cr
&  \big ( \vep - \wL_{\smash{{\hB},\hrho}} \big  ) \C_{ILJK}
- {\hat \Omega}_{IL}{}^M \B_{MJK} - {\hat \Omega}_{JK}{}^M \B_{MIL}
+ \K_{IL} \cdot \epsilon_{JK} + \K_{JK}\cdot \epsilon_{IL} \cr
&\quad {} +( \pr_I \hrho_{(J} - \pr_{(J} \hrho_I) \cdot \P_{K)L}
+( \pr_L \hrho_{(J} - \pr_{(J} \hrho_L) \cdot \P_{K)I} = C_{ILJK} \, ,
}
}
and also for terms involving $\pr_\mu \sigma$,
\eqn\STrel{\eqalign{
{\hat \Psi}_I{}^K \A_{KJ} + \B_{JIK} \hB^K - \J_J \cdot \theta_I = {}&
S_{IJ} + \G_{IJ} \, , \cr
{\hat \Psi}_I{}^L \B_{LJK} + \C_{ILJK}\hB^L - \K_{JK} \cdot \theta_I
- \hrho_{(J} \cdot \P_{K)I} = {}& T_{IJK} + \Gamma^{(\G)}{\!}_{IJK} \, .
}
}
This determines
\eqn\resST{\eqalign{
S_{IJ} = {}& - (k_I+\half ) \,  \A_{IJ,1}
- {\ts {\sum_K}} \, \B_{JIK ,1} k_K g^K  \, , \cr
T_{IJK}  = {}& - (k_I+\half ) \,  \B_{IJK,1}
- {\ts {\sum_L}} \, \C_{ILJK ,1} k_L g^L \, . 
}
}
Since $(\vep - \wL_{\smash{{\hB},\hrho}}) \Gamma^{(\G)}{\!}_{IJK}
- {\hat \Omega}_{\smash{JK}}{}^L \G_{IL} + ((\pr_I \hrho_{\smash{(J}} - 
\pr_{\smash{(J}} \hrho_I)g)^L \,\G_{K)L} = \Gamma^{(G)}{\!}_{IJK}$ then
applying ${\vep - \wL_{\smash{{\hB},\hrho}}}$ to \STrel\ and using
\rgeqA\ gives finite relations which, after dropping ${\rm O}(\vep)$
contributions, are identical to \GArelb\ and \relGBCa.

Furthermore, eliminating $\A_{IJ},\B_{IJK},\C_{IJKL}$ from \STrel\ gives
\eqn\fineq{
L_{[I} \cdot \theta_{J]} + \half \, \hrho_{[I} \cdot Q_{J]} - 
{\tilde S}_{IJ} = \Gamma^{(\G)}{\!}_{[IJ]K}\hB^K - \big ( {\hat \Psi}_{[I}{}^K
- \half ( \hrho_{[I} g)^K\big )  \G_{J]K} = \pr_{[I} W_{J]} \, ,
}
where  
${\tilde S}_{[IJ]} = - {\hat \Psi}_{[I}{\!}^K  S_{J] K} + T_{[IJ]K} \hB^K
= - \Psi_{[I}{\!}^K  S_{J] K} + T_{[IJ]K} B^K$ and
$W_I$ is determined by \defW. Hence \WSa\ is recovered.

\newsec{Calculations for a Scalar Fermion Theory}

For the theory defined by \Lag, where $\Phi=(\phi, \psi , \chi)$, 
$g^I= \{ y^i , \by_i , \lambda_{ij}{}^{kl}\}$, then the kinetic symmetry
group ${\rm G}_K  =U(n_\phi) \times U(n_\psi) \times U(n_\chi)$ and
for $\omega \in {\frak g}_K$ then
\eqn\omegaeq{
\omega = - \omega^\dagger = \{ \omega_{\phi \,i}{}^j , \omega_\psi,
\omega_\chi \} \, , \quad \omega\cdot \omega' = 
\omega_{\phi \,i}{}^j  \omega'{\! }_{\phi \,j}{}^i 
+\tr (\omega_\psi \, \omega'{\!}_\psi ) 
+\tr (\omega_\chi \, \omega'{\!}_\chi ) \, .
}
To allow application of the formalism of section 2 it is necessary to extend 
the theory to include background gauge fields 
$a_\mu =  \{ a_{\phi\mu\,i}{}^j, a_{\psi \mu}, a_{\chi \mu}\}
= - a_\mu{\!}^\dagger\in {\frak g}_K$ and a scalar field mass term 
\eqn\Lagg{\eqalign{
\L = {}& - D \bphi^i \cdot D \phi_i - \bpsi \, i \sigma \cdot D \, \psi
- \bchi \, i \bsi \cdot D \, \chi - {\bchi} \, y^i \phi_i \, \psi  - 
{\bpsi} \,  \bphi^i \,\by_i   \, \chi \cr
&{} - {M}_i{}^j \bphi^i \phi_j 
- \quar \,\lambda_{ij}{\!}^{kl}\, \bphi^i\bphi^j\phi_k\phi_l  \, ,
}
}
where the covariant derivatives depending on the background gauge fields
are
\eqn\covarD{
D_\mu \phi_i = \pr_\mu \phi_i + a_{\phi\mu\,i}{}^j \phi_j \, , \quad
D_\mu \psi = \pr_\mu \psi + a_{\psi \mu} \psi \, , \quad
D_\mu \chi = \pr_\mu \chi+ a_{\chi \mu} \chi \, .
}
Acting on the local couplings, in accord with \deg, the covariant derivative
is determined by using \upg\ for $(a_\mu g)^I$.
For this theory the minimal subtraction ${\hat \beta}$-functions 
are expressible as in \Bhatr\ in the form
\eqn\betle{
{\hat \beta}_{\lambda\, ij}{}^{kl} = - \vep \, \lambda_{ij}{}^{kl}+
\beta_{\lambda\, ij}{}^{kl} \, , \quad 
{\hat \beta}_y{\!}^i = - \half \vep \, y^i +  \beta_y{\!}^i \, , \quad
{\hat \beta}_{\by\,  i}= - \half \vep \, \by_i  +  \beta_{\by\,  i} \, .
}

To obtain counterterms involving derivatives of the couplings
when they are $x$-dependent the methods described in 
\refs{\JackOne,\JackO}, which avoid momentum space,  may be adapted.
For the theory defined by \Lag, neglecting mass terms and background
gauge fields, the propagators are given by
\eqn\prop{
\l \psi(x) \, \bpsi(y) \r = S(s)= - i \, \bsi \cdot \pr G_0(s) \, , \quad
\l \chi(x) \, \bchi(y) \r = {\bar S}(s) =- i \, \sigma \cdot \pr G_0(s) \, ,
}
and
\eqn\propB{
\l \phi_i(x) \, \bphi^j(y) \r = \delta_i{\!}^j G_0(s) \, ,
}
with
\eqn\Gzero{
G_0(s) = {1\over (d-2)S_d} \, (s^2)^{1-{1\over 2}d} \, , \qquad
S_d = {2 \pi^{{1\over 2}d}\over \Gamma(\half d)} \, , \qquad s = x-y \, ,
}
so that $-\pr^2 G_0(s) = \delta^d(s)$. For graphs involving two vertices
the $\vep$ poles may be determined by using
\eqn\Gdiv{
G_0(s)^n \sim
{2\over \vep} \, {1\over (16\pi^2)^{n-1}} \, {1\over (n-1)!^2} \,
(\pr^2)^{n-2} \delta^d(s) \quad  \hbox{for} \quad n=2,3,\dots \, ,
}
and various extensions involving derivatives \JackOne. At one loop
it is sufficient to use \Gdiv\ for $n=2$ since
\eqn\oneloop{
\tr_\sigma \big ( S(s)\, {\bar S} (-s) \big )  = - \pr^2 G_0(s)^2 \, , \qquad
S(s)G_0(s) = - \half i \, \bsi \cdot \pr\, G_0(s)^2 \, .
}
This formalism may also be extended to allow for mass terms and gauge
fields in a manifestly gauge covariant fashion.

With these results it is straightforward to obtain
\eqn\onelK{
\L^{(1)}_{\rm c.t.} = {1\over \vep} \, 
\big ( 2\, \bphi^i \, \tr(\by_i {\overleftarrow \pr}{\!} \cdot \pr \, 
y^j) \, \phi_j  + \bpsi\, \by_i  \, i \sigma \cdot \olr \pr \, y^i \,\psi
+ \bchi \, y^i \, i \bsi \cdot \olr \pr \; \by_i \, \chi \big ) \, ,
}
for $\olr \pr = \half ( \pr - {\overleftarrow \pr})$ and also
rescaling the couplings as in \rescale. At two loops the corresponding
contribution to $\L^{(2)}_{\rm c.t.}$ involving $\chi$ is given by
\eqn\twolK{\eqalign{
\L^{(2)}_{\rm c.t.\chi} = {}& {1\over 4\vep^2}(1-\quar \vep)  \, \big (
\bchi \, y^i \by_j \, y^j i \bsi \cdot \pr \; \by_i \, \chi
-  \bchi \, y^i \,  i \bsi \cdot {\overleftarrow \pr}  \by_j \, y^j
 \by_i \, \chi \big ) \cr
&{} - {1\over 2\vep^2}(1- {\ts {5\over 4}} \vep) \,  \bchi \, y^i \, 
i \bsi \cdot (\by_j \olr \pr \, y^j) \,  \by_i \, \chi \cr
{}& + {1\over 2\vep^2}(1- {\ts {3\over 4}} \vep)  \, \big (
 \bchi \, y^i \, \tr(\by_i \, y^j) \,  i \bsi \cdot \pr \; \by_j \, \chi
-  \bchi \, y^i \,  i \bsi \cdot {\overleftarrow \pr}  \, \tr( \by_i \, y^j)
\,  \by_j \, \chi \big ) \cr
&{} - {1\over \vep^2}(1 + \quar \vep) \,  \bchi \, y^i \,
i \bsi \cdot \tr (\by_i \olr \pr \, y^j) \,  \by_j \, \chi \, ,
}
}
and similarly for $\L^{(2)}_{\rm c.t.\psi}$ obtained from \twolK\
with $\chi \to \psi$, $y \leftrightarrow \by$. Furthermore the two
loop scalar field counterterm is given by
\eqnn\twolKp$$\eqalignno{
\L^{(2)}_{\rm c.t.\phi} = {}& {1\over 4 \vep} \, \bphi^i \lambda_{ik}{}^{mn}
{\overleftarrow \pr} \cdot \pr \lambda_{mn}{}^{kj}\phi_j \cr
&{}+ {2\over \vep^2} (1-{\ts{1\over 4}} \vep)  \,  \bphi^i \lambda_{ik}{}^{lj}
\tr \big ( \pr \by_l \cdot  \pr y^k \big )  \phi_j 
- {1\over 2 \vep}\,  \bphi^i \lambda_{ik}{}^{lj} \big ( 
\tr ( \pr^2 \by_l \, y^k ) + \tr ( \by_l \, \pr^2 y^k ) \big ) \phi_j 
\cr
& {} + \bigg (
 {1\over \vep^2}(1-{\ts{3\over 4}} \vep)  \, 
\bphi^i  \, \tr \big ( \by_i   {\overleftarrow \pr}  \cdot 
y^k \by_k \, \pr \, y^j \big ) \phi_j \cr
&\qquad {} +{1\over \vep^2}(1- \quar \vep) \,  \Big ( \bphi^i \, 
\tr \big ( \by_i \,
( y^k \olr \pr \, \by_k ) \cdot \pr \, y^j \big ) \phi_j 
- \bphi^i \, \tr \big ( \by_i \,  {\overleftarrow \pr}  \cdot 
( y^k \olr \pr \, \by_k ) \, y^j \big ) \phi_j \Big )  \cr
&\qquad {}  - {1\over 2\vep^2}(1- {\ts {5\over 4}} \vep)  \, 
\bphi^i  \, \tr \big ( \by_i \, \pr^2( y^k \, \by_k )\,  y^j \big ) \phi_j 
- {1\over \vep} \, \bphi^i  \, \tr \big ( \by_i \, \pr y^k \cdot \pr \by_k \,  
y^j \big ) \phi_j  \cr
& \qquad {} + \bphi \leftrightarrow \phi \, , \ y \leftrightarrow \by \,  \bigg )  \, .
& \twolKp }
$$

The result \onelK\ then determines
\eqn\diffA{\eqalign{
\Z^{(1)} ={}&  -{1\over \vep} \big \{
 \tr  ( \by_i  \, y^j ) , \, \half\, 
\by_i \, y^i , \, \half \, y^i  \by_i \big \} \, , \cr
a_0{\!}^{(1)}{\!}_\mu = \fN_I{\!}^{(1)} \pr_\mu g^I = {}& -{1\over \vep} \big \{
2 \, \tr \big ( \by_i  \olr \pr_\mu \, y^j \big ) , \, 
\by_i  \olr \pr_\mu \, y^i , \, y^i \olr \pr_\mu \, \by_i \big \} \, , 
}
}
as well as the required  contributions to $M_0{\!}^{(1)}$
\eqn\Mzeroone{
M_0{\!}^{(1)}{\!}_i{}^j = {1\over \vep} \Big ( 2\, \lambda_{ik}{}^{jl} M_l{}^k 
+  \tr(\by_i \, y^k) M_k{}^j +M_i{}^k \, \tr(\by_k \, y^j) 
- 2 \, \tr \big ( \pr^\mu \by_i \, \pr_\mu y^j \big ) \Big ) \, .
}
In consequence at one loop from \diffA\ using \Nloop
\eqn\oneN{
 \rho_I{\!}^{(1)}  \del g^I = - \big \{
 \tr ( \by_i\,   \del y^j  - \del \by_i\,  y^j  ) , \, \half (
\by_i  \, \del y^i - \del \by_i \, y^i )  , \, \half ( y^i \, \del \by_i
- \del y^i \, \by_i) \big \} \, .
}
From \Mzeroone\ using \Mzero\ and \gmm
\eqn\oneM{
\delta_I {\!}^{(1)} \del g^I  = 0 \, , \qquad
\big ( \epsilon_{IJ}{}^{(1)}  \del g^I \del g^J \big ){}_i{}^j
= 2 \, \tr \big ( \del \by_i \, \del  y^j \big )  \, ,
}
and also
\eqn\oneMt{
\big ( \theta_{I}{\!}^{(1)}  \del  g^I \big ){}_i{}^j
= \half \big (   \tr (\by_i \, \del y^j  ) + 
\tr ( \del  \by_i \, y^j  ) \big ) \, .
}

From the two loop result \twolK\ we may obtain, as well as $Z^{(2)}{\!}_\chi$,
\eqnn\achi$$\eqalignno{
a_0{\!}^{(2)}{\!}_{\chi\mu} = {}& - {1\over 4\vep^2}(1-\quar \vep) \,
\big (  y^i \by_j \, y^j \, \pr_\mu \by_i 
-  \pr_\mu y^i \, \by_j \, y^j \by_i \big ) 
+ {1\over 2\vep^2}(1- {\ts {5\over 4}} \vep) \, y^i \,
(\by_j \olr \pr_\mu  \, y^j) \,  \by_i \cr
{}& - {1\over \vep^2}(1- {\ts {3\over 4}} \vep)  \; 
\tr(\by_i \, y^j) \,  (y^i \olr \pr_\mu \,\by_j)
+ {1\over \vep^2}(1 + \quar \vep)  \; y^i \by_j \,  
\tr (\by_i \olr \pr_\mu \, y^j) \cr
&{} - \Z^{(1)}{\!}_\chi \, \olr \pr_\mu \,\Z^{(1)}{\!}_\chi  
- \Z^{(1)}{\!}_\chi \, a_0{\!}^{(1)}{\!}_{\chi\mu} 
- a_0{\!}^{(1)}{\!}_{\chi\mu}\, \Z^{(1)}{\!}_\chi  \, . & \achi
}
$$
Also from \twolKp
\eqn\aphia{\eqalign{
a_0{\!}^{(2)}{\!}_{\phi\mu \, i}{}^j  
=  {}& - {1\over 4\vep} \, \lambda_{ik}{}^{mn}
{\olr \pr}_\mu \, \lambda_{mn}{}^{kj} \cr
{}& - {1\over 2\vep^2}(1- {\ts {3\over 4}}  \vep)\,  \big ( 
\tr ( \by_i \, y^k \by_k \, \pr_\mu y^j ) -  
\tr ( \pr_\mu \by_i \, y^k \by_k \, y^j ) \cr
\noalign{\vskip -4pt}
& \hskip 2.5cm {}+ \tr ( \by_k \, y^k \by_i \, \pr_\mu y^j ) -
\tr ( \by_k \, y^k \,\pr_\mu \by_i \,  y^j )  \big ) \cr
&{} + {1\over \vep^2}(1- \quar  \vep)\,  \big (
\tr ( \by_i \, ( y^k \olr \pr_\mu \, \by_k ) \, y^j ) 
- \tr (( y^k \olr \pr_\mu \, \by_k ) \, \by_i \, y^j ) \big ) \cr
&{} - \Z^{(1)}{\!\!}_{\phi\,  i}{}^k  \, \olr \pr_\mu \, 
\Z^{(1)}{\!\!}_{\phi \, k}{}^j 
- \Z^{(1)}{\!\!}_{\phi\, i}{}^k \, a_0{\!}^{(1)}{\!}_{\phi\mu  k}{}^j
- a_0{\!}^{(1)}{\!}_{\phi\mu  i}{}^k \, \Z^{(1)}{\!\!}_{\phi\, k}{}^j\, .
}
}
Furthermore
\eqnn\Mzerot$$\eqalignno{
M_0{\!}^{(2)}{\!}_i{}^j = {}&- {1\over 4 \vep} \, \pr \lambda_{ik}{}^{mn}
 \cdot \pr \lambda_{mn}{}^{kj} \cr
&{} - {2\over \vep^2} (1-{\ts{1\over 4}} \vep)  \,  \lambda_{ik}{}^{lj}
\tr \big ( \pr \by_l \cdot  \pr y^k \big )  
+ {1\over 2 \vep}\,  \lambda_{ik}{}^{lj} \big ( 
\tr ( \pr^2 \by_l \, y^k ) + \tr ( \by_l \, \pr^2 y^k ) \big ) \cr
& {} -
 {1\over \vep^2}(1-{\ts{3\over 4}} \vep)  \, 
\Big ( \tr \big (\pr  \by_i     \cdot y^k \by_k \, \pr y^j \big ) + 
 \tr \big ( \by_k \,  y^ k  \, \pr  \by_i     \cdot  \, \pr y^j \big ) \Big ) 
\cr
&- {1\over \vep^2}(1- \half  \vep) \,  \Big ( 
\tr \big ( \by_i \,  y^k \,  \pr  \by_k  \cdot \pr  y^j \big ) +
\tr \big ( \pr \by_i \cdot  \pr y^k \,  \by_k  \,   y^j \big )\cr
\noalign{\vskip -5 pt}
& \hskip 2.5cm {}  +
\tr \big ( \pr \by_k  \cdot  y^k \,   \by_i  \, \pr  y^j \big )  +
\tr \big ( \by_k  \,  \pr  y^k \cdot   \pr \by_i \,  y^j \big )  \Big ) \cr
&+ {1\over 4\vep}\,  \Big ( 
\tr \big ( \by_i \, \pr  y^k \cdot  \by_k  \,  \pr  y^j \big ) +
\tr \big ( \pr \by_i \cdot  y^k \,  \pr \by_k  \,   y^j \big )\cr
\noalign{\vskip -5 pt}
& \hskip 2 cm {}  +
\tr \big ( \by_k  \, \pr  y^k \cdot   \by_i  \, \pr  y^j \big )  +
\tr \big ( \pr \by_k  \cdot    y^k \,  \pr \by_i \,   y^j \big )  \Big ) \cr
&{}  - {1\over 4\vep} \Big ( 
\tr \big ( \by_i \, \pr^2( y^k \, \by_k )\,  y^j \big ) +
\tr \big (\pr^2(\by_k \, y^k )\,  \by_i \,  y^j \big ) \Big )  \cr
&{}  + {1\over \vep} \Big ( 
\tr \big ( \by_i \, \pr y^k \cdot \pr  \by_k \,  y^j \big ) +
\tr \big (\pr\by_k \cdot \pr  y^k \,  \by_i \,  y^j \big ) \Big )  \cr
\noalign{\vskip 2 pt}
&{} - \big ( \pr \Z^{(1)}{\!\!}_{\phi\,  i}{}^k  - 
a_0{\!}^{(1)}{\!}_{\phi \,  i}{}^k \big )
\cdot  \big (  \pr \Z^{(1)}{\!\!}_{\phi \, k}{}^j + 
a_0{\!}^{(1)}{\!}_{\phi\, k}{}^j \big )\cr
\noalign{\vskip 2 pt}
&{} - \Z^{(1)}{\!\!}_{\phi\,  i}{}^k  M_0{\!}^{(1)}{\!}_k{}^j 
- M_0{\!}^{(1)}{\!}_i{}^k \Z^{(1)}{\!\!}_{\phi \, k}{}^j  + {\rm O}(M) \, .
& \Mzerot
}
$$

Letting in \diffA, using \upg,
\eqnn\repD$$\eqalignno{
y^i \olr \pr_\mu \, \by_i  \to  {}& y^i  {\olr D}_\mu \, \by_i  =
 y^i \olr \pr_\mu \, \by_i + y^i \, a_{\psi \mu} \, \by_i + y^i \by_j \,
a_{\phi \mu i}{}^j - \half \, a_{\chi \mu} \, y^i \by_i  - \half \, 
y^i \by_i \, a_{\chi \mu} \, , \cr
\tr \big ( \by_i  \olr \pr_\mu \, y^j \big ) \to {}&
\tr \big ( \by_i  \olr D_\mu \, y^j \big ) = 
\tr \big ( \by_i  \olr \pr_\mu \, y^j \big ) +
\tr \big ( \by_i  \, a_{\chi \mu} \, y^j \big ) -
\tr \big ( a_{\psi\mu}\, \by_i \, y^j \big ) \cr
&\hskip 2.5cm {} - \half \, \tr(\by_i \, y^k) \, a_{\phi\mu k}{}^j -
\half \,  a_{\phi\mu i}{}^k \, \tr(\by_k \, y^j) \, , &\repD
}
$$
we may verify that the RG equations \rga\ are consistent with the 
double pole terms in \achi\ and \aphia\ with $\omega=0$ to this order.
The double $\vep$-poles in \Mzerot\ are also determined by \rgM.

The two loop results \achi\ and \aphia\ then entail
\eqn\twoN{\eqalign{
\big ( \rho_I{\!}^{(2)} \del g^I \big ){}_\chi = {}&
{\ts {1\over 8}}\,
\big (  y^i \by_j \, y^j \, \del \by_i
-  \del y^i \, \by_j \, y^j \by_i \big )
- {\ts {5\over 8}}  \, y^i \,
(\by_j \, \del y^j -\del \by_j \, y^j ) \,  \by_i \cr
&{}+  {\ts {3\over 4}} \,
\tr(\by_i \, y^j) \,  (y^i  \, \del \by_j -\del y^i \, \by_j )
+ \quar  \, y^i \by_j \, \tr (\by_i \, \del  y^j - \del \by_i \, y^j ) \, , \cr
\big ( \rho_I{\!}^{(2)} \del g^I \big ){}_{\phi}{}_i{}^j = {}&
- {\ts {1\over 4}} \, ( \lambda_{ik}{}^{mn}
\, \del \lambda_{mn}{}^{kj}   -\del \lambda_{ik}{}^{mn}\, 
\lambda_{mn}{}^{kj} ) \cr
{}& + {\ts {3\over 4}} \, 
\tr \big ( y^k \by_k \, (\del  y^j \by_i - y^j \del \by_i ) +
\by_k \, y^k \, ( \by_i \, \del y^j - \del \by_i \, y^j ) \big ) \cr
&{} + \quar  \, \tr \big ( ( 
\del y^k \, \by_k - y^k \, \del  \by_k   ) \, y^j \by_i 
+ ( \by_k \, \del y^k  - \del \by_k \, y^k ) \, \by_i \, y^j )\big )  \, .
}
}
A related calculation, which was extended to three loops, was described 
in \FortinC. Also from \rga\ we obtain
\eqn\onetu{
\upsilon^{(1)} = \upsilon^{(2)} = 0 \, .
}

A useful check is to restrict \oneN\ and \twoN\ to the 
supersymmetric case \red, where, with a similar notation to that in
section 7,
\eqn\SusyN{\eqalign{
\big ( \rho_I{\!}^{(1)} \del g^I \big ){}_{\rm Susy}{}_i{}^j = {}& \half 
\big ( - (\bY \del Y)_i{}^j + (\del \bY Y )_i{}^j \big ) \, , \cr
\big ( \rho_I{\!}^{(2)} \del g^I \big ){}_{\rm Susy}{}_i{}^j = {}& 
\half \big( ( \bY Y\bY \del Y)_i{}^j - ( \del \bY Y\bY Y)_i{}^j  
+ (\bY \del Y \bY Y)_i{}^j - (\bY Y \del \bY Y )_i{}^j \big ) \, .
}
}
These results are in accord with \Susyall, using \otl.

The condition \Brho, which links different loop orders,  provides a further
verification of the results \oneN\ and \twoN. It is easy to check that 
$ \rho_I{\!}^{(1)}\beta^{(1)I} =0$ and also
\eqn\twolch{\eqalign{
\big (  \rho_I{\!}^{(2)} \beta^{I(1)} \big ){}_\psi = {}& 
- \big (  \rho_I{\!}^{(1)}\beta^{I(2)} \big ){}_\psi
 =  {\ts{1\over 16}} \, \by_i\, y^j\by_j\, y^i \by_k\,  y^k 
 + {\ts {3\over 8}}\,  \by_i\,  y^j \by_k \, y^k \, \tr(y^i\by_j ) - 
 \hbox {conjugate} \, , \cr
 \big (  \rho_I{\!}^{(2)} \beta^{I(1)} \big ){}_\phi{}_i{}^j = {}& 
- \big (  \rho_I{\!}^{(1)}\beta^{I(2)} \big ){}_\phi{}_i{}^j 
=  2 \, \lambda_{ik}{}^{mn} \, \tr(\by_m \, y^k\, \by_n \, y^j ) -
\quar \, \lambda_{ik}{}^{mn}\lambda_{mn}{}^{kl }  \, \tr(\by_l \, y^j ) \cr
&\hskip 1.5 cm {}+ {\ts{3\over 4}} \big ( \tr (\by_i \, y^k\by_k \, y^l ) 
+ \tr(\by_k \, y^k \by_i \, y^l) \big ) \, \tr(\by_l \, y^j) ) - 
 \hbox {conjugate} \, .
 }
 }

From \Mzerot\ we may also read off
\eqnn\twoM$$\eqalignno{
\big (\delta_I & {\!}^{(2)} \del g^I \big ){} _i{}^j 
= - \lambda_{ik}{}^{lj} \big ( \tr ( \del \by_l \, y^k ) 
+ \tr ( \by_l \, \del y^k )  \big ) \cr
&\hskip 2.2 cm {} +  \half \, 
\tr \big ( \by_i \,(  \del y^k \, \by_k  + y^k \, \del \by_k )\,  y^j  \big )  
+ \half\,  \tr \big ((\del \by_k \, y^k  
+ \by_k\,  \del y^k )\,  \by_i \,  y^j \big )  \, , \cr
\big (  
\epsilon_{I J} & {\!}^{(2)} \del g^I \del g^J\big ){} _i{}^j 
= \half\, \del \lambda_{ik}{}^{mn} \,  \del \lambda_{mn}{}^{kj} 
- \lambda_{ik}{}^{lj} \,  \tr \big ( \del \by_l \,  \del y^k \big )  \cr
& {} - {\ts{3\over 2}}  \,
\big ( \tr \big (\del  \by_i     \, y^k \by_k \, \del y^j \big ) +
 \tr \big ( \by_k \,  y^ k  \, \del \by_i    \, \del y^j \big ) \big )\cr
& {}- 
\tr \big ( \by_i \,  y^k \,  \del  \by_k  \, \del  y^j \big ) -
\tr \big ( \del  \by_i \,  \del y^k \,  \by_k  \,   y^j \big )
- \tr \big ( \del \by_k  \,  y^k \,   \by_i  \, \del  y^j \big ) - 
\tr \big ( \by_k  \,  \del  y^k \,  \del  \by_i \,  y^j \big )  \cr
&-  \half \big (
\tr \big ( \by_i \, \del  y^k \,  \by_k  \,  \del  y^j \big ) +
\tr \big ( \del \by_i \, y^k \,  \del \by_k  \,   y^j \big )
+ \tr \big ( \by_k  \, \del  y^k \,  \by_i  \, \del  y^j \big )  +
\tr \big ( \del \by_k  \,   y^k \,  \del \by_i \,   y^j \big )  \big ) \cr
&{} - \tr \big (  \by_i \, \del y^k \, \del \by_k \,  y^j \big ) 
 - \tr \big (   \del  \by_k \,  \del  y^k \,  \by_i \,  y^j \big ) 
 \, . &\twoM
}
$$
Reducing  \oneM\  and \twoM\ to the supersymmetric case 
\eqn\SusyM{\eqalign{
\big ( \epsilon_{IJ}{}^{(1)} \del g^I  \del g^J 
\big ){}_{\rm Susy}{}_i{}^j = {} &
( \del \bY \del Y)_i{}^j  \, , \cr
\big ( \epsilon_{IJ}{}^{(2)} \del g^I \del
g^J\big ){}_{\rm Susy}{}_i{}^j = {}& 
- ( \del \bY Y\bY \del Y)_i{}^j
- ( \del \bY \del Y\bY Y)_i{}^j \cr 
\noalign{\vskip -1pt}
&{} - (\bY \del Y \del \bY Y )_i{}^j - (\bY Y \del \bY \del Y)_i{}^j\, ,
}
}
which agrees with \SusyMall.

The results are compatible with the consistency relation \derel\ or \deth.
Assuming \oneMt, and for simplicity $\del g^I = ( \del y^i , 0 ,0)$,
then
\eqn\cloop{\eqalign{
\big ( \L_{\beta^{(1)}} \theta_I{\!}^{(1)}\, \del g^I \big  ){}_i{}^j = {}&
\half \, \tr(\by_i\, y^k \by_k \, \del y^j ) 
+ \half \, \tr ( \by_k \, y^k \by_i \, \del y^j)\cr
\noalign{\vskip - 1pt}
&{}  + \quar \, \tr(\by_i\, \del y^k \by_k \, y^j ) 
+ \quar \, \tr ( \by_k \, \del y^k \by_i \, y^j)  \cr
\noalign{\vskip - 1pt}
&{} + \tr(\by_i \, y^k ) \, \tr ( \by_k \, \del y^j) + \half \, 
 \tr (\by_i \,\del  y^k ) \, \tr ( \by_k \, y^j) \, , \cr
\big ( \gamma_M  {\!}^{(1)}
\theta_I{\!}^{(1)}\, \del g^I \big ){}_i{}^j  = {}&
\lambda_{ik}{}^{lj}\, \tr( \by_l\, \del y^k)
+ \half \, \tr(\by_i \, y^k ) \, \tr ( \by_k \, \del y^j) 
+ \half \,  \tr (\by_i \,\del  y^k ) \, \tr ( \by_k \, y^j) \, , \cr
 \big ( (\rho_I {\!}^{(1)}g)^J
  \theta_J{\!} ^{(1)}\, \del g^I \big ){}_i{}^j  = {}&
 \quar \, \tr(\by_i\, \del y^k \by_k \, y^j ) 
+ \quar \, \tr ( \by_k \, \del y^k \by_i \, y^j)  
+ \half \, \tr(\by_i \, y^k ) \, \tr ( \by_k \, \del y^j) \, .
}
}
The sum is then equal to 
$( \delta_I{\!}^{(2)} + \epsilon_{IJ}{\!}^{(1)} \beta^{J(1)})_i{}^j\, \del g^I$,
as required by \derel\ to this order.

Similar calculations determine $\X$. At one loop there is no dependence on
the couplings and
\eqn\Lone{
\X(a,M)^{(1)} =
{1\over 6 \vep}\Big ( \tr \big ( f_{\phi}{\!}^{\mu \nu} f_{\phi \mu \nu} \big )
+ 2\, \tr \big ( f_{\psi}{\!}^{\mu \nu} f_{\psi \mu \nu} \big )
+ 2\, \tr \big ( f_{\chi}{\!}^{\mu \nu} f_{\chi \mu \nu} \big ) \Big ) 
+  {1\over \vep} \, {M}_i{}^j  {M}_j{}^i \, ,
}
giving $\L_f{\!}^{(1)}$ and $\L_M{\!}^{(1)}$ in \counter.
 Two loop contributions to $\X $, which determine the leading contributions
 to $\A_{IJ},\P_{IJ},\J_I,\K_{IJ}$ in \counter, may also be 
undertaken within the framework of \JackOne.
For the scalar/fermion theory determined by \Lagg\ there is just one two 
loop graph involving only the Yukawa couplings. For zero $a_\mu,M$
this gives
\eqn\Wtwo{
W^{(2)} = - \int \d^d x \, \d^d y \; \tr \big ( y^i(x)\, \by_i(y) \big )
\; \tr_\sigma \big ( S(s)\, {\bar S} (-s) \big ) \, G_0 (s) \, .
}
Since $ \tr_\sigma \big ( S(s)\, {\bar S} (-s) \big ) G_0(s)
= - {1\over 3}\, \pr^2 G_0(s)^3$
the divergent part of \Wtwo\  is determined by using  \Gdiv\
and gives, after rescaling according to \rescale, 
$ \X(g)^{(2)}  = {1\over 6\vep} \, 
\tr \big ( \pr^2 y^i \, \pr^2 \by_i \big )$ as was obtained in \Analog.

Extending this two loop calculation  to include the additional contributions
involving the background gauge fields $a_\mu$ and also $M$ gives
\eqnn\WtwoM$$\eqalignno{
\X(g, {} & a , M) ^{(2)}\cr
 ={}&  {1\over \vep} \; {1\over 6} \, 
\tr \big ( D^2 y^i \, D ^2 \by_i \big ) +{2\over 3 \vep^2} 
(1+{\ts{5\over 12}}\vep )
\, \tr \big (D_\mu y^i \, D_\nu \by_j ) \, f_\phi{\!}^{\mu\nu}{\!}_i{}^j \cr
&{}+ {2\over 3 \vep^2} (1-{\ts{7\over 12 }}\vep )\, \Big ( 
\tr \big ( D_\mu y^i  f_\psi{\!}^{\mu\nu} D_\nu \by_i \big ) 
- \tr \big ( D_\mu y^i \, D_\nu \by_i \,  f_\chi{\!}^{\mu\nu} \big ) \Big ) 
\cr 
&{} - {2\over \vep^2 }(1-\quar \vep)  \, 
M_i{}^j\, \tr \big ( D^\mu \by_j \, D_\mu y^i \big )  
+{1\over 2\vep} \, M_i{}^j  
\big ( \tr (  \by_j\, D^2  y^i ) +  \tr (D^2  \by_j \,  y^i ) \big )  \cr 
&{} - {1\over 4\vep} \, \Big (  \tr \big ( y^i f_\psi{\!}^{\mu\nu}
f_{\psi\, \mu\nu}\, \by_i  \big ) + \tr \big ( y^i \by_i \, f_\chi{\!}^{\mu\nu} 
f_{\chi\, \mu\nu} \big ) \Big ) -  {1\over 3\vep^2} (1- {\ts{1\over 12}}\vep) \,
\tr \big ( (f^{\mu\nu} y)^i \, (f_{\mu\nu}\by)_i \big ) \cr
&{}+ {1\over 6\vep} \, \tr \big ( (f_{\mu\nu} y)^i  \by_j \big )  
f_\phi{\!}^{\mu\nu}{\!}_i{}^j 
+ {2\over \vep^2} \, \lambda_{ij}{}^{kl} M_k{}^i M_l{}^j
+ {2\over \vep^2} (1-\half \vep) \, M_i{}^j \, \tr(\by_j \, y^k) \, M_k{}^i \, ,
& \WtwoM
}
$$
which is consistent with the general form \counter.
The RG equation \rgeq\ provides
a non trivial check of the double poles in $\vep$ present in $\X^{(2)}$
which are determined in terms of
\Lone\  and the one loop results \oneN\ and \oneM.  In the $ {\rm O} ( f^2 ) $
terms it is useful to note $\tr \big ( (f^{\mu\nu} y)^i  f_{\psi\, \mu \nu} \by_i \big )
= - \tr \big ( y^i f_{\psi\, \mu \nu }(f^{\mu\nu} \by)_i \big )$, with similar relations
for $f_\psi \to f_\chi, f_\phi$. For $(\omega y)^i , (\omega \by)_i=0$ the $ {\rm O} ( f^2 ) $
contributions are just the  two loop Yukawa contribution to the 
gauge beta function \refs{\JackO,\Mach}.

The two loop contributions to $X$ and $Y^\mu$  
are determined as in \rgLa. This gives using from \rgeqA\ and \resST
\eqn\twoAres{\eqalign{
G_{IJ}{\!}^{(2)} \del g^I \del g^J = {}& 
A_{IJ}{\!}^{(2)} \del g^I \del g^J =  {\ts {2\over 3}} \,
\tr (\del \by_i \,  \del y^i) \, ,  \cr 
S_{IJ}{\!}^{(2)} \del g^I \del'  g^J = {}& -{\ts {1\over 6}} \,
\big ( \tr (\del \by_i \,  \del' y^i) + \tr (\del' \by_i \,  \del y^i) \big )  \, .
}
}
For terms involving $M$ using \resLJK,
\eqn\twores{\eqalign{
\big ( J_I {\!}^{(2)} \del g^I \big ){}_i{}^j = {}& \tr ( \by_i \, \del y^j) 
+ \tr(\del \by_i \, y^j ) 
\, , \quad \big ( K_{IJ}{\!}^{(2)} \del g^I \del g^J \big ){}_i{}^j=  
\tr (\del \by_i \, \del y^j ) \, , \cr
\big ( L_I {\!} ^{(2)} \del g^I \big ){}_i{}^j = {}& -{\ts {5\over 8}}\, \big (
\tr ( \by_i \, \del y^j) + \tr(\del \by_i \, y^j ) \big ) \, , 
}
}
while for the $f_{\mu\nu}$ terms, if $P_{IJ},Q_I$ are decomposed as in 
\omegaeq, 
\eqn\PQres{\eqalign{
& P_{IJ}{\!}^{(2)} \del g^I \del' g^J =   \big \{
-{\ts {5\over 18}} \, \tr (\del \by_i \, \del'  y^j- \del'\by_i\, \del y^j ) , \cr
&\hskip 3cm {\ts {7\over 18}} \,  (  \del \by_i  \,  \del'  y^i - 
\del'  \by_i\, \del y^i ) , \,  {\ts {7\over 18}} \,  
(  \del y^i \, \del'\by_i -  \del' y^i \, \del\by_i ) \big \} \, ,  \cr
& Q_{I}{\!}^{(2)} \del g^I  =   \big \{
{\ts {5\over 72}} \, \tr ( \by_i \, \del  y^j- \del \by_i\,   y^j ) ,  
- {\ts {7\over 72}} \, ( \by_i  \, \del  y^i - \del \by_i\,  y^i ) , \,
{\ts {7\over 72}} \, (\del y^i \,\by_i - y^i \, \del\by_i ) \big \} \, . 
}
} 
It is easy to see that $J_I {\!}^{(2)} = 2 \, \theta_I{\!}^{(1)} = 
\theta_I{\!}^{(1)}  \cdot \beta_M{\!}^{(1)}$ in accord with \relJKL\ at this
order. From \Lone\ and \WtwoM
\eqn\betaf{\eqalign{
\omega \cdot \beta_f{\!}^{(1)}\! \cdot \omega  = {}& {\ts {2\over 3}} \big ( 
 \omega_{\phi \, i}{}^j  \omega_{\phi \, j}{}^i + 2\, \tr ( \omega_\psi{\!}^2 )  + 2\, \tr ( \omega_\chi{\!}^2 ) 
\big ) \, , \cr
\omega \cdot \beta_f{\!}^{(2)} \! \cdot \omega = {}& - 2\, 
 \tr ( \by_i \, y^i \, \omega_\psi{\!}^2  + y^i\by_i \,  \omega_\chi{\!}^2 ) 
 + {\ts {2\over 9}}\, \tr \big ( (\omega y)^i (\omega \by_i ) \big ) 
 + {\ts {4\over 3}}\, \tr \big ( (\omega y)^i\by_j  \big ) \omega_{\phi \, i}{}^j \, .
}
}
It is easy to check that $G_{IJ}{\!}^{(2)} (\omega g)^J = - \half \, \omega \cdot \beta_f{\!}^{(1)}
\! \cdot \rho_I{\!}^{(1)}$, as required by \WQ{b}.

At three loops we determine for simplicity just contributions independent of
$a_\mu,M$.  For the quartic scalar coupling there is a single vacuum graph
\eqn\Wthree{
W_a^{(3)} = {1\over 8} \int \d^d x \, \d^d y \; \lambda_{ij}{}^{kl}(x) 
\lambda_{kl}{}^{ij}(y) \, G_0(s)^4 \, ,
}
which gives, using \Gdiv\ for $n=4$,
\eqn\Wthreea{
\X_a( \lambda)^{(3)} = {1\over \vep} \; {1\over 144} \,
 \pr^2 \lambda_{ij}{}^{kl}\, \pr^2 \lambda_{kl}{}^{ij} \, .
}
At four-loop  order the vacuum graphs involving just the
scalar couplings also give
\eqnn\Wfour$$\eqalignno{
W^{(4)} =  - {1\over 48} \int \d^d x \, \d^d y \, \d^d z\, {}&
\big ( \lambda_{ij}{}^{mn}(x)
\lambda_{mn}{}^{kl}(y)\lambda_{kl}{}^{ij}(z)
+  \lambda_{ij}{}^{mn}(y)
\lambda_{mn}{}^{kl}(x)\lambda_{kl}{}^{ij}(z)\cr
\noalign{\vskip -6pt} 
&\  \ {} + 8\,  \lambda_{m i}{}^{n k}(x)
\lambda_{n j}{}^{m l}(y)\lambda_{kl}{}^{ij}(z) \big ) \cr
&{}\times  \R G_0(x-z)^2\, \R G_0(z-y)^2 \, \R G_0(x-y)^2 \, ,
& \Wfour
}
$$
where
\eqn\RGO{
\R G_0(s)^2  = G_0(s)^ 2 - {1\over 8\pi^2 \vep} \, \delta^d(s) \, .
}
The additional pole term in $\vep$ is necessary 
to ensure  subtraction of one loop sub-divergences
and would be generated by appropriate counterterms
consistent with minimal subtraction.  Using results from
appendix D the divergent part of \Wfour\ determines
\eqn\Wfoura{\eqalign{
\X( \lambda)^{(4)}  = {} & {1\over \vep^2}(1+{\ts{11\over 12}}\vep) \; 
{1\over 288} \, \pr^2 \big ( \lambda_{ij}{}^{mn}\, \lambda_{mn}{}^{kl} 
+ 4 \,   \lambda_{im}{}^{kn}\, \lambda_{jn}{}^{l m}  \big ) 
\pr^2 \lambda_{kl}{}^{ij} \cr
&{}  - {1\over \vep} \; {1\over 96} \,
\big ( \lambda_{ij}{}^{mn}\,\pr^2  \lambda_{mn}{}^{kl} + 4 \,
\lambda_{im}{}^{kn}\,\pr^2  \lambda_{jn}{}^{l m}  \big ) 
\pr^2 \lambda_{kl}{}^{ij} \, .
}
}
It is easy to check that \rgeq\ determines the double poles in 
\Wfoura\ using \betl\  and \betle. \Wfoura\ gives \Gfoura\ with 
$\uG = -7/216$.

At three-loop order there are also further vacuum graphs involving 
solely the Yukawa couplings. There are just two relevant graphs
which contain two and one fermion loops giving at this 
order in addition to \Wthree
\eqn\Wthreeb{\eqalign{
W_b^{(3)} =  {}& {1\over 2} \int \d^d x \, \d^d y \; 
\tr \big ( \by_i(x)\, \pr_x{\!}^2 Y_{y^j}(x,y) \big )
\; \tr \big ( \by_j(y)\, \pr_y {\!}^2 Y_{y^i} (y,x) \big ) \, , \cr
W_c^{(3)} =  {}& - {1\over 8} \int \d^d x \, \d^d y \;
\Big ( \tr\, \tr_\sigma \big ( \by_i(x)\, \sigma{\cdot \, \pr_x} Y_{y^i}(x,y)\, 
\bsi {\cdot \!  {\overleftarrow \pr}}{\!}_y
\;\by_j(y)\,\sigma{\cdot \, \pr_y} Y_{y^j}(y,x)\, 
\bsi {\cdot  \! {\overleftarrow \pr}}{\!}_x \big ) \cr
\noalign{\vskip -6pt}
& \hskip 2.4cm {}+ \tr \, \tr_\sigma\big ( y^i(x)\, \bsi{\cdot \, \pr_x} 
Y_{\by_i}(x,y)\, \sigma {\cdot \! {\overleftarrow \pr}}{\!}_y
\; y^j(y)\,\bsi{\cdot \, \pr_y} Y_{\by_j} (y,x)\,
\sigma {\cdot \! {\overleftarrow \pr}}{\!}_x \big ) \Big )  \, , 
}
}
using \oneloop\ with
\eqn\defY{\eqalign{
Y_f(x,y) = {}& \int \d^d z \; \R G_0(x-z)^2  \,
f(z) \, G_0(z-y)  \, . 
}
}
From  \defY\ it is easy to obtain
\eqn\deY{
Y_{f}(x,y)(-{\overleftarrow\pr}{\!}_y{\!}^2) = \R G_0(s)^2 f(y) \, .
}

The analysis of \Wthreeb\ is more involved than obtaining \WtwoM\ or
\Wthreea\ and is described in appendix C by obtaining formulae for the local
$\vep$-poles which arise from products of $Y_f$ with derivatives. Thus
\eqnn\chib$$\eqalignno{
\chi_b(y,\by)^{(3)} = {}& {2\over 9\vep^2}\big ( 1 + {\ts {5\over 12}} \vep \big )
\, \tr(\by_i\, \pr^2 y^j) \, \tr( \pr^2\by_j \, y^i ) 
+ { 1\over 9\vep^2}\big ( 1-{\ts {25\over 12}}\vep \big )
\, \tr(\pr^2 \by_i\, \pr^2 y^j) \, \tr( \by_j \, y^i )  \cr
&{} 
 + { 1\over 36\vep} \, \big ( \tr(\pr^2 \by_i\,  y^j) \, \tr( \pr^2 \by_j \, y^i ) 
+  \tr( \by_i\, \pr^2 y^j) \, \tr( \by_j \,  \pr^2 y^i )  \big ) \cr
\noalign{\vskip 2pt}
&{}+ {2\over 9\vep^2}\big ( 1-{\ts{7\over 12}}\vep \big )
 \big ( \tr(\pr^\mu \by_i\, \pr^2 y^j) \, \tr( \pr_\mu \by_j \, y^i ) +
 \tr(\pr^2 \by_i\, \pr^\mu y^j) \, \tr( \by_j \, \pr_\mu y^i )\big )  \cr
\noalign{\vskip 2pt}
&{}- {2\over 9\vep^2}\big ( 1+{\ts{5\over 12}}\vep \big )\, 
\tr(\pr^\mu \by_i\, \pr_\mu y^j)  \big ( \tr( \pr^2 \by_j \, y^i ) +
 \tr( \by_j \, \pr^2  y^i )\big )  \cr
&{}- {1\over 9\vep} \big (
\tr(\pr^\mu \by_i\, \pr^2 y^j) \, \tr( \by_j \,  \pr_\mu y^i ) +
 \tr(\pr^2 \by_i\, \pr^\mu y^j) \, \tr( \pr_\mu \by_j \, y^i )\big )  \cr
 &{} + {4\over 9 \vep^3} \big (1 + {\ts {5\over 12}} \vep -{\ts {35\over 144}}\vep^2\big ) \, 
 \tr(\pr^\mu \by_i\, \pr^\nu y^j)  \big ( \tr( \pr_\mu \by_j \, \pr_\nu y^i ) - 
 \tr(\pr_\nu  \by_j \, \pr_\mu  y^i )\big )  \cr
&{}+   {4\over 3\vep^3} \big (
1-\quar \vep -{\ts {19\over 48}}\vep^2 \big ) \,
 \tr(\pr^\mu \by_i\, \pr_\mu y^j) \, \tr( \pr^\nu \by_j \,  \pr_\nu y^i ) \cr
&{} + {1\over 18\vep} \,  \tr(\pr^\mu \by_i\, \pr^\nu y^j) 
 \big ( \tr( \pr_\mu \by_j \, \pr_\nu y^i ) +  \tr(\pr_\nu  \by_j \, \pr_\mu  y^i )\big )  \, , 
 & \chib
}
$$
and
\eqnn\chic$$\eqalignno{
\chi_c(y,\by)^{(3)} = {}& {1\over 18\vep^2}
\big ( 1 - {\ts {13\over 12}} \vep \big ) \, 
\tr \big ( \pr^2 \by_i \, \pr^2  y^i \, \by_j \, y^j +  
\pr^2 y^i \pr^2\by_i  \, y^j \by_j  \big )\cr 
&{} + { 1\over 9\vep^2}\big ( 1 - {\ts {7 \over 12}}\vep \big )\, \tr
 \big ( \by_i\, \pr^2 y^i \, \pr^2 \by_j \, y^j  + 
\pr^2 y^i \,  \by_i\, y^j \, \pr^2 \by_j  \big )\cr
&{} - { 1\over 72\vep} \, \tr \big ( \pr^2 \by_i \,y^i \,\pr^2 \by_j  \, y^j  
+ \by_i\, \pr^2 y^i \, \by_j \, \pr^2 y^j 
+ \pr^2 y^i\, \by_i \, \pr^2 y^j\,  \by _j  
+ y^i \pr^2\by_i \,  y^j \pr^2 \by_j \big ) \cr
\noalign{\vskip 2pt}
&{}+ {1\over 9\vep^2}\big ( 1-{\ts{1\over 12}}\vep \big ) \, \tr
\big (\pr^\mu \by_i \, \pr^2 y^i \, \pr_\mu \by_j \,  y^j 
+ \pr^2 \by_i \, \pr^\mu y^i \, \by_j \, \pr_\mu y^j \cr
\noalign{\vskip - 3pt}
& \hskip 3cm {} + \pr^\mu y^i  \, \pr^2 \by_i \, \pr_\mu y^j  \, \by_j  +
\pr^2 y^i \, \pr^\mu \by_i  \, y^j \, \pr_\mu \by_j \big )  \cr
\noalign{\vskip 2pt}
&{}+ {1\over 18\vep^2}\big ( 1+{\ts{5\over 12}}\vep \big )\big (
 \pr^\mu \by_i\,  \pr_\mu y^i \,   (\pr^2  \by_j \, y^j + \by_j \, \pr^2 y^j ) +
 \pr^\mu y^i \pr_\mu \by_i \, ( \pr^2 y^j \by_j + y^j \pr^2 \by_j  )  \big )\cr
\noalign{\vskip 2pt}
&{}
+ {2\over 9 \vep^3} \big (1 - {\ts {7\over 12}} \vep 
- {\ts {41\over 144}}\vep^2\big ) \,
\tr \big ( \pr^\mu \by_i \, \pr^\nu y^i \, (  \pr_\mu \by_j \, \pr_\nu y^j
-\pr_\nu \by_j \, \pr_\mu y^j ) \cr
\noalign{\vskip - 3pt}
&{} \hskip 4.3cm {}+ \pr^\mu y^i\, \pr^\nu \by_i \, (  
\pr_\mu y^j \, \pr_\nu \by_j - \pr_\nu y^j \, \pr_\mu \by_j ) \big )\cr 
& {} - {1\over 18\vep} \, \tr \big ( \pr^\mu \by_i \, \pr_\mu y^i \,  \pr^\nu \by_j \, \pr_\nu y^j
+  \pr^\mu y^i\, \pr_\mu \by_i \,   \pr^\nu y^j \, \pr_\nu \by_j \big  ) \cr
&{}
+ {1\over 18 \vep} \,
\tr \big ( \pr^\mu \by_i \, \pr^\nu y^i \, (  \pr_\mu \by_j \, \pr_\nu y^j
+\pr_\nu \by_j \, \pr_\mu y^j ) \cr
\noalign{\vskip - 3pt}
&{} \hskip 4.3cm {}+ \pr^\mu y^i\, \pr^\nu \by_i \, (  
\pr_\mu y^j \, \pr_\nu \by_j + \pr_\nu y^j \, \pr_\mu \by_j ) \big )
\, . &\chic
}
$$
The double and triple $\vep$-poles are determined by \rgeq\ starting from
\WtwoM\ using the one loop results \oneN\ and \oneM. 

From \Wthreea, \chib\ and \chic\ we may determine $X(g)^{(3)}$ and
$Y^\mu(g)^{(3)}$. In particular the $\pr^2 g^I \pr^2 g^J$ terms give
the three loop contribution to $A_{IJ}$ which involves both the scalar
and Yukawa couplings
\eqnn\Athree$$\eqalignno{
A_{IJ}{\!}^{(3)} \d g^I \d g ^J 
= {}&  {\ts{1\over 24}} \,
\d \lambda_{ij}{}^{kl}\, \d \lambda_{kl}{}^{ij}\cr
&{\!}-  {\ts {13\over 36}}  \,
\tr \big ( \d \by_i \, \d  y^i \, \by_j \, y^j +
\d y^i \d \by_i  \, y^j \by_j  \big )
- {\ts {7 \over 18}} \, \tr \big ( \by_i\, \d y^i \, \d \by_j \, y^j  +
\d y^i \,  \by_i\, y^j \, \d \by_j  \big ) \cr
&{\!}-  {\ts {1\over 12}}  \,
\tr \big ( \d \by_i \,  y^i \, \d \by_j \, y^j +
\d y^i  \by_i  \, \d y^j \by_j  + \by_i\, \d y^i \, \by_j \, \d y^j  +
\d y^i \,  \by_i\, \d y^j \, \by_j  \big ) \cr
&{} - { \ts {25\over 18}} \, \tr(\d \by_i\, \d y^j) \, \tr( \by_j \, y^i )
+ {\ts {5\over 9}} \, \tr(\by_i\, \d y^j) \, \tr( \d\by_j \, y^i ) \cr
&{\!}
+ {\ts{ 1\over 6}} \, \big ( \tr(\d \by_i\,  y^j) \, \tr( \d \by_j \, y^i )
+  \tr( \by_i\, \d y^j) \, \tr( \by_j \,  \d y^i )  \big )\, .  &\Athree
}
$$

At this order there are extra terms  necessary to calculate
$G_{IJ}$.  Using results in \oneN, \oneMt\ and \twoAres, \twores\ 
then, since $ S_{(IJ)}{\!}^{(2)}+ \half A_{IJ}{\!}^{(2)} = 0$, \GArelb\ gives
\eqnn\Athree$$\eqalignno{
G_{IJ}{\!}^{(3)} \d g^I \d g ^J = {}& 
\big ( A_{IJ}{\!}^{(3)} - \half \, (\rho_I{\!}^{(1)} g)^K A_{KJ}{\!}^{(2)}
- J_I{\!}^{(2)} \cdot \theta_J{\!}^{(1)} \big )\, \del g^I \del g^J \cr
= {}&  {\ts{1\over 24}} \,
\d \lambda_{ij}{}^{kl}\, \d \lambda_{kl}{}^{ij}\cr
&{\!}-  {\ts {13\over 36}}  \, 
\tr \big ( \d \by_i \, \d  y^i \, \by_j \, y^j +  
\d y^i \d \by_i  \, y^j \by_j  \big ) 
- {\ts {5 \over 9}} \, \tr \big ( \by_i\, \d y^i \, \d \by_j \, y^j  + 
\d y^i \,  \by_i\, y^j \, \d \by_j  \big ) \cr
&{} - { \ts {25\over 18}} \, \tr(\d \by_i\, \d y^j) \, \tr( \by_j \, y^i ) 
 -   {\ts {7\over 9}} \, \tr(\by_i\, \d y^j) \, \tr( \d\by_j \, y^i ) \cr
&{\!}
- {\ts{ 1\over 6}} \, \big ( \tr(\d \by_i\,  y^j) \, \tr( \d \by_j \, y^i ) 
+  \tr( \by_i\, \d y^j) \, \tr( \by_j \,  \d y^i )  \big )\, .  &\Athree 
}
$$
This gives the results in \Resabc. We have verified that the three loop 
result \Athree\ is precisely determined by the one and two loop results
in \oneN, \twoN, \PQres\  and \betaf,  according to \WQ{b}.

For 
\eqn\gmet{
g_{IJ} \, \d g^I \d g^J = \tr (\d  \by_i \, \d y^i ) \, ,
}
then
\eqn\Lbg{\eqalign{
\big ( \wL_{\beta^{(1)},\rho^{(1)}}\, g_{IJ} \big )\, 
\del g^I \del g^J = {}&
\tr \big ( \d \by_i \, \d  y^i \, \by_j \, y^j +  
\d y^i \d \by_i  \, y^j \by_j  \big ) \cr
&{} + 2  \, \tr \big ( \by_i\, \d y^i \, \d \by_j \, y^j  +
\d y^i \,  \by_i\, y^j \, \d \by_j  \big ) \cr
&{}+ 2  \, \tr(\d \by_i\, \d y^j) \, \tr( \by_j \, y^i )
+ 4  \, \tr(\by_i\, \d y^j) \, \tr( \d\by_j \, y^i ) \, , 
}
}
which determines the possible freedom in $G_{IJ}{\!}^{(3)}$ shown in
\equivabc.

Using \defW\ we may determine from \twoAres\ and \Athree\ the
two and three loop contributions to $W_I$. This gives
\eqn\Wrestt{\eqalign{
W_I{\!}^{(2)} \d g^I = {}& \d \, {\ts {1\over 12}}\, \tr ( \by_i \, y^i ) \, , \cr
W_I{\!}^{(3)} \d g^I = {}& \d \big ( {\ts {1\over 144}} \, \lambda_{ij}{}^{kl}
\lambda_{kl}{}^{ij} - {\ts {5\over 48}} \, \tr (\by_i \, y^j) \, \tr(\by_j \, y^i)  \cr
 \noalign{\vskip - 2pt}
 & \quad {} 
- {\ts {11\over 288}}\, \tr ( \by_i \, y^i \,\by_j \, y^j + y^i \by_i \, y^j  \by_j) \big )\, ,
}
} 
which is compatible with \WQ{a}\ since $Q_I{\!}^{(2)} B^{(1)I}=0$.

Restricting to the supersymmetric case according to the prescription
described in section 5 then \WtwoM, neglecting gauge fields and $M$ terms, gives
\eqn\susyXt{
X(Y,\bY)_{\rm {Susy}}^{(2)} = {\ts{1\over 6}} \,(\pr^2 \bY \, \pr^2 Y)_i{}^i \, ,
}
while from \Wthreea, \chib\ and \chic
\eqnn\susyX$$\eqalignno{
X (Y,\bY)_{\rm{Susy}}^{(3)} = {}& - {\ts {5\over 16}} 
\, (\bY Y)_i{}^j \, (\pr^2 \bY \, \pr^2 Y)_j{}^i  - {\ts {1\over 8}} 
\,  (\bY  \, \pr^2 Y)_i{}^j \, (\pr^2 \bY \, Y)_j{}^i \cr
&{}  - {\ts {1\over 8}} \, \big  ( (\bY\,   \pr^\mu Y)_i{}^j \, 
(\pr^2 \bY \, \pr_\mu Y)_j{}^i
+ (\pr^\mu \bY  \, Y)_i{}^j \, (\pr_\mu \bY \, \pr^2 Y)_j{}^i \big ) \cr
&{} -{\ts {1\over 16}} \, 
 ( \pr^\mu \bY  \pr^\nu Y)_i{}^j \, (\pr_\mu \bY \pr_\nu Y  )_j{}^i
 +{ \ts{9\over 16}} \,  ( \pr^\mu \bY  \pr^\nu Y)_i{}^j \, (
\pr_\nu \bY \pr_\mu Y )_j{}^i \cr 
&{}- {\ts{9\over 16}} \, ( \pr^\mu \bY  \pr_\mu Y)_i{}^j \, 
(\pr^\nu \bY \pr_\nu Y)_j{}^i \, . &\susyX
}
$$
This three loop result has been verified  by an independent superspace
calculation. 

In the supersymmetric case the gauge field contributions at two loops
may be obtained from the calculations in the scalar/fermion model
by letting $a_\psi = - a_\chi{\!}^T = a_\phi$ so that the results in \PQres\
may be added to give
\eqn\PQsusy{\eqalign{
\big ( P_{IJ}{\!}^{(2)}\d g^I \d' g^J \big ){}_{\rm Susy}{}_i{}^j = {}&
\quar ( \d \bY \d'Y - \d' \bY \d Y )_i{}^j \, , \cr
\big ( Q_{I}{\!}^{(2)}\d g^I \big ){}_{\rm Susy}{}_i{}^j = {}&
{\ts{1\over 16}} ( \d \bY Y - \bY \d Y )_i{}^j \, .
}
}
Assuming \gone\ then \equivWQ\ gives
$ ( \Delta Q_{I}{\!}^{(2)}\d g^I  ){}_{\rm Susy}{}= 
{\ts{3\over 2}} z ( \d \bY Y - \bY \d Y )$ so that $Q^{(2)} \to 0$ 
if $z=-{1\over 24}$.  If this is done, from \varab\ and \Resab,
$a\to - {3\over 4},\,  b \to -1$.
Furthermore from \betaf
\eqn\betaf{\eqalign{
\big (\omega \cdot \beta_f{\!}^{(1)} \! \cdot \omega \big ){}_{\rm Susy} 
={}&  2\, \tr(\omega^2) \, ,\cr
\big (\omega \cdot \beta_f{\!}^{(2)}\! \cdot \omega \big ){}_{\rm Susy} 
={}&-   2\, \tr\big (\omega^2 (\bY Y)\big ) - {\ts {1\over 3}} \,
(Y*\omega) \cirk (\omega * \bY)   \, .
}
}
These results \betaf\ together with \PQsusy\ are sufficient to check \WQ{b}
with the three loop $G_{IJ}$ given by \susyGt\ and \Resab. The one and two
loop expressions for $\beta_f$ are compatible with an extension of the
NSVZ formula for the matter contributions to $\N=1$ gauge $\beta$-function of the form
$(\omega \cdot \beta_f \cdot \omega)_{\rm {Susy}} = 2\, 
\tr \big ( \omega^2 ( 1-2\gamma )\big ) - (Y*\omega)\cirk {\tilde G} \cirk
(\omega * \bY)$.

\newsec{Conclusion}

In this paper we have endeavoured to show that the existence
of  a metric on the space of couplings, for renormalisable theories
at least, and the associated equations, which are related to 
gradient flow,  provide significant constraints on $\beta$-functions
and anomalous dimensions. These results are applicable in the
context of the standard model in that their application here provides
a partial check of the three-loop Yukawa $\beta$-function in \Chet. 
For supersymmetric theories there are additional constraints such
as the metric being hermitian and K\"ahler which might follow
from an extension of the present discussion to  superspace.

A critical issue which has not been analysed in any detail here
is the role of anomalies which render the assumption of invariance
under arbitrary gauge transformations G${}_K$ invalid. This is
crucial for a more complete analysis of supersymmetric theories
where careful analysis of anomaly matching links IR and UV limits
under RG flow \AnselmiAM.

In this paper we have avoided perturbative calculations on curved space
backgrounds. Nevertheless the techniques described here for three
loop calculations of vacuum graphs with local couplings should allow
an extension to arbitrary metrics following \JackSK\ 
although as always the calculational
details are non-trivial.
\bigskip
\medskip
\vfil\eject
\noindent
{\bf Note Added}

\medskip
It has been pointed out to us by Zohar Komargodski
that the requirement that the metric be K\"ahler
in supersymmetric theories should only be possible for strictly marginal
operators.   This suggests that \Kahler\ be modified to
$$
\d Y \! \cirk G \cirk \d \bY = \d_Y \d_{\smash \bY} F
+  \d Y \! \cirk H \cirk \d \bY \, ,
$$
where $H$ vanishes if $\beta_Y$ or $\beta_\bY$ are zero.

\bigskip
\noindent
{\bf Acknowledgements}

\medskip

We are grateful to Konstantin Chetyrkin and and Max Zoller for informing about
some details of their work on three loop Yukawa $\beta$-functions.
HO is grateful to Boaz Keren-Zur for several
discussions about the details of their approach described in \Other\ and its 
relation to results in this paper. This allowed some simplifications in
the expressions obtained in sections 2 and 3 and later correction of some
errors. IJ is grateful to John Gracey for showing him how to
apply MINCER \Mincer, \Form\ and its relevance in three loop calculations. HO would also
like to express his thanks 
for hospitality at the University of Crete, CERN and Syddansk Universitet
during the slow progress of this work and also to KITP, UCSB for the
final revision.

\vfil\eject

\appendix{A}{Higher Loops in the Wess-Zumino Supersymmetric Theory}

In higher loop  calculations of the anomalous dimension $\gamma$  
transcendental numbers, such as 
$\zeta(3)$ in three or more loops, arise. These numbers are associated  
to diagrams with particular topologies which are possible initially
only at some minimal loop order $\ell$.  The connection between particular
transcendental numbers and a particular graph topology is valid only
up to scheme-dependent contributions to $\gamma$ and these need to
be considered separately.  
For each such non scheme dependent term $\gamma_\zeta$ 
contributing to $\gamma^{(\ell)}= {\rm O}(Y^\ell \bY^\ell)$,  
which is proportional to a  transcendental number $\zeta$ and 
 corresponds to diagrams involving a  topology which are not
 present at lower loop orders, the equations simplify. 
It is only necessary then to consider the lowest order $T^{(2)}$ and also
$T^{(\ell+1)}$ to determine the associated contribution to ${\tilde A}^{(\ell+2)}$.

The simplest case is when $\tr(\gamma_\zeta)$ corresponds to a connected 
symmetric graph with $\ell+1$ loops and $\ell$ $Y$-vertices linked to 
$\ell$  $\bY$-vertices. Such graphs are edge transitive so that all $3\ell$  
lines  are related by an  automorphism and are therefore equivalent. 
In this case $\gamma_\zeta$ may be recovered from  $\tr(\gamma_\zeta)$ 
by cutting any line. This implies the identity, for any $\omega_i{}^j$ and with
notation as in section 7,
\eqn\trgam{
\tr ( \omega \, \gamma_\zeta) = {1\over 3\ell} \,
(\omega * \bY)\cirk \pr_\bY\; \tr(\gamma_\zeta) =  {1\over 3\ell} \,
(Y * \omega )\cirk \pr_Y\; \tr(\gamma_\zeta) \, .
}
With $\beta_{\zeta \bY} = (\gamma_\zeta * \bY)$ and $T^{(2)}$ 
given by \susyG
\eqn\Tgam{
\d Y \cirk T^{(2)} \cirk \beta_{\zeta\bY} = 
\tr \big ( (\bY \d Y) \, \gamma_\zeta \big ) \, .
}
To ensure integrability in \ST\ it is necessary to assume 
$T^{(\ell+1)}$ contains a term proportional to $\zeta$ of the form
\eqn\TgamL{
\d Y \cirk T_\zeta \cirk \d \bY = {2\over 3\ell} \, \d_{\smash \bY} \, \d_Y \, 
\tr (\gamma_\zeta) \, ,
}
since then
\eqn\integAL{
\d Y \cirk T^{(2)} \cirk \beta_{\zeta \bY} + 
\d Y \cirk T_\zeta \cirk \beta_\bY{\!}^{(1)} = 
\d_Y \, \tr \big ( (\bY Y) \, \gamma_\zeta  \big ) \, .
}
In consequence there is an associated contribution ${\tilde A}_\zeta$ to 
${\tilde A}^{(\ell+2)}$ given by
\eqn\AL{
{\tilde A}_\zeta = \half \, \tr \big ( (\bY Y) \, \gamma_\zeta \big ) 
= {1\over 3\ell} \, \beta_\bY{\!}^{(1)} \cirk \pr_\bY\; \tr(\gamma_\zeta) =  
{1\over 3\ell } \,\beta_Y{\!}^{(1)} \cirk \pr_Y\; \tr(\gamma_\zeta)\, .
}
For this case
\eqn\lowest{
- \tr ( \gamma^{(1)} \gamma_\zeta ) + 
\Lambda^{(2)} \cirk \beta_{\zeta\bY} = 0 \, ,
}
so that we must take in \Aexact
\eqn\ALL{
{\tilde A}_\zeta =  \Lambda_\zeta \cirk \beta_\bY{\!}^{(1)} \, ,
}
where $\Lambda_\zeta$  is part of $\Lambda^{(\ell+1)}$. Hence from \AL\
\eqn\GAML{
\Lambda_\zeta \cirk \d \bY = {1\over 3\ell} \; \d_\bY \, \tr(\gamma_\zeta) 
\quad \Rightarrow \quad 3\, (\bY \Lambda_\zeta ) = \gamma_\zeta \, ,
}
in accord with \detL. In this case the metric $G_\zeta = T_\zeta$ so that
\TgamL\ ensures \Kahler\ is satisfied in this case with
\eqn\Fzeta{
F_\zeta =   {2\over 3\ell} \,  \tr (\gamma_\zeta) \, .
}

These results apply when $\ell=3$ for the term proportional to
$\zeta(3)$, $\gamma_{\zeta(3)} = D \, \gamma_D$ where 
 $\tr(\gamma_{\zeta(3)}) = {3\over 2} \zeta(3)\, 
(Y^3 \bY^3 )_{K_{3,3}}$, and also when  $\ell=4$, according to \FerreiraRC,
 for the term proportional to $\zeta(5)$, which satisfies
$\tr(\gamma_{\zeta(5)}) = - 10 \zeta(5)\, (Y^4 \bY^4 )_{M_8}$, with the 
vertices contracted as in the symmetric non planar graphs $K_{3,3}$, with 
6 vertices 9 edges, and $M_8$ forming a cube respectively.

At the next order
there are additional non planar contributions to $\gamma^{(4)}$ which are
proportional to $\zeta(3)$. These are determined by the corresponding
term at three loops. To show this we consider a contribution to $\gamma$,
in addition to the $\ell$-loop $\gamma_\zeta$ satisfying \trgam, 
at $\ell+1$ loops which is expressed  in terms of $\gamma_\zeta$. 
It is sufficient to assume that
the relevant term in $\gamma^{(\ell+1)}$ has the form
\eqn\gampr{
\gamma'{\!} _\zeta{}_i{}^j = A\, \bY_{ikm} Y^{lmj} \gamma_\zeta{\,}_l{}^k \, ,
}
with an undetermined coefficient $A$. As usual $\gamma'{\!} _\zeta$ determines 
$\beta'{\!}_{\zeta \bY} = (\gamma'{\!}_\zeta * \bY)$ and hence we may obtain
$\d Y \cirk T^{(2)} \cirk \beta'{\!}_{\zeta \bY} + 
\d Y \cirk T^{(3)} \cirk \beta_{\zeta \bY}$ which is part of
$\d_Y {\tilde A}'{\!}_\zeta$. There are also contributions 
$\d Y \cirk T_\zeta \cirk \beta_\bY{\!}^{(2)}$, determined by \TgamL, 
but it is necessary also to allow corresponding terms in $T^{(\ell+1)}$
and $K^{(\ell+1)}$. Assuming these must contain the subgraph associated
with $\gamma_\zeta$ they can have the general form proportional to
$\gamma_\zeta$
\eqn\Tellone{\eqalign{
\d Y \cirk T'{\!}_\zeta \cirk \d \bY 
= {}& \alpha_1 \, \tr \big ( (\d \bY \d Y ) \, \gamma_\zeta \big )
+ \alpha_2 \, \tr \big ( (\d \bY Y ) \,  \d_Y \gamma_\zeta \big )  +
\alpha_3 \, \tr \big ( (\bY \d Y ) \, \d_\bY \gamma_\zeta \big ) \cr
&{}+ \alpha_4 \, \tr \big ( (\bY Y ) \, 
\d_{\smash{\bY}} \d_Y \gamma_\zeta \big ) \,, \cr
\d' Y \cirk K'{\!}_\zeta \cirk \d Y = {}& \beta \, 
\tr \big ( ( \bY \d' Y ) \, \d_Y \gamma_\zeta \big ) - 
\d'Y \leftrightarrow \d Y \, .
}
}
To calculate $\d Y \cirk T'{\!}_\zeta \cirk \beta_\bY{\!}^{(1)} 
+ \beta_Y{\!}^{(1)}  \cirk K'{\!}_\zeta \cirk \d Y$ we use the identities
\eqn\idone{
\big ( \beta_\bY{\!}^{(1)} \cirk \pr_\bY - \beta_Y{\!}^{(1)} \cirk \pr_Y
\big ) \, \gamma_\zeta = \half \, \big [ ( \bY Y) , \gamma_\zeta \big ] \, ,
}
a special case of \conga\ valid for any $\gamma_\zeta$, and
\eqn\idtwo{
\tr \big ( ( \bY \d Y) \, \beta_\bY{\!}^{(1)} \cirk \pr_\bY \, \gamma_\zeta
\big ) =
\tr \big ( ( \bY Y) \, ( \d_Y \beta_\bY{\!}^{(1)}) \cirk \pr_\bY \, 
\gamma_\zeta \big ) + \half \, \tr \big ( \big [  ( \bY \d Y) , (\bY Y) 
\big ] \, \gamma_\zeta \big ) \, ,
}
which may be derived from \trgam\ and reflects that all lines in the
graph for $\tr( \gamma_\zeta)$ are equivalent. Combining all contributions
to $\d_Y  {\tilde A}'{\!}_\zeta$ gives finally
\eqn\Apgam{
2\, {\tilde A}'{\!}_\zeta = 
Y_1 \, \tr \big ( (\bY Y \bY Y ) \, \gamma_\zeta \big )
+  Y_2 \, \tr \big ( (\bY Y )^2 \, \gamma_\zeta \big )
+ Y_3 \, 
\tr \big ( (\bY Y ) \,  \beta_\bY{\!}^{(1)} \cirk \pr_\bY \gamma_\zeta \big ) 
\, ,
}
where 
\eqn\Yeqs{\eqalign{
Y_1 = {}& 2a + \alpha_1 = 2b + A = -1 + \alpha_2 +\beta \, , \cr
Y_2 = {}& a + \quar ( \alpha_3 + \beta ) = b - \quar ( \alpha_3 + \beta ) 
+ \half \alpha_1 = \half ( \alpha_2 + \beta ) \, , \cr
Y_3 ={}& \alpha_4 = \half ( \alpha_3 - \beta ) \, .
}
}
The equations for $Y_1,Y_2$ give rise to integrability conditions once more
so that we may eliminate $\alpha_1, \alpha_2+\beta,
\alpha_3 + \beta$ and then determine
\eqn\Ares{
A= - 2 \, ,
}
independent of $a,b$.
Remarkably this agrees with the non planar $\zeta(3)$ term in $\gamma^{(4)}$,
after subtracting scheme-dependent terms,
obtained in \FerreiraRC{}. Subject to \Ares
\eqn\Yres{
Y_1 = 4a-1 \, , \qquad Y_2= 2 a \, .
}

At this order there is the freedom due to \varKT\ 
arising from taking $g = w \, g_\zeta$ for
\eqn\gzeta{
\d Y \cirk g_\zeta \cirk \d  \bY = {1\over 3\ell}\,
\d_Y \d_{\smash \bY} \tr (\gamma_\zeta) \, ,
}
which leads to an arbitrariness under variations
\eqn\varabw{
\Delta \alpha_2 = \Delta \alpha_3 =\Delta \alpha_4 =-\Delta \beta = w \, ,
}
giving $\Delta Y_1 = \Delta Y_2 =0$, $\Delta Y_3 = w$. The corresponding
variation in ${\tilde A}'{\!}_\zeta$ follows from
\eqn\varAW{
\beta_Y{\!}^{(1)}  \cirk g_\zeta \cirk \beta_\bY{\!}^{(1)}  = \half \,
\tr \big ( (\bY Y ) \,\beta_\bY{\!}^{(1)} \cirk \pr_\bY \gamma_\zeta \big ) \, .
 }
For \gone, $\d Y \cirk g \cirk \d \bY = z\, \d Y \cirk \d \bY=  z  \, 
\tr \big ( (\d \bY \d Y ) \big ) $, leading to \varab\  
$\Delta a = 3z  , \ \Delta b = 6 z $ then also it is necessary that
\eqn\varabz{
\Delta \alpha_1 = \Delta \alpha_2 =\Delta \alpha_3  = \Delta \beta = 6 z \, ,
}
so that $\Delta Y_1 = 12z, \ \Delta Y_2 = 6z, \ \Delta Y_3 = 0$. In this case
\eqn\varAz{
{\ts {1\over 3}} \big ( \beta_Y{\!}^{(1)} \cirk \beta_{\smash{\zeta\, \bY}} 
+ \beta_{\zeta \, Y} \cirk \beta_\bY{\!}^{(1)} \big ) =  
2\, \tr \big ( (\bY Y \bY Y ) \, \gamma_\zeta \big )
+  \tr \big ( (\bY Y )^2 \, \gamma_\zeta \big ) \, .
}

If $a=b=-\half$, as in \resab,  \Yeqs\ has
the solution
\eqn\resabz{
\alpha_1=\alpha_2+\beta =\alpha_3 + \beta = - 2 \, , 
}
but the metric $G'{\!}_\zeta$ obtained then from \Tellone\ 
cannot be written in the K\"ahler form  \Kahler\ for any choice of 
$\alpha_4 = \half(\alpha_3-\beta)$ making use of the
the freedom under \varabw. However if we also allow a change of
scheme as in \varKTn\ with $T\to T^{(1)}$ and $h\to - \gamma_\zeta$
so that 
\eqn\Telltwo{
\d Y \cirk \delta  T'{\!}_\zeta \cirk \d \bY 
=  \tr \big ( (\d \bY \d Y ) \, \gamma_\zeta \big )
+ \tr \big ( (\d \bY Y ) \,  \d_Y \gamma_\zeta \big )  +
\tr \big ( (\bY \d Y ) \, \d_\bY \gamma_\zeta \big ) \, ,
}
then, taking $\beta=0, \alpha_4=-1$, 
$\d Y \cirk (T'{\!}_\zeta +  \delta  T'{\!}_\zeta)\cirk \d\bY
 = \d Y \cirk G'{\!}_\zeta \cirk \d \bY = \d_Y
\d_{\smash \bY} F'{\!}_\zeta$  with
\eqn\Fzeta{
F'{\!}_\zeta = -  \tr \big ( (\bY Y) \, \gamma_\zeta \big ) \,. 
}

The result \Apgam\ with \Yres\ may also be expressed in the form
\Aexact.  To solve \detL\ it is sufficient to take
\eqn\lamprf{
 \Lambda'{\!}_\zeta \cirk \d \bY =  u\, 
\tr \big ( (\d \bY Y ) \, \gamma_\zeta \big ) + v \, 
\tr \big ( ( \bY Y ) \, \d_\bY \gamma_\zeta \big ) \, .
}
Using $\big [ (\omega' * \bY ) \cirk \pr_\bY ,  (\omega * \bY ) \cirk \pr_\bY \big ]
=  ([\omega,\omega' ]* \bY ) \cirk \pr_\bY $ then from \trgam\ we
may derive
\eqn\omeq{
\tr \big ( \omega \,  (\omega' * \bY ) \cirk \pr_\bY \gamma_\zeta \big )
= \tr \big ( \omega' \,  (\omega * \bY ) \cirk \pr_\bY \gamma_\zeta \big )
+ \tr \big ( [ \omega , \omega'] \, \gamma_\zeta \big ) \, ,
}
and hence obtain, with $\Theta^{(2)}$ as in \GHtwo,
\eqn\Ylam{
3\, ( \bY  \Lambda'{\!}_\zeta ) - \Theta^{(2)} \cirk \beta_{\zeta \bY} = \gamma'{\!}_\zeta 
+ (u-v) \, (\bY Y) \,  \gamma_\zeta
+(v-\theta)\, \gamma_\zeta \, (\bY Y) +v\,  (\bY Y) * \pr_\bY \, \gamma_\zeta  \, ,
}
for $\gamma'{\!}_\zeta $ as in \gampr\ so long as
\eqn\uvA{
2(u-\theta) = A \, .
}
Hence  $3\, ( \bY  \Lambda'{\!}_\zeta )  - \Theta^{(2)} \cirk \beta_{\zeta \bY} 
- \Theta_\zeta \cirk \beta_{\bY}{\!}^{(1)}
= \gamma'{\!}_\zeta - \gamma^{(1)} \gamma_\zeta
- \gamma_\zeta \gamma^{(1)} $ if we take
\eqn\Thetazeta{
\Theta_\zeta \cirk \d \bY = 2v\,  \d_\bY \gamma_\zeta \, ,
}
and
\eqn\resuv{
u-v = v-\theta =  - \half \, .
}
Applying \resuv\ in \uvA\ gives \Ares\ once more.

Using \lamprf\ and \Thetazeta\ in \resGK\ gives a metric of the form
\Tellone\ with
\eqn\azeta{
\alpha_1 = 2u=-2+2\theta \, , \quad \alpha_2 = 
\alpha_3 =4v = -2 +4\theta \, , \quad \alpha_4 =2v =   -1 +2\theta
\, , \quad \beta = 0 \, .
}
These results satisfy \Yeqs\ for $a,b$ given by \ablt\ so that
\eqn\Yres{
Y_1 = -3 + 4\theta \, , \qquad Y_2= Y_3 = -1 + 2\theta  \, .
}

Since, with $\Lambda^{(2)},\Lambda^{(3)}$ given in \Wlam,
\eqn\gamLam{\eqalign{
& - \tr \big ( \gamma^{(1)} \, \gamma'{\!}_\zeta \big )
+  \Lambda^{(2)} \cirk \beta'{\!}_{\zeta\bY}
= - \tr \big ( \gamma^{(2)} \, \gamma_\zeta \big )+
 \Lambda_\zeta \cirk \beta_\bY{\!}^{(2)} = 0 \, , \cr
&  \tr \big ( \gamma^{(1)}{}^2 \, \gamma_\zeta \big ) =   
\quar \, \tr \big ( (\bY Y)^2 \, \gamma_\zeta \big )\, , \cr
&  \Lambda^{(3)} \cirk \beta_{\zeta\bY} = 
2\lambda \,  \tr \big ( (\bY Y \bY Y ) \, \gamma_\zeta \big ) +
\lambda  \, \tr \big ( (\bY Y)^2 \, \gamma_\zeta \big ) \, , \cr
& \Lambda'{\!}_\zeta \cirk \beta_\bY{\!}^{(1)} =
u\,  \tr \big ( (\bY Y \bY Y ) \, \gamma_\zeta \big )
+ \half u \,  \tr \big ( (\bY Y)^2 \, \gamma_\zeta \big )\big ) +
v \, \tr \big ( (\bY Y) \, \beta_\bY{\!}^{(1)} \cirk \pr_\bY 
\gamma_\zeta \big ) \, , 
}
}
we may verify
\eqn\Azeta{\eqalign{
 {\tilde A}'{\!}_\zeta ={}& 
- \tr \big ( \gamma^{(1)} \, \gamma'{\!}_\zeta \big )
 - \tr \big ( \gamma^{(2)} \, \gamma_\zeta \big ) 
 + \tr \big ( \gamma^{(1)2} \gamma_\zeta \big ) \cr
&{} +  \Lambda^{(2)} \cirk \beta'{\!}_{\zeta\bY}
+  \Lambda^{(3)} \cirk \beta_{\zeta\bY}
+  \Lambda_\zeta \cirk \beta_\bY{\!}^{(2)} 
+  \Lambda'{\!}_\zeta \cirk \beta_\bY{\!}^{(1)} \, ,
}
}
as required by \Aexact\ with $H=0$.

\appendix{B}{Derivation of Local RG Equations}

Usually RG equations are derived by considering the response
to a change of cut-off scale or using dimensional regularisation
variations in the arbitrary mass scale $\mu$ which is necessary
for dimensions $d\ne 4$. For the equations in section 8, which
are related to broken conformal symmetry, a slightly different 
approach is required. For renormalisable scalar fermion theories 
in $d$ dimensions $ \L(\Phi_0,{\bar \Phi}_0, g_0 , a_{0} , M_0 )$
can be chosen to be  conformal primary under
conformal transformations so long as $g_0,a_0,M_0$ transform
appropriately as well as $\Phi_0,{\bar \Phi}_0$. The generator
of conformal transformations for this theory is then,
for any conformal Killing vector $v^\mu$,
\eqn\conDO{\eqalign{
-\D_{0,v}={}&  \big  ( \L_v \Phi_0 - \half \vep\, \sigma_v \,  \Phi_0  \big )\cdot  
{\pr \over \pr \Phi_0}
+ \big  ( \L_v  {\bar \Phi}_0  - \half \vep  \, \sigma_v \, {\bar \Phi}_0  
\big ) \cdot {\pr \over \pr {\bar \Phi}_0 } \cr
\noalign{\vskip -4pt}
& {}+   \big ( v^\mu \pr_\mu \,  g_0{\!}^I + \vep\, \sigma_v \, k_I  g_0{\!}^I \big )
 \cdot {\pr \over \pr g_0{\!}^I} 
+\L_v a_{0\mu} \cdot {\pr \over \pr a_{0\mu} } + 
\L_v M_0 \cdot  {\pr \over \pr M_0} \, ,
}
}
for
\eqn\LppaM{\eqalign{
\L_v \Phi_0 = {}& \big ( v^\mu  \pr_\mu - \half \, \omega_v{\!}^{\mu\nu} 
s_{\Phi \mu\nu} + \sigma_v \, \Delta_\Phi \big ) \Phi_0 \, , \cr
\L_v {\bar \Phi}_0 = {}& \ v^\mu \pr_\mu\, {\bar \Phi}_0 +
 {\bar \Phi}_0 \big ( \half \, \omega_v{\!}^{\mu\nu} s_{{\bar \Phi}\mu\nu}
+ \sigma_v \, \Delta_{\bar \Phi} \big )  \, , \cr
\L_v a_{0\mu} ={}&  v^\nu\pr_\nu\, a_{0\mu} + \pr_\mu v^\nu a_{0\nu} \, ,\cr
\L_v M_0 = {}& v^\mu\pr_\mu\, M_0 + \sigma_v \big ( 4 \, M_0 -
\Delta_{\bar \Phi} M_0 - M_0 \Delta_\Phi \big ) \, ,
}
}
where $\omega_v{\!}^{\mu\nu} = \pr^{[\mu} v^{\nu]}$ 
and $s_{\Phi\mu\nu}, {s}_{{\bar \Phi}\mu\nu}$ are the appropriate spin matrices.
$\Delta_\Phi, \Delta_{\bar \Phi}$ are the canonical dimension matrices for 
$\Phi,{\bar \Phi}$ when $d=4$ and in consequence $\L_v$ has no explicit 
dependence on $\vep$ for  each case in \LppaM. It is easy to verify
\eqn\DOalg{
\big [  \D_{0,v} , \D_{0,v'} \big ] = \D_{\smash{0,[v,v']}} \, .
}
The crucial assumption is then that $\L$ satisfies\foot{This is
the condition for $\L$ to be a conformal primary, it dictates
the form of the scalar kinetic term so that $\L_{K0} = - \pr \bphi_0 \cdot
\pr \phi_0 + {1\over 2} \,\pr^2 ( \bphi_0 \, \phi_0 )$.
}
\eqn\bareW{
- \D_{0,v}\,  \L(\Phi_0,{\bar \Phi}_0, g_0 , a_{0} , M_0 ) 
= \big ( v^\mu  \pr_\mu + d \, \sigma_v \big ) 
\L(\Phi_0,{\bar \Phi}_0, g_0 , a_{0} , M_0 )  \, . 
}

The derivation of finite local RG equations depends  on the detailed
form of the relation between
$\Phi_0,{\bar \Phi}_0, g_0{\!}^I , a_{0\mu} , M_0$ and the corresponding finite
$\Phi,{\bar \Phi}, g^I , a_\mu, M$ implicitly defined by \bareL. Defining
\eqn\defDv{
\D_v =  - v^\mu \pr_\mu \,  g^I  \cdot {\pr \over \pr g^I}
- \L_v a_{\mu} \cdot {\pr \over \pr a_{\mu} } -
\L_v M \cdot  {\pr \over \pr M} -  \L_v  {\Phi}   \cdot {\pr \over \pr {\Phi} }
-  \L_v  {\bar \Phi}   \cdot {\pr \over \pr {\bar \Phi} } \, ,
}
then $\D_{0,v}$ may be expressed in terms of $g^I,a_\mu,M,\Phi,{\bar \Phi}$
in the form
\eqn\DDvspp{
\D_{0,v} = \D_v + \D_{\sigma_v} + \D_{\smash{\sigma_v,\Phi,{\bar \Phi}}} \, ,
}
with $\D_\sigma ,\D_{\smash{\sigma,\Phi,{\bar \Phi}}}$ as in \defDsi.
The commutation relation \DOalg\  ensures that the coefficients in $\D_\sigma$
obey the required consistency conditions.

Since  $-\D_v \Z = v^\mu \pr_\mu \Z, \, - \D_v D_\mu  g^I = \L_v D_\mu g^I$ 
then
\eqn\relzero{
 - \D_v \, \Phi_0 = \L_v \Phi_0 \, , \qquad
- \D_v \, {\bar \Phi}_0 = \L_v {\bar \Phi}_0 \, , \qquad
- \D_v \, a_{0\mu} = \L_v a_{0\mu} \, .
}
However
\eqn\LDsq{
- \D_v \, D^2 g^I = ( v^\mu \pr_\mu + 2 \, \sigma_v ) D^2 g^I
+ \pr^2 v_\mu \, D^\mu g^I \, , \qquad  \pr^2 v_\mu = - (d-2)\, \pr_\mu \sigma_v \, .
}
As $M_0$ may contain counterterms involving $D^2 g^I$ in general
$- \D_v M_0 \ne \L_v M_0$ but taking this into account
\eqn\bareWt{\eqalign{
  - \bigg (&  \D_v 
+  (d-2) \, \pr_\mu \sigma_v \, D^{\mu} g^I {\pr \over \pr D^2 g^I } \bigg  ) 
\L(\Phi_0,{\bar \Phi}_0, g_0 , a_{0} , M_0 ) \cr
 & {}\sim  \big ( v^\mu \pr_\mu + 4 \, \sigma_v \big ) 
\L(\Phi_0,{\bar \Phi}_0, g_0 , a_{0} , M_0 ) \, ,
}
}
where $\sim$ denotes equality up to total derivatives.
Subtracting  \bareWt\ from \bareW\ then gives
\eqn\DbareW{
 \bigg ( 
 \vep \, \sigma - \D_\sigma - \D_{\smash{\sigma,\Phi,{\bar \Phi}}} 
-  (2-\vep ) \, \pr_\mu \sigma \, D^{\mu} g^I {\pr \over \pr D^2 g^I }  
\bigg  ) \L(\Phi_0,{\bar \Phi}_0, g_0 , a_{0} , M_0 ) \sim 0 \, ,
}
for $\sigma$ linear in $x$, which is identical to \rgflow\ for a suitable 
choice of total derivative contributions. As shown in section 8
\DbareW\ is sufficient to determine the various
contributions to $\D_\sigma$, in particular
\eqn\defbet{
{\hat \beta}^J {\pr \over \pr g^J} \, g_0{\!}^I  =  - \vep \, k_I g_0{\!}^I \, .
}
 This is equivalent to the 
standard definition $\mu {\d \over \d \mu} g^I |_{g_0} = {\hat \beta}^I$
when $ g_0{\!}^I = \mu^{k_I \vep}(g^I + L^I(g))$, with $L^I$ containing
just poles in $\vep$ and gives the standard form \Bhatr.

We assume also that \bareW\ with \DDvspp\ extend also to $\L_0$
including also the field independent counterterms  so that
\eqn\wardL{
\big( \D_v + \D_{\sigma_v} 
+ \D_{\smash{\sigma_v,\Phi,{\bar \Phi}}} + v^\mu \pr_\mu + d \, \sigma_v \big  )
\L_0 \sim  {1\over 16\pi^2}
\big ( \sigma_v X - 2 \, \pr_\mu \sigma_v \, Y \big ) \, ,
}
In a similar fashion to the above this leads to \rgeq. \wardL\ directly
implies the broken conformal Ward identities discussed in section 4.


\appendix{C}{Calculations}

Assuming continuation to a Euclidean metric
the short distance divergent parts in  \Wthreeb\ may be obtained by using
 the integral formula
\eqn\intF{\eqalign{
{1\over \pi^{{1\over 2}d} } \int \d^d z \; {}& 
\big ( (x-z)^2 \big )^{-\lambda} \, \big ( (y-z)^2 \big )^{-\mu} \, f(z) \cr
= {}& {1\over \Gamma(\lambda) \Gamma(\mu)} \sum_{n\ge 0} {1\over n!}\, 
\Gamma(\lambda+\mu - \half d - n) \, (s^2)^{{1\over 2}d-\lambda-\mu +n } \,
b_n(x,y) \cr
&{}+  \hbox{terms analytic in} \ s \, , 
}
}
for
\eqn\defbn{
b_n ( x,y) = \int_0^1 \! \d t \, t^{{1\over 2}d-\lambda+n-1}
(1-t)^{{1\over 2}d-\mu +n-1} \, (\quar \pr^2)^n f(x - ts) \, .
}
To verify \intF\ it is sufficient to consider Fourier transforms with
respect to $x,y$ where
\eqn\Fxl{
\int \d^d x \; e^{i k \cdot x} \, (x^2)^{-\lambda} = \pi^{{1\over 2}d} \, 
{ \Gamma( \half d - \lambda) \over \Gamma(\lambda) } \, \big (
\quar k^2 \big )^{\lambda-{1\over 2}d} \, ,
}
and on the right hand side the sum over $n$ reproduces the left hand
side within an appropriate region of convergence. For generic $\lambda,\mu$,
$b_n$ satisfies
\eqn\bnrel{
 ( s \cdot \pr_x +d -\lambda-\mu + n-1) b_n(x,y)  - 
n  \, \quar \pr_x{\!}^2 b_{n-1}(x,y) 
= ( \half d -\lambda -1 ) b_n(x,y) \big |_{\lambda \to \lambda+1} \, ,
}
as well as the similar equation obtained by $x \leftrightarrow y,
\lambda \leftrightarrow \mu$. The $t$-integration in $b_n$ is 
convergent when $\lambda,\mu < \half d +n$ but it may be extended by
analytic continuation. $b_n(x,y)$ are smooth functions for $y$ in the
neighbourhood of $x$ but there are poles
for $\lambda,\mu = \half d +n + p$, $p=0,1,\dots$, which 
reflect short distance sub-divergences. The poles present in the
expansion \intF\ at $\lambda + \mu = \half d + n$ are 
generated by divergences for large $z$ which should be cancelled by the
analytic terms assuming $f(z)$ falls off sufficiently fast as $z\to \infty$.

For calculations here the divergent $\vep$-poles are obtained by
using, for $\mu$ an arbitrary scale mass,
\eqn\divs{\eqalign{
& \sum_{i=0}^p \alpha_i \, (s^2)^{-{1\over 2}d-n+{1\over 2}\delta_i} = 
\sum_{i=0}^p \alpha_i \, {\mu^{\delta_i}\over \delta_i } \, 
S_d \, {1\over (\half d)_n \, n! } \;
(\quar \pr^2 )^n \delta^d(s)  + {\rm O}(1)  \, , 
}
}
as  $\vep \to 0$ where $\alpha_i, \delta_i$ are assumed to
depend on $\vep$ such that in this limit
\eqn\condit{
\delta_i = {\rm O}(\vep) \, , \ \ \alpha_i = {\rm O}(\vep^{-p}) \, , \quad
\sum_{i=0}^p \alpha_i \, \delta_i{\!}^r = {\rm O}(1) \, , \  
r=0, \dots, p-1 \, .
}
The conditions \condit\ are necessary and sufficient for the left hand
side of \divs\ to have a finite limit as $\vep\to 0$ and also ensure
that the pole terms on the right hand side, of ${\rm O}(\vep^{-r}), \, 
r=1,\dots, p$, have no $\mu$ dependence. The result \Gdiv\ is a special
case of \divs.

The results given in \intF\ and \divs\ may be used to obtain the
$\vep$-poles reflecting short distance divergences in products
involving $Y_f(x,y)$, as defined in \defY, and also
\eqn\defYt{
{\tilde Y}_f(x,y) = \int \d^d z \; G_0(x-z)  \, f(z) \, G_0(z-y)  \, .
}
\intF\ gives the expansion, up to terms which are regular as
$s\to 0$,
\eqnn\Yres$$\eqalignno{
Y_{f} (x,y) \sim {}& {1 \over 4 S_d{\!}^2}\,
{1\over (d-2)^2(d-3)} \, b_{f,0}(x,y) \, (s^2)^{3-d} 
- {1\over \vep}\, {1 \over 4 S_d S_4} \, {1\over d-2}\, f(x)\,
(s^2)^{1-{1\over 2}d}   \cr
{}& - {1 \over 4 S_d{\!}^2}\,
{1\over (d-2)^2(d-3)} \, b_{f,1}(x,y) \, {1\over \vep} \big (
(s^2)^{4-d} - 1 \big ) \, ,& \Yres
}
$$
where
\eqn\bnj{
b_{f,n} ( x,y) = \int_0^1 \! \d t \,
t^{n+1-{1\over 2}d} \, (1-t)^{n} \, (\quar \pr^2)^n f(x - ts) \, ,
}
and also
\eqn\Ytres{
{\tilde Y}_f(x,y)  \sim - {1 \over 4 S_d}\, {1\over d-2} \,  {2\over \vep} \Big ( 
{\tilde b}_{f,0}(x,y) - {1\over 3-\half d} \, {\tilde b}_{f,1}(x,y)\,  s^2
\Big )  \,\big ( (s^2)^{2-{1\over 2}d}  -1 \big ) \, ,
}
for
\eqn\bnt{
{\tilde b}_{f,n}( x,y) = \int_0^1 \! \d t \,  t^n(1-t)^n \, (\quar \pr^2)^n f(x - ts) \, .
}
In both \Yres\ and \Ytres\ terms which are regular as $s\to 0$ have been 
subtracted to cancel an IR divergence at $\vep=0$. The terms omitted in 
\Yres, \Ytres\ are then  without any $\vep$-poles. 
In consequence
\eqn\Yxx{\eqalign{
Y_{f} (x,x) \sim {}& {1\over \vep} \,  {1 \over  S_d{\!}^2}
{1\over (d-2)^2(d-3)(6-d)(8-d)} \, \quar \pr^2 f(x) \, , \cr
{\tilde Y}_{f} (x,x) \sim {}& {1\over \vep} \,  {1 \over 2 S_d}\, {1\over d-2} 
\, f(x) \, . 
}
}
There is also a UV sub-divergence present in $b_{f,0}$ since
\eqn\subb{
b_{f,0} ( x,y) \sim  {2\over \vep} \, f(x) \, .
}

The various results in the text can be obtained from analysing the
singularities in products involving ${\tilde Y}_f , Y_f$ using \divs.
For two loop graphs relevant for calculating \achi, \aphia\ we used
\eqn\YtY{\eqalign{
(16\pi^2)^2& {\tilde Y}_f(x,y) {\overleftarrow\pr}{\!}_{\mu y}\, Y_g(y,x)
{\overleftarrow\pr}{\!}_{\nu x} \sim -  (16\pi^2)^2 
\pr_{\nu x} {\tilde Y}_f(x,y) {\overleftarrow\pr}{\!}_{\mu y}\, Y_g(y,x) \cr
={}& {1\over 2 \vep^2}\, (1-\quar \vep) \, f(x) g(x) \, \delta_{\mu\nu} 
\delta^d(s) \, ,
}
}
and
\eqn\YGpr{\eqalign{
(16\pi^2)^2 {Y}_f(x,y) \, G_0(s)
\sim {}&-   {2\over \vep^2}\, (1-\half  \vep) \,  f(x) \,  \delta^d(s) \, ,\cr 
(16\pi^2)^2 {Y}_f(x,y) {\overleftarrow\pr}{\!}_{\mu y}\, G_0(s)
\sim {}& {1\over \vep^2}\, (1-\quar \vep) \,  f(x) \, \pr_\mu \delta^d(s) 
+ {1\over 2\vep} \, \pr_\mu f(x) \, \delta^d(s) \, ,\cr 
(16\pi^2)^2 G_0(s)  {\overleftarrow\pr}{\!}_{\mu y}\, Y_f(y,x)
\sim {}& {1\over \vep^2}\, (1-{\ts{3\over 4}} \vep) \,  f(y) \,
\pr_\mu \delta^d(s) + {1\over 2\vep} \, \pr_\mu f(x) \, \delta^d(s) \, . 
}
}

For the three loop integrals in \Wthreeb\ it is necessary 
to determine the $\vep$-poles in various products involving $Y_f$
with $Y_g$ or $G_0{\!}^2$. These can be reduced to
\eqna\YYo$$\eqalignno{
(16\pi^2)^3\, Y_{f}(x,y) \, Y_{g}(y,x) \sim{}&
{8\over 3\vep^3}\, \big (1 - \half \vep - \quar \vep^2 \big ) \, f(x)g(x) \, 
\delta^d(s) \, ,  & \YYo{a} \cr
(16\pi^2)^3\, Y_{f}(x,y) \, \R G_0(s)^2 \sim{}&
- {1\over 3\vep^2}\, \big ( 1 - {\ts {1\over 4}} \vep  \big ) 
\big ( f(x) \, \, \pr^2 \delta^d(s) + \pr^2 f(x) \, \delta^d(s) \big ) \cr
&{} + {1\over 3 \vep} \,  f(y) \, \pr^2 \delta^d(s) \, , &\YYo{b}
}
$$
and with one derivative
\eqna\YYy$$\eqalignno{
(16\pi^2)^3\, Y_{f}(x,y){\overleftarrow\pr}{\!}_{\mu y} \, Y_{g}(y,x) \sim{}&
- {4\over 3\vep^3}\, \big (1 - \half \vep - \quar \vep^2 \big ) \, f(x)g(y) \,
\pr_\mu \delta^d(s) \cr
&{} - {1\over 3\vep^2}\, \big (1 - \quar \vep \big )\,
\pr_\mu \big ( f(x)g(x) \big ) \, \delta^d(s) \, ,  &  \YYy{a} \cr
(16\pi^2)^3 \, Y_{f}(x,y){\overleftarrow\pr}{\!}_{\mu y}\,
\R G_0(s)^2 \sim{}&
 {1\over 9\vep^2}\, \big ( 1 - {\ts {7\over 12}} \vep  \big )
\big ( f(x) \,\pr_\mu \pr^2 \delta^d(s) -\pr_\mu \pr^2 f(x) \, \delta^d(s) 
\big ) \cr
&{} + {2\over 9 \vep} \, \big ( \pr_\mu f(y) \, \pr^2 \delta^d(s) 
- \pr_\nu  f(y) \, \pr_\mu \pr_\nu  \delta^d(s) \big ) \, , & \YYy{b}
}
$$
and with two derivatives
\eqn\YYdiv{\eqalign{
& (16\pi^2)^3 \, Y_{f}(x,y){\overleftarrow\pr}{\!}_{\mu y}\,
Y_{g}(y,x){\overleftarrow\pr}{\!}_{\nu x} \cr
& {} \sim  - \Big (  
{2\over 9\vep^3} 
\big ( 1 - {\ts{1\over 12}}\vep - {\ts{83\over 144}} \vep^2 \big )\, f(x)g(y)
- {1\over 18\vep^2} 
\big ( 1 - {\ts{13\over 12}}\vep \big )\,  
\big ( f(x) g(x)+f(y)g(y) \big ) \Big ) \cr
&{} \hskip 1cm {}\times 
\, \big ( 2 \pr_\mu \pr_\nu + \delta_{\mu\nu} \pr^2 \big )  \delta^d(s) \cr
& \hskip 0.5cm {} +  {1\over 3\vep^2} 
\big ( 1- \quar \vep \big ) \Big( f(x) g(y) 
- \half \big ( f(x) g(x)+f(y)g(y) \big ) 
\Big ) \, \pr_\mu \pr_\nu  \delta^d(s) \cr
& \hskip 0.5cm {} + {1\over 18\vep^2}\big ( 1 -  {\ts{1\over 12}} \vep \big)
\, \delta_{\mu\nu} \, \big ( f(x)\, \pr^2 g(x) + \pr^2 f(x) \, g(x)
\big ) \, \delta^d(s) \, \cr
& \hskip 0.5cm {} - {1\over 18\vep^2}\big ( 1 -  {\ts{7\over 12}} \vep \big)
\, \big ( f(x)\, \pr_\mu \pr_\nu g(x) +  \pr_\mu \pr_\nu f(x) \, g(x)\big ) \, 
\delta^d(s) \, \cr
& \hskip 0.5cm {} +  {1\over 3\vep^2} 
\big ( 1- \quar \vep ) \, \pr_{[\mu} \delta^d(s) \,
\big ( \pr_{\nu]} f(x)\, g(y) + f(x)  \, \pr_{\nu]} g(y) \big ) \cr
& \hskip 0.5cm {} +  {1\over 9\vep}  
\big ( \delta_{\mu\nu} \, \pr f(x) \cdot \pr g(x)
+ \pr_\mu f(x)\,  \pr_\nu g(x) + 
\pr_\nu f(x)\,  \pr_\mu g(x) \big ) \delta^d(s) \, .
}
}

By integrating $\pr_x{\!}^2, \pr_y{\!}^2$ by parts and using
\deY\ with \YYo{a,b}, \YYy{a,b} and \YYdiv\ we may obtain
\eqnn\YYdivhk$$\eqalignno{
(16\pi^2)^3 \,  &  {1\over 2}  \int   \d^dx \, \d^d y \; h(x) \,
\pr_x{\!}^2 Y_{f}(x,y) \, k(y) \,  \pr_y{\!}^2 Y_{g}(y,x) \cr
\sim  \int &  \! \d^dx \, \bigg (  {4\over 3\vep^3} \big (
1-\quar \vep -{\ts {19\over 48}}\vep^2 \big ) \, \pr_\mu h \, \pr_\mu f \, \pr_\nu k\, \pr_\nu g
+ {1\over 18\vep} \, \pr_\mu h\, \pr_\nu f \, \big (  \pr_\mu k \, \pr_\nu g 
+\pr_\nu k \, \pr_\mu g \big ) \cr 
&{} + {4\over 9 \vep^3} \big (1 + {\ts {5\over 12}} \vep - {\ts {35\over 144}}\vep^2\big ) \, 
\pr_\mu h\, \pr_\nu f \, \big (  \pr_\mu k \, \pr_\nu g 
-\pr_\nu k \, \pr_\mu g \big ) \cr
&{}+ {1\over 18\vep^2}\big ( 1 - {\ts {25\over 12}} \vep \big )
\big ( \pr^2 h \, \pr^2\!  f\, k \, g + h\, f \, \pr^2 k \, \pr^2 g \big )\cr 
&{} + { 1\over 9\vep^2}\big ( 1+{\ts {5\over 12}}\vep \big )
 \big ( h\, \pr^2\!  f\, \pr^2 k \, g  + \pr^2 h \,  f\, k \, \pr^2 g  \big )
 + { 1\over 36\vep}
 \big ( \pr^2 h \,  f\,  \pr^2 k \,  g  + h\, \pr^2\!  f\, k \, \pr^2 g \big ) \cr
\noalign{\vskip 2pt}
&{}+ {1\over 9\vep^2}\big ( 1-{\ts{7\over 12}}\vep \big )
 \big (\pr_\mu h \, \pr^2 \! f\, \pr_\mu k \, g+ \pr^2 h \, \pr_\mu f \, k \, \pr_\mu g
+ h \, \pr_\mu f \, \pr^2 k \, \pr_\mu g + 
\pr_\mu h \, f \, \pr_\mu k \, \pr^2 g \big )  \cr
\noalign{\vskip -2pt}
&{}- {1\over 9\vep^2}\big ( 1+{\ts{5\over 12}}\vep \big )\big (
 \pr_\mu h\,  \pr_\mu f \,   \pr^2  k \, g + \pr_\mu h\,  \pr_\mu f \,  k \, \pr^2 g  +
\pr^2h  \,  f \, \pr_\mu k \, \pr_\mu g + h \, \pr^2 \! f \,  \pr_\mu k \, \pr_\mu g  \big ) \cr
&{}- {1\over 18\vep} \big (
\pr^2h  \, \pr_\mu  f \, \pr_\mu k \,  g + \pr_\mu h \,  f\, \pr^2 k \, \pr_\mu g
+ \pr_\mu h \, \pr^2 \! f \,  k \,\pr_\mu  g +
h \, \pr_\mu f \, \pr_\mu k \, \pr^2  g \big ) \bigg ) \, . &\YYdivhk
}
$$
In a similar fashion, neglecting possible $\epsilon$-tensor contributions,
\eqnn\YYdivhks$$\eqalignno{
- (16\pi^2)^3 \,  &  {1\over 4}  \int   \d^dx \, \d^d y \; 
\tr_\sigma \big ( h(x)\, \sigma{\cdot \, \pr_x} Y_{f}(x,y)\, 
\bsi {\cdot {\overleftarrow \pr}}{\!}_y \; k(y) \,
\sigma{\cdot \, \pr_y} Y_{g}(y,x)\, 
\bsi {\cdot {\overleftarrow \pr}}{\!}_x \big )  \cr
\sim  \int &{}   \d^dx \, \bigg (  
{4\over 9 \vep^3} \big (1 - {\ts {7\over 12}} \vep -
{\ts {41 \over 144}}\vep^2\big ) \,
\pr_\mu h\, \pr_\nu f \, \big (  \pr_\mu k \, \pr_\nu g
-\pr_\nu k \, \pr_\mu g \big ) \cr
&{}- {1\over 9\vep} \, 
\pr_\mu h \, \pr_\mu f \, \pr_\nu k\, \pr_\nu g
+ {1\over 9\vep} \, \pr_\mu h\, \pr_\nu f \, \big (  \pr_\mu k \, \pr_\nu g 
+\pr_\nu k \, \pr_\mu g \big ) \cr 
&{}+ {1\over 18\vep^2}\big ( 1 - {\ts {13\over 12}} \vep \big )
\big ( \pr^2 h \, \pr^2\!  f\, k \, g + h\, f \, \pr^2 k \, \pr^2 g \big )\cr 
&{} + { 1\over 9\vep^2}\big ( 1 - {\ts {7 \over 12}}\vep \big )
 \big ( h\, \pr^2\!  f\, \pr^2 k \, g  + \pr^2 h \,  f\, k \, \pr^2 g  \big )
-  { 1\over 36\vep} \big ( \pr^2 h \,  f\,  \pr^2 k \,  g  + 
h\, \pr^2\!  f\, k \, \pr^2 g \big ) \cr
\noalign{\vskip 2pt}
&{}+ {1\over 9\vep^2}\big ( 1-{\ts{1\over 12}}\vep \big )
\big (\pr_\mu h \, \pr^2 \! f\, \pr_\mu k \, g+ 
\pr^2 h \, \pr_\mu f \, k \, \pr_\mu g + 
h \, \pr_\mu f \, \pr^2 k \, \pr_\mu g + 
\pr_\mu h \, f \, \pr_\mu k \, \pr^2 g \big )  \cr
\noalign{\vskip -2pt}
&{}+  {1\over 18\vep^2}\big ( 1+{\ts{5\over 12}}\vep \big )\big (
\pr_\mu h\,  \pr_\mu f \,   \pr^2  k \, g + 
\pr_\mu h\,  \pr_\mu f \,  k \, \pr^2 g  +
\pr^2h  \,  f \, \pr_\mu k \, \pr_\mu g + 
h \, \pr^2 \! f \,  \pr_\mu k \, \pr_\mu g  \big ) \bigg ) \, . \cr
&{} &\YYdivhks
}
$$

\appendix{D}{Four Loop Calculations for Scalar Fields}

The additional counterterms necessary for $x$-dependent couplings may
be extended to four loops for purely scalar field theories. For
simplicity we assume here a single component real scalar field $\phi$
with interaction $V(\phi)= {1\over 24} \lambda \phi^4$. For arbitrary
$\lambda(x)$ the first relevant vacuum graph is at three loops giving
\eqn\WAthree{
W^{(3)} = {1\over 48} \int \d^d x \, \d^d y \; \lambda(x)
\lambda(y) \, G_0(s)^4 \sim 
{1\over (16\pi^2)^3} \, {1\over \vep} \; {1\over 864}
\int \d^d x \; \pr^2 \lambda \, \pr^2 \lambda  \, .
}

At four loops there is also just one vacuum graph which generates simple poles
in $\vep$ and therefore contributes to $A_{IJ}$ and other terms in \counter,
\eqn\WAfour{
W^{(4)} = - {1\over 48} \int \d^d x \, \d^d y \, \d^d z\; \lambda(x)
\lambda(y)\lambda(z)\, \R G_0(x-z)^2\, \R G_0(z-y)^2 \, \R G_0(x-y)^2 \, ,
}
for $\R G_0{\!}^2$ as in \RGO.  Letting
\eqn\Yfour{
Y(x,y) = \int \d^d z\; \R G_0(x-z)^2\, \lambda(z) \, \R G_0(z-y)^2 \, ,
}
then using \intF, in order to determine just the contributions containing
poles in $\vep$ it is sufficient to replace
\eqn\Yred{
Y(x,y) \to Y_0(x,y)  +  Y_1(x,y) + Y_2(x,y) \, ,
}
where
\eqn\YYY{\eqalign{
Y_0(x,y) = {} & {1\over 2(d-2)^4 S_d{\!}^3} \, {\Gamma(\half d)
\Gamma({\ts{3\over 2}}d-4)\over \Gamma(d-2)^2} \, b_0(x,y) \, 
(s^2)^{4-{3\over 2}d} \cr
&{} - {1\over \vep} \, {1\over 4(d-2)^2 S_d{\!}^2 S_4}
\, \big ( \lambda(x)+\lambda(y) \big )\, (s^2)^{2-d}
+ {1\over \vep^2} \, {1\over 16 S_4{\!}^2} \, \lambda(x) \, \delta^d(s) \, ,\cr
Y_1(x,y) = {} & {1\over 2(d-2)^4 S_d{\!}^3} \, {\Gamma(\half d)
\Gamma({\ts{3\over 2}}d-5)\over \Gamma(d-2)^2} \, b_1(x,y) \, 
(s^2)^{5-{3\over 2}d} \, , \cr
Y_2(x,y) = {}& {1\over 4(d-2)^4 S_d{\!}^3} \, {\Gamma(\half d)
\Gamma({\ts{3\over 2}}d-6)\over \Gamma(d-2)^2} \, b_2(x,y) \, 
\big ( (s^2)^{6-{3\over 2}d} - 1 \big )  \, , 
}
}
where now
\eqn\defbn{
b_n(x,y) = \int_0^1 \! \d t \, t^{1-{1\over 2}d +n}
(1-t)^{1-{1\over 2}d +n } \, (\quar \pr^2)^n \lambda (x - ts) \, .
}
$b_0$ has the expansion
\eqn\expb{\eqalign{
b_0(x,y) = {\Gamma(3-\half d)^2 \over \Gamma(5-d)} \bigg ( & 
{2\over \vep} \big ( \lambda(x)+\lambda(y) \big )-
{1\over 5-d} \, {1\over 4} \big ( (s \cdot \pr)^2 \lambda(x)+
(s \cdot \pr)^2  \lambda(y) \big ) \cr
\noalign{\vskip -2pt}
{} + {}& { 34-5d \over (5-d)(7-d)} \, {1\over 192} 
\big ( (s \cdot \pr)^4 \lambda(x) +  (s \cdot \pr)^4 \lambda(y) \big )
+ {\rm O} (s^6) \bigg ) \, .
}
}

Applying \divs\ gives
\eqnn\Ydiv$$\eqalignno{
(16\pi^2)^4 \,Y_0(x,y) G_0(s)^2 \sim {}& 
{-{1\over 36 \vep^2}} \big (1-{\ts{7\over 12}}\vep \big) 
\big ( \lambda(x)+\lambda(y) \big ) \, \pr^2 \pr^2 \delta^d(s) \cr
& {} - {1\over 24\vep} \big ( \pr_\mu \pr_\nu\lambda(x)+  
\pr_\mu \pr_\nu\lambda(y) \big ) \big ( 2 \pr_\mu \pr_\nu + \delta_{\mu\nu}
\pr^2 \big )  \delta^d(s)  \cr
& {} + {7\over 144\vep}\, \pr^2\pr^2 \lambda(x) \,\delta^d(s) \, ,  \cr
(16\pi^2)^4 \, Y_1(x,y) G_0(s)^2 \sim {}&  {1\over 8\vep} \Big (
\big   ( \pr^2 \lambda(x)+ \pr^2 \lambda(y) \big ) \, \pr^2 \delta^d(s)
- {\ts{1\over 3}} \,  \pr^2 \pr^2 \lambda(x) \, \delta^d(s) \Big ) \, , \cr
(16\pi^2)^4 \, Y_2(x,y) \R G_0(s)^2 \sim {}& - {1\over 36\vep^2}
\big (1-{\ts{1\over 3}}\vep \big)\, \pr^2 \pr^2 \lambda(x) \, \delta^d(s)
\, , & \Ydiv
}
$$
and hence
\eqn\Ydiva{\eqalign{
(16\pi^2)^4 \,Y(x,y) \R G_0(s)^2 \sim {}& 
{-{1\over 36 \vep^2}} \big (1+{\ts{11\over 12}}\vep \big) \Big (
\big ( \lambda(x)+\lambda(y) \big ) \, \pr^2 \pr^2 \delta^d(s) 
+ \pr^2 \pr^2 \lambda(x) \, \delta^d(s) \Big )  \cr
& {} + {1\over 12\vep} \Big (
\big   ( \pr^2 \lambda(x)+ \pr^2 \lambda(y) \big ) \, \pr^2 \delta^d(s)
+\pr_x{\!}^2 \pr_y{\!}^2 \big ( \lambda(x) \, \delta^d(s) \big )  \Big ) \, .
}
}
This gives
\eqn\WAfourdiv{
W^{(4)} \sim
{1\over (16\pi^2)^4} \, {1\over \vep^2} \; {1\over 576}
\int \d^d x \; \Big ( \big ( 1 + {\ts {11\over 12}}\vep \big ) \,
\lambda^2 \, \pr^2\pr^2 \lambda -3 \vep \, \lambda 
\big (\pr^2\lambda \big )^2 \Big ) \, .
}

Using \Yred\ with \YYY\ we may further find
\eqn\YdivG{
(16\pi^2)^3 \,Y(x,y) G_0(s) \sim 
-{1\over 3 \vep^2} (1- \quar\vep ) \,
\big ( \lambda(x)+\lambda(y) \big ) \, \pr^2 \delta^d(s) 
+ {1\over 3 \vep} \, \pr^2 \lambda(x) \,   \delta^d(s) \, ,
}
which is equivalent to \YYo{b}, and  to a result obtained  in \Analog,
and also
\eqn\YhdivG{
(16\pi^2)^3 \,{\tilde Y}_f (x,y) \R G_0(s)^3 \sim 
-{1\over 3 \vep^2} (1- {\ts {3\over 8}} \vep ) \,
\big ( f(x)+ f(y) \big ) \, \pr^2 \delta^d(s) 
+ {1\over 3\vep^2}(1-{\ts {7\over 8}}\vep)  \, \pr^2 f(x) \,   \delta^d(s) \, .
}

\listrefs

\bye

\eqn\Lfour{\eqalign{
\Lambda^{(4)} \cirk \d \bY = {}& 
{\ts{1\over 2}}A\, \tr \big ( ( \d \bY Y ) \, (\bY Y)^2 \big ) 
+ B \, \tr \big ( ( \bY Y ) \, (\d \bY Y \bY Y ) \big )\cr
&{} +  
{\ts{1\over 2}}C \, \tr \big ( ( \d \bY Y ) \, (\bY Y \bY Y ) \big )
+ D \, {\ts {1\over 9}} d_\bY \tr \big ( \gamma_D \big ) \, ,
}
}
then
\eqn\Lathree{\eqalign{
3\, (\bY \Lambda^{(4)})  = {}&  A \,\big (\gamma_A + \half\, (\bY Y )^3 \big ) 
+ B \, \big ( \gamma_B +  2 \,   (\bY Y \bY Y) (\bY Y )  \big ) \cr 
&{} + C \, \big (\gamma_C +  \half \,  (\bY Y) (\bY Y \bY Y ) \big ) 
+  D \, \gamma_D \cr
={}& \gamma^{(3)}-  \gamma^{(1)} \gamma^{(2)} - \gamma^{(2)} \gamma^{(1)}
+ \Theta^{(3)} \cirk \beta_\bY{\!}^{(1)} \, , 
}
}
which is consistent  with \detL\ so long as
\eqn\Thett{
\Theta^{(3)} \cirk \d \bY =  (2B-\quar ) \, (\d \bY  Y )( \bY Y )
+ (\half C -\quar ) \, ( \bY Y ) (\d \bY  Y )\, , \quad
A-2B-\half C  = - \half \,  .
} 
Using \Lfour\ and \Thett\ in \resGK\ gives contributions to $T^{(4)},K^{(4)}$ of the
form \Tfour, \Kfour\ with
\eqn\resabc{\eqalign{
&{} a_1=a_2=a_3=C\, , \quad a_4=2B\, , \ a_5 = 4B \, , \quad e = f =0 \, , \cr
&{} b_1 = A \, , \ b_2 = A +\half C- \quar \, , \ b_3 = A+2B - \quar \, ,
}
}
which is compatible with \integfive\ for $X_1 = 2A+\half, \, X_2 = C+\quar, \,
X_3=B, \, X_4 = \half A$ so long as $a=b=-\half$.

Using \Lfour\ allows the verification of consistency 
of the results obtained in \Afive\
along with \integfive, so long as $A,B,C$  in the 
general expression for  $\gamma^{(3)}$, given in \sthree,
satisfy the numerical constraint  in \XXXX.
By subtracting from ${\tilde A}^{(5)}$ the relevant $\tr(\gamma^2),
\tr(\gamma^3)$ contributions, the corresponding terms arising
from \GHone\ and \GHtwo\ in \Aexact, the remainder has to be written as
\eqn\Hthree{\eqalign{
2\,\big ( \Lambda^{(4)} \cirk  \beta_\bY{\!}^{(1)} + {}&
 \beta_Y{\!}^{(1)} \cirk H^{(3)} \cirk \beta_\bY{\!}^{(1)} \big ) \cr
= {}&  {\hat X}_1\, \tr \big (  (\bY Y)^2 \, (\bY Y \bY Y ) \big )
+ {\hat X}_2 \, \tr \big ( (\bY Y \bY Y )^2 \big )\cr
&{}  +  X_3 \, \tr \big (  (\bY Y) \, \gamma_B  \big ) 
+ X_4\; \tr \big (  (\bY Y)^4 \big ) 
+ D \; \tr \big (  (\bY Y) \, \gamma_D  \big ) \, ,
}
}
for
\eqn\XXrev{
{\hat X}_1 =X_1 + a -l + {\ts {1\over 4}}\, , \qquad 
 {\hat X}_2 = X_2 + 2 (a-l) + {\ts {1\over 4}}\, .
}
and using \integfive\ to eliminate $X_5$.  Since
$2{\hat X}_1 - {\hat X}_2 - 4 X_3 -4X_4 = 0$ the right hand
side of \Hthree\ vanishes when either $\beta_Y{\!}^{(1)}$ or
$\beta_\bY{\!}^{(1)}$ are zero. From \Lfour\
\eqn\LfourB{\eqalign{
2 \,\Lambda^{(4)} \cirk \beta_\bY{\!}^{(1)} = {}&
 (A+2B+\half C) \, \tr \big (  (\bY Y)^2 \, (\bY Y \bY Y ) \big )
+ C  \, \tr \big ( (\bY Y \bY Y )^2 \big )\cr
&{}  +  B \, \tr \big (  (\bY Y) \, \gamma_B  \big ) 
+ \half A \; \tr \big (  (\bY Y)^4 \big ) 
+ D \; \tr \big (  (\bY Y) \, \gamma_D  \big ) \,  ,
}
}
which ensures, with $l$ given by \resl, that the remaining contributions
can be absorbed into $H^{(3)}$ by taking
\eqn\Hthres{
2 \, \del Y \cirk H^{(3)}  \cirk \del \bY   = (X_3 -B) 
\; \tr \big (  (\d \bY \d Y ) \, ( \bY Y ) \big ) +
\half ({\hat X}_2 - C ) \; 
\tr \big (  (\d \bY  Y ) \, ( \bY \d Y ) \big )  \, .
}

\bye